\documentclass[runningheads,citeauthoryear,envcountchap]{svmult}

\usepackage{makeidx}   
\usepackage{graphicx}  
\usepackage{subeqnar}  
\usepackage{multicol}  
\usepackage{physmubb}  
\makeindex             

\def\gax{\mathrel{\raise.3ex\hbox{$>$}\mkern-14mu\lower0.6ex\hbox{$\sim$}}}
\def\lax{\mathrel{\raise.3ex\hbox{$<$}\mkern-14mu\lower0.6ex\hbox{$\sim$}}}
\def\gtorder{\mathrel{\raise.3ex\hbox{$>$}\mkern-14mu
             \lower0.6ex\hbox{$\sim$}}}
\def\ltorder{\mathrel{\raise.3ex\hbox{$<$}\mkern-14mu
             \lower0.6ex\hbox{$\sim$}}}

\def\labelprint#1{\label{#1}}
\def\farcs{\hbox{$.\!\!^{\prime\prime}$}}

\input psfig.sty

\begin{document}

\setcounter{chapter}{1}
\renewcommand{\thechapter}{\Alph{chapter}}

\title{ The Saas Fee Lectures on \protect\newline 
          Strong Gravitational Lensing}
\titlerunning{Strong Gravitational Lensing}

\author{C.S. Kochanek}
\institute{
  Department of Astronomy, The Ohio State University \protect\newline
  email: ckochanek@astronomy.ohio-state.edu}
\authorrunning{C.S. Kochanek}

\def\avgm{\langle M \rangle}
\def\avgmhat{\langle M/M_\odot \rangle}
\def\kbar{\langle\kappa\rangle}
\def\thx{\hat{\theta}_1}
\def\thy{\hat{\theta}_2}
\def\rbar{\langle \theta \rangle}
\def\dr{\delta \theta}
\def\kmsmpc {~km~s$^{-1}$~Mpc$^{-1}$}
\def\vsis{48\pm3}
\def\vml{71\pm3}

\def\partintro{Part~1}
\def\partstrong{Part~2}
\def\partweak{Part~3}
\def\partmicro{Part~4}

\maketitle

\section{Introduction}

The objective of these lectures is to provide a practical introduction to strong 
gravitational lensing including the data, the theory, and the application of 
strong lensing to other areas of astrophysics. 
This is \partstrong\, of the complete Saas Fee lectures on gravitational
lensing.  \partintro\, (Schneider 2004) provides a basic introduction, \partstrong\, (Kochanek 2004)
examines strong gravitational lenses, \partweak\, (Schneider 2004) explores cluster lensing
and weak lensing, and \partmicro\, (Wambsganss 2004) examines microlensing.\footnote{For
 astro-ph users, the lectures should be referenced as: Kochanek, C.S., Schneider, P.,
  Wambsganss, J., 2004, Gravitational Lensing: Strong, Weak \& Micro,
  Proceedings of the 33$^{rd}$ Saas-Fee Advanced Course, G. Meylan,
  P. Jetzer \& P. North, eds. (Springer-Verlag: Berlin).}  The complete
lectures provide an updated summary of the field from Schneider, Ehlers \& 
Falco~(\cite{Schneider92}).  There are also many earlier (and shorter!) reviews of 
strong lensing (e.g. Blandford \&
Kochanek~\cite{Blandford1987p133}, Blandford \& Narayan~\cite{Blandford1992p311},
Refsdal \& Surdej~\cite{Refsdal1994p117}, Wambsganss~\cite{Wambsganss1998p12},
Narayan \& Bartelmann~\cite{Narayan1999p360},  Courbin, Saha \&
Schechter~\cite{Courbin2002p1},
Claeskens \& Surdej~\cite{Claeskens2002p263}).

It is not my objective in this lecture to provide a historical
review,  carefully outlining the genealogy of every development in gravitational 
lensing, but to focus on current research topics.  \partintro\ of these lectures
summarizes the history of lensing and introduces most of the basic equations of 
lensing.  
The present discussion is divided into 9 sections.  We start in \S\ref{sec:data} with an introduction
to the observational data.  In \S\ref{sec:basics} we outline the basic principles of strong
lenses, building on the general theory of lensing from \partintro.
In \S\ref{sec:mass} we discuss modeling gravitational lenses and the
determination of the mass distribution of lens galaxies.    In \S\ref{sec:time} we discuss
time delays and the Hubble constant.  In \S\ref{sec:stat} we discuss gravitational lens 
statistics and the cosmological model.  In \S\ref{sec:cluster} we discuss the differences
between galaxies and clusters as lenses.  In \S\ref{sec:substruc} we discuss the effects
of substructure or satellites on gravitational lenses.  In \S\ref{sec:optical} we discuss
the optical properties of lens galaxies and in \S\ref{sec:hosts} we discuss extended sources
and quasar host galaxies.  Finally in \S\ref{sec:future} we discuss the future of strong
gravitational lensing.

It will be clear to readers already familiar with the field that these are my lectures 
on strong lensing rather than an attempt to achieve a mythical consensus.  
I have tried to make clear what matters (and what does not), what lensing can do (and
cannot do) for astrophysics, where the field is serving the community well
(and poorly), and where non-experts have understood the consequences 
(or have failed to do so).  Doing so requires having definite opinions 
with which other researchers may well disagree.  We will operate on the 
assumption that anyone who disagrees sufficiently violently will have an
opportunity to wreak a horrible revenge at a later date by spending 
six months doing their own set of lectures.

\begin{itemize}
\item B.1 Introduction \hfill 1
\item B.2 An Introduction to the Data  \hfill 3
\item B.3 Basic Principles   \hfill 8
\begin{enumerate}
\item Some Nomenclature \hfill 10
\item Circular Lenses \hfill 12
\item Non-Circular Lenses  \hfill 24
\end{enumerate}
\item B.4 The Mass Distributions of Galaxies   \hfill 34
\begin{enumerate}
\item Common Models for the Monopole  \hfill 38
\item The Effective Single Screen Lens  \hfill 42
\item Constraining the Monopole   \hfill 43
\item The Angular Structure of Lenses  \hfill 49
\item Constraining Angular Structure   \hfill 53
\item Model Fitting and the Mass Distribution of Lenses   \hfill 56
\item Non-Parametric Models   \hfill 62
\item Statistical Constraints on Mass Distributions   \hfill 65
\item Stellar Dynamics and Lensing   \hfill 72
\end{enumerate}
\item B.5 Time Delays   \hfill 77
\begin{enumerate}
\item A General Theory of Time Delays   \hfill 79
\item Time Delay Lenses in Groups or Clusters  \hfill 82
\item Observing Time Delays and Time Delay Lenses  \hfill 84
\item Results: The Hubble Constant and Dark Matter  \hfill 88
\item The Future of Time Delay Measurements   \hfill 95
\end{enumerate}
\item B.6 Gravitational Lens Statistics   \hfill 96
\begin{enumerate}
\item The Mechanics of Surveys  \hfill 96
\item The Lens Population  \hfill 100
\item  Cross Sections  \hfill 106
\item Optical Depth  \hfill 108
\item  Spiral Galaxy Lenses  \hfill 110
\item Magnification Bias  \hfill 111
\item Cosmology With Lens Statistics   \hfill 119
\item The Current State \hfill 120
\end{enumerate}
\item B.7 What Happened to The Cluster Lenses?   \hfill 124
\begin{enumerate}
\item The Effects of Halo Structure and the Power Spectrum \hfill 132
\item Binary Quasars   \hfill 133
\end{enumerate}
\item B.8 The Role of Substructure   \hfill 135
\begin{enumerate}
\item Low Mass Dark Halos  \hfill 145
\end{enumerate}
\item B.9 The Optical Properties of Lens Galaxies   \hfill 147
\begin{enumerate}
\item The Interstellar Medium of Lens Galaxies   \hfill 153
\end{enumerate}
\item B.10 Extended Sources and Quasar Host Galaxies   \hfill 158
\begin{enumerate}
\item An Analytic Model for Einstein Rings  \hfill 159
\item Numerical Models of Extended Lensed Sources  \hfill 163
\item Lensed Quasar Host Galaxies   \hfill 167
\end{enumerate}
\item B.11 Does Strong Lensing Have A Future?   \hfill 170
\end{itemize}

\section{An Introduction to the Data \labelprint{sec:data}}

There are now 82 candidates for multiple image lenses besides those found
in rich clusters.  Of these candidates, there is little doubt about 74 of them.
The ambiguous candidates consist of faint galaxies with nearby
arcs and no spectroscopic data.  Indeed, the absence of complete spectroscopic 
information is the bane of most astrophysical applications of lenses.  Less
than half (38) of the good candidates have both source and lens redshifts --
43 have lens redshifts, 64 have source redshifts, and 5 have neither
redshift.  Much of this problem could be eliminated in about 5 clear nights of 8m time, 
but no TAC seems
willing to devote the effort even though lens redshifts probably provide more 
cosmological information per redshift than any other sparsely distributed source.  
Of these 74 lenses, 10 have had their central velocity dispersions measured and 10 have
measured time delays.  
A reasonably complete summary of the lens data is available at the
CASTLES WWW site http://cfa-www.harvard.edu/castles/, although lack
of manpower means that it is updated only episodically.

Fig.~\ref{fig:basic1} shows the distribution of the lenses in image separation
and source redshift.
The separations of the images range from 0\farcs35 to 15\farcs9 (using either half
the image separations or the mean distance of the images from the lens).  The
observed distribution combines both the true separation distribution
and selection effects.  For example, in simple statistical models using standard
models for galaxy properties  we would expect to find that 
the logarithmic separation distribution $dN/d\ln\Delta\theta$ is nearly constant
at small separations (i.e. $dN/d\Delta\theta \propto \Delta\theta$,  \S\ref{sec:stat}), while the
raw, observed distribution shows a cutoff due to the finite resolution of lens
surveys (typically 0\farcs25 to 1\farcs0 depending on the survey).  The cutoff
at larger separations is real, and it is a consequence of the vastly higher 
lensing efficiency of galaxies relative to clusters created by the cooling of the
baryons in galaxies (see \S\ref{sec:cluster}).  

Fig.~\ref{fig:basic3}
 shows the distribution in image separation and lens galaxy redshift.  
There is no obvious trend in the typical separation with redshift, as might
be expected if there were rapid evolution in the typical masses of galaxies.
Unfortunately, there is also an observational bias to measure the redshifts
of large separation lenses, where the lens galaxies tend to be brighter and
less confused with the images, which makes quantitative interpretation of 
any trends in separation with redshift difficult.  There is probably also
a bias against finding large separation, low lens redshift systems because the flux
from the lens galaxy will more easily mask the flux from the source.  
We examine the correlations
between image separations and lens magnitudes in \S\ref{sec:optical}. 

\begin{figure}[ph]
\begin{center}
\centerline{\psfig{figure=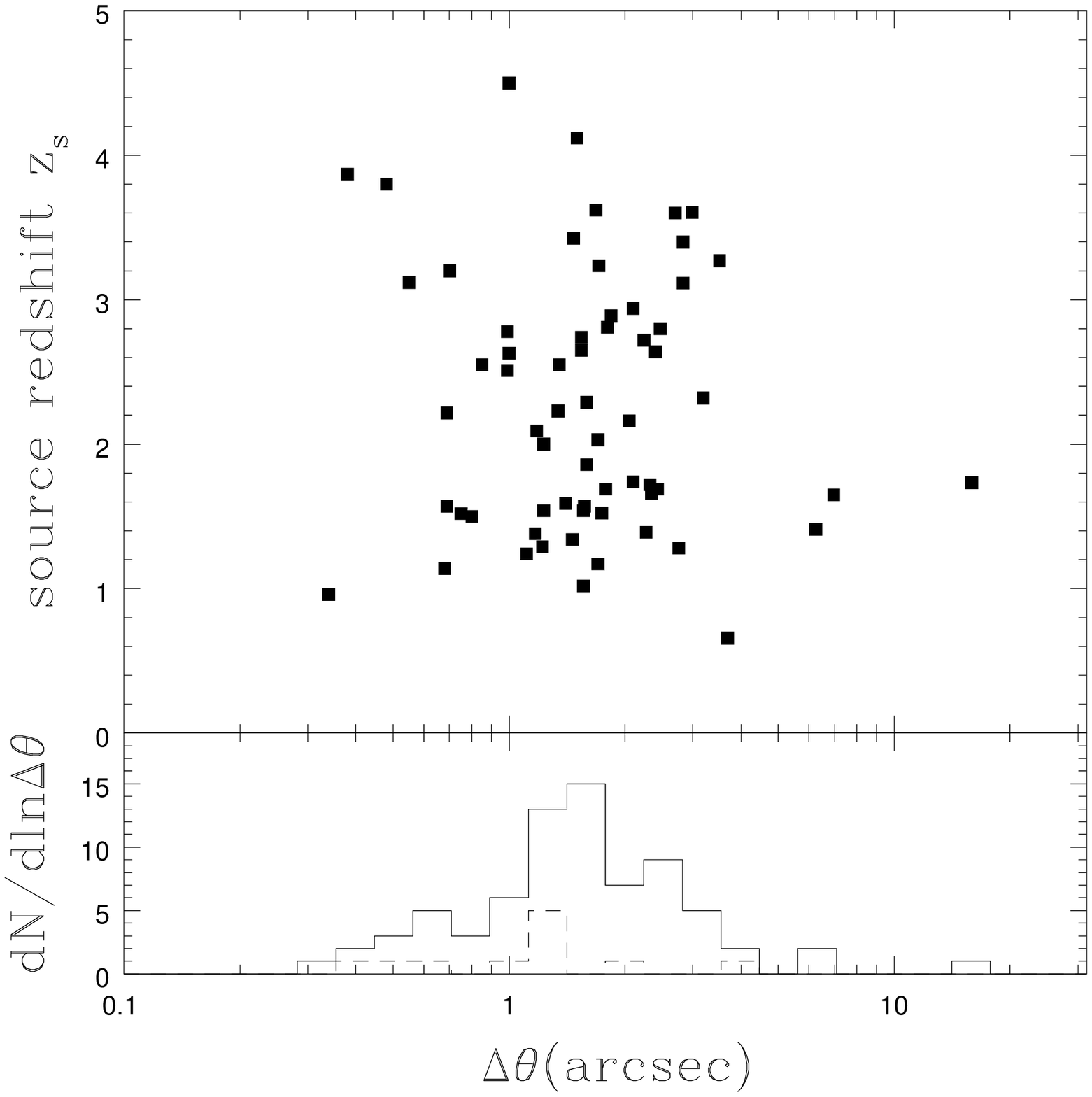,width=3.0in}}
\end{center}
\caption{
   The distribution of lens galaxies in separation $\Delta\theta$ and source
   redshift $z_s$.  
   The solid histogram shows the distribution in separation
   for all the lenses while the dashed histogram shows the distribution of
   those with unmeasured source redshifts.
   }
\labelprint{fig:basic1}
\begin{center}
\centerline{\psfig{figure=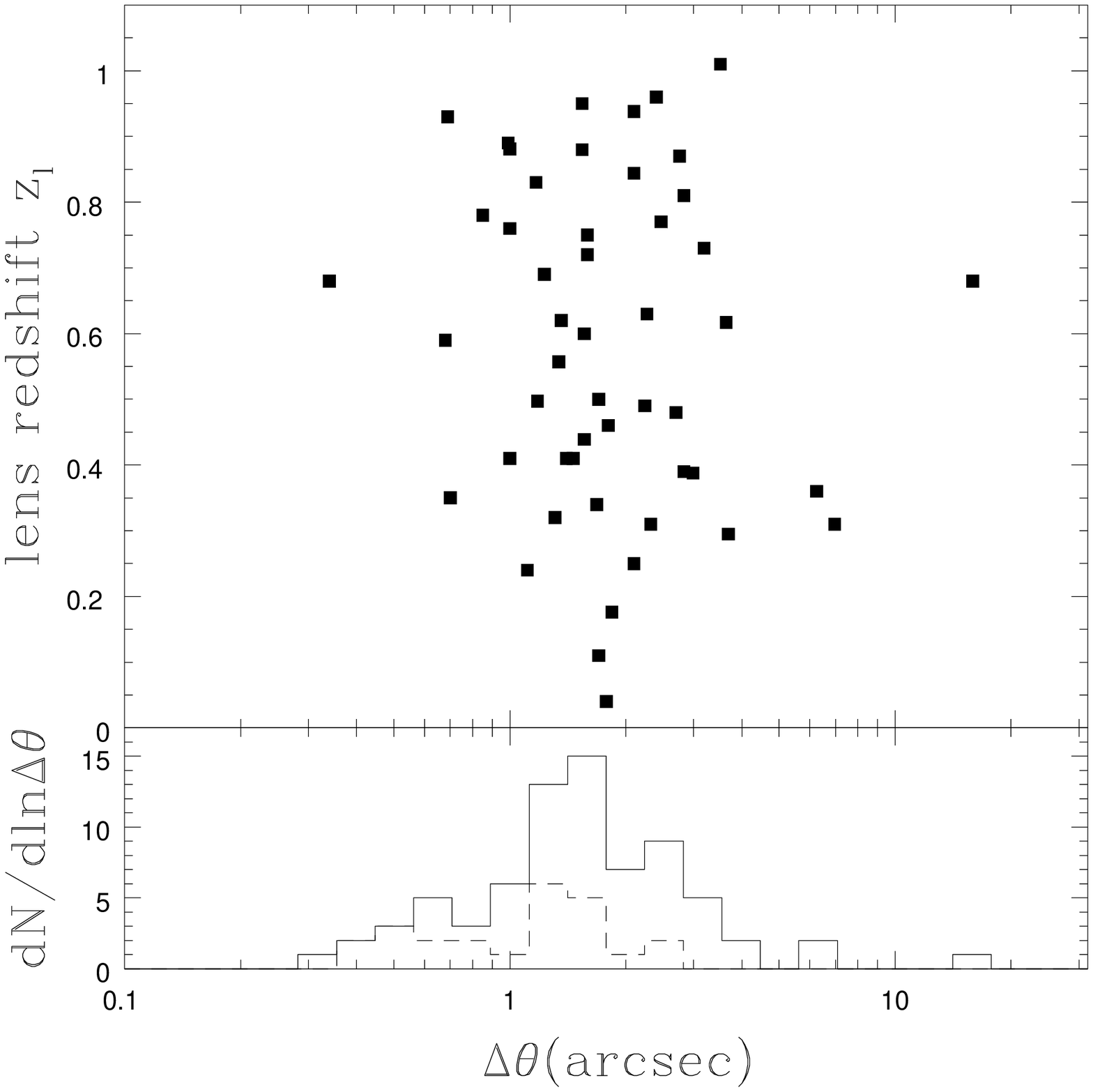,width=3.0in}}
\end{center}
\caption{
   The distribution of lens galaxies in separation $\Delta\theta$ and lens
   redshift $z_l$.  
   The solid histogram shows the distribution in separation
   for all the lenses while the dashed histogram shows the distribution of
   those with unmeasured lens redshifts.
   There are no obvious correlations between lens redshift $z_l$ and separation
   $\Delta\theta$, but the strong selection bias that small separation lenses are
   less likely to have measured redshifts makes this difficult to interpret.
   There may also be a deficit of low redshift, large separation lenses, which
   may be a selection bias created by the difficulty of finding quasar lenses 
   embedded in bright galaxies.
   }
\labelprint{fig:basic3}
\end{figure}

\begin{figure}[ph]
\begin{center}
\centerline{\psfig{figure=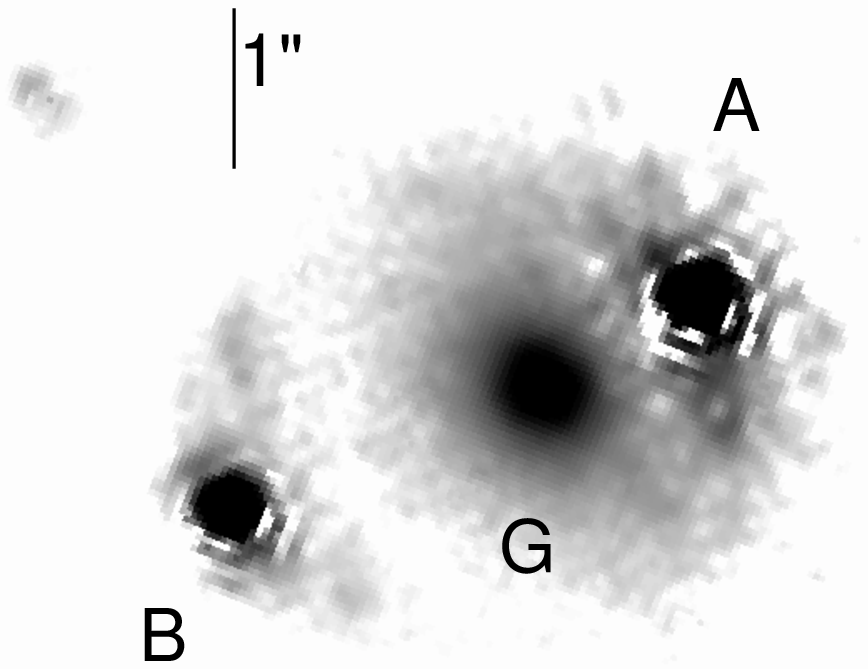,width=3.2in}}
\end{center}
\begin{center}
\centerline{\psfig{figure=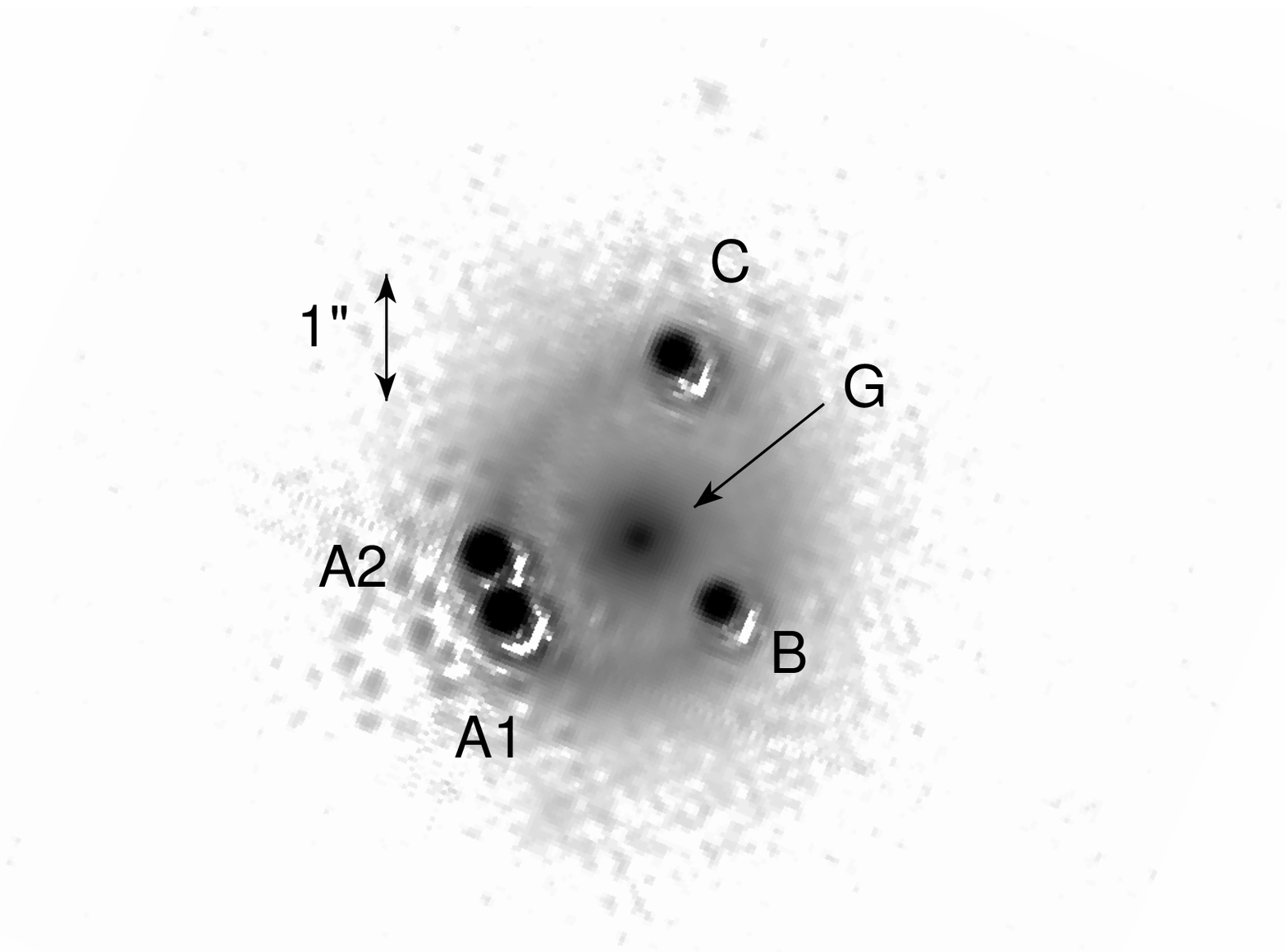,width=3.2in}}
\end{center}
\caption{Standard image geometries.  
   (Top) The two-image lens HE1104--1805. 
     G is the lens galaxy and A and B are the quasar images.  We also see arc images
     of the quasar host galaxy underneath the quasar images.
   (Bottom) The four-image lens PG1115+080 showing the bright A$_1$ and A$_2$ images
     created by a fold caustic.
   (Top, next page) The four-image lens RXJ1131--1231
     showing the bright A, B and C images
     created by a cusp caustic.
   (Bottom, next page) The four-image lens HE0435--1223, showing the cruciform geometry 
     created by a source near the center of the lens.
   For each lens we took the CASTLES H-band image, subtracted the bright quasars
   and then added them back as Gaussians with roughly the same FWHM as the real
   PSF.  This removes the complex diffraction pattern of the HST PSF and makes
   it easier to see low surface brightness features.  
   }
\labelprint{fig:basic4a}
\end{figure}

\begin{figure}[ph]
\vspace{0.1in}
\begin{center}
\centerline{\psfig{figure=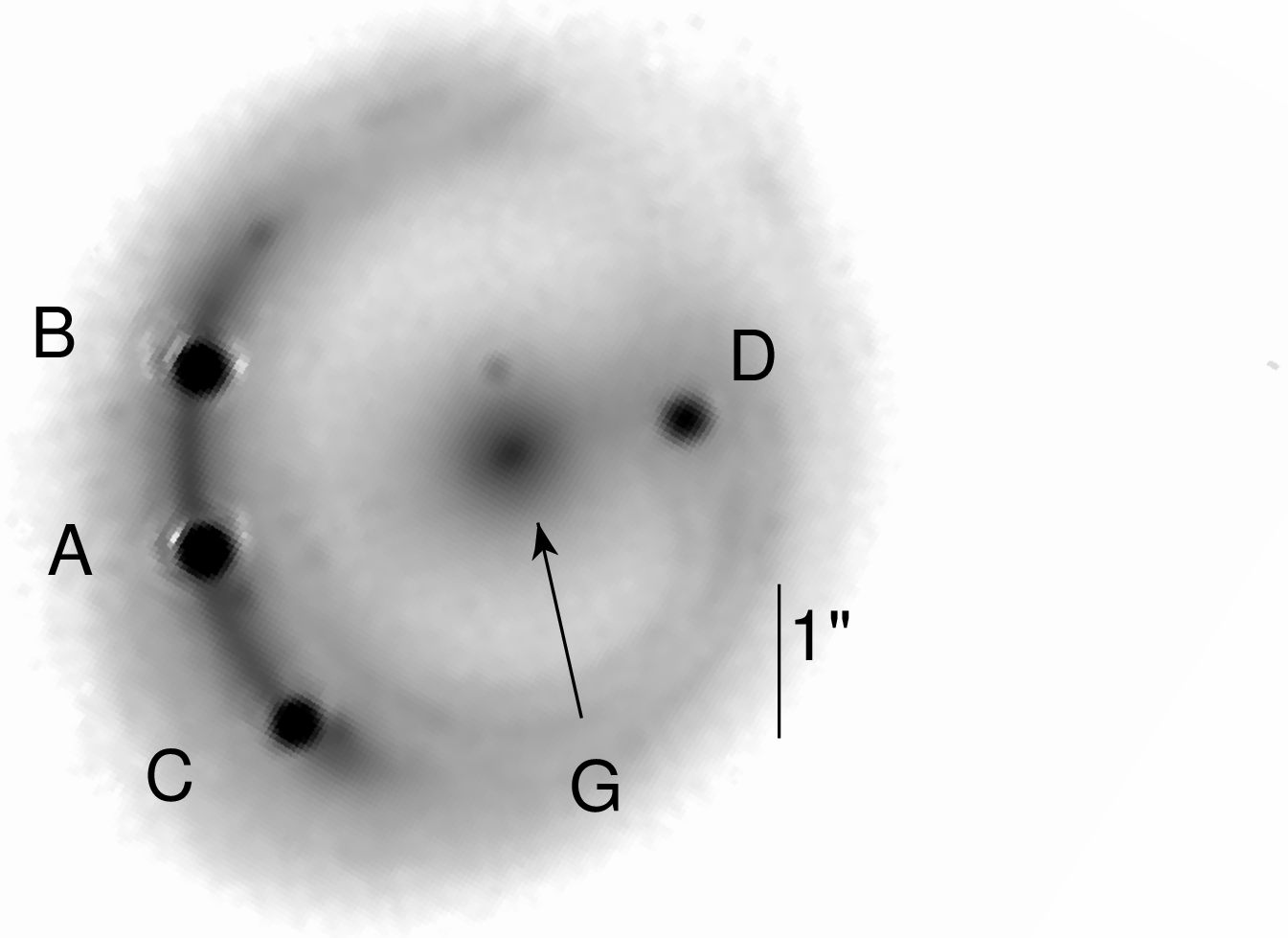,width=3.1in}}
\end{center}
\begin{center}
\centerline{\psfig{figure=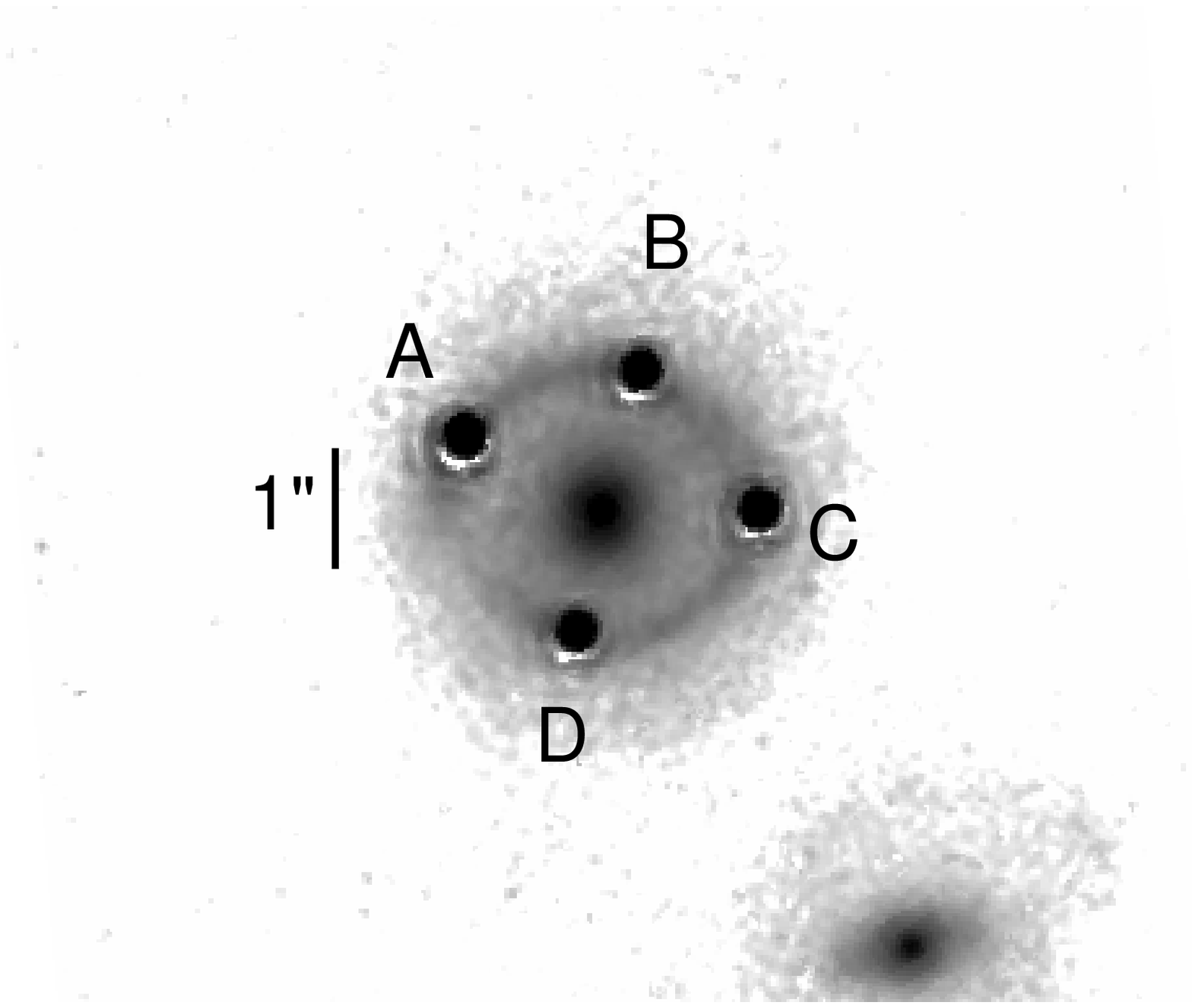,width=3.1in}}
\end{center}
\vspace{0.1in}
\caption{Standard image geometries continued. See the caption for Fig.~\ref{fig:basic4a}. }
\labelprint{fig:basic4b}
\end{figure}

\begin{figure}[t]
\begin{center}
\centerline{\psfig{figure=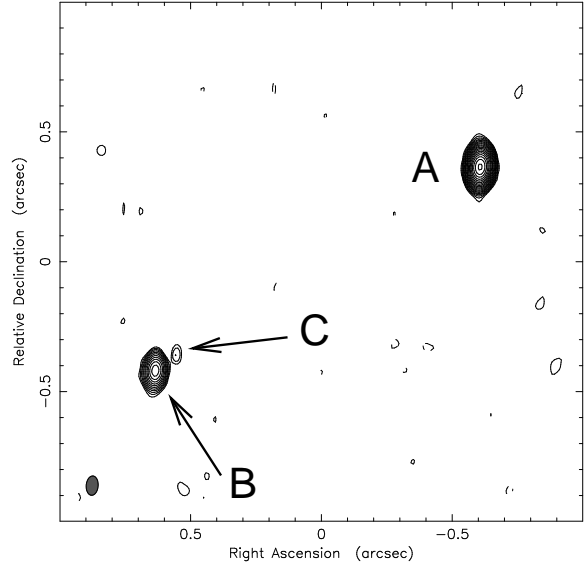,width=3.0in}}
\end{center}
\caption{PMN1632--0033 is the only known lens with a ``classical'' third
  image in the core of the lens galaxy.  The center of the lens galaxy
  is close to the faint C image.  Images A, B and C have identical radio
  spectra except for the longest wavelength flux of C, which can be 
  explained by absorption in the core of the lens galaxy.  Time delay
  measurements would be required to make the case absolutely secure.
  A central black hole in the lens galaxy might produce an additional
  image with a flux about 10\% that of image C. (Winn et al.~\cite{Winn2004p613})
  }
 \labelprint{fig:basic5}
\end{figure}

\begin{figure}[p]
\begin{center}
\centerline{\psfig{figure=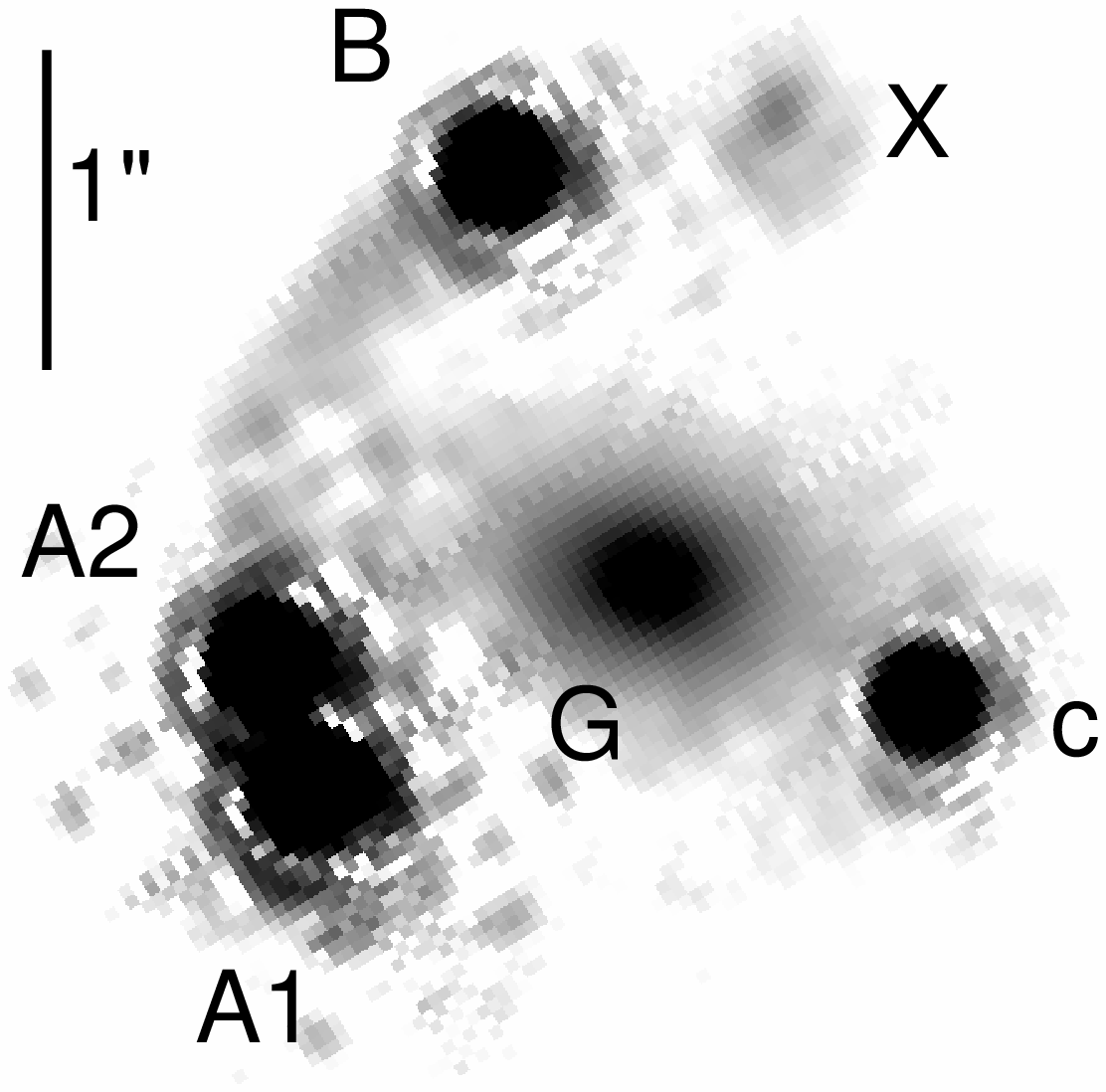,width=4.0in}}
\end{center}
\begin{center}
\centerline{\psfig{figure=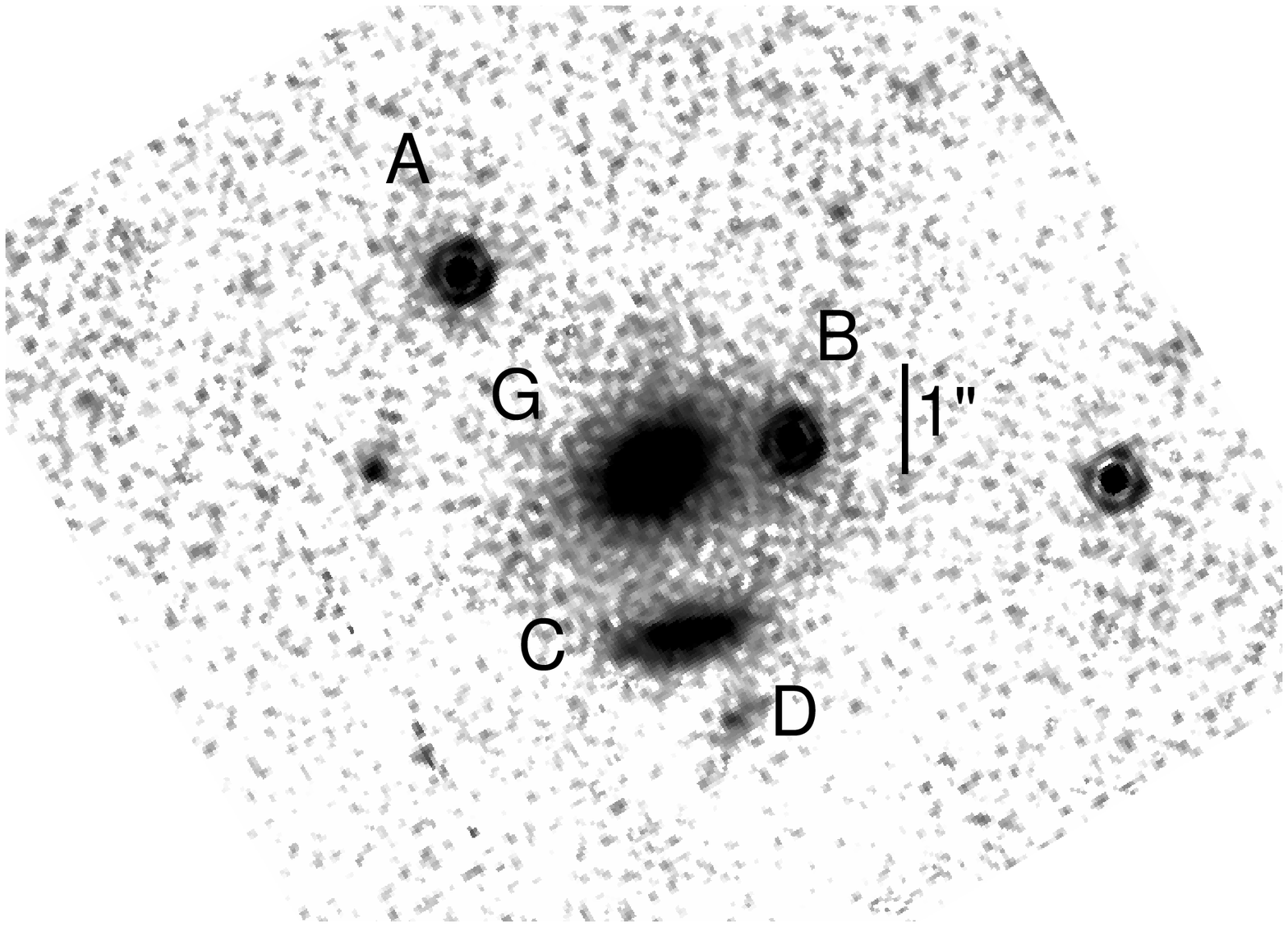,width=4.0in}}
\end{center}
\vspace{-0.3in}
\caption{H-band images of two lenses with small companions that are crucial
  for successful models.  The upper image shows ``Object X'' in MG0414+0534, and
  the lower image shows component D of MG2016+112.  MG2016+112 has the additional
  confusion that only A and B are images of the quasar
  (Koopmans et al.~\cite{Koopmans2002p39}).   Image C is some combination of
  emission from the quasar jet (it is an extended X-ray source, Chartas
  et al.~\cite{Chartas2001p163}) and the
  quasar host galaxy.  Object D is known to be at the same redshift as the
  primary lens galaxy G (Koopmans \& Treu~\cite{Koopmans2002p5}).
  }
 \labelprint{fig:mg0414h}
\end{figure}

In almost all cases the lenses have geometries that are ``standard'' for
models in which the angular structure of the gravitational potential
is dominated by the quadrupole moments of the density distribution, either 
because the lens is ellipsoidal or because
the lens sits in a strong external (tidal) shear field.  Of the 60 lenses 
where a compact component (quasar or radio core) is clearly identifiable, 
36 are doubles, 2 are triples, 20 are quads, 1 has five images and 1 has
six images.  The doubles and quads are the standard geometries produced
by standard lenses with nearly singular central surface densities.  
Examples of these basic patterns are shown in Figs.~\ref{fig:basic4a}
and \ref{fig:basic4b}.

In a two-image lens like HE1104--1805 (Wisotzki et al.~\cite{Wisotzki1993p15}),
the images usually lie at markedly different distances from the lens galaxy
because the source must be offset from the lens center to avoid
producing four images.  The
quads show three generic patterns depending on the location of the source
relative to the lens center and the caustics.  There are cruciform quads 
like HE0435--1223 (Wisotzki et al.~\cite{Wisotzki2002p17}), where the
images form a cross pattern bracketing the lens. These are created when
the source lies almost directly behind the lens.  There are fold-dominated quads
like PG1115+080 (Weymann et al.~\cite{Weymann1980p641}), where
the source is close to a fold caustic and we observe a close pair of highly
magnified images.  Finally, there are 
cusp-dominated quads like RXJ1131--1231 (Sluse et al.~\cite{Sluse2003p43}),
where the source is close to a cusp caustic and we observe a close
triple of highly magnified images.  These are all generic 
patterns expected from caustic theory, as we discuss in \partintro\, and \S\ref{sec:basics}.  
We will discuss the relative numbers of doubles and quads in \S\ref{sec:stat}.

The lenses with non-standard geometries all have differing origins.
One triple, APM08279+5255 (Irwin et al.~\cite{Irwin1998p529}, Ibata et al.~\cite{Ibata1999p1922},
Mu\~noz, Kochanek \& Keeton~\cite{Munoz2001p657}), is probably an 
example of a disk or exposed cusp lens (see \S\ref{sec:basics}), while the 
other, PMNJ1632--0033 (Winn et al.~\cite{Winn2002p10}, 
Winn, Rusin \& Kochanek~\cite{Winn2004p613}), appears to be a classical
three image lens with the third image in the core of the lens (Fig.~\ref{fig:basic5}).  
The system with five images, PMNJ0134--0931 (Winn et al.~\cite{Winn2002p143},
Keeton \& Winn~\cite{Keeton2003p39}, Winn et al.~\cite{Winn2003p26}), is 
due to having two lens galaxies, while the system with six images, 
B1359+154 (Myers et al.~\cite{Myers1999p2565}, Rusin et al.~\cite{Rusin2001p594}), 
is a consequence of having
three lens galaxies inside the Einstein ring.  Many lenses have luminous
satellites that are required in any successful lens model, such as the
satellites known as  ``Object X'' in MG0414+0534 (Hewitt et al.~\cite{Hewitt1992p968},
Schechter \& Moore \cite{Schechter1993p1}) and object D  in MG2016+112 
(Lawrence et al.~\cite{Lawrence1984p46}  ) shown in
Fig.~\ref{fig:mg0414h}.  These satellite galaxies can be crucial parts of lens
models, although there has been no systematic study of their properties in
the lens sample.

\begin{figure}[t]
\vspace{-0.3in}
\begin{center}
\centerline{\psfig{figure=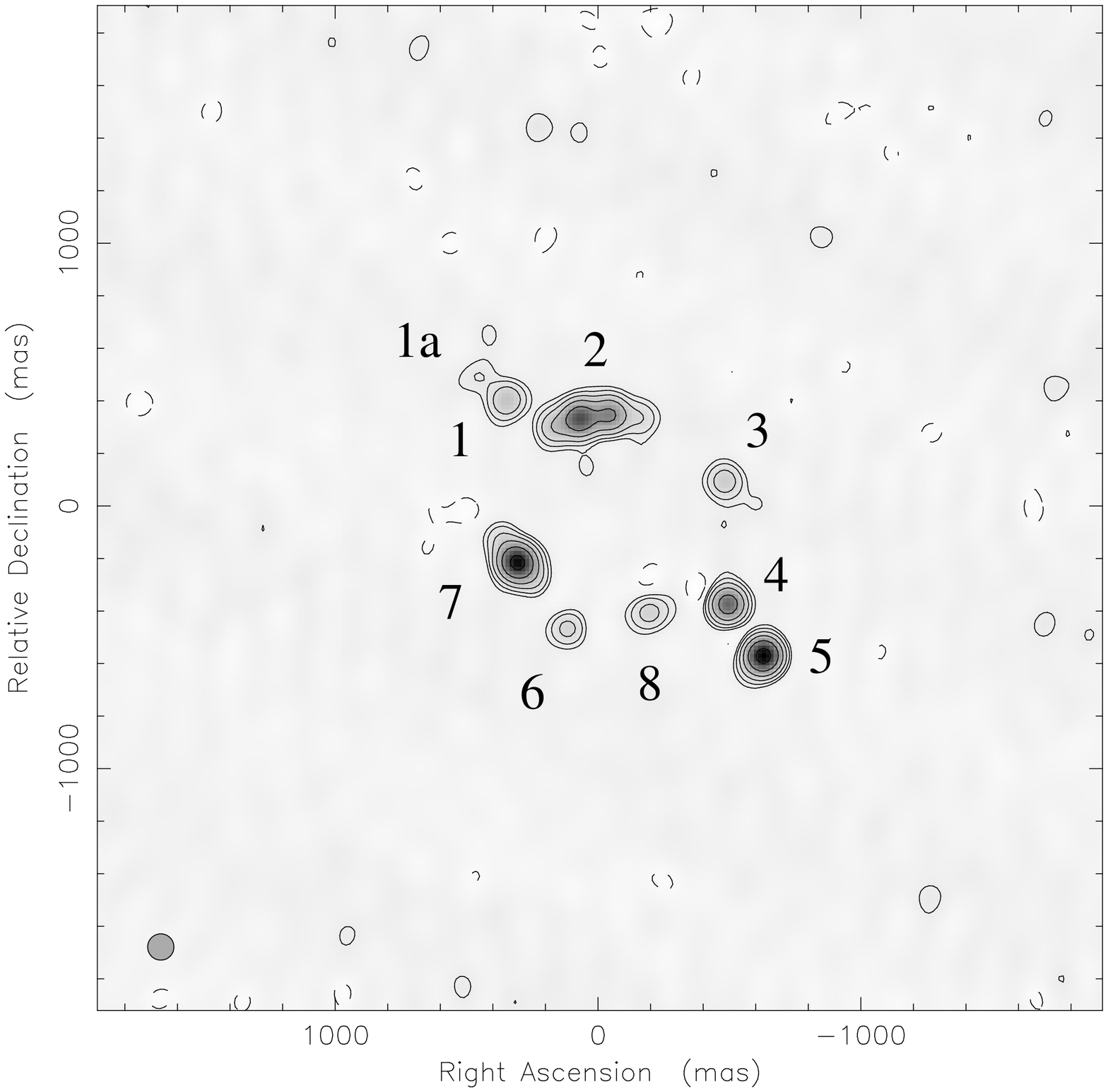,width=3.0in}}
\end{center}
\vspace{-0.3in}
\caption{A Merlin map of B1933+503 showing the 10 observed images of
  the three component source (Marlow et al.~\cite{Marlow1999p15}).
  The flat radio spectrum core is lensed into images 1, 3, 4 and 6.
  One radio lobe is lensed into images 1a and 8, while the other is
  lensed into images 2, 7 and 5.  Image 2 is really two images 
  merging on a fold.
  }
 \labelprint{fig:b1933merlin}
\end{figure}

\begin{figure}[p]
\begin{center}
\centerline{\psfig{figure=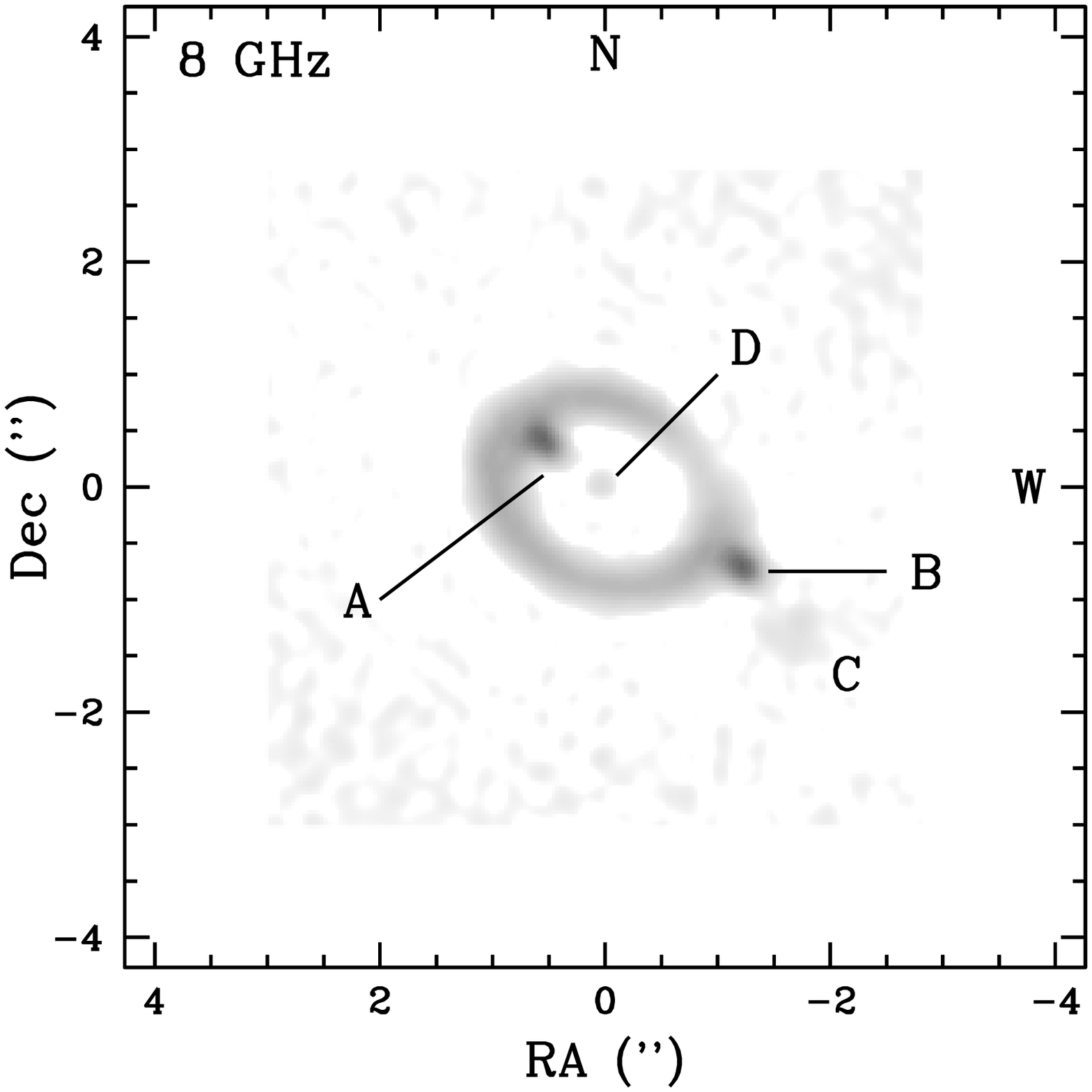,width=3.75in}}
\end{center}
\begin{center}
\centerline{\psfig{figure=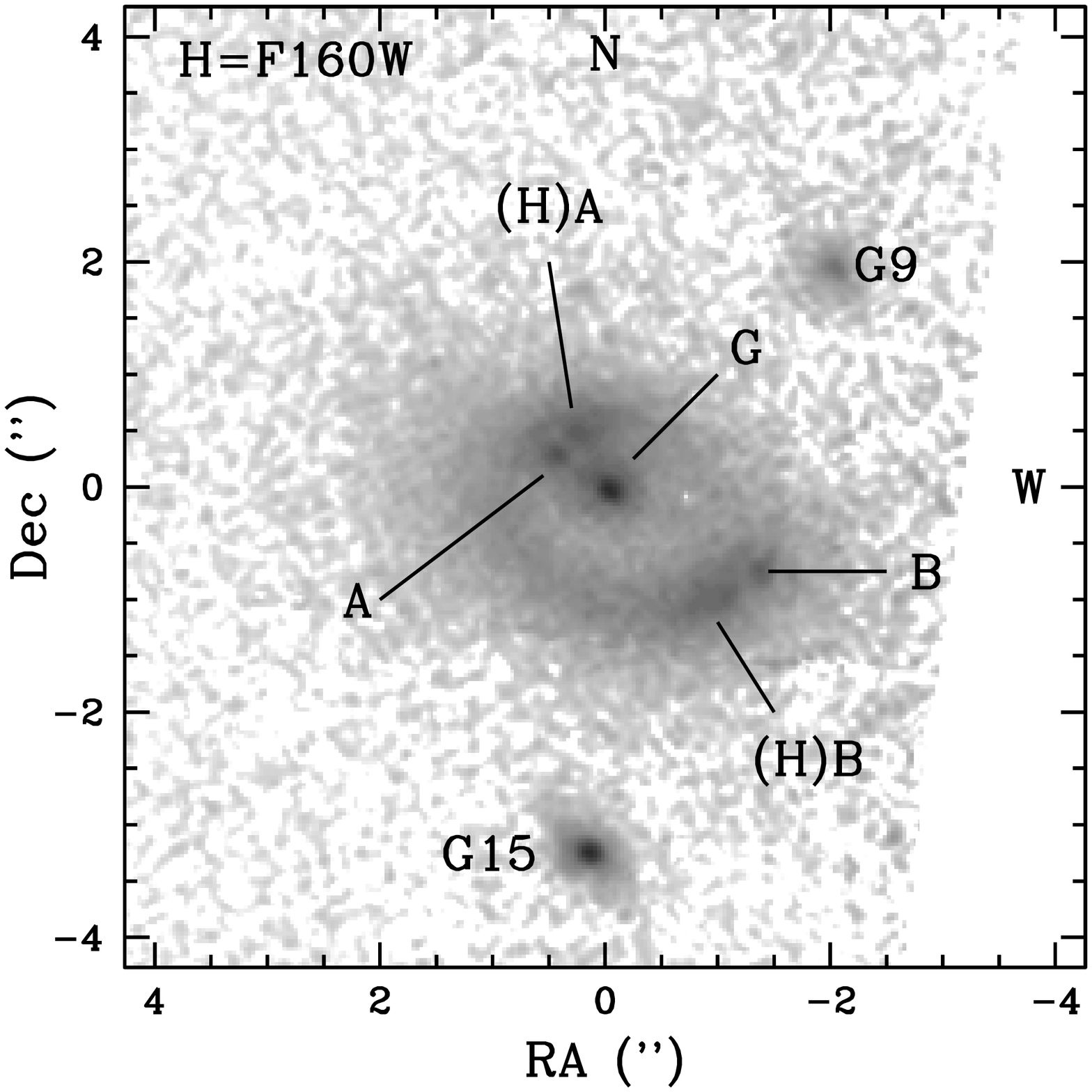,width=3.75in}}
\end{center}
\caption{The radio (top) and H-band (bottom) rings in MG11131+0456.  The
  radio map was made at 8~GHz by Chen \& Hewitt~(\cite{Chen1993p1719}), while
  the H-band image is from Kochanek et al.~(\cite{Kochanek2000p692}).  The
  radio source D is probably another example of a central odd image, but
  the evidence is not as firm as that for PMN1632--0033.  Note the
  perturbing satellite galaxies (G9 and G15) in the H-band image.
  }
 \labelprint{fig:mg1131}
\end{figure}

If the structure of the source is more complicated, then the resulting
image geometries become more complicated.  For example, the source of
the radio lens B1933+503 (Sykes et al.~\cite{Sykes1998p310})
consists of a radio core and two radio lobes,
leading to 10 observable images because the core and one lobe are 
quadruply imaged and the other lobe is doubly imaged (Fig.~\ref{fig:b1933merlin}).  
If instead
of discrete emission peaks there is a continuous surface brightness
distribution, then we observe arcs or rings surrounding the lens galaxy.
Fig.~\ref{fig:mg1131}
shows examples of Einstein rings for the case of MG1131+0456 (Hewitt et al.~\cite{Hewitt1988p537})
in both the radio (Chen \& Hewitt~\cite{Chen1993p1719}) and the infrared
(Kochanek et al.~\cite{Kochanek2000p692}).  The radio ring is formed
from an extended radio jet, while the infared ring is formed from the
host galaxy of the radio source.  We also chose
most of the examples in Figs.~\ref{fig:basic4a} and \ref{fig:basic4b}
to show prominent arcs and
rings formed by lensing the host galaxy of the source quasar.  We discuss arcs
and rings in \S\ref{sec:hosts}.

\section{Basic Principles \labelprint{sec:basics} }

Most gravitational lenses have the standard configurations we illustrated
in \S\ref{sec:data}.  These configurations are easily understood in terms 
of the caustic structures generic to simple lens models.  In this section
we illustrate the origin of these basic geometries using simple mathematical
examples.  We build on the general outline of lensing theory from \partintro.

\subsection{Some Nomenclature}

Throughout this lecture we use comoving angular diameter distances (also known
as proper motion distances) rather than the 
more familiar angular diameter distances  because almost every
equation in gravitational lensing becomes simpler.  The distance between
two redshifts $i$ and $j$ is 
\begin{equation}
    D_{ij} = { r_H \over |\Omega_k|^{1/2} } \hbox{sinn}\left[ \int_i^j
 { |\Omega_k|^{1/2} dz \over \left[ (1+z)^2 (1 + \Omega_M z) - z(2+z)\Omega_\Lambda \right]^{1/2} 
   }\right]
\end{equation}
where $\Omega_M$, $\Omega_\Lambda$ and $\Omega_k=1-\Omega_M-\Omega_\Lambda$ are the present
day matter density, cosmological constant and ``curvature'' density respectively,
$r_H=c/H_0$ is the Hubble radius, and the function $\hbox{sinn}(x)$ becomes 
$\sinh(x)$, $x$ or $\sin(x)$ for open ($\Omega_k > 0$), flat ($\Omega_k=0$) and
closed ($\Omega_k <0$) models (Carroll, Press \& Turner~\cite{Carroll1992p30}).  
We use $D_d$, $D_s$ and $D_{ds}$ for the
distances from the observer to the lens, from the observer to the source and 
from the lens to the source.  These distances are trivially related to the
angular diameter distances, $D_{ij}^{ang} = D_{ij}/(1+z_j)$, 
and luminosity distances, $D_{ij}^{lum} = D_{ij}(1+z_j)$.  In a flat
universe, one can simply add comoving angular diameter distances ($D_s = D_d + D_{ds}$),
which is not true of angular diameter distances.  The comoving volume element is
\begin{equation}
     dV = { 4\pi D_d^2 dD_d\over \left( 1+\Omega_k r_H^{-2} D_d^2 \right)^{1/2} }
      \rightarrow 4\pi D_d^2 dD_d 
      \labelprint{eqn:volume}
\end{equation}   
for flat universes.  
We denote angles on the lens plane by $\vec{\theta}=\theta(\cos\chi, \sin\chi)$ and
angles on the source plane by $\vec{\beta}$.  Physical lengths on the lens plane
are $\vec{\xi} = D_d^{ang} \vec{\theta}$.  The lensing potential, denoted by
$\Psi(\vec{\theta})$, satisfies the Poisson equation $\nabla^2 \Psi = 2\kappa$
where $\kappa=\Sigma/\Sigma_c$ is the surface density $\Sigma$ in units of the
critical surface density $\Sigma_c = c^2 (1+z_l)D_s/ (4\pi G D_d D_{ds})$.  For a 
more detailed review of the basic physics, see \partintro.

\subsection{Circular Lenses}

While one of the most important lessons about modeling gravitational lenses in the
real world is that you can never (EVER!)\footnote{AND I MEAN EVER! DON'T 
EVEN THINK OF IT!}  safely neglect the angular structure
of the gravitational potential, it is still worth starting with circular
lens models.  They provide a basic introduction to many of the elements which
are essential to realistic models without the need for numerical calculation. 
In a circular lens, the effective lens potential (\partintro) is a function
only of the distance from the lens center $\theta=|\vec{\theta}|$.  Rays 
are radially deflected by the angle
\begin{equation}
    \alpha(\theta) = { 2 \over \theta } \int_0^\theta \theta d\theta \kappa(\theta)
            = { 4 G M(<\xi) \over c^2 \xi } { D_{ds} \over D_{s}} 
  \labelprint{eqn:aaa}
\end{equation}
where we recall from \partintro\, that $\kappa(\theta)=\Sigma(\theta)/\Sigma_c$ is 
the surface density in units of the critical surface density, $D_{ds}$ and $D_s$
are the lens-source and observer-source comoving distances and $\xi = D_d^{ang} \theta$ 
is the proper distance from the lens. 
The bend angle is simply twice the Schwarzschild radius of the  enclosed mass, 
$4GM(<\xi)/c^2$, divided by the impact 
parameter $\xi$ and scaled by the distance ratio $D_{ds}/D_s$. 

The lens equation (see \partintro) becomes
\begin{equation}
   \vec{\beta} = \vec{\theta} \left[ 1 -  \alpha(\theta)/\theta \right]
               = \vec{\theta} \left[ 1 - \langle \kappa(\theta)\rangle \right]
  \labelprint{eqn:aab}
\end{equation}
where 
\begin{equation}
    \langle \kappa (\theta)\rangle = 
    { 2 \over \theta^2 } \int_0^\theta \theta d\theta \kappa(\theta)  
     = \alpha(\theta)/\theta
  \labelprint{eqn:aac}
\end{equation}
is the average surface density interior to $\theta$ in units of the 
critical density.   Note that there must be a region with 
$\langle\kappa\rangle > 1$ to have solutions on both sides
of the lens center.  Because of the circular symmetry, all images
will lie on a line passing through the source and the lens center.

The inverse magnification tensor (or Hessian, see \partintro) also has a simple form, with
\begin{equation}
   M^{-1} =  { d\vec{\beta} \over d\vec{\theta} } =
    \left(1-\kappa\right)
    \left( \begin{array}{cc}
              1 &0 \\
              0 &1 
           \end{array}
    \right)
     +\gamma 
    \left( \begin{array}{rr}
               \cos 2\chi  & \sin 2\chi \\
               \sin 2\chi  & -\cos 2\chi
           \end{array}
    \right)
  \labelprint{eqn:aad}
\end{equation}
where $\vec{\theta}=\theta(\cos\chi,\sin\chi)$. The convergence (surface
density) is 
\begin{equation}
  \kappa={ 1 \over 2 } \left( { \alpha \over \theta} +{ d\alpha\over d\theta } \right) 
  \labelprint{eqn:aae1}
\end{equation}
and the shear is
\begin{equation}
  \gamma={ 1\over 2} 
   \left( {\alpha \over \theta } - { d\alpha \over d\theta } \right)=\kbar-\kappa.
  \labelprint{eqn:aae2}
\end{equation}
The eigenvectors of $M^{-1}$ point in the radial and tangential 
directions, with a radial
eigenvalue of $\lambda_+=1-\kappa+\gamma=1-d\alpha/d\theta$ and a tangential
eigenvalue of $\lambda_-=1-\kappa-\gamma=1-\alpha/\theta=1-\kbar$.  If either one of
these eigenvalues is zero, the magnification diverges and we are on
either the radial or tangential critical curve.   If we can resolve
the images, we will see the images radially magnified near the radial
critical curve and tangentially magnified near the tangential critical
curve.  For example, all the quasar host galaxies seen in Figs.~\ref{fig:basic4a}
and \ref{fig:basic4b}
lie close to the tangential critical line and are stretched tangentially
to form partial or complete Einstein rings.  
The signs of the eigenvalues $\lambda_\pm$ give the parities of the images
and the type of time delay extremum associated with the images. 
If both eigenvalues are positive, the image is a minimum.  If both
are negative, the image is a maximum.  If one is positive and the
other negative, the image is a saddle point.  The inverse of the 
total magnification $\mu^{-1}=|M^{-1}|$ is the product of the eigenvectors, 
so it is positive for minima and maxima and negative for saddle points.  
The signs of the eigenvalues are referred to as the partial parities of
the images, while the sign of the total magnification is referred to as
the total parity.

It is useful to use simple examples to illustrate the behavior of circular
lenses for different density profiles.  In most previous lensing reviews,
the examples are based on lenses with finite core radii.  However, most 
currently popular models of galaxies and clusters have central density cusps
rather than core radii, so we will depart from historical practice and
focus on the {\it power-law lens} (e.g. Evans \& Wilkinson~\cite{Evans1998p800}).  
Suppose, in three dimensions, that
the lens has a density distribution $\rho \propto r^{-n}$.  Such a lens
will produce deflections of
\begin{equation}
   \alpha(\theta) = b \left( { \theta \over b } \right)^{2-n}
  \labelprint{eqn:aaf}
\end{equation}
as shown in Fig.~\ref{fig:bendang}, with convergence and shear profiles
\begin{equation}
  \kappa(\theta) = { 3-n \over 2 } \left( { \theta \over b } \right)^{1-n} 
  \quad\hbox{and}\quad
  \gamma(\theta) = { n-1 \over 2 } \left( { \theta \over b } \right)^{1-n}. 
  \labelprint{eqn:aag}
\end{equation}
The power law lenses cover most of the simple, physically interesting models.
The {\it point mass} lens is the limit $n\rightarrow 3$, with deflection
$\alpha=b^2/\theta$, convergence $\kappa=0$ (with a central singularity)
and shear $\gamma=b^2/r^2$.  The {\it singular isothermal sphere (SIS)}
is the case with $n=2$. It has a constant deflection $\alpha=b$, and
equal convergence and shear $\kappa=\gamma=b/2\theta$.  A {\it uniform
critical sheet} is the limit $n \rightarrow 1$ with $\alpha=\theta$, $\kappa=1$ 
and $\gamma=0$.  Models with $n\rightarrow 3/2$ have the cusp exponent of
the Moore (\cite{Moore1998p5}) halo model.  
The popular $\rho \propto 1/r$ NFW (Navarro, Frenk \& White~\cite{Navarro1996p563},
see \S\ref{sec:massmono}) density cusps are not 
quite the same as the $n\rightarrow 1$ case because the projected
surface density of a $\rho \propto 1/r$ cusp has $\kappa \propto \ln\theta$ rather 
than a constant.  Nonetheless, the behavior of the power law models as $n\rightarrow1$
will be very similar to the NFW model if the lens is dominated by the central
cusp.  The central regions of galaxies probably act like cusps with 
$1 \ltorder n \ltorder 2$.

The tangential magnification eigenvalue of these models is
\begin{equation}
   \lambda_- = 1-\kappa-\gamma=1-{\alpha\over \theta}=1-\kbar=1-(\theta/b)^{1-n}
  \labelprint{eqn:aah}
\end{equation}
 which is always equal to zero at $\theta=b\equiv\theta_E$.  This circle defines the 
tangential critical curve or
Einstein (ring) radius of the lens.  We normalized the models in this fashion 
because the Einstein radius is usually the best-determined parameter 
of any lens model, in the sense that all successful models will find
nearly the same Einstein radius (e.g. Kochanek~\cite{Kochanek1991p354},
Wambsganss \& Paczynski~\cite{Wambsganss1994p1156}).  The source position corresponding
to the tangential critical curve is the origin ($\beta=0$), and the
reason the magnification diverges is that a point source at the 
origin is converted into a ring on the tangential critical curve
leading to a divergent ratio between the ``areas'' of the source
and the image.  The other important point to notice is that the mean
surface density  inside the tangential critical radius is
$\kbar \equiv 1$ independent of the model.  This is true of any
circular lens.  With the addition of angular structure it is not
strictly true, but it is a very good approximation unless the mass
distribution is very flattened. 
 The definition of $b$ in terms of the properties of the
lens galaxy will depend on the particular profile.  For example, in a point
mass lens ($n\rightarrow 3$), $b^2=(4GM/c^2 D_d^{ang})(D_{ds}/D_s)$ where $M$
is the mass, while in an SIS lens ($n=2$), $b=4\pi(\sigma_v/c)^2 D_{ds}/D_s$ where 
$\sigma_v$ is the (1D) velocity dispersion of the lens.  For the other profiles, $b$
can be defined in terms of some velocity dispersion or mass estimate for
the lens, as we will discuss later in \S\ref{sec:dynamics} and \S\ref{sec:stat}. 
The radial magnification eigenvalue of these models is
\begin{equation}
   \lambda_+ = 1-\kappa+\gamma=1-{ d\alpha \over d\theta} = 1-(2-n)(\theta/b)^{1-n}
  \labelprint{eqn:aai}
\end{equation}
which can be zero only if $n<2$.  If $n<2$ the deflection goes to zero
at the origin and the lens has a
radial critical curve at $\theta = b(2-n)^{1/(n-1)} < b$ interior
to the tangential critical curve.  Models with $n\geq 2$ have constant
($n=2$) or rising deflection profiles as we approach the lens center
and have negative derivatives $d\alpha/d\theta$ at all radii.  

\begin{figure}[t]
\begin{center}
\centerline{\psfig{figure=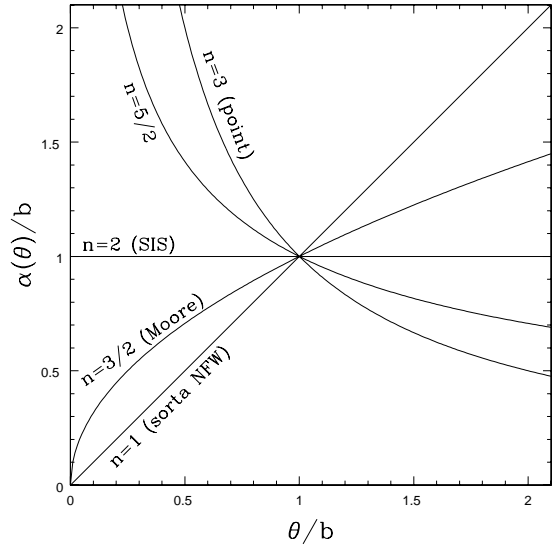,width=3.0in}}
\end{center}
\caption{
   The bending angles of the power law lens models.  Profiles more centrally
   concentrated ($n>2$) than the SIS ($n=2$), have divergent central deflections,
   while profiles more extended ($n<2$) than SIS have deflection profiles that
   become zero at the center of the lens.  The $n=1$ model is not quite an NFW
   model because the surface density is constant rather than logarithmic.
   }
\labelprint{fig:bendang}
\end{figure}

\begin{figure}[ph]
\begin{center}
\centerline{\psfig{figure=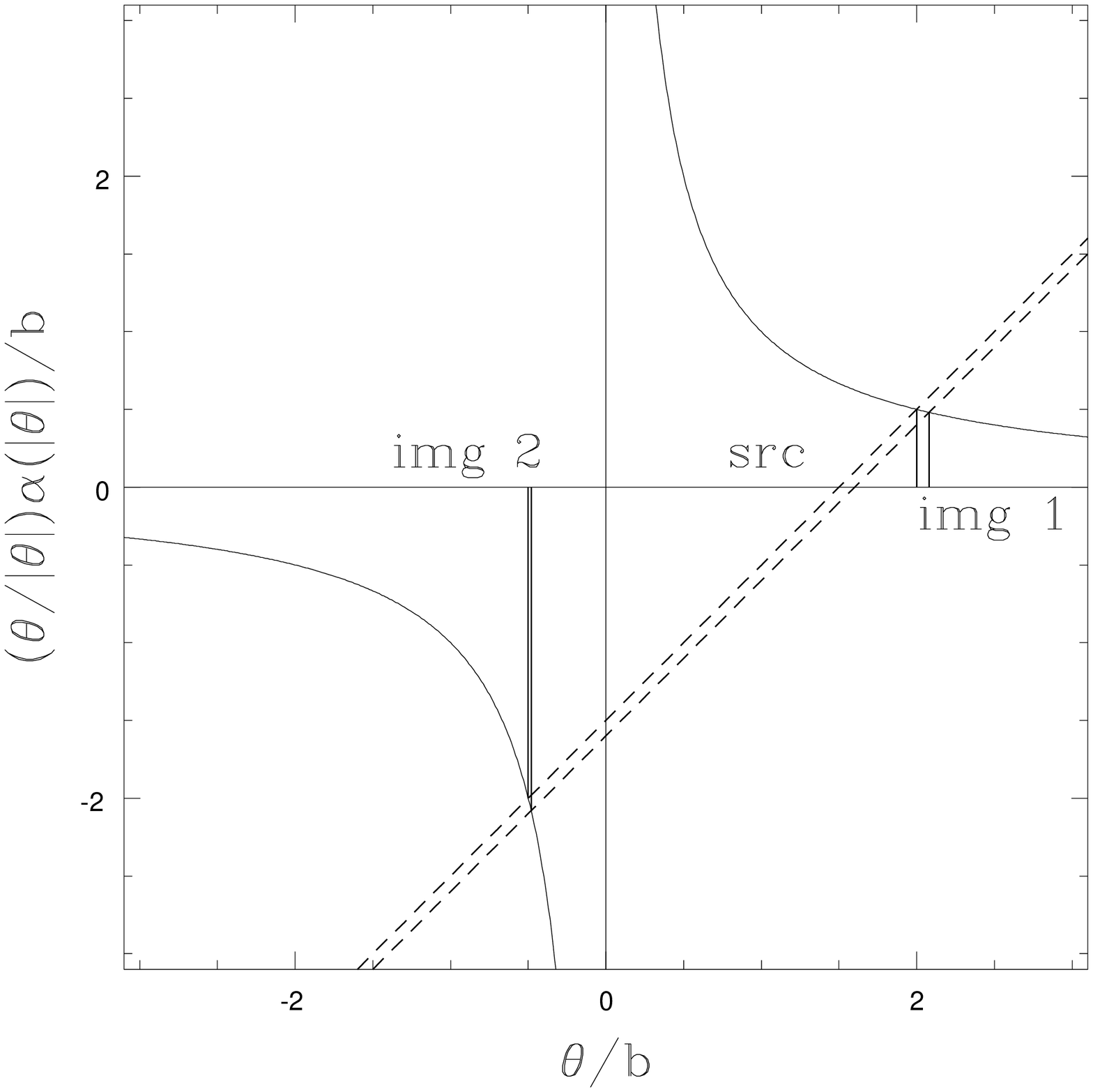,width=3.0in}}
\end{center}
\begin{center}
\centerline{\psfig{figure=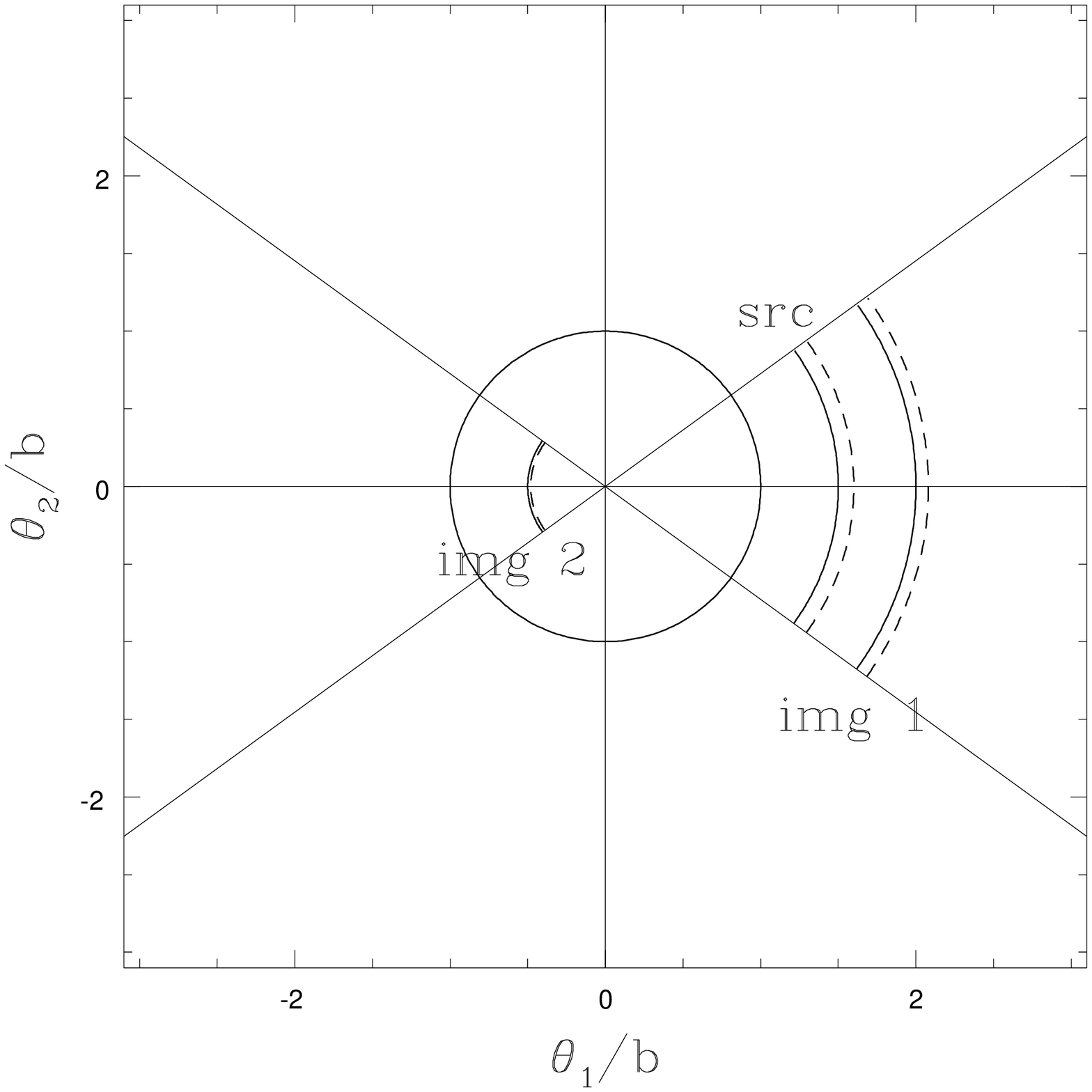,width=3.0in}}
\end{center}
\caption{
   Graphical solutions for the point mass ($n=3$) lens.  The top panel shows the
   graphical solution for the radial positions of the images, and the bottom
   panel shows the graphical solution for the image structure.  Note the strong
   radial demagnification of image 2 produced by the falling deflection profile.
   }
\labelprint{fig:pntmass}
\end{figure}

\begin{figure}[ph]
\begin{center}
\centerline{\psfig{figure=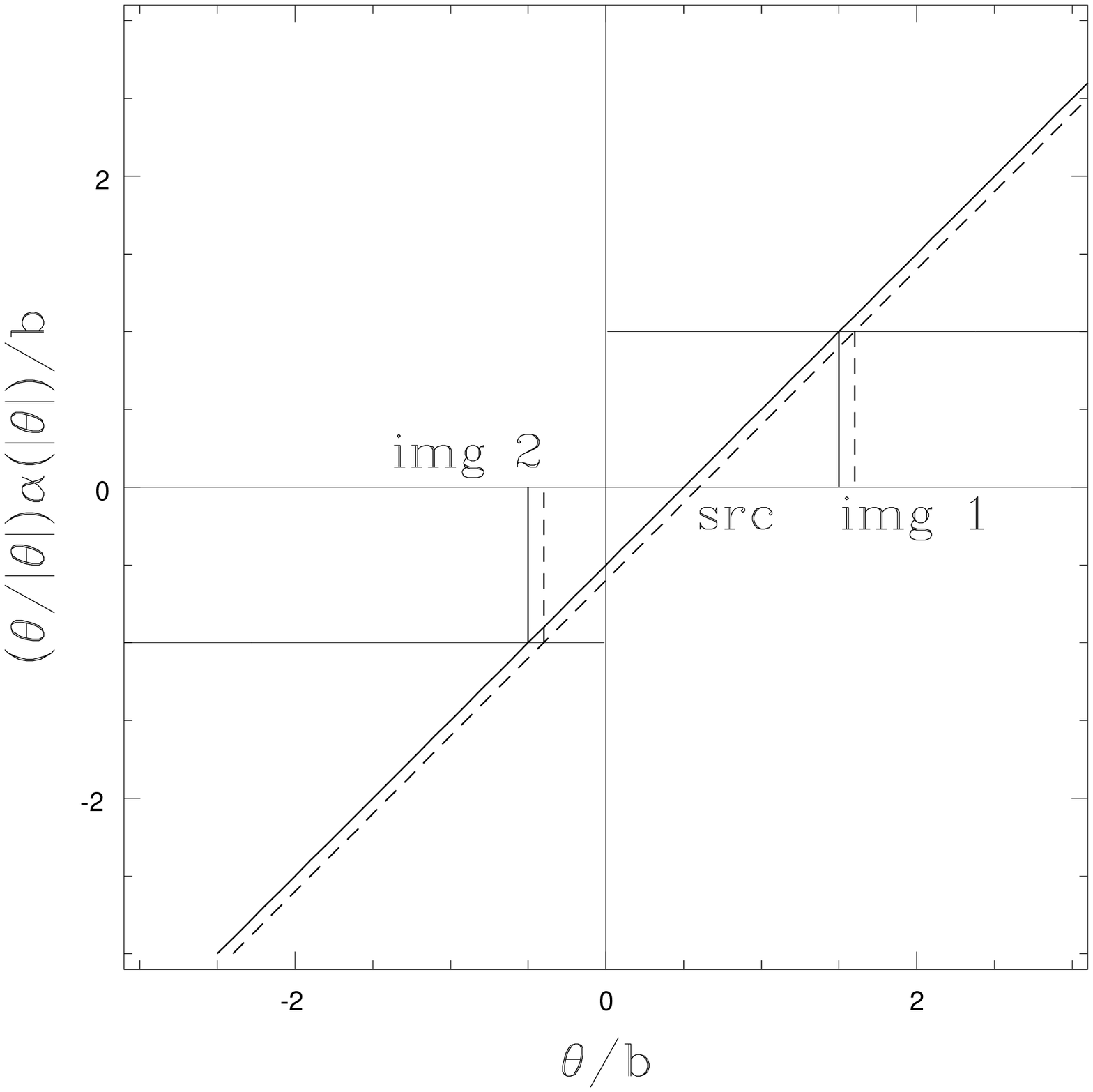,width=3.0in}}
\end{center}
\begin{center}
\centerline{\psfig{figure=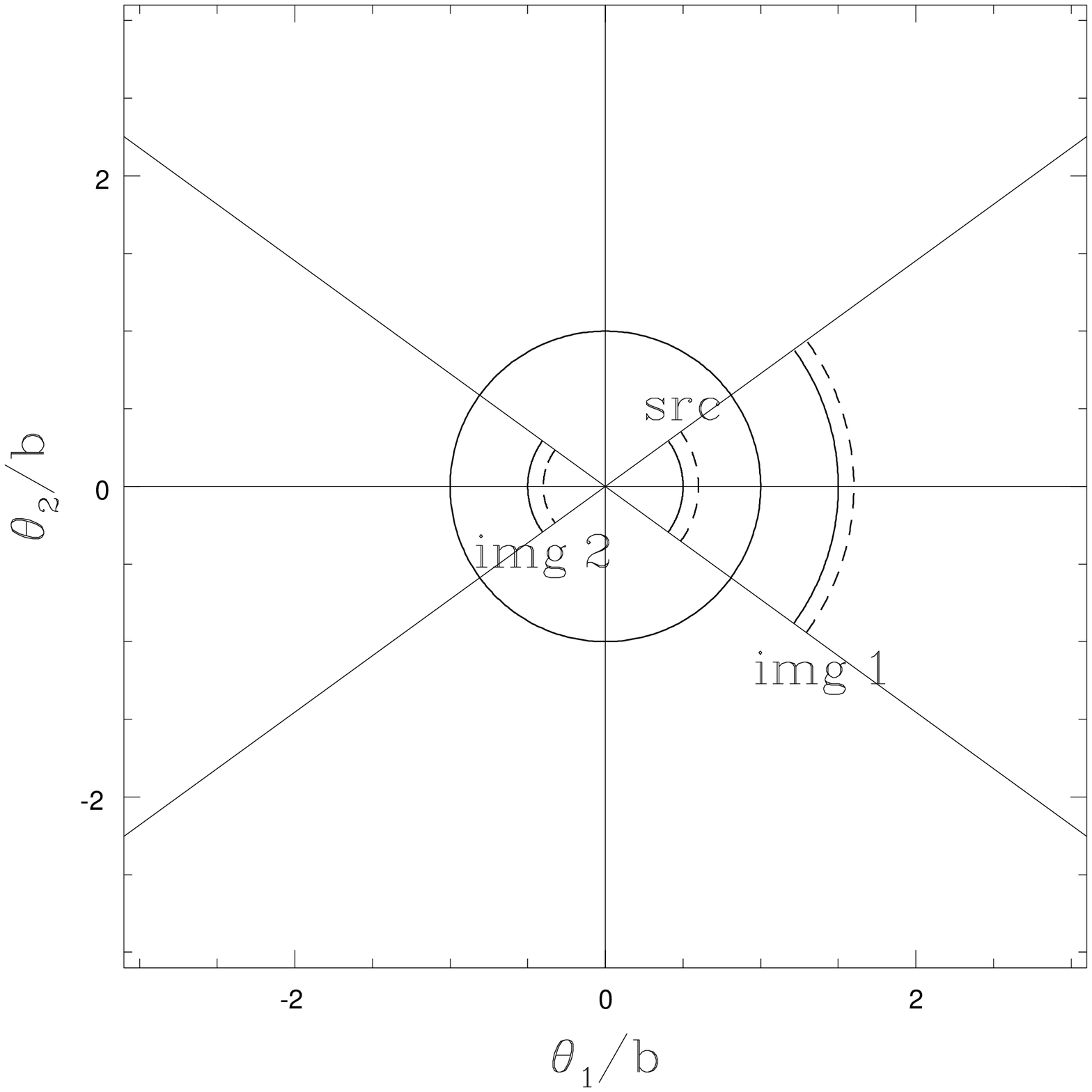,width=3.0in}}
\end{center}
\caption{
   Graphical solutions for the SIS ($n=2$) lens when $\beta<b$ and
   there are two images. 
   }
\labelprint{fig:sis1}
\end{figure}

\begin{figure}[ph]
\begin{center}
\centerline{\psfig{figure=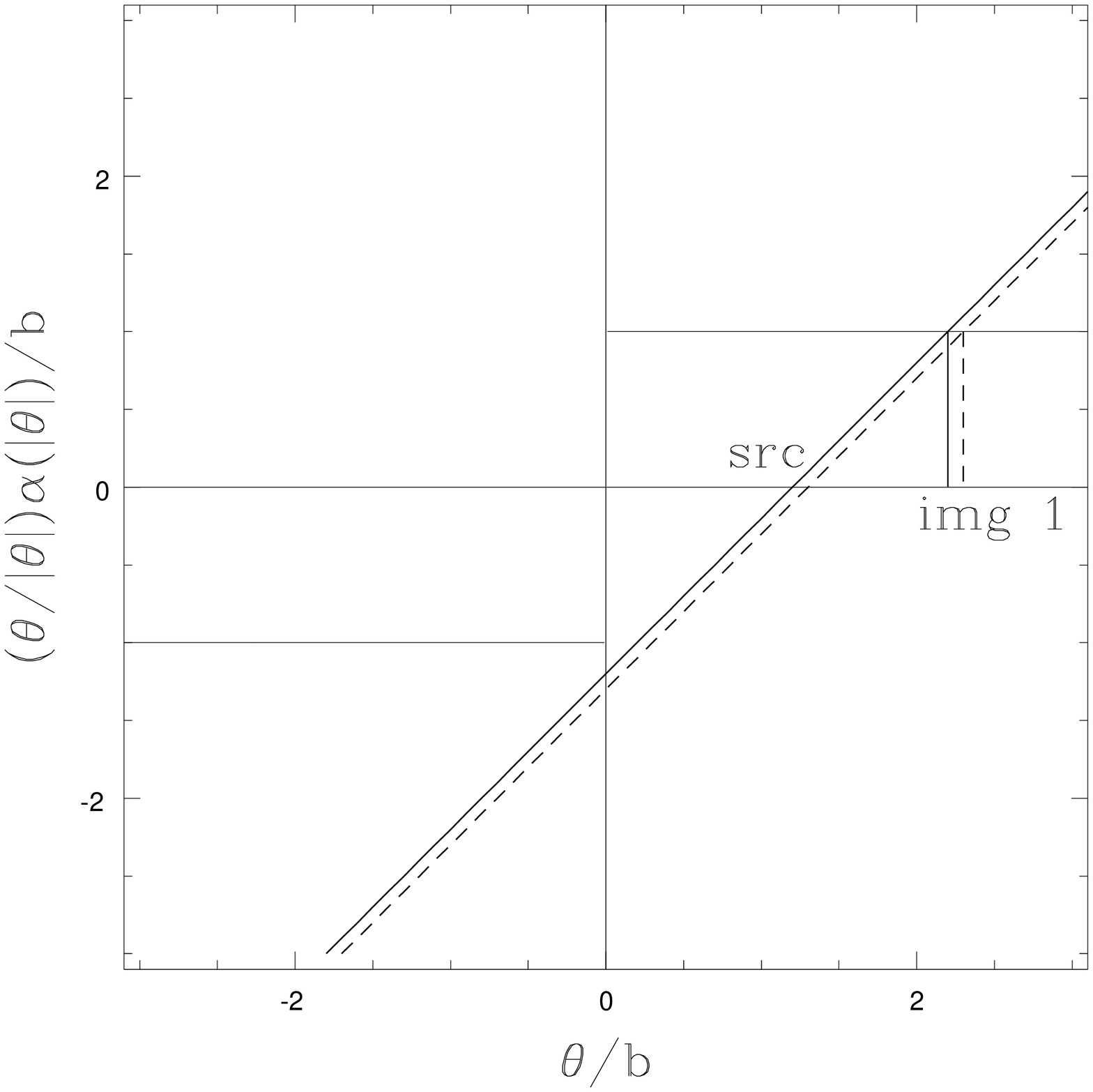,width=3.0in}}
\end{center}
\begin{center}
\centerline{ \psfig{figure=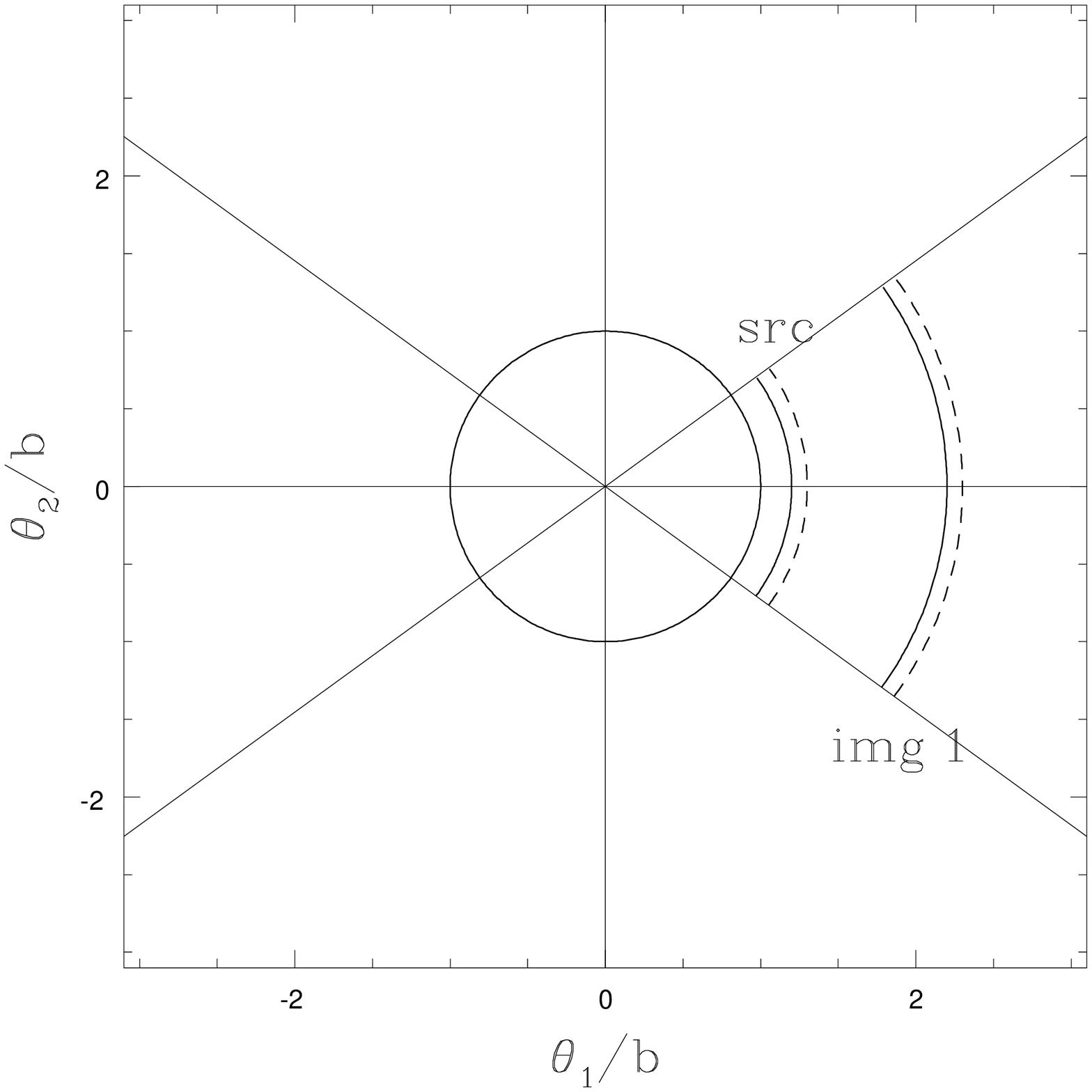,width=3.0in}}
\end{center}
\caption{
   Graphical solutions for the SIS ($n=2$) lens when $\beta>b$ and
   there is only one image.
   }
\labelprint{fig:sis2}
\end{figure}

\begin{figure}[ph]
\begin{center}
\centerline{\psfig{figure=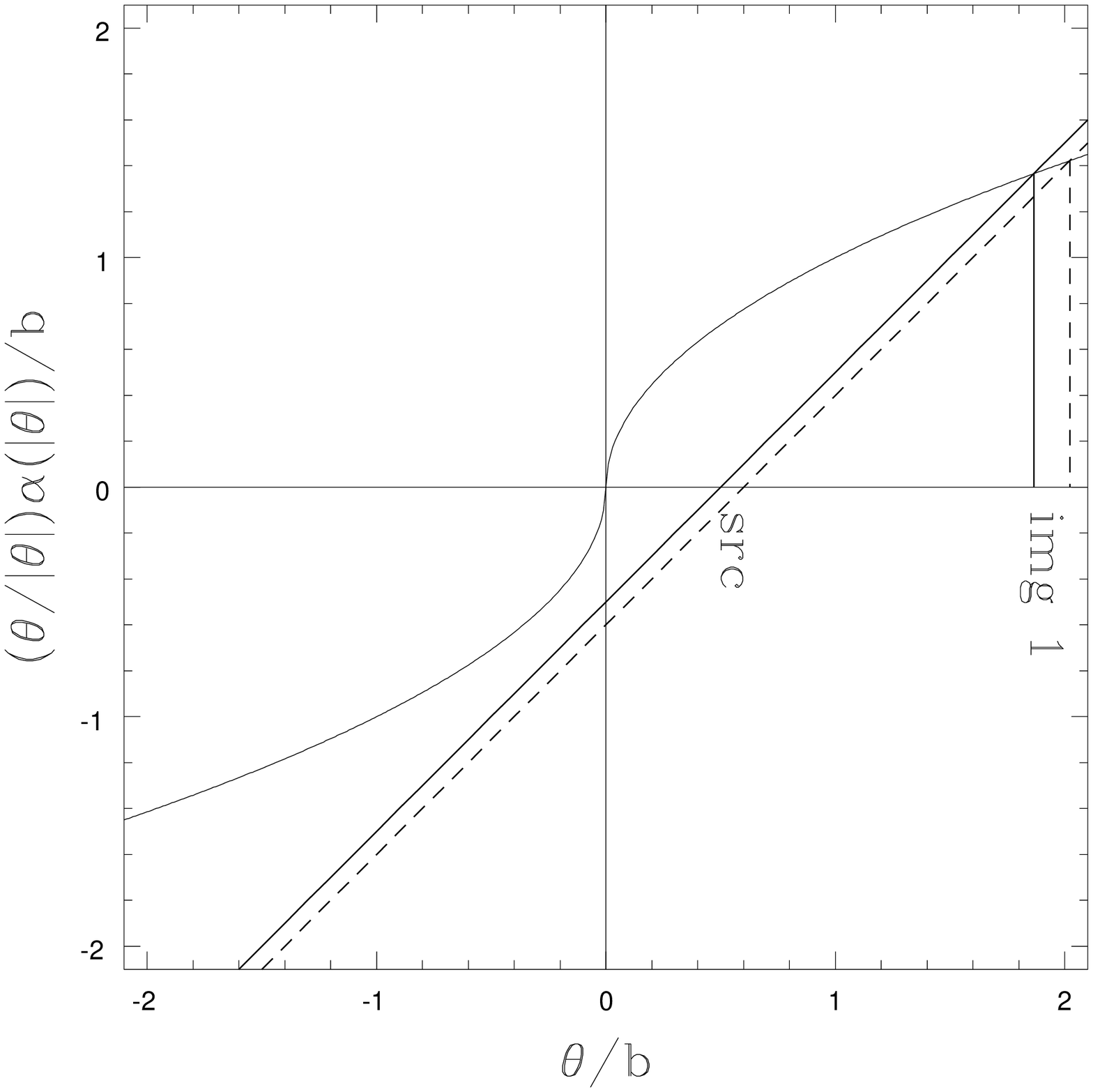,width=3.0in}}
\end{center}
\begin{center}
\centerline{\psfig{figure=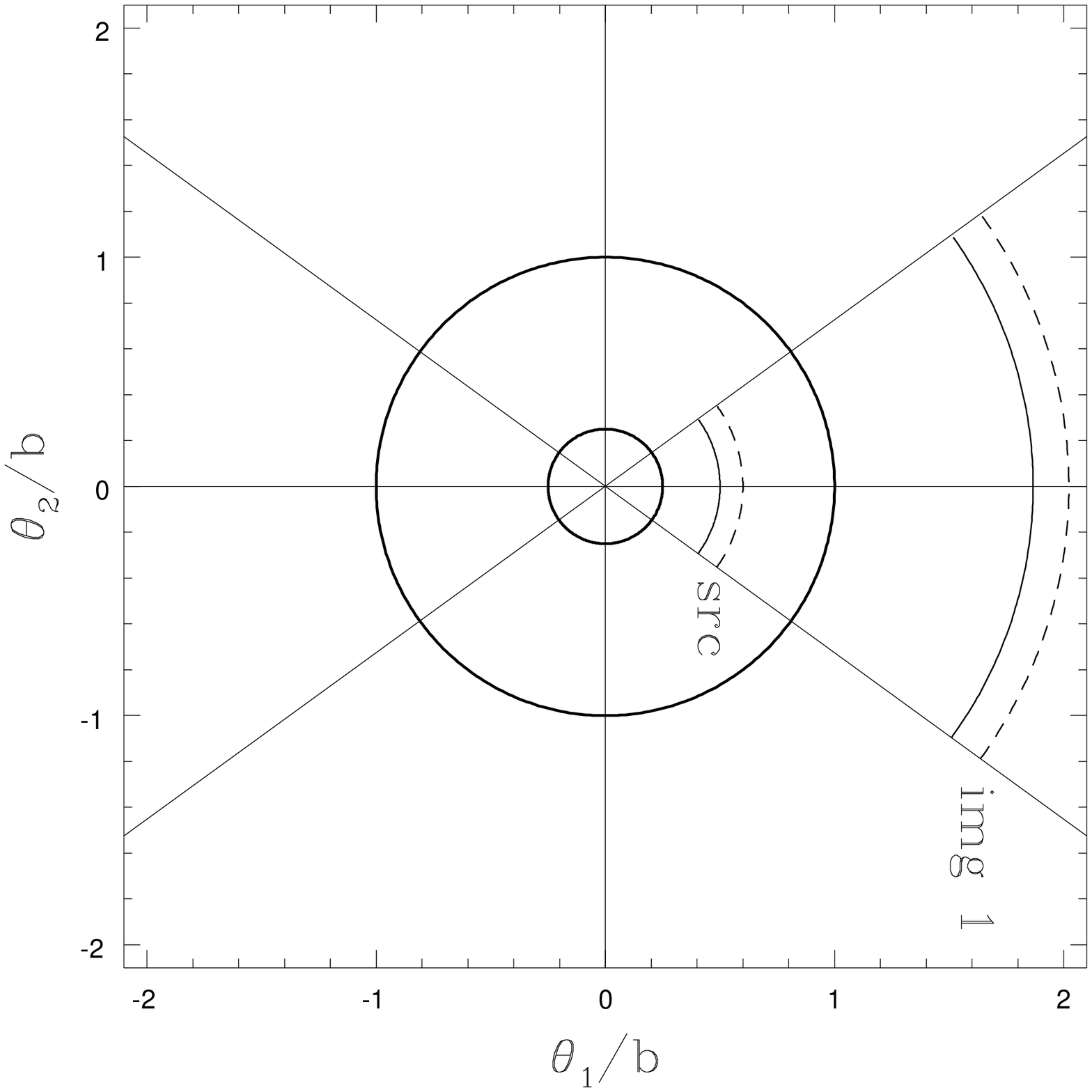,width=3.0in}}
\end{center}
\caption{
   Graphical solutions for the Moore profile cusp ($n=3/2$) lens when $\beta>b/4$ and
   there is only one image.
   }
\labelprint{fig:moore1}
\end{figure}

\begin{figure}[ph]
\begin{center}
\centerline{\psfig{figure=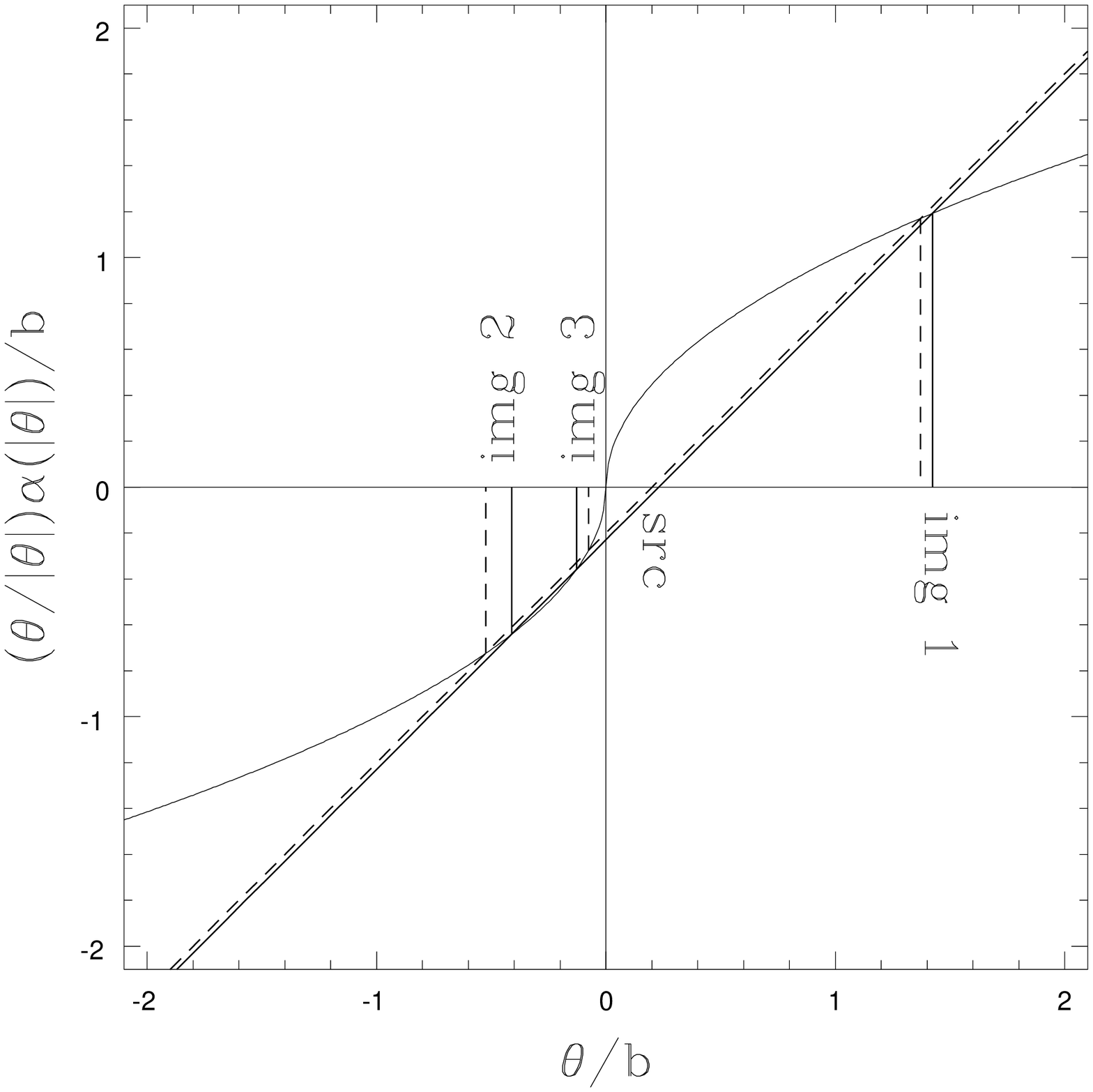,width=3.0in}}
\end{center}
\begin{center}
\centerline{\psfig{figure=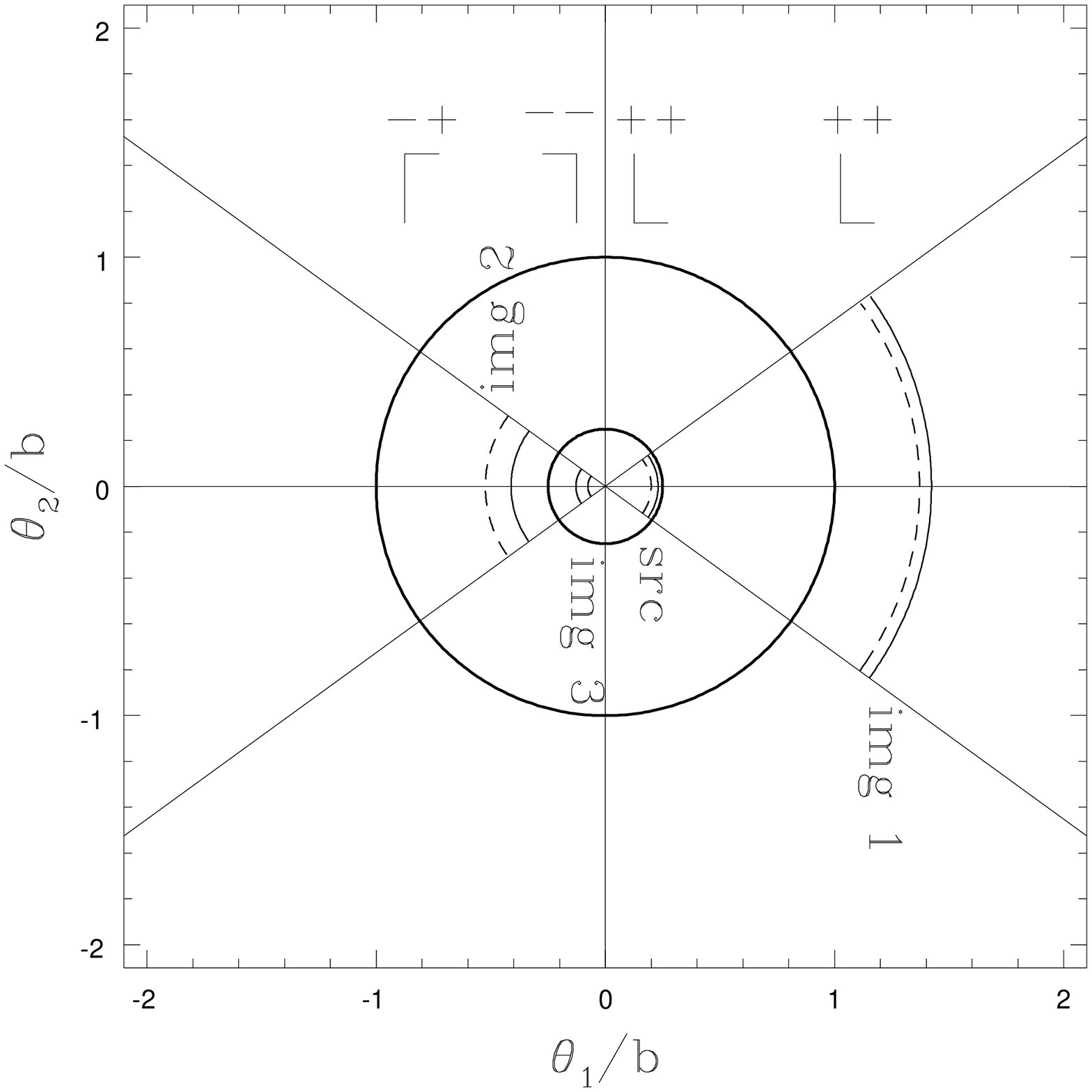,width=3.0in}}
\end{center}
\caption{
   Graphical solutions for the Moore profile cusp ($n=3/2$) lens when $\beta<b/4$ and
   there are three images.  At the top of the lower panel we illustrate the geometric
   meaning of the image partial parities defined by the signs of the magnification
   tensor eigenvalues (see text).
   }
\labelprint{fig:moore2}
\end{figure}

A nice property of circular lenses is that they allow simple graphical solutions
of the lens equation for arbitrary deflection profiles.  There are two parts
to the graphical solution -- the first is to determine the radial positions
$\theta_i$ of the images given a source position $\beta$, and the second is
to determine the magnification by comparing the area of the images to the
area of the source.  Recall first, that by symmetry, all the images must lie 
on a line passing through the source and the lens.  Let $\theta$ now be 
a signed radius that is positive along this line on one side of the lens
and negative on the other.  The lens equation (Eqn.~\ref{eqn:aab}) along the line is
simply
\begin{equation}
     { \theta \over |\theta| } \alpha(|\theta|) = \theta - \beta
  \labelprint{eqn:aaj}
\end{equation}
where we have rearranged the terms to put the deflection on one side and the
image and source positions on the other.  One side of the equation is 
the bend angle (Fig.~\ref{fig:bendang}), while the other side  
of the equation, $\theta-\beta$, is simply a line of unit slope passing
through the source position $\beta$. The solutions to the lens equation
for any source position $\beta$ are the radii $\theta_i$ where the line 
crosses the curve.

For understanding any observed lens, it is always useful to first sketch where
the critical lines must lie.  Recall from the discussion of caustics in 
\partintro, that images are always created and destroyed on critical lines
as the source crosses a caustic, so the critical lines and caustics define
the general structure of the lens.  All our power law models have a 
tangential critical line at $\theta=b$, which is the solution $\alpha(b)=b$
and corresponds to the source position $\beta=0$.  The origin, as the 
projection of the critical curve onto the source plane, is the tangential
caustic (strictly speaking a degenerate pseudo-caustic) corresponding to
the critical line.  A point source at the
origin is transformed into an Einstein ring of radius $\theta_E=b$.  

The second step of the graphical construction is to determine the angular
structure of the image.  For simplicity, suppose the source is an arc with
radial width $\Delta\beta$ and angular width $\Delta\chi$.  By symmetry,
the angle subtended by an image relative to the lens center must be
the same as that subtended by the source.  For an image at $\theta_i$
and a source at $\beta$, the tangential extent of the image is 
$|\theta_i| \Delta\chi$ while that of the source is $\beta \Delta\chi$.
The tangential magnification of the image is simply 
$|\theta_i|/\beta=(1-|\alpha(\theta_i)/\theta_i|)^{-1}$ after making
use of the lens equation (Eqn.~\ref{eqn:aaj}), and this is identical to the tangential
magnification eigenvalue (Eqn.~\ref{eqn:aah}).  The thickness of the arc requires finding
the image radii for the inner and outer edges of the source,
$\theta_i(\beta)$ and $\theta_i(\beta+\Delta\beta)$.  The ratio
of the thickness of the two arcs is the radial magnification,
\begin{equation}
 { \theta_i(\beta+\Delta\beta)-\theta_i(\beta) \over \Delta\beta }
  \simeq { d \theta \over d\beta } = 
  \left( 1 - { d \alpha \over d\theta } \right)^{-1},
  \labelprint{eqn:aak}
\end{equation}
and this is simply the inverse of the radial eigenvalue of the magnification matrix 
(Eqn.~\ref{eqn:aai}) where 
we have taken the derivative of the lens equation (Eqn.~\ref{eqn:aaj}) with respect to
the source position to obtain the final result.  Thus, the tangential
magnification simply reflects the fact that the angle subtended by the
source is the angle subtended by the image, while the radial magnification
depends on the slope of the deflection profile with declining
deflection profiles ($d\alpha/d\theta<0$) demagnifying the source and rising profiles
magnifying the source.       
 
In Fig.~\ref{fig:pntmass} we illustrate this for the point mass lens ($n\rightarrow 3$).
From the shape of the deflection profile, it is immediately obvious
that there will be only two images, one on each side of the lens.  If
we assume $\beta >0$, the first image is a minimum located at
\begin{equation}
   \theta_1 = { 1\over 2} \left( \beta + \sqrt{\beta^2+4b^2} \right)
  \labelprint{eqn:aal}
\end{equation} 
with $\theta_1>\theta_E$ and positive magnification
\begin{equation}
    \mu_1 = { 1 \over 4 } \left( { \beta \over \sqrt{\beta^2+4b^2} }
              + { \sqrt{\beta^2+4b^2} \over \beta } +2 \right) >0,
  \labelprint{eqn:aam1}
\end{equation}
while the second image is a saddle point located at 
\begin{equation}
  \theta_2 = { 1\over 2} \left( \beta - \sqrt{\beta^2+4b^2} \right)
  \labelprint{eqn:aan1}
\end{equation}
with $-\theta_E < \theta_2 <0$ and negative magnification
\begin{equation}
    \mu_2 = -{ 1 \over 4 } \left( { \beta \over \sqrt{\beta^2+4b^2} }
              + { \sqrt{\beta^2+4b^2} \over \beta } -2 \right) <0.
  \labelprint{eqn:aam2}
\end{equation}
As the source approaches the tangential caustic ($\beta \rightarrow 0$) the
magnifications of both images diverge as $\beta^{-1}$ and the image radii
converge to $\theta_E$.  As the source moves to infinity, the magnification
of the first image approaches unity and its position approaches that of the
source, while the second image is demagnified by the factor $(1/2)(b/\beta)$ 
and converges to the position of the lens.  The image separation
\begin{equation}
   \Delta\theta = | \vec{\theta}_1 -\vec{\theta}_2| = 2 b \sqrt{1+\beta^2/4b^2} \geq 2b
  \labelprint{eqn:aan2}
\end{equation}
is always larger than the diameter of the Einstein ring and the total
magnification
\begin{equation}
   | \mu_1| + |\mu_2| = { 2 b^2 + \beta^2 \over \beta \sqrt{\beta^2+4b^2} } \geq 1
  \labelprint{eqn:aao}
\end{equation}
is the characteristic light curve expected for isolated Galactic microlensing
events (see \partmicro).  
The point mass lens has one peculiarity that makes it different from extended
density distributions like galaxies in that it has two images 
independent of the impact parameter of the source and no radial caustic.
 This is a characteristic of any density distribution with a divergent central 
deflection ($n>2$). 

The SIS ($n=2$) model is the ``standard'' lens model for galaxies.
Figs.~\ref{fig:sis1} and \ref{fig:sis2}
show the geometric constructions for the images of an SIS lens.  
If $0 < \beta < b$, then the SIS lens also produces two images (Fig.~\ref{fig:sis1}).  
The first image is a minimum located at 
\begin{equation}
   \theta_1 = \beta + b \quad\hbox{with $\theta_1>b$ and positive magnification}\quad 
   \mu_1 = 1+b/\beta 
  \labelprint{eqn:aap}
\end{equation}
and the second image is a saddle point located at
\begin{equation}
   \theta_2 = \beta - b \quad\hbox{with $-b<\theta_2<0$ and negative magnification} \quad
   \mu_2 = 1-b/\beta.
  \labelprint{eqn:aaq}
\end{equation}
The image separation $|\theta_1-\theta_2|=2b$ is constant, and the total
magnification $|\mu_1|+|\mu_2|=2b/\beta$ is a simple power law.  
The magnification produced by an SIS lens is purely tangential
since the radial magnification is unity.  If, however, $\beta > b$,
then there is only one image, corresponding to the minimum located
on the same side of the lens as the source (see Fig.~\ref{fig:sis2}).  This
boundary on the source plane at $\beta=b$ between having two images
at smaller radii and only one image at larger radii is a
radial (pseudo)-caustic that can be thought of as being associated
with a radial critical curve at the origin.  It is a pseudo-caustic
because there are neither images nor a divergent magnification associated
with it.

Historically the next step is to introduce a core radius to have a model
with a true radial critical line and caustic (see \partintro, Blandford \& Kochanek~\cite{Blandford1987p658}, Kochanek \& Blandford~\cite{Kochanek1987p676}, Kovner~\cite{Kovner1987p22},
Hinshaw \& Krauss~\cite{Hinshaw1987p468}, Krauss \& White~\cite{Krauss1992p385},
Wallington \& Narayan~\cite{Wallington1993p517}, Kochanek~\cite{Kochanek1996p638}).  
Instead we will consider 
the still softer power law model with $n=3/2$, which would correspond to
the central exponent of the ``Moore'' profile proposed for CDM halos (Moore et al. 
\cite{Moore1998p5}).  As 
Fig.~\ref{fig:moore1} shows, there is only one solution for $|\beta| > b/4$, a 
minimum located at
\begin{equation}
  \theta_1  = { 1 \over 2 } \left( b + 2\beta + \sqrt{b+4\beta}\right)
  \labelprint{eqn:aar}
\end{equation} 
and with $\theta_1 > b$ assuming $\beta$ is positive.
The magnification expressions are too complex to be of much use,
but the magnification $\mu_1$ diverges at $\theta=b$ when the source is
on the tangential pseudo-caustic at $\beta=0$.  As Fig.~\ref{fig:moore2} 
shows, we
find two additional images once $|\beta|<b/4$.  The first additional
image is a saddle point located at
\begin{equation}
 \theta_2  = { 1 \over 2 } \left( -b + 2\beta - \sqrt{b+4\beta}\right) 
  \labelprint{eqn:aas}
\end{equation}
with $-b<\theta_2 < -b/4$,
which has a negative magnification that diverges at both $\theta_2=-b$ (the
tangential critical curve) and $\theta_2=-b/4$.  This latter radius defines the
radial critical curve where the magnification diverges because the
radial magnification eigenvalue $1-\kappa+\gamma=1-d\alpha/d\theta=0$
at radius $\theta=b/4$.
The third image is a maximum located at
\begin{equation}
 \theta_3  = { 1 \over 2 } \left( -b + 2\beta + \sqrt{b+4\beta}\right)
  \labelprint{eqn:aat}
\end{equation}
with $-b/4<\theta_2<0$ and 
a positive magnification that diverges on the radial critical curve.
As we move the source outward from the center we would see images 2 and 3
approach each other, merging on the radial critical line where they would
have divergent magnifications, and then vanishing to leave only image 1.
We would see the same pattern if instead of softening the exponent we had
followed the traditional path and added a core radius to the SIS model.
With a finite core radius the central deflection profile would pass through
zero, and this would introduce a radial critical curve and a third image 
which would be a maximum of the time delay surface.  

In Fig.~\ref{fig:moore2} we also illustrate the geometric meaning of the partial 
parities (the signs of the magnification eigenvalues). 
A source structure (the L) defines the
   reference shape.  Image 1 is a minimum with positive partial parities ($++$)
   defined by the signs of the tangential and radial eigenvalues. The 
   orientation of image 1 is the same as the source.  Image 2 is a saddle point with
   mixed partial parities ($-+$) because the tangential eigenvalue is negative
   while the radial eigenvalue is positive.  This means that the image is
   inverted in the tangential direction relative to the source.  Image 3 is a
   maximum with negative partial parties ($--$), so the image is inverted
   in both the radial and tangential directions relative to the source.
The total parity, the product of the partial parities, is positive for maxima
and minima so the orientation of the image can be produced by rotating the
source.  The total parity of the saddle point image is negative, so its 
orientation cannot be produced by a rotation of the source.

\subsection{Non-Circular Lenses }

The tangential pseudo-caustic at the origin producing Einstein ring images
is unstable to the introduction of any angular structure into the gravitational
potential of the lens.  There are two generic sources of angular perturbations.
The first source of angular perturbations is 
the ellipticity of the lens galaxy. What counts here is the
ellipticity of the gravitational potential rather than of the surface density.
For a lens with axis ratio $q$, ellipticity $\epsilon=1-q$, or 
eccentricity $e=(1-q^2)^{1/2}$, the 
ellipticity of the potential is usually $\epsilon_\Psi \sim \epsilon/3$ --
potentials are always rounder than densities. 
The second source of angular perturbations is tidal perturbations from any
nearby objects.  This is frequently called the ``external shear" or the
``tidal shear" because it can be modeled as a linear shearing of the
deflections.
In all known lenses, quadrupole perturbations (i.e. $\Psi \propto \cos(2\chi)$
where $\chi$ is the azimuthal angle)
dominate -- higher order multipoles are certainly present and they can
be quantitatively important, but they are smaller.  For example, in an
ellipsoid the amplitude of the $\cos 2m\chi$ multipole scales as
$\epsilon_\Psi^m$ (see \S\ref{sec:massquad} and \S\ref{sec:substruc}).

Unfortunately, there is no example of a non-circular lens 
that can be solved in full generality unless you view the nominally analytic
solutions to quartic equations as helpful.
 We can make the greatest progress 
for the case of an SIS in an external (tidal) shear field.  Tidal shear is due
to perturbations from nearby objects and its amplitude can be determined by
Taylor expanding its potential near the lens (see \partintro\, and \S\ref{sec:mass}).  
Consider a lens with Einstein
radius $\theta_E$ perturbed by an object with effective lens potential $\Psi$
a distance $\theta_p$ away.  For $\theta_E \ll \theta_p$ we can Taylor 
expand the potential of the nearby object about the center of the primary
lens, dropping the leading two terms.\footnote{The first term,
a constant, gives an equal contribution to the time delays of all the images,
so it is unobservable when all we can measure is relative delays.  The 
second term is a constant deflection, which is unobservable when all we
can measure is relative deflections.} This leaves, as the first term with
observable consequences,
\begin{equation}
   \Psi(\theta) \simeq { 1 \over 2 } \vec{\theta} \cdot \vec{\nabla}
    \vec{\nabla} \Psi \cdot \vec{\theta}
    = { 1 \over 2 } \kappa_p \theta^2 
      - { 1 \over 2 } \gamma_p \theta^2 \cos 2 (\chi-\chi_p)
  \labelprint{eqn:aau}
\end{equation}
where $\kappa_p$ is the surface density of the perturber at the center of the
lens galaxy and $\gamma_p>0$ is the tidal shear from the perturber.  If the
perturber is an SIS with critical radius $b_p$ and distance $\theta_p$ from
the primary lens, then $\kappa_p=\gamma_p=b_p/2\theta_p$.  With this
normalization, the angle $\chi_p$ points toward the perturber.  For a
circular lens, the shear $\gamma_p=\kbar-\kappa$ can be expressed in terms
of the surface density of the perturber, and it is larger (smaller) than the
convergence if the density profile is steeper (shallower) than isothermal.

The effects of $\kappa_p$ are observable only if we measure a time delay
or have an independent estimate of the mass of the lens galaxy, while the
effects of the shear are easily detected from the relative positions of the
lensed images (see \partintro).  Consider, for example, one component of the
lens equation including an extra convergence,
\begin{equation}
    \beta_1 = \theta_1(1-\kappa_p) - d\Psi/d\theta_1
  \labelprint{eqn:aav}
\end{equation}
and then simply divide by $1-\kappa_p$ to get
\begin{equation}
    \beta_1/(1-\kappa_p) = \theta_1 - (d\Psi/d\theta_1)/(1-\kappa_p).
  \labelprint{eqn:aaw}
\end{equation}
The rescaling of the source position $\beta_1/(1-\kappa_p)$ has no
consequences since the source position is not an observable 
quantity, while the rescaling of the deflection is simply a change
in the mass of the lens.  This is known as the ``mass sheet degeneracy''
because it corresponds to adding a constant surface density sheet to the
lens model (Falco, Gorenstein \& Shapiro~\cite{Falco1985p1}, see \partintro), and it is
an important systematic problem for both strong lenses and cluster
lenses (see \partweak).  

Thus, while the extra convergence can be important for the quantitative
understanding of time delays or lens galaxy masses, it is only the shear
that introduces qualitatively new behavior to the lens equations.
The effective potential of an SIS lens in an external shear is 
$\Psi = b\theta + (\gamma/2)\theta^2\cos 2\chi$ leading to the lens equations  
\begin{equation}
   \begin{array}{cc}
     \beta_1 &= \theta_1 (1-\gamma) - b  \theta_1 / |\vec{\theta}|  \\
     \beta_2 &= \theta_2 (1+\gamma) - b  \theta_2 / |\vec{\theta}|  
    \end{array}
  \labelprint{eqn:aax}
\end{equation}
where for $\gamma>0$ the perturber is due North (or South) of the lens.  The
inverse magnification is
\begin{equation}
   \mu^{-1} = 1 -\gamma^2 - { b \over \theta } \left(1-\gamma\cos 2\chi\right)
  \labelprint{eqn:aay}
\end{equation}
where $\vec{\theta}=(\theta_1,\theta_2)=\theta(\cos\chi,\sin\chi)$.

The first step in any general analysis of a new lens potential is to locate
the critical lines and caustics.  In this case we can easily solve $\mu^{-1}=0$
to find that the tangential critical line
\begin{equation}
     \theta = b { 1 - \gamma \cos 2\chi \over 1-\gamma^2 }
  \labelprint{eqn:aaz}
\end{equation}
is an ellipse whose axis ratio is determined by the amplitude of the shear
$\gamma$ and whose major axis points toward the perturber.
We call it the tangential critical line because the associated 
magnifications are nearly tangential to the direction to the lens galaxy 
and because it is a perturbation to the Einstein ring of a circular lens.
The tangential caustic, the image of the critical line on the source plane,
is a curve called an astroid (Fig.\ref{fig:cuspminor}, it is not a ``diamond'' despite
repeated use of the term in the literature).  The parametric expression for
the astroid curve is
\begin{equation}
   \beta_1 =  -{ 2 b \gamma \over 1+\gamma} \cos^3 \chi
           =  - \beta_+ \cos^3 \chi
   \qquad
   \beta_2 =  +{ 2 b \gamma \over 1-\gamma} \sin^3 \chi
           = \beta_- \sin^3 \chi
  \labelprint{eqn:aba}
\end{equation}
where the parameter $\chi$ is the same as the angle  appearing in the critical
curve (Eqn.~\ref{eqn:aaz}) and we have defined  $\beta_\pm=2b\gamma/(1\pm\gamma)$
for the locations of the cusp tips on the axes. The astroid consists of 4 cusp 
caustics on the symmetry axes
of the lens connected by fold caustics with a major axis pointing
toward the perturber.  Like the SIS model without
any shear, the origin plays the role of the radial critical line and
there is a circular radial pseudo-caustic at $\beta=b$.  

As mentioned earlier, there is no useful general solution for the image
positions and magnifications.  We can, however, solve the equations for
a source on one of the symmetry axes of the lens.  For example, consider
a solution on the minor axis of the lens ($\beta_2=0$ for $\gamma>0$).
There are two ways of solving the lens equation to satisfy the 
criterion.  One is to put the images on the same axis ($\theta_2=0$)
and the other is to place them on the arc defined by $0=1+\gamma-b/\theta$.
The images with $\theta_2=0$ are simply the SIS solutions corrected
for the effects of the shear.  Image 1 is defined by
\begin{equation}
  \theta_1 = { \beta_1 + b \over 1-\gamma} \quad\hbox{with}\quad
    \mu^{-1} =  \left( 1-\gamma^2 \right) { \beta_+ + \beta_1 \over b + \beta_1 }
  \labelprint{eqn:abb}
\end{equation}
and image 2 is defined by 
\begin{equation}
  \theta_1 = { \beta_1 - b \over 1-\gamma} \quad\hbox{with}\quad
    \mu^{-1} =  \left( 1-\gamma^2 \right) { \beta_+ - \beta_1 \over b - \beta_1 }
  \labelprint{eqn:abc}
\end{equation}
Image 1 exists if $\beta_1 > -b$, it is a saddle point for $-b<\beta_1 <-\beta_+$ 
and it is a minimum for $\beta_1 > -\beta_+$.  Image 2 has the reverse ordering.
It exists for $\beta_1<b$, it is a saddle point for $\beta_+ < \beta_1 < b $
and it is a minimum for $\beta_1 < \beta_+$.   The magnifications of both
images diverge when they are on the tangential critical line ($\beta_1=-\beta_+$
for image 1 and $\beta_1=+\beta_+$ for image 2) and approach zero as
they move into the core of the lens ($\beta_1\rightarrow -b$ for image 1
and $\beta_1 \rightarrow +b$ for image 2).  These two images
shift roles as the source moves through the origin.
The other two solutions are both saddle points, and they exist only
if the source lies inside the astroid ($|\beta_1| < \beta_+$ along
the axis).  The positions of images 3 ($+$) and 4 ($-$) are 
\begin{equation}
  \theta_1=-{ \beta_1 \over 2 \gamma} \qquad
  \theta_2= \pm { b \over 1+\gamma} 
    \left[ 1- \left( { \beta_1 \over \beta_+}\right)^2\right]^{1/2}
  \labelprint{eqn:abd}
\end{equation}
and they have  equal magnifications 
\begin{equation}
    \mu^{-1} = -2\gamma(1+\gamma)  \left[ 1- \left( { \beta_1 \over \beta_+}\right)^2\right].
  \labelprint{eqn:abe}
\end{equation}
The magnifications of the images diverge when the source reaches the cusp
tip ($|\beta_1|=\beta_+$) and the image lies on the tangential critical
curve.  

Thus, if we start with a source at the origin we can follow the changes in
the image structure (see Fig.~\ref{fig:cuspminor}, \ref{fig:cuspmajor}).  
With the source at the origin we see 4 images on the
symmetry axes with reasonably high magnifications, 
$\sum|\mu_i| = (2/\gamma)/(1-\gamma^2) \sim 10$.
It is a generic result that the least magnified four-image system is found for an
on-axis source, and this configuration has a total magnification of order the inverse of the 
ellipticity of the gravitational potential.
As we move the source toward the tip of the cusp ($ \beta \rightarrow \beta_+$,
Fig.~\ref{fig:cuspminor}),
image 1 simply moves out along the symmetry axis with slowly dropping
magnification, while images 2, 3 and 4 move toward a merger on the 
tangential critical curve at $\vec{\theta}=(-\beta_+,0)$.  Their
magnifications steadily rise and then diverge when the source reaches
the cusp.  If we move the source further outward we find only images
1 and 2 with 1 moving outward and 2 moving inward toward the origin.
As it approaches the origin, image 2 becomes demagnified and vanishes
when $\beta \rightarrow b$.  
Had we done the same calculation on the major axis (Fig.~\ref{fig:cuspmajor}), 
there is a qualitative
difference.  As we moved image 1 outward along the $\beta_2$ axis, image 3 and
4 would merge with image 1 when the source reaches the tip of the cusp
at $\beta_2=\beta_-$ rather than with image 2.   

Unfortunately once we move the source off a symmetry axis, there is no simple
solution.  It is possible to find the locations of the remaining images given
that two images have merged on the critical line, and this is useful for
determining the mean magnifications of the lensed images, a point we will
return to when we discuss lens statistics in \S\ref{sec:stat}.  Here we simply illustrate
(Fig.~\ref{fig:foldlens}) the behavior of the images when we move the source radially 
outward from the origin away from the symmetry axes.  Rather than three
images merging on the tangential critical line as the source approaches
the tip of a cusp, we see two images merging as the source approaches the
fold caustic of the astroid.  This difference, two images merging versus
three images merging, is a generic difference between folds and cusps 
as discussed in \partintro.  All images in these four-image configurations are
restricted to an annulus of width $\sim \gamma b$ around the critical line,
so the mean magnification of all four image configurations is also 
of order $\gamma^{-1}$ (see Finch et al.~\cite{Finch2002p51}).

\begin{figure}[ph]
\begin{center}
\centerline{\psfig{figure=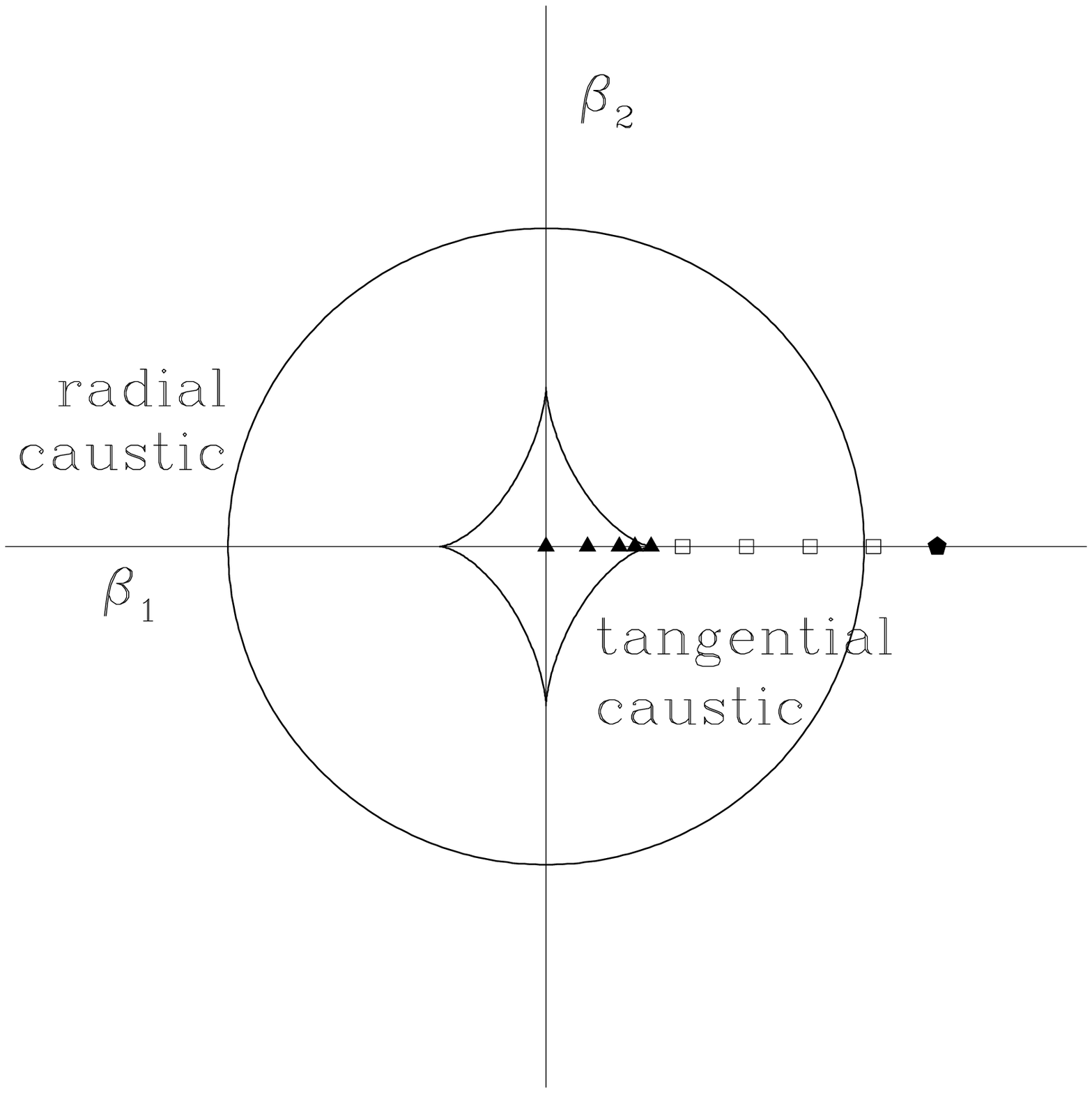,width=2.6in}}
\end{center}
\begin{center}
\centerline{\psfig{figure=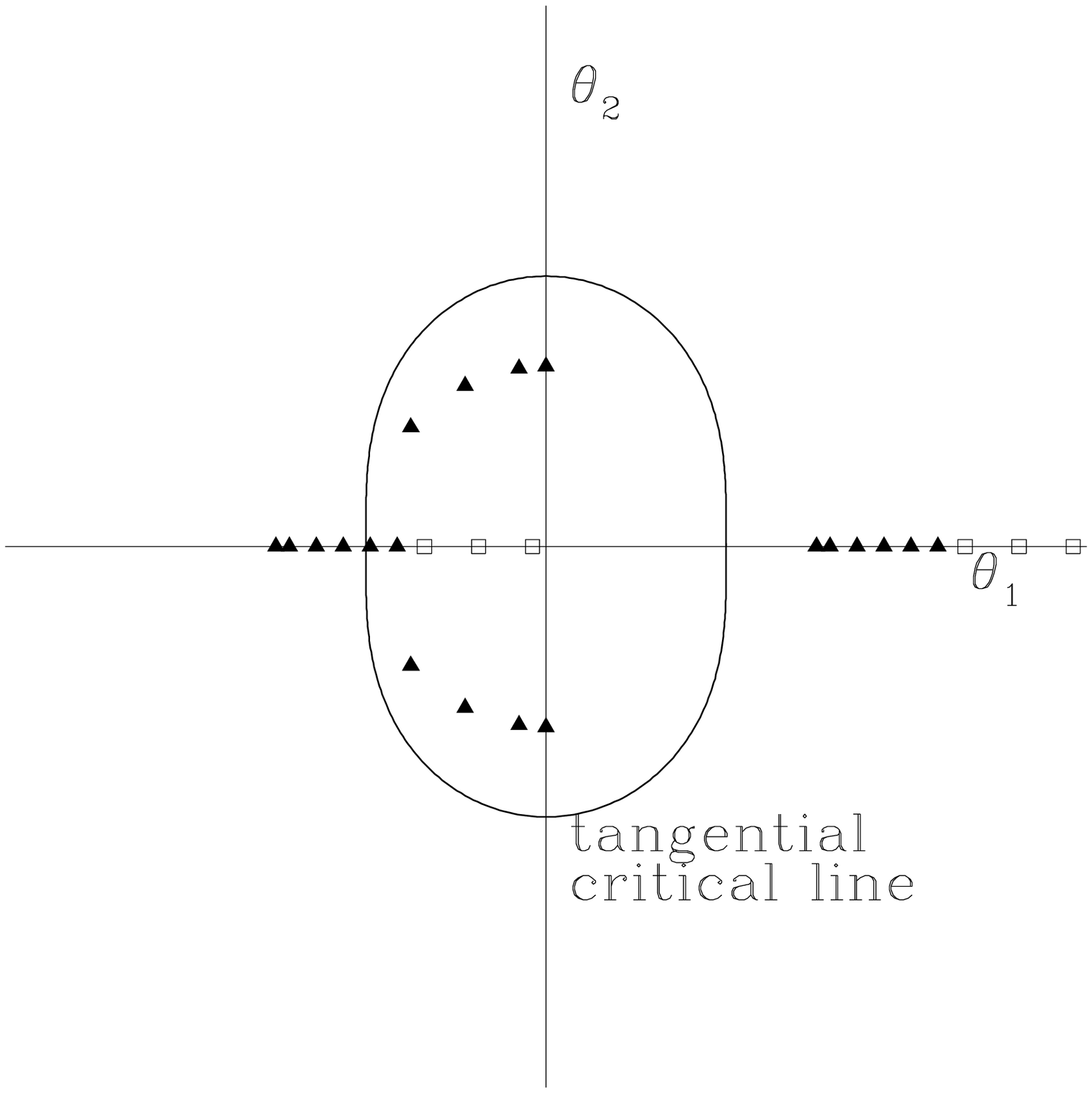,width=2.6in}}
\end{center}
\caption{
   Example of a minor axis cusp on the source (top) and image (bottom) planes.
   When the source is inside both the radial and tangential caustics (triangles)
   there are four images.  As the source moves toward the cusp, three of the 
   images head towards a merger on the critical line and become highly magnified to
   leave only one image once the source crosses the cusp and lies between the
   two caustics (open squares).  In a minor axis cusp, the image surviving the 
   cusp merger is a saddle point interior to the critical line.
   As the source approaches the radial caustic,
   one image approaches the center of the lens and then vanishes as the it
   crosses the caustic to leave only one image (pentagons).
   }
\labelprint{fig:cuspminor}
\end{figure}

\begin{figure}[ph]
\begin{center}
\centerline{\psfig{figure=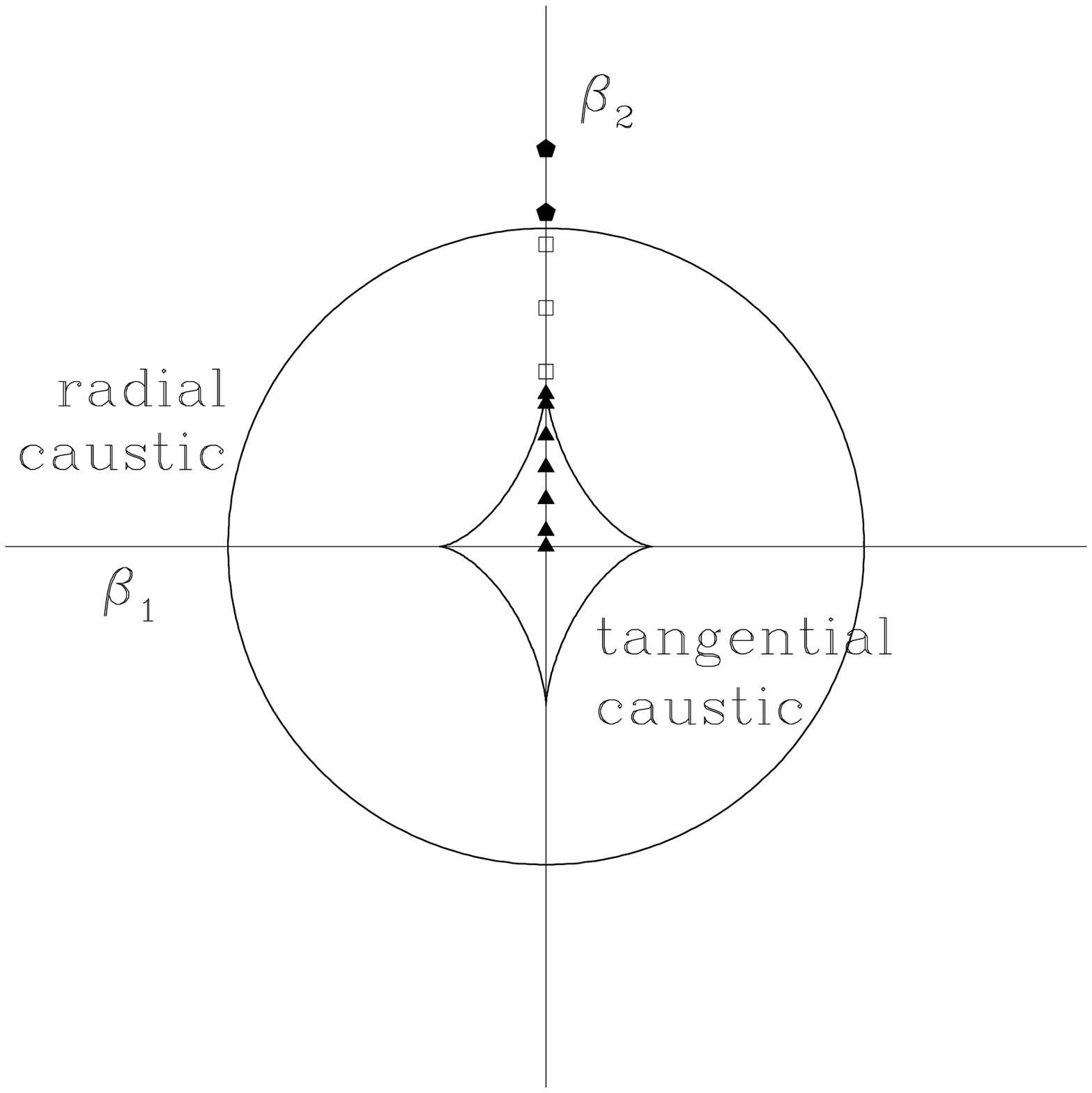,width=2.6in}}
\end{center}
\begin{center}
\centerline{\psfig{figure=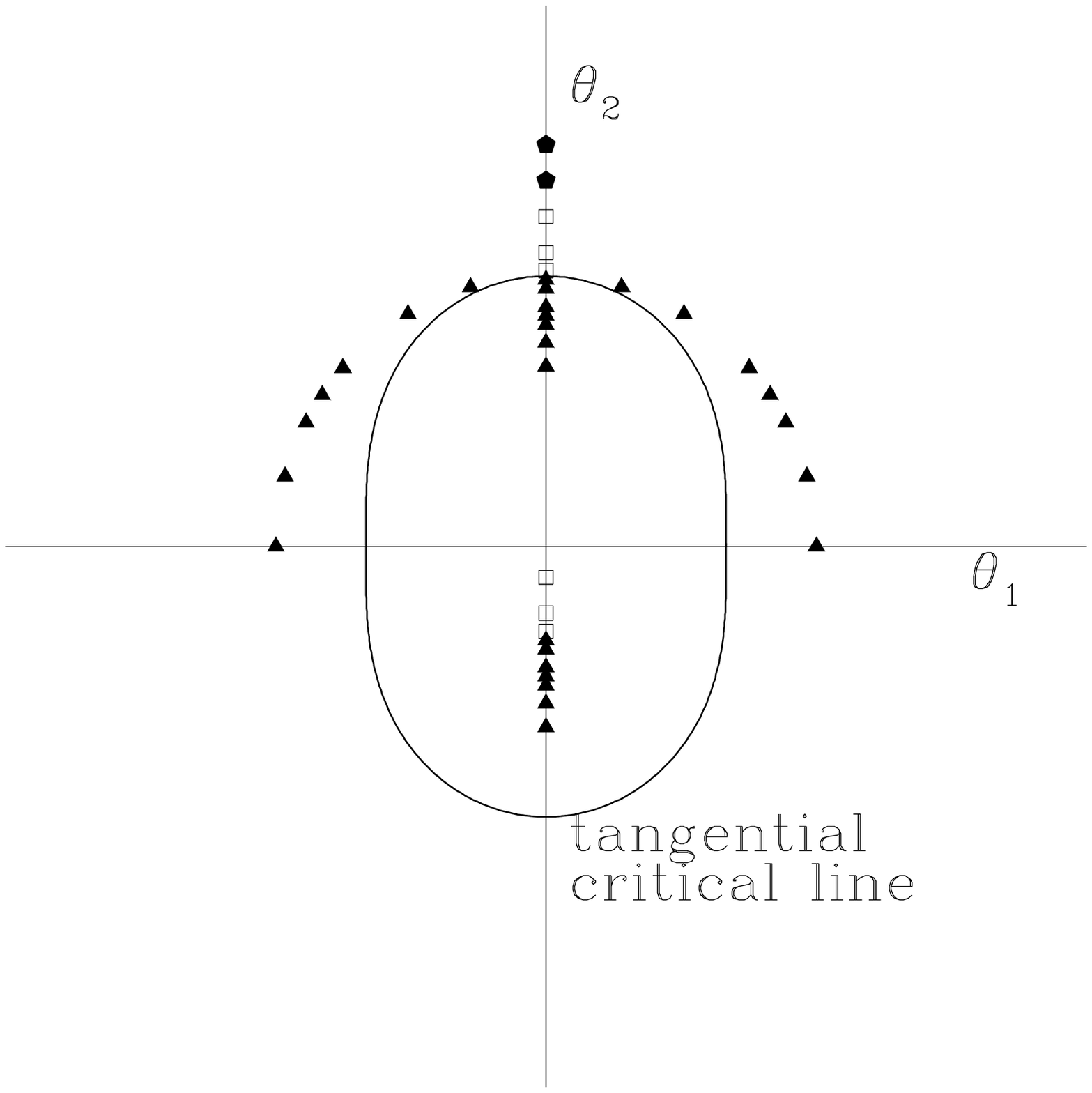,width=2.6in}}
\end{center}
\caption{
   Example of a major axis cusp on the source (top) and image (bottom) planes.
   When the source is inside both the radial and tangential caustics (triangles)
   there are four images.  As the source moves toward the cusp, three of the 
   images head towards a merger on the critical line and become highly magnified to
   leave only one image once the source crosses the cusp and lies between the
   two caustics (open squares).  In a major axis cusp, the image surviving the
   cusp merger is the minimum corresponding to the image we would see in the
   absence of a lens. As the source approaches the radial caustic,
   one image approaches the center of the lens and then vanishes as the source
   crosses the caustic to leave only one image (pentagons).
   }
\labelprint{fig:cuspmajor}
\end{figure}

\begin{figure}[ph]
\begin{center}
\centerline{\psfig{figure=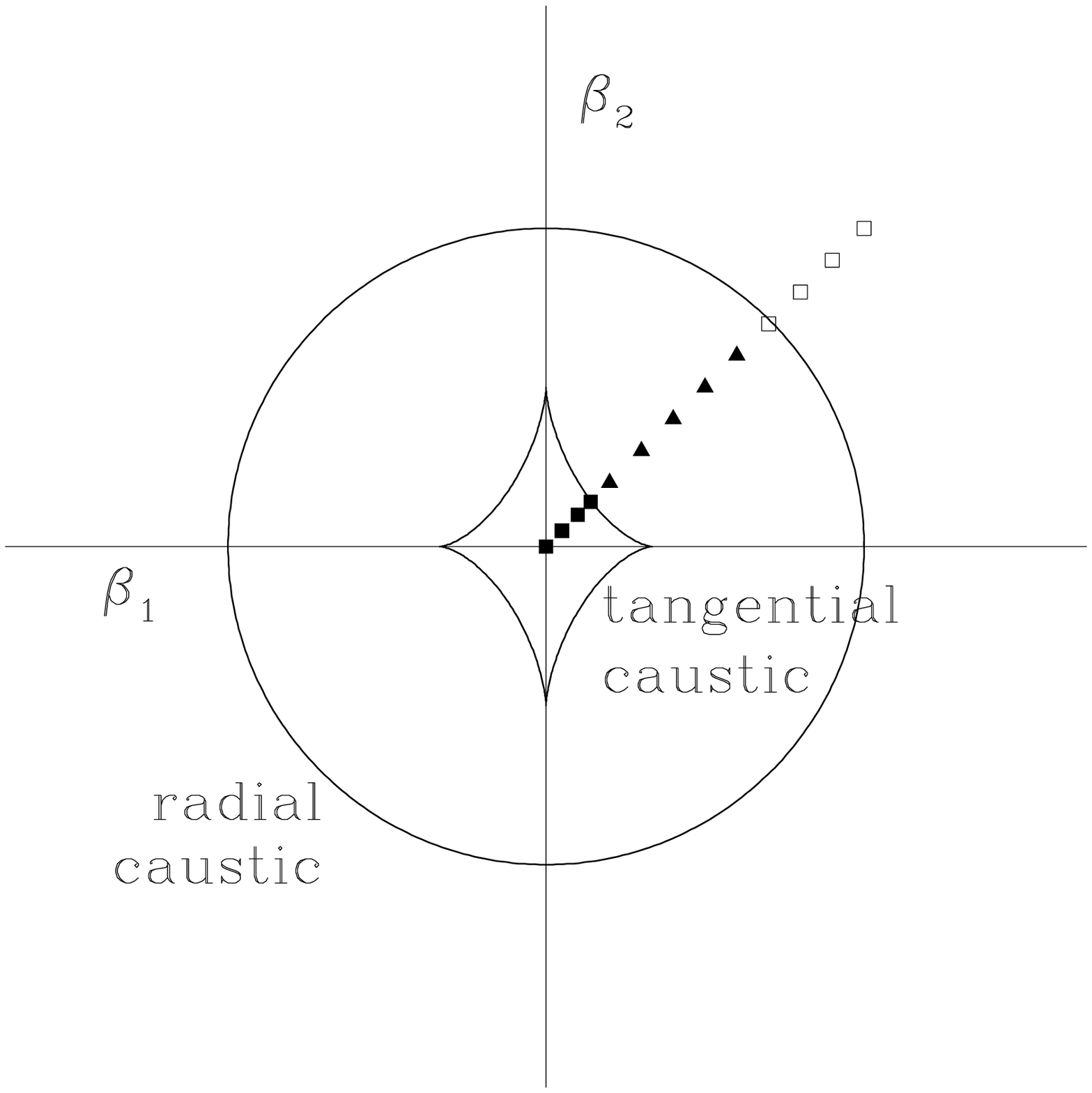,width=2.6in}}
\end{center}
\begin{center}
\centerline{\psfig{figure=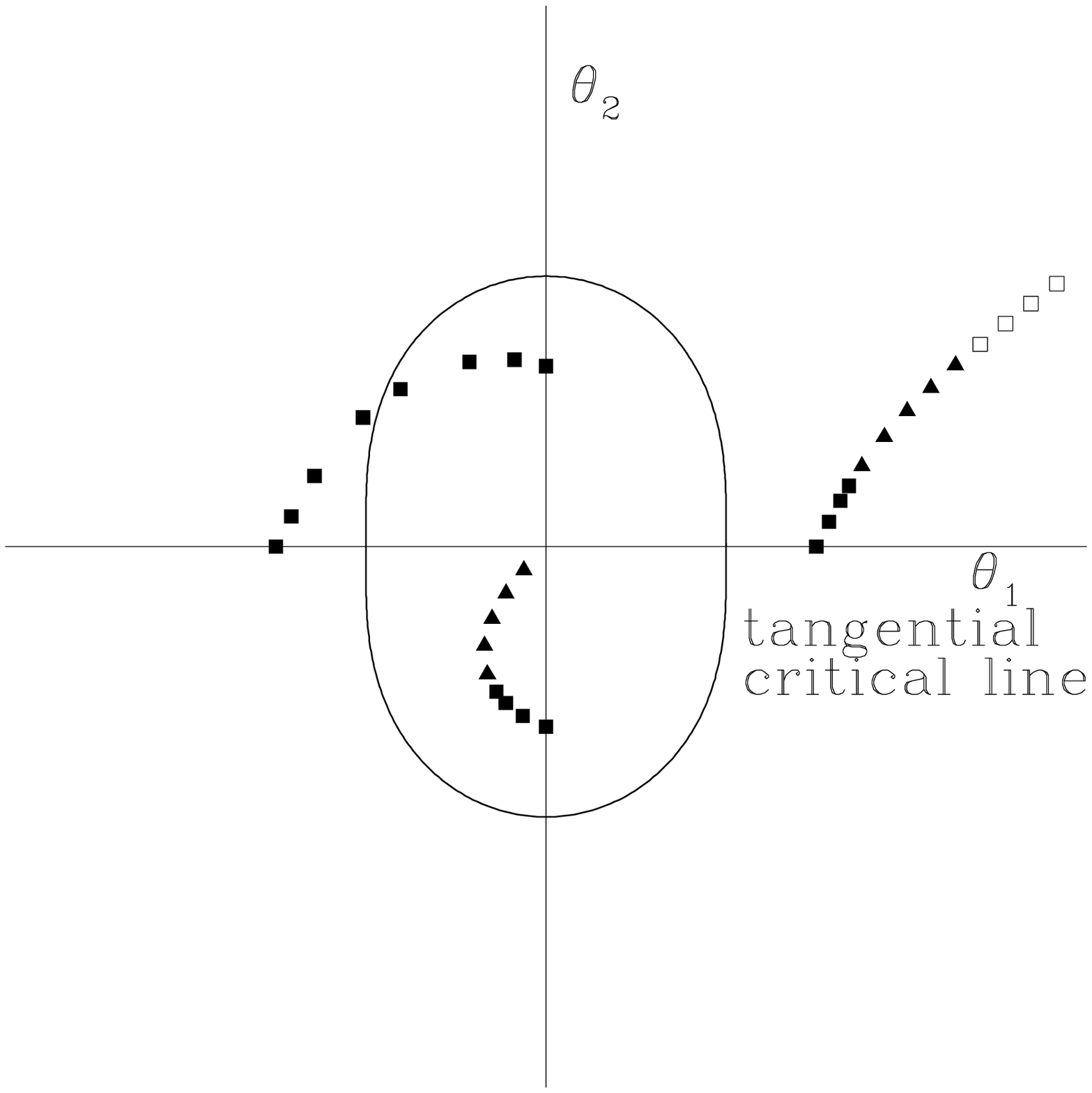,width=2.6in}}
\end{center}
\caption{
   Example of a fold merger on the source (top) and image (bottom) planes. When
   the source is inside both the radial and tangential caustics (filled squares)
   there are four images.  As the source crosses the tangential caustic, two
   images merge, become highly magnified and then vanish, leaving only two images
   (triangles) when the source is outside the tangential caustic but inside the
   radial caustic.  As the source approaches the radial caustic, one image moves
   into the center of the lens and then vanishes when the source crosses the
   radial caustic to leave only one image when the source is outside both
   caustics (open squares).    
   }
\labelprint{fig:foldlens}
\end{figure}

\begin{figure}[ph]
\begin{center}
\centerline{\psfig{figure=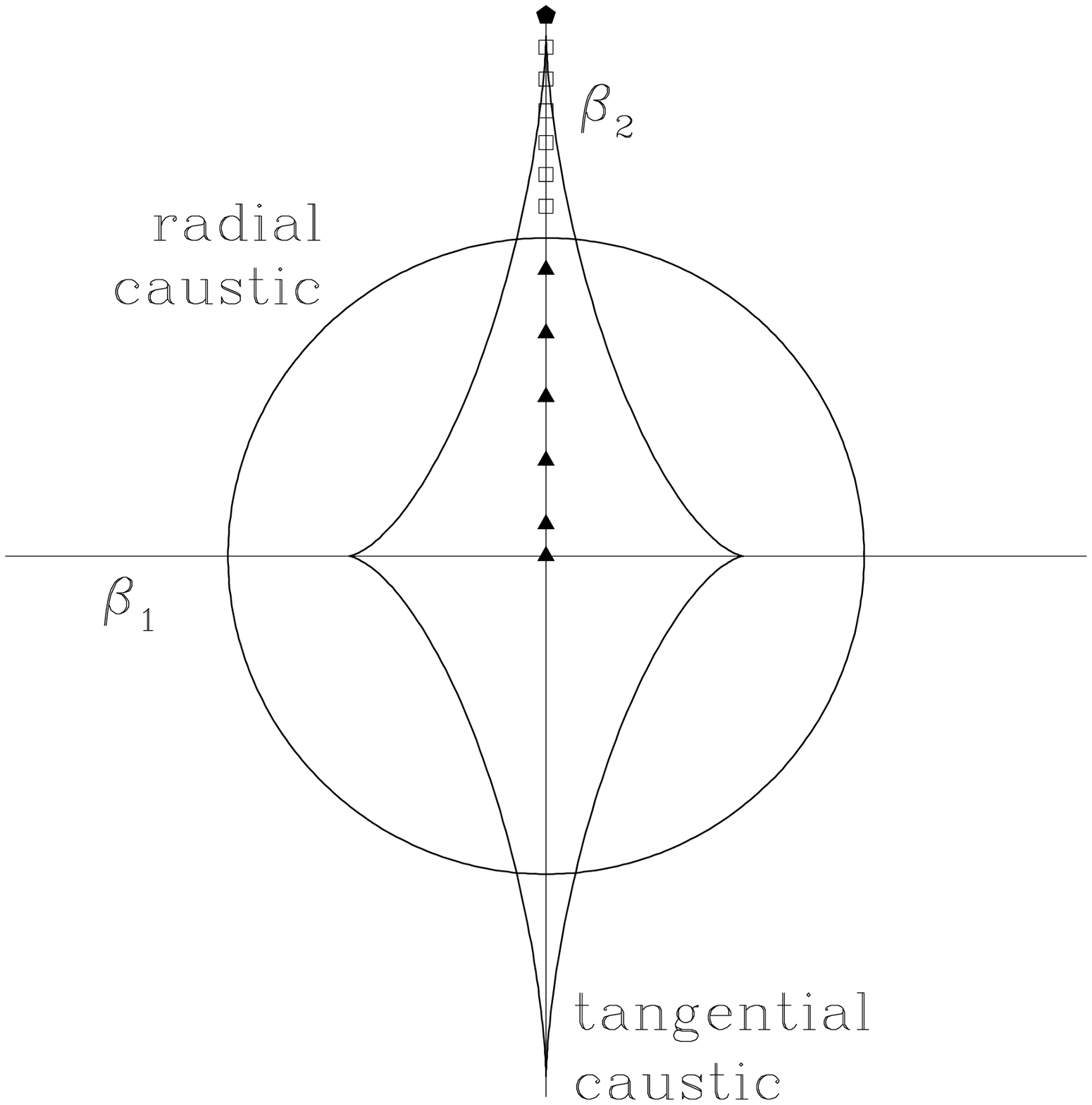,width=2.6in}}
\end{center}
\begin{center}
\centerline{\psfig{figure=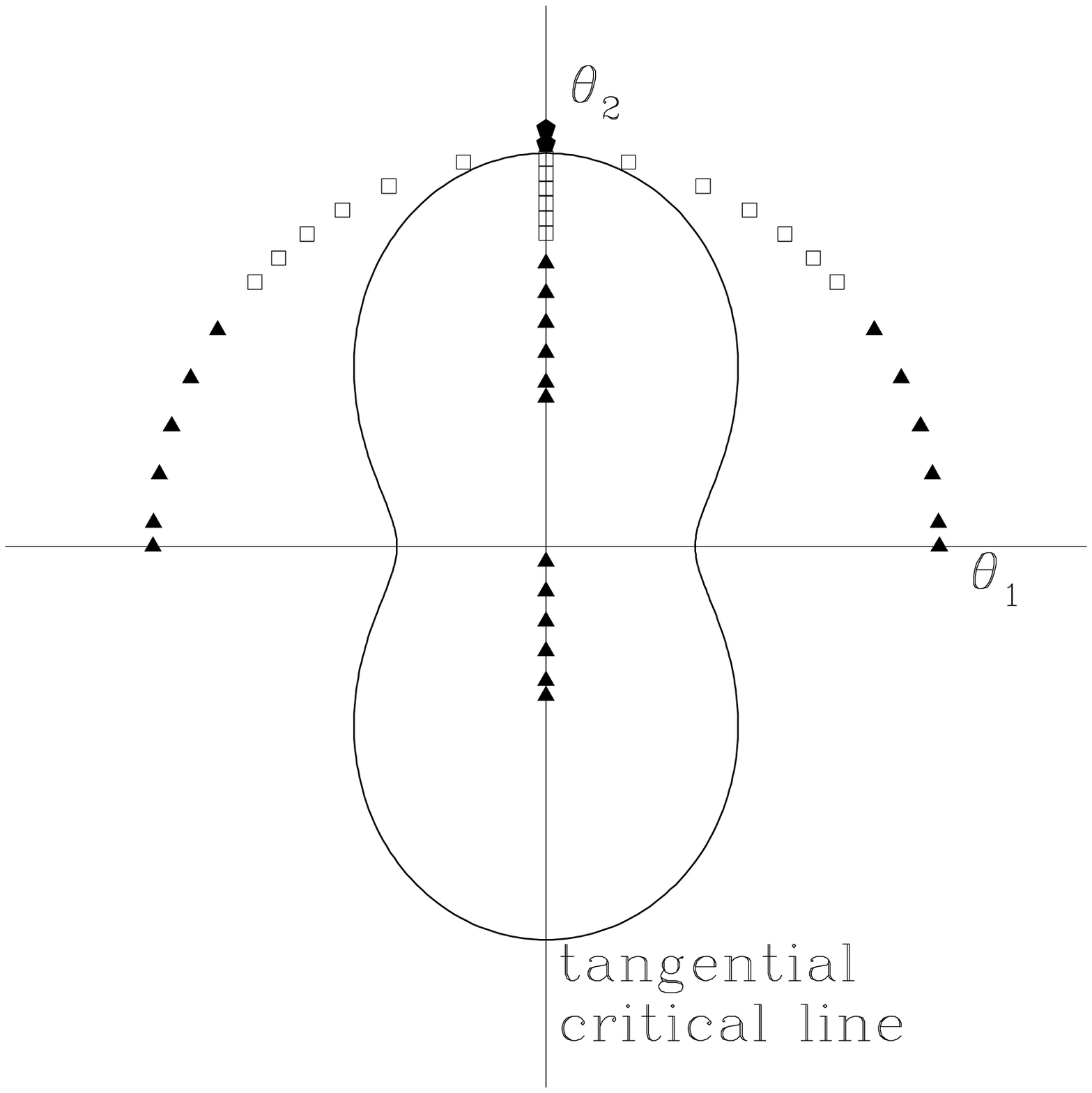,width=2.6in}}
\end{center}
\caption{
   Example of a cusp or disk image geometry on the source (top) and image (bottom)
   planes.  The shear is high enough to make the tangential caustic extend outside
   the radial caustic.  For a source inside both caustics (triangles) we see a 
   standard four-image geometry as in Fig.~\ref{fig:cuspmajor}.  
  However, for a source outside the
   radial caustic but inside the tangential caustic (squares) we have three images
   all on one side of the lens.  This is known as the cusp geometry because it is
   always associated with cusps, and the disk geometry because flattened disks are
   the only natural way to produce them.  Once the source is outside the cusp tip
   (pentagon), a single image remains.  
   }
\labelprint{fig:cusplens}
\end{figure}

There is one more possibility for the caustic structure of the lens if the
external shear is large enough.  For $1/3<|\gamma|<1$, the tip of the 
astroid caustic extends outside the radial caustic, as shown if Fig.~\ref{fig:cusplens}.
This allows a new image geometry, known as the cusp or disk geometry,
where we see three images straddling the major axis of a very flattened
potential.  It is associated with the caustic region inside the astroid
caustic but outside the
radial caustic.  This configuration appears to be rare for lenses produced
by galaxies, with APM08279+5255 as the only likely candidate, but relatively
more common in clusters.  The difference is that clusters tend to have 
shallower density profiles than galaxies, which shrinks the radial 
caustics relative to the tangential caustics to allow more cross section
for this image configuration and lower ellipticity thresholds before it
becomes possible (Oguri \& Keeton~\cite{Oguri2004p1} most recently, but
also see Kochanek \& Blandford~\cite{Kochanek1987p676}, Kovner~\cite{Kovner1987p22},
Wallington \& Narayan~\cite{Wallington1993p517}). 

In general, it is far more difficult to analyze ellipsoidal lenses, in part
because few ellipsoidal lenses have analytic expressions for their deflections.
The exception is the isothermal ellipsoid (Kassiola \& Kovner~\cite{Kassiola1993p450}, 
Kormann, Schneider \& Bartelmann~\cite{Kormann1994p285}, 
Keeton \& Kochanek~\cite{Keeton1998p157}), including a core radius $s$, 
which is both analytically tractable and generally viewed as the most likely 
average mass distribution for gravitational lenses.  The surface density
of the isothermal ellipsoid 
\begin{equation}
   \kappa = { 1 \over 2 } { b \over \omega} 
  \quad\hbox{where}\quad \omega^2=q^2(\theta_1^2+s^2)+\theta_2^2
  \labelprint{eqn:abf}
\end{equation}
depends on the axis ratio $q$ and the core radius $s$.  For $q=1-\epsilon<1$
the major axis is the $\theta_1$ axis and $s$ is the major axis core
radius.  The deflections produced by this lens are remarkably simple,
\begin{equation}
   \alpha_1 = { b \over \sqrt{1-q^2} } \tan^{-1} 
       \left[ { \theta_1 \sqrt{1-q^2} \over \omega + s } \right]
   \quad\hbox{and}\quad 
   \alpha_2 = { b \over \sqrt{1-q^2} } \hbox{tanh}^{-1} 
       \left[ { \theta_2 \sqrt{1-q^2} \over \omega + q^2 s } \right].
  \labelprint{eqn:abg}
\end{equation}
The effective lens potential is cumbersome but analytic,
\begin{equation}
   \Psi = \vec{\theta}\cdot \vec{\alpha} 
   - b s \ln\left[ (\omega+s)^2 + (1-q^2)\theta_1^2 \right]^{1/2},
  \labelprint{eqn:abh}
\end{equation}
the magnification is simple
\begin{equation}
    \mu^{-1} = 1 - { b \over \omega} -
     { b^2 s \over \omega \left[  (\omega+s)^2 + (1-q^2)\theta_1^2 \right] }
  \labelprint{eqn:abi}
\end{equation}
and becomes even simpler in the limit of a singular isothermal ellipsoid
(SIE) with $s=0$ where $\mu^{-1}\rightarrow 1-b/\omega$.  In this case, 
contours of surface density $\kappa$ are also contours of the magnification,
and the tangential critical line is the $\kappa=1/2$ isodensity contour
just as for the SIS model.  The critical radius scale $b$ can be related
to the circular velocity in the plane of the galaxy relatively easily.  For an isothermal
sphere we have that $b_{SIS} = 4\pi (\sigma_v/c)^2 D_{ds}/D_s$ where the
circular velocity is $v_c=\sqrt{2}\sigma_v$.  For the projection of a 
three-dimensional (3D) oblate ellipsoid of axis ratio $q_3$ and inclination $i$, so that
$q^2 = q_3^2 \cos^2 i + \sin^2 i$, the deflection scale is
$b = b_{SIS} (e_3/\sin^{-1}e_3)$ where $e_3=\sqrt{1-q_3^2}$ is the eccentricity
of 3D mass distribution.  In the limit that $q_3\rightarrow 0$ the 
model becomes a Mestel (\cite{Mestel1963p553})
disk, the infinitely thin disk producing a flat rotation 
curve, and $b=2b_{SIS}/\pi$ (see \S\ref{sec:dynamics} and
Keeton, Kochanek \& Seljak~\cite{Keeton1997p604},
Keeton \& Kochanek~\cite{Keeton1998p157}, Chae~\cite{Chae2003p746}).  
At least for the case of a face-on disk,
at fixed circular velocity you get a smaller Einstein radius as you make
the 3D distribution flatter because a thin disk requires less mass to 
produce the same circular velocity. 

We can generate several other useful models from the isothermal ellipsoids.
For example, steeper ellipsoidal density distributions can be derived
by differentiating with respect to $s^2$.  The most useful of these
is the first derivative with $\kappa \propto \omega^{-3/2}$ which is
related to the Kuzmin (\cite{Kuzmin1956p27}) disk 
(see Kassiola \& Kovner~\cite{Kassiola1993p450}, Keeton \& Kochanek~\cite{Keeton1998p157}).   
It is also easy to generate models with
flat inner rotation curves and truncated halos by taking the difference
of two isothermal ellipsoids.  In particular if $\kappa(s)$ is an
isothermal ellipsoid with core radius $s$, the model 
\begin{equation}
   \kappa = \kappa(s) - \kappa(a)
  \labelprint{eqn:abj}
\end{equation}
with $a > s$ has a central core region with a rising rotation curve for
$\theta \ltorder s$, a flat rotation curve for $s \ltorder \theta \ltorder a$
and a dropping rotation curve for $\theta \gtorder a$.  In the singular limit
($s\rightarrow 0$), 
it becomes the ``pseudo-Jaffe model'' corresponding to a 3D density 
distribution $\rho \propto (r^2+s^2)^{-1} (r^2+a^2)^{-1}$ whose name derives
from the fact that it is very similar the Jaffe model with 
$\rho \propto r^{-2} (r+a)^{-2}$ (Kneib et al.~\cite{Kneib1996p643},
Keeton \& Kochanek~\cite{Keeton1998p157}).  We will discuss other common lens
models in \S\ref{sec:massmono}. 

The last simple analytic models we mention are the generalized singular isothermal
potentials of the form $\Psi = \theta F(\chi)$ with surface density 
$\kappa(\theta,\chi) = (1/2)(F(\chi)+F''(\chi))/\theta$.  Both the SIS and SIE are 
examples of this model.  The generalized isothermal sphere has a number
of useful analytic properties.  For example, the magnification contours 
are isodensity contours
\begin{equation}
    \mu^{-1} = 1 - { 1 \over \theta } \left[ F(\chi) + F''(\chi) \right]
             = 1 - 2\kappa(\theta,\chi) \label{eqn:wevans}
\end{equation}
with the tangential critical line being the contour with $\kappa=1/2$, and
the time delays between images depend only on the distances from the images
to the lens center (see Witt, Mao \& Keeton~\cite{Witt2000p98}, Kochanek, 
Keeton \& McLeod~\cite{Kochanek2001p50}, Wucknitz~\cite{Wucknitz2002p332}, Evans \& Witt~\cite{Evans2003p1351}).

\section{The Mass Distributions of Galaxies \labelprint{sec:mass} }

Contrary to popular belief, the modeling of gravitational lenses to determine the
mass distribution of a lens is not a ``black art.'' It is, however, an area
in which the lensing community has communicated results badly.  There are two
main problems.  First,  many modeling results seem almost deliberately obfuscatory 
as to what models were actually used, what data were fit and what was actually
constrained.    Not only 
do many lens papers insist on taking well known density distributions from the 
dynamical literature and assigning them new names simply because they have been 
projected into two dimensions, but they then assign them a plethora of bizarre 
acronyms.  Sometimes the model used is not actually the one named, for example 
using tidally truncated halos but calling them isothermal models.  Second, there
is a steady confusion between the parameters of models and the aspects of the
mass distribution that have actually been constrained.  Models with 
apparently very different parameters may be in perfect accord as to the 
properties of the mass distribution that are actually relevant to what is observed.
Discussions of non-parametric mass models then confuse the issue further by
conflating differences in parameters with differences in what is actually
constrained to argue for non-parametric models when in fact they also are 
simply matching the same basic properties with lots of extra noise from the
additional and uninteresting degrees of freedom.  In short, the problem with lens
modeling is not that it is a ``black art,'' but that the practitioners try to
make it seem to be a ``black art'' (presumably so that people will believe they
need wizards).  The most important point to take from
this section is that any idiot can model a lens and interpret it properly with
a little thinking about what it is that lenses constrain.

There are two issues to think about in estimating the mass distributions
of gravitational lenses.  The first issue is how to model the mass distribution
with a basic choice between parametric and non-parametric models.  In
\S\ref{sec:massmono} we summarize the most commonly used radial mass 
distributions for lens models.  Ellipsoidal versions of these profiles
combined with an external (tidal) shear are usually used to describe the
angular structure, but there has been recent interest in deviations from
ellipsoidal distributions which we discuss in \S\ref{sec:massquad} and
\S\ref{sec:substruc}.  In \S\ref{sec:nonparam} we summarize the most 
common approaches for non-parametric models of the mass distribution.  Since
this is my review, I will argue that the parametric models are all that
is needed to model lenses and that they provide a better basis for understanding
the results than non-parametric models (but the reader should be warned that
if Prasenjit Saha was writing this you would probably get a different opinion).

The second issue is to determine the aspects of the lens data that actually constrain the mass
distribution.  Among the things that can be measured for a lens are the
relative positions of the components (the astrometric constraints), the
relative fluxes of the images, the time delays between the images, the
dynamical properties of the lens galaxy, and the microlensing of the 
images.  Of these, the most important constraints are the positions.  
We can usually measure the relative
positions of the lensed components very accurately (5~mas or better)
compared to the arc second scales of the component separations.
Obviously the accuracy diminishes when components are faint, and the
usual worst case is having very bright lensed quasars that make it
difficult to detect the lens galaxy.
As we discuss in \S\ref{sec:substruc}, substructure and/or satellites of the lens galaxy
set a lower limit of order 1--5~mas with which it is safe to impose astrometric
constraints independent of the measurement accuracy.
When the source is extended, the resulting arcs and rings discussed in \S\ref{sec:hosts}
provide additional constraints.  These are essentially astrometric in nature,
but are considerably more difficult to use than multiply imaged point sources.
Our general discussion of how lenses constrain the radial (\S\ref{sec:monofit})
and angular structure (\S\ref{sec:massquad}) focus on the use of astrometric
constraints, and in \S\ref{sec:modelfit} we discuss the practical details of 
fitting image positions in some detail.

The flux ratios of the images are one of the most easily measured constraints, but
are cannot be imposed stringently enough to constrain radial density profiles because
of systematic uncertainties.  Flux ratios measured at a single epoch are
affected by time variability in the source (\S\ref{sec:time}), microlensing by the 
stars in the lens galaxy in the optical continuum (see \partmicro), magnification perturbations
from substructure at all wavelengths (see \S\ref{sec:substruc}), absorption by the ISM
of the lens (dust in the optical, free-free in the radio) and scatter broadening in the
radio (see \S\ref{sec:substruc} and \S\ref{sec:optical}).  Most applications of flux 
ratios have focused on using them to probe these perturbing effects rather than for 
studying the mean mass distribution of the lens.  Where radio sources have small
scale VLBI structures, the changes in the relative astrometry of the components
can constrain the components of the relative magnification tensors without needing
to use any flux information (e.g. Garrett et al.~\cite{Garrett1994p457}, Rusin et 
al.~\cite{Rusin2002p205}).

Two types of measurements, time delays (\S\ref{sec:time}) and microlensing by the
stars or other compact objects
in the lens galaxy (\partmicro) constrain the surface density near the lensed 
images.  Microlensing also constrains the fraction of that surface density that 
can be in the form of stars.  To date, time delays have primarily been used to
estimate the Hubble constant rather than the surface density, but if we 
view the Hubble constant as a known quantity, consider only time delay
ratios, or simply want to compare surface densities between lenses, then 
time delays can be used to constrain the mass distribution.  We discuss time 
delays separately because of their close association with attempts to measure the 
Hubble constant.  Using microlensing variability to constrain the mass distribution
is presently more theory than practice due to a lack of microlensing light
curves for almost all lenses.  However, the light curves of the one well monitored 
lens, Q2337+0305, appear to require a surface density composed mainly of stars
as we would expect for a lens where we see the images deep in the bulge of a nearby
spiral galaxy (Kochanek~\cite{Kochanek2004p58}).   We will not discuss this approach further
in \partstrong.

Any independent measurement of the mass of a component will also help to 
constrain the structure of the lenses.  At present this primarily means
making stellar dynamical measurements of the lens galaxy and comparing the
dynamical mass estimates to those from the lens geometry.  We discuss this
in detail in \S\ref{sec:dynamics}.  For lenses associated with
clusters, X-ray, weak lensing or cluster velocity dispersion measurements
can provide estimates of the cluster mass.  While this has been done in a 
few systems (e.g. X-rays, Morgan et al.~\cite{Morgan2001p1}, 
Chartas et al.~\cite{Chartas2002p96}; weak lensing,  Fischer et al.~\cite{Fischer1997p521};
velocity dispersions, Angonin-Willaime, Soucail \& Vanderriest~\cite{Angonin1994p411}),  
the precision of these mass
estimates is not high enough to give strong constraints on lens models.
X-ray observations are probably more important for locating the positions
of groups and clusters relative to the lens than for estimating their masses.

The most useful way of thinking about lensing constraints on mass distributions
is in terms of multipole expansions (e.g. Kochanek~\cite{Kochanek1991p354}, Trotter, Winn \& 
Hewitt~\cite{Trotter2000p671}, Evans \& Witt~\cite{Evans2003p1351}, Kochanek \& 
Dalal~\cite{Kochanek2004}).   An arbitrary surface density $\kappa(\vec{\theta})$
can be decomposed into multipole components,
\begin{equation}
    \kappa(\vec{\theta}) = \kappa_0(\theta) +
   \sum_{m=1}^\infty 
   \left[ \kappa_{cm}(\theta)\cos(m\chi) +\kappa_{sm}(\theta)\sin(m\chi) \right] 
\end{equation}
where  the individual components are angular averages over the surface density
\begin{equation}
   \kappa_0(\theta)={1 \over 2 \pi} \int_0^{2\pi} d\chi \kappa(\vec{\theta}),
   \quad\hbox{and}\quad
   { \kappa_{cm}(\theta) \choose  \kappa_{sm}(\theta) }
    ={1 \over \pi} \int_0^{2\pi} d\chi 
     { \kappa(\vec{\theta}) cos(m\chi) \choose  \kappa(\vec{\theta})  \sin(m\chi) }.
\end{equation}
The first three terms are the monopole ($\kappa_0$), the dipole ($m=1$)
and the quadrupole ($m=2$) of the lens.
The Poisson equation $\nabla^2\Psi=2\kappa$ is separable in polar coordinates, so 
a multipole decomposition of the effective potential
\begin{equation}
    \Psi(\vec{\theta}) = \Psi_0(\theta) +
   \sum_{m=1}^\infty 
   \left[ \Psi_{cm}(\theta)\cos(m\chi) +\Psi_{sm}(\theta)\sin(m\chi) \right] 
\end{equation}
will have terms that depend only on the corresponding multipole of the surface
density, $\nabla^2 \Psi_{cm}(\theta)\cos(m\chi)=2\kappa_{cm}(\theta)\cos(m\chi)$. 
The monopole of the potential is simply
\begin{equation}
    \Psi_0(\theta) = 2 \log(\theta) \int_0^\theta u du \kappa_0(u) +
                 2 \int_\theta^\infty u du \log(u) \kappa(u)
\end{equation}
and its derivative is the bend angle for a circular lens,
\begin{equation}
     \alpha_0(\theta) = { d\Psi_0 \over d \theta} = 
             { 2 \over \theta} \int_0^\theta u du \kappa_0(u),
\end{equation}
just as we derived earlier (Eqn.~\ref{eqn:aaa}).  The higher
order multipoles are no more complicated, with 
\begin{equation}
    { \Psi_{cm}(\theta) \choose \Psi_{sm}(\theta) }
  = -{ 1 \over m\theta^m} \int_0^\theta u^{1+m} du 
   { \kappa_{cm}(u) \choose \kappa_{sm}(u) }-
              { \theta^m \over m } \int_\theta^\infty u^{1-m} du 
   { \kappa_{cm}(u) \choose \kappa_{sm}(u) }.
   \labelprint{eqn:ajk}
\end{equation}
The angular multipoles are always composed of two parts.  There is an interior
pole $\Psi_{cm,int}(\theta)$ due to the multipole surface density interior to
$\theta$ (the integral from $0<u<\theta$) and an exterior pole $\Psi_{cm,ext}(\theta)$
due to the multipole surface density exterior to $\theta$ (the integral from
$\theta<u<\infty$).  
The higher order multipoles produce deflections in both the radial 
\begin{equation}
    \alpha_{cm,rad}=
    {d\hphantom{\theta}\over d\theta}\left[\Psi_{cm}\cos (m\chi) \right]
          = { d \Psi_{cm} \over d\theta } \cos (m\chi)
\end{equation}
and tangential
\begin{equation}
    \alpha_{cm,tan}=
    {1\over \theta } {d\hphantom{\theta}\over d\chi}\left[\Psi_{cm}\cos (m\chi)\right] 
          = - {m \over \theta}  \Psi_{cm} \sin (m\chi)
\end{equation}
directions, where the radial deflection depends on the derivative of $\Psi_{cm}$ and 
the tangential deflection depends only on $\Psi_{cm}$.
This may seem rather formal, but the multipole expansion provides the basis
for understanding which aspects of mass distributions will matter for lens 
models.  Obviously it is the lowest order angular multipoles which are most
important.  The most common angular term added to lens models is the external shear 
\begin{equation}
    \Psi_{2,ext}=
  {1\over 2} \gamma_c \theta^2 \cos 2(\chi-\chi_\gamma)+
  {1\over 2} \gamma_s \theta^2 \sin 2(\chi-\chi_\gamma)
  \labelprint{eqn:acc}
\end{equation}
with dimensionless amplitudes $\gamma_c$ and $\gamma_s$ and axis $\chi_\gamma$.
The external (tidal) shear and any accompanying mean convergence are the lowest
order perturbations from any object near the lens that have measurable effects
on a gravitational lens (see Eqn.~\ref{eqn:aau}).
While models usually consider only external (tidal) shears where these coefficients
are constants, in reality $\gamma_c$, $\gamma_s$ and $\chi_\gamma$ are functions of
radius (i.e. Eqn.~\ref{eqn:ajk}).  Along with the external shear, there is an
internal shear  
\begin{equation}
    \Psi_{2,int}=
  {1\over 2} \Gamma_c {\rbar^4 \over \theta^2 } \cos 2(\chi-\chi_\Gamma)+
  {1\over 2} \Gamma_s {\rbar^4 \over \theta^2}  \sin 2(\chi-\chi_\Gamma).
  \labelprint{eqn:adl}
\end{equation}
due to the quadrupole moment of the mass interior to a given radius.  We introduce
the mean radius of the lensed images $\rbar$ to make $\Gamma_c$ and 
$\Gamma_s$ dimensionless with magnitudes that can be easily compared to the
external shear amplitudes $\gamma_c$ and $\gamma_s$.  Arguably the critical radius of
the lens is a better physical choice, but the mean image radius will be close to
the critical radius and using it avoids any trivial covariances between the
internal shear strength and the monopole mass.  Usually the internal quadrupole
is added as part of an ellipsoidal model for the central lens galaxy,
but it is useful in analytic studies to consider it separately.

\subsection{Common Models for the Monopole \labelprint{sec:massmono}}

Most attention in modeling lenses focuses on the monopole or radial mass distribution
of the lenses.  Unfortunately, much of the lensing literature uses an almost impenetrable
array of ghastly non-standard acronyms to describe the mass models even though many
of them are identical to well-known families of density distributions used in 
stellar dynamics.  Here we summarize the radial mass distributions which are most
commonly used and will keep reappearing in the remainder of \partstrong.

The simplest possible choice for the mass distribution
is to simply trace the light.  The standard model for early-type galaxies or
the bulges of spiral galaxies is the de Vaucouleurs (\cite{de_Vaucouleurs1948p247})
profile with surface density 
\begin{equation}
     \Sigma(R) = I_e \exp\left[ -7.67 \left[ \left(R/R_e\right)^{1/4} -1 \right]\right],
  \labelprint{eqn:abk}
\end{equation}
where the effective radius $R_e$ encompasses half the total mass (or light) of
the profile.  Although the central density of a de Vaucouleurs model is finite,
it actually acts like a rather cuspy density distribution and will generally
fit the early-type lens data with no risk of producing a detectable central
image (e.g. Leh\'ar et al.~\cite{Lehar2000p584}, Keeton~\cite{Keeton2003p17}).  
The simplest model for a disk galaxy is an exponential disk,
\begin{equation}
     \Sigma(R) = I_0 \exp\left[-R/R_d\right]
  \labelprint{eqn:abl}
\end{equation}
where $R_d$ is the disk scale length.  An exponential disk by itself is rarely
a viable lens model because it has so little density contrast between the center
and the typical radii of images that detectable central images are almost always
predicted but not observed.  Some additional component, either a de Vaucouleurs bulge or a 
cuspy dark matter halo, is always required.  This makes spiral galaxy lens models 
difficult because they generically require two stellar components (a bulge and a disk) 
and a dark matter halo, while the photometric data are rarely good enough to
constrain the two stellar components (e.g. Maller, Flores \& Primack~\cite{Maller1997p681},
Koopmans et al.~\cite{Koopmans1998p534},
Maller et al.~\cite{Maller2000p194}, Trott \& Webster~\cite{Trott2002p621},
Winn, Hall \& Schechter~\cite{Winn2003p672}).  Since spiral lenses are already relatively
rare, and spiral lens galaxies with good photometry are rarer still, less attention
has been given to these systems.  The de Vaucouleurs and exponential disk
models are examples of Sersic~(\cite{Sersic1968}) profiles 
\begin{equation}
     \Sigma(R) = I_0 \exp\left[ - b_n \left[ \left(R/R_e(n)\right)^{1/n}\right]\right]
  \labelprint{eqn:abm}
\end{equation}
where the effective radius $R_e(n)$ is defined to encompass half the light and 
$n=4$ is a de Vaucouleurs model and $n=1$ is an exponential disk.  These profiles
have not been used as yet for the study of lenses except for some quasar host
galaxy models (\S\ref{sec:hosts}).
The de Vaucouleurs model can be approximated (or the reverse) by
the Hernquist (\cite{Hernquist1990p359}) model with the 3D density distribution
\begin{equation}
      \rho(r) = { M \over \pi r } { a \over (a+r)^3 }
  \labelprint{eqn:abn}
\end{equation}
and $a\simeq 0.55 R_e$ if matched to a de Vaucouleurs model.
For lensing purposes, the Hernquist model has one major problem.
Its $ \rho \propto 1/r$ central density cusp is shallower than the effective 
cusp of a de Vaucouleurs model, so Hernquist models tend to predict detectable 
central images even when  the
matching de Vaucouleurs model would not.  As a result, the Hernquist model 
is more often used as a surrogate for dynamical normalization of the de Vaucouleurs 
model than as an actual lens model (see below).  

Theoretical models for lenses started with simple, softened power laws of the form
\begin{equation}
     \kappa (R) \propto \left(R^2+s^2 \right)^{-(n-1)/2} \rightarrow R^{1-n}
  \labelprint{eqn:abo}
\end{equation}
in the limit where there is no core radius.  We are using these simple power law
lenses in all our examples (see \S\ref{sec:basics}).  These models include many well known
stellar dynamical models such as the singular isothermal sphere (SIS, $n=2$,
$s=0$), the modified Hubble profile ($n=3$) and the Plummer model ($n=5$).
Since we only see the projected mass, these power laws are also related 
to common models for infinitely thin disks.  The Mestel (\cite{Mestel1963p553})
disk ($n=2$, $s=0$) is the disk that produces a flat rotation curve, and the 
Kuzmin (\cite{Kuzmin1956p27}) disk ($n=3$) can be used to mimic the rising and 
then falling rotation curve of an exponential disk.  
The softened power-law models have generally fallen out of favor other than as
simple models for some of the visible components of lenses because the
strong evidence for stellar and dark matter cusps makes models with core
radii physically unrealistic.  While ellipsoidal versions of these models
are not available in useful form, there are fast series expansion methods
for numerical models (Chae, Khersonsky \& Turnshek~\cite{Chae1998p80},
Barkana~\cite{Barkana1998p531}).  

Most ``modern'' discussions of galaxy density distributions are based on
sub-cases of the density distribution
\begin{equation}
     \rho(r) \propto { 1 \over r^n } 
              { 1 \over \left( a^\alpha + r^\alpha \right)^{(m-n)/\alpha} },
  \labelprint{eqn:abp}
\end{equation}
which has a central density cusp with $\rho \propto r^{-n}$, asymptotically
declines as $\rho \propto r^{-m}$ and has a break in the profile near $r\simeq a$
whose shape depends on $\alpha$ (e.g. Zhao~\cite{Zhao1997p525}).  The most common cases are the
Hernquist model ($n=1$, $m=4$, $\alpha=1$) mentioned above, the
Jaffe (\cite{Jaffe1983p995}) model ($n=2$, $m=4$, $\alpha=1$), the NFW (Navarro, Frenk
\& White~\cite{Navarro1996p563}) model
($n=1$, $m=3$, $\alpha=1$) and the Moore (\cite{Moore1998p5}) model 
($n=3/2$, $m=3$, $\alpha=1$).  We can view the power-law models
either as the limit $n \rightarrow0$ and $\alpha=2$, or we could 
generalize the $r^{-n}$ term to $(r^2+s^2)^{-n/2}$ and consider
only regions with $r$ and $s \ll a$.   Projections of these models are
similar to surface density distributions of the form 
\begin{equation}
     \kappa(R) \propto { 1 \over R^{n-1} } 
              { 1 \over \left( a^\alpha + R^\alpha \right)^{(m-n)/\alpha} }
  \labelprint{eqn:abq}
\end{equation}
(although the definition of the break radius $a$ changes) with the
exception of the limit $n \rightarrow 1$ where the projection of a 
3D density cusp $\rho \propto 1/r$ produces surface density terms 
$\kappa \propto \ln R$ that cannot be reproduced by the broken surface
density power law.  This surface density model is sometimes called the Nuker law 
(e.g. Byun et al.~\cite{Byun1996p1889}).
A particularly useful case for lensing is the pseudo-Jaffe model with
$n=2$, $m=4$ and $\alpha=2$ (where the normal Jaffe model has
$\alpha=1$) as the only example of a broken power law with simple analytic
deflections even when ellipsoidal because the density distribution is
the difference between two isothermal ellipsoids (see Eqn.~\ref{eqn:abj}).   
These cuspy models
also allow fast approximate solutions for their ellipsoidal counterparts
(see Chae 2002).

The most theoretically important of these cusped profiles is the NFW
profile (Navarro et al.~1996) because it is the standard model for dark matter halos.  Since it
is such a common model, it is worth discussing it in a little more detail,
particularly its peculiar normalization.  The NFW profile is normalized by
the mass $M_{vir}$ inside the virial radius $r_{vir}$, with
\begin{equation}
   \rho_{NFW}(r) = { M_{vir} \over 4 \pi f(c) } { 1 \over r (r+a)^2 }
   \quad\hbox{and}\quad
   M_{NFW}(<r) = { f(r/r_{vir}) \over f(c) }
  \labelprint{eqn:abr}
\end{equation}
where $f(c)=\ln(1+c)-c/(1+c)$ and the concentration $c=r_{vir}/a\sim 5$ for
clusters and $c\sim 10$ for galaxies.  The concentration is a function
of mass whose scaling is determined from N-body simulations.  A typical
scaling for a halo at redshift $z$ in an $\Omega_M=0.3$ flat cosmological 
models is (Bullock et al. 2001)
\begin{equation}
    c(M) = { 9 \over 1+z } \left( { M_{vir} \over 8 \times 10^{12} h M_\odot }\right)^{-0.14}
  \labelprint{eqn:abs}
\end{equation}
with a dispersion in $\log c$ of $\sigma_{\log(c)} \simeq 0.18$~dex.  Because
gravitational lensing is very sensitive to the central density of the lens,
including the scatter in the concentration is quantitatively important for 
lensing by NFW halos (Keeton~\cite{Keeton2001p46}).  
The
virial mass and radius are related and determined by the overdensity $\Delta_{vir}(z)$ required
for a halo to collapse given the cosmological model and the redshift.  This can
be approximated by
\begin{equation}
   M_{vir} = { 4 \pi \over 3 } \Delta_{vir}(z) \rho_u(z) r_{vir}^3 
     \simeq 0.23 \times 10^{12} h 
          \left( { (1+z) r_{vir} \over 100 h^{-1}\hbox{kpc} } \right)^3
          \left( { \Omega_M \Delta_{vir} \over 200 } \right) M_\odot
  \labelprint{eqn:abt}
\end{equation}
where $\rho_u(z)=3H_0^2 \Omega_M (1+z)^3/8\pi G$ is the mean matter density when the
halo forms and $\Delta_{vir}\simeq (18\pi^2 +82 x-39 x^2)/\Omega(z)$ with 
$x=\Omega-1$ is the overdensity needed for a halo to collapse.  There
are differences in normalizations between authors and with changes in the 
central cusp exponent $\gamma$, but models of this type are what we presently
expect for the structure of dark matter halos around galaxies.  

For most lenses, HST imaging allows us to measure the spatial
distribution of the stars, thereby providing us with a model for the 
distribution of stellar mass with only the stellar mass-to-light ratio
as a parameter.  For present purposes, gradients in the stellar mass-to-light
ratio are unimportant compared to the uncertainties arising from the dark
matter.  Unless we are prepared to abandon the entire paradigm for modern
cosmology, the luminous galaxy is embedded in a dark matter halo and we
must decide how to model the overall mass distribution.  The most common
approach, as suggested by the rich variety of mass profiles we introduced
in \S\ref{sec:massmono}, is to assume a parametric form for the total 
mass distribution rather than attempting to decompose it into luminous
and dark components.  The alternative is to try to embed the stellar component 
in a dark matter halo.  Operationally, doing so is trivial -- the lens
is simply modeled as the sum of two mass components.  However, there
are theoretical models for how CDM halos should be combined with 
the stellar component.

Most non-gravitational lensing applications focus
on embedding disk galaxies in halos because angular momentum conservation
provides a means of estimating a baryonic scale length (e.g. Mo, Mao
\& White~\cite{Mo1998p319}).  The spin parameter of the halo sets the
angular momentum of the baryons, and the final disk galaxy is defined
by the exponential disk with the same angular momentum.  As the baryons
become more centrally concentrated, they pull the dark matter inwards
as well through a process known as adiabatic contraction
(Blumenthal et al.~\cite{Blumenthal1986p27}).   The advantage of
this approach, which in lensing has been used only by Kochanek
\& White~(\cite{Kochanek2001p531}), is that it allows a full 
{\it ab initio} calculation of lens statistical properties when
combined with a model for the cooling of the baryons (see \S\ref{sec:cluster}).
It has the major disadvantage that most lens galaxies are 
early-type galaxies rather than spirals, and that there is no analog
of the spin parameter and angular momentum conservation to set the
scale length of the stellar component in a model for an early-type
galaxy.  

Models of early-type galaxies embedded in CDM halos have to start with
an empirical estimate of the stellar effective radius.  In models of
individual lenses this is a measured property of the lens galaxy
(e.g. Rusin et al.~\cite{Rusin2003p29}, \cite{Rusin2004p1},
Koopmans \& Treu~\cite{Koopmans2002p5}, Kochanek~\cite{Kochanek2003p49}).  
Statistical models must
use a model for the scaling of the effective radius with luminosity
or other observable parameters of early-type galaxies
(e.g. Keeton~\cite{Keeton2001p46}).  From the luminosity, a
mass-to-light ratio is used to estimate the stellar mass.  If
all baryons have cooled and been turned into stars, then the 
stellar mass provides the total baryonic mass of the halo, 
otherwise the stellar mass sets a lower bound on the baryonic
mass.  Combining the baryonic mass with an estimate of the
baryonic mass fraction yields the total halo mass to be fed
into the model for the CDM halo.

In general, there is no convincing evidence favoring either approach --
for the regions over which the mass distributions are constrained by
the data, both approaches will agree on the overall mass distribution.
However, there can be broad degeneracies in how the total mass distribution
is decomposed into luminous and dark components (see \S\ref{sec:modelfit}).

\begin{figure}[t]
\begin{center}
\vspace{0.5in}
\centerline{\psfig{figure=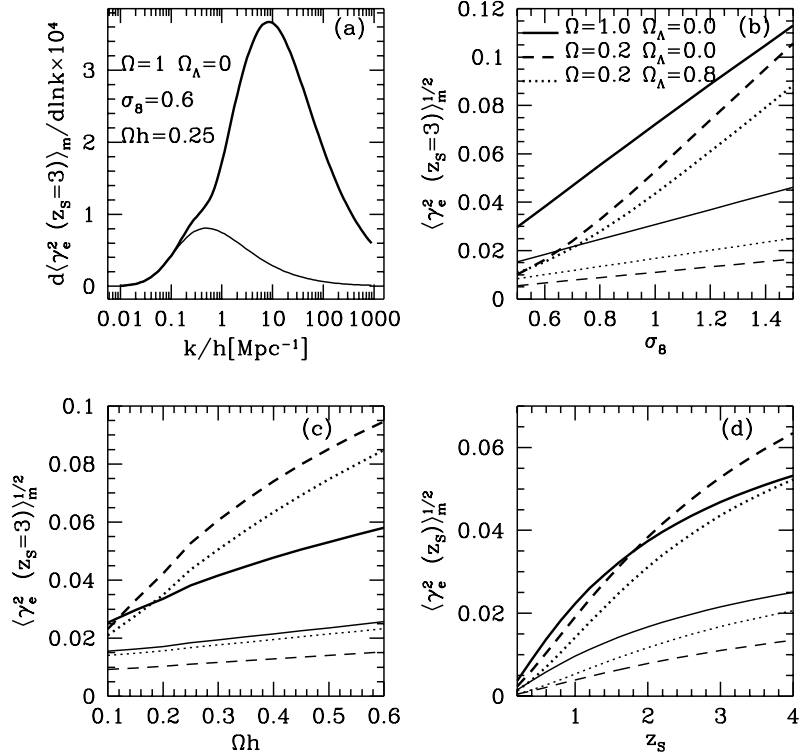,width=3.5in}}
\end{center}
\caption{
   Dependence of the shear generated by other objects along the line of sight 
   for both linear (light lines) and non-linear (heavy lines) power spectra.
   (a) Shows the logarithmic contribution to the rms effective shear for a 
   source at redshift $z_s=3$ as a function of wave vector $k$.  (b) Shows
   the dependence on $\sigma_8$ for a fixed power spectrum shape 
   $\Omega_M h=0.25$.  (c) Shows the dependence on the shape $\Omega_M h$
   with $\sigma_8=0.6$ for $\Omega_M=1$ and $\sigma_8=1.0$ for $\Omega_M <1$.
   (d) Shows the variation in the shear with source redshift for the models
   in (c) with $\Omega_M h=0.25$.
   }
\labelprint{fig:los}
\end{figure}

\subsection{The Effective Single Screen Lens \labelprint{sec:singlescreen}}

Throughout these notes we will treat lenses as if all the lens components lay at a
single redshift (``the single screen approximation'').  The lens equations
for handling multiple deflection screens (e.g.  Blandford \& 
Narayan~\cite{Blandford1986p568}, Kovner~\cite{Kovner1987p52}, Barkana~\cite{Barkana1996p17})
 are known but little used except
for numerical studies (e.g. Kochanek \& Apostolakis~\cite{Kochanek1988p1073},
Moller \& Blain~\cite{Moller2001p339})  in large part because few lenses
require multiple lens galaxies at different redshifts with the 
exception of B2114+022 (Chae, Mao \& Augusto~\cite{Chae2001p1015}).  
In fact, we are not being as cavalier in making this approximation as 
it may seem.

The vast majority of strong lenses consist of a single lens galaxy 
perturbed by other objects.  We can divide these objects into those near
the primary lens, where a single screen is clearly appropriate, and those distributed
along the line of site for which a single screen may be inappropriate.
Because the correlation
function is so strong on small scales, the perturbations are dominated
by objects within a correlation length of the lens galaxy (e.g. Keeton, 
Kochanek \& Seljak~\cite{Keeton1997p604}, Holder \& Schechter~\cite{Holder2003p688}).  
The key to the relative safety of the single screen model is that weak
perturbations from objects along the line of site, in the sense that 
in a multi-screen lens model they could be treated as a convergence and
a shear, can be reduced to a single ``effective''
lens plane in which the true amplitudes of the convergence and shear are
rescaled by distance ratios to convert them from their true redshifts
to the redshift of the single screen (Kovner~\cite{Kovner1987p52}, 
Barkana~\cite{Barkana1996p17}).  The lens equation on the effective
single screen takes the form 
\begin{equation}
   \vec{\beta} = \left(I + F_{OS}\right) \vec{\theta} - 
   \left(I+F_{LS}\right)\vec{\alpha}\left[\left(I+F_{OL}\right)\vec{\theta}\right]
     \labelprint{eqn:efflens1}
\end{equation}
where $F_{OS}$, $F_{LS}$ and $F_{OL}$ describe the shear and convergence
due to perturbations between the observer and the source, the lens and
the source and the observer and the lens respectively.  For statistical
calculations this can be simplified still further by making the coordinate
transformation $\vec{\theta}'=(I+F_{OL})\vec{\theta}$ and 
$\vec{\beta}'=(I+F_{LS})\vec{\beta})$ to leave a lens equation, 
\begin{equation}
   \vec{\beta}' = \left(I + F_e\right) \vec{\theta}' - 
   \vec{\alpha}\left[\vec{\theta}'\right],
     \labelprint{eqn:efflens2}
\end{equation}
identical
to a single screen lens 
in an effective convergence and shear of $F_e=F_{OL}+F_{LS}-F_{OS}$ (to linear
order).  In practice it will usually be safe to neglect the differences between
Eqns.~\ref{eqn:efflens1} and \ref{eqn:efflens2} because the shearing terms 
affecting the deflections in Eqn~\ref{eqn:efflens1} are easily mimicked by 
modest changes in the ellipticity and orientation of the primary lens.  
The rms amplitudes of these perturbations depend on the cosmological model
and the amplitude of the non-linear power spectrum, but the general scaling
is that the perturbations grow as $D_s^{3/2}$ with source redshift, and 
increase for larger $\sigma_8$ and $\Omega_M$ as shown in Fig.~\ref{fig:los}
from Keeton et al.~(\cite{Keeton1997p604}).   The importance of these 
effects is very similar to concerns about the effects of lenses along the
line of sight on the brightness of high redshift supernova being used to
estimate the cosmological model (e.g. Dalal et al.~\cite{Dalal2003p11}).

\subsection{Constraining the Monopole \labelprint{sec:monofit} }

The most frustrating aspect of lens modeling is that it is very difficult to
constrain the monopole.  If we take a simple lens and fit it with any
of the parametric models from \S\ref{sec:massmono}, it will be possible
to obtain a good fit provided the central surface density of the model is high
enough to avoid the formation of a central image.
As usual, it is simplest to begin understanding the problem with a circular,
two-image lens whose images lie
at radii $\theta_A$ and $\theta_B$ from the lens center (Fig.~\ref{fig:geometry}).
The lens equation (\ref{eqn:aab}) constrains the deflections so that 
the two images correspond to the same source position, 
\begin{equation}
   \beta = \theta_A-\alpha(\theta_A)=-\theta_B+\alpha(\theta_B),
  \labelprint{eqn:abu}
\end{equation}
where the sign changes appear because the images are on opposite 
sides of the lens.  Recall that for the power-law lens model,
$\alpha(\theta) = b^{n-1} \theta^{2-n}$ (Eqn.~\ref{eqn:aaf}), so we can easily solve
the constraint equation to determine the Einstein radius of the
lens,
\begin{equation}
    b = \left[ { \theta_A + \theta_B \over \theta_A^{2-n}+\theta_B^{2-n} } 
    \right]^{1/(n-1)}
  \labelprint{eqn:abv}
\end{equation}
in terms of the image positions.  In the limit of an SIS ($n=2$) the
Einstein radius is the arithmetic mean, $b=(\theta_A+\theta_B)/2$,
and in the limit of a point source ($n\rightarrow 3$), it is the 
geometric mean, $b=(\theta_A\theta_B)^{1/2}$, of the image radii.
More generally, for any deflection profile $\alpha(\theta)=b f(\theta)$, the two
images simply determine the mass scale 
$b=(\theta_A+\theta_B)/(f(\theta_A)+f(\theta_B))$.

There are two important lessons here.
First, the location of the tangential critical line is determined
   fairly accurately independent of the mass profile.  We may only
   be able to determine the mass scale, but it is the most accurate
   measurement of galaxy masses available to astronomy.  The dependence
   of the mass inside the Einstein radius on the shape of the
   deflection profile is weak, with fractional differences between 
   profiles being of order $(\dr/\rbar)^2/8$ where $\dr=\theta_A-\theta_B$
  and $\rbar=(\theta_A+\theta_B)/2$ (i.e. if the images
   have similar radii, the difference beween the arthmetic
   and geometric mean is small).
Second, it is going to be very difficult to determine radial mass 
   distributions.  In this example there is a perfect degeneracy
   between the exact location of the tangential critical line $b$ and 
   the exponent $n$.  In theory, this is broken by the flux ratio
   of the images. However, a simple two-image lens has too few
   constraints even with perfectly measured flux ratios because a realistic
   lens model must also include some freedom in the angular structure of the lens.
   For a simple four-image lens, there begin to be enough constraints but the images
   all have similar radii, making the flux ratios relatively insensitive to changes 
   in the monopole.  Combined with the systematic uncertainties in flux ratios, they
   are not useful for this purpose.

\begin{figure}[t]
\begin{center}
\centerline{\psfig{figure=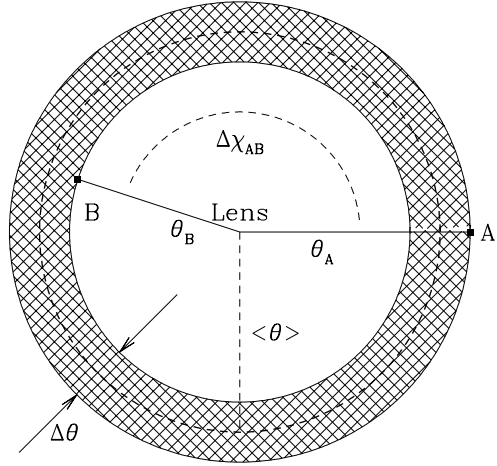,width=3.0in}}
\end{center}
\caption{
   A schematic diagram of a two-image lens.  The lens galaxy lies at the origin
   with two images A and B at radii $\theta_A$ and $\theta_B$ from the lens center.
   The images define an annulus of average radius $\rbar=(\theta_A+\theta_B)/2$ and
   width $\dr=\theta_A-\theta_B$, and they subtend an angle $\Delta\chi_{AB} $
   relative to the lens center.  For a circular lens $\Delta\chi_{AB}=180^\circ$
   by symmetry.
   }
\labelprint{fig:geometry}
\end{figure}

\begin{figure}[p]
\begin{center}
\centerline{\psfig{figure=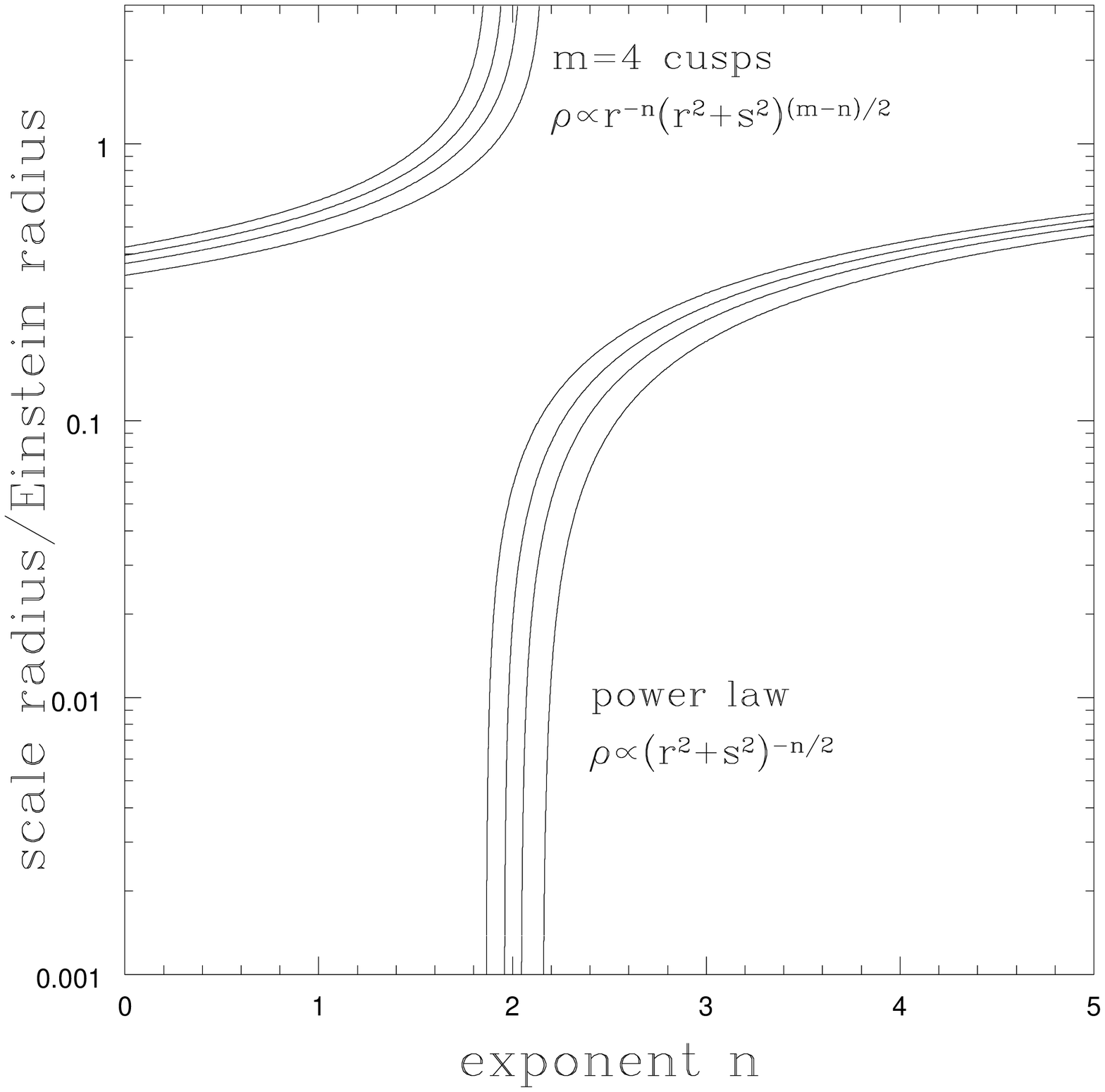,width=3.3in}}
\end{center}
\begin{center}
\centerline{\psfig{figure=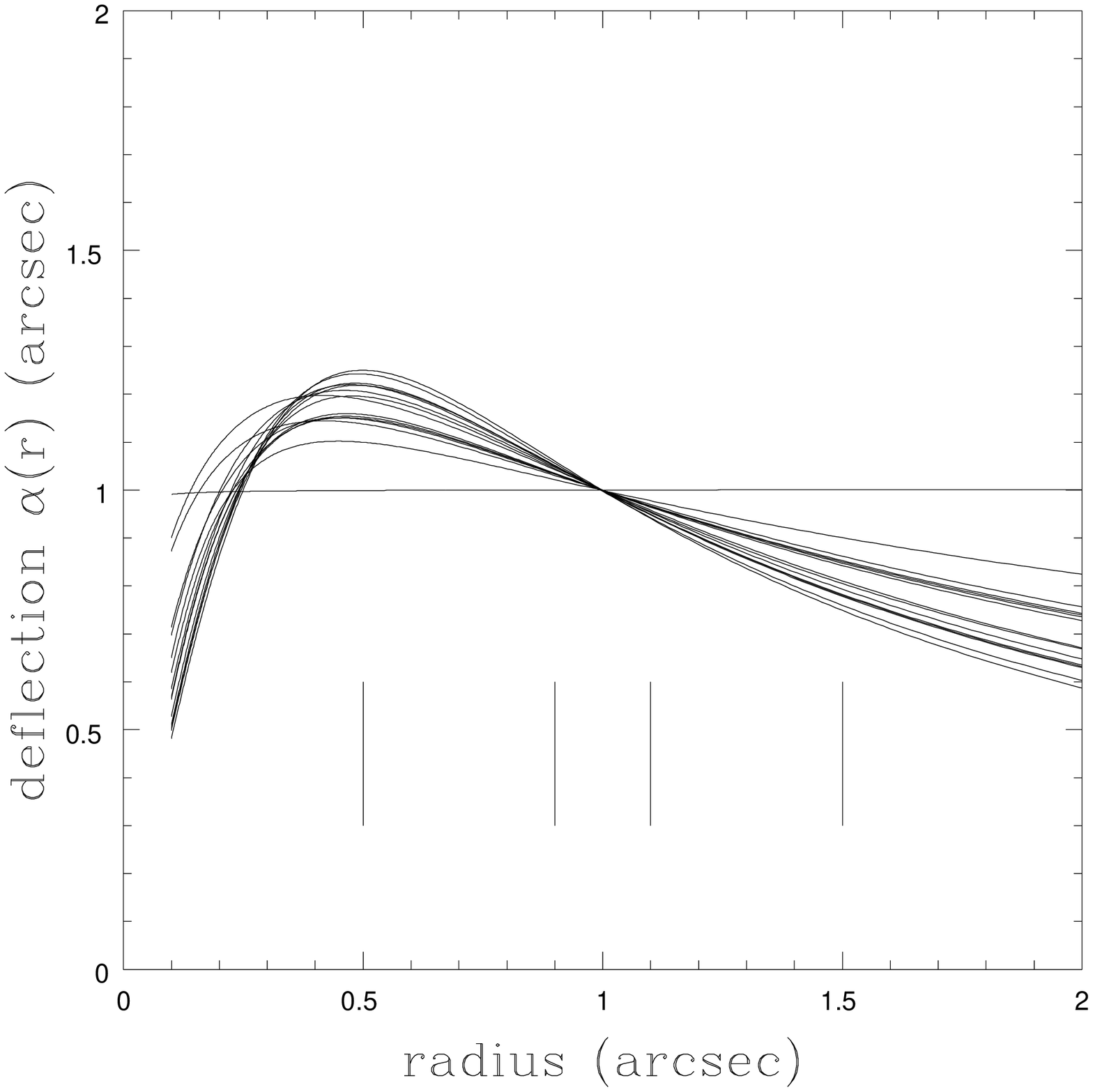,width=3.3in}}
\end{center}
\caption{
   Softened power law and cusped model fits to the images produced by an SIS
   lens with Einstein radius $b=1\farcs0$ and two source components located
   $0\farcs1$ and $0\farcs5$ from the lens center.  In the top panel, the contours 
   show the regions with astrometric fit residuals per image of $0\farcs003$
   and $0\farcs010$.  Models with $m=3$ cusps so closely overly the $m=4$
   models that their error contours were not plotted.  The bottom panel 
   shows the deflection profiles of the best models at half-integer increments
   in the exponent $n$.  The SIS model has a constant deflection, and the
   power-law and cusp models approach it in a sequence of slowly falling
   deflection profiles.  All models agree with the SIS Einstein radius at
   $r=1\farcs0$.  The positions of the images are indicated by the vertical
   bars.
   }
\labelprint{fig:cusp}
\end{figure}

This example also leads to the major misapprehension about lens 
models and radial mass distributions, in that the constraints 
appear to lead to a degeneracy related to the global structure of
the potential (i.e. the exponent $n$).  This is not correct.  The degeneracy
is a purely local one that depends only on the structure of the
lens in the annulus defined by the images, $\theta_B < \theta < \theta_A$,
as shown in Fig.~\ref{fig:geometry}.
To see this we will rewrite the expression 
for the bend angle (Eqn.~\ref{eqn:aaa}) as
\begin{equation}
   \alpha(\theta)= { 2 \over \theta } 
      \left[ \int_0^{\theta_B} u du \kappa(u)
           + \int_{\theta_B}^\theta  u du \kappa(u)
      \right]
    = { 1 \over \theta }\left[ b_B^2 + (\theta^2-\theta_B^2)\kbar(\theta,\theta_B)
     \right]
  \labelprint{eqn:abw}
\end{equation}
where $b_B^2=2\int_0^{\theta_B} u du \kappa(u)$ is the
Einstein radius of the total mass interior to image $B$, and 
\begin{equation}
  \kbar(\theta,\theta_B)={ 2 \over \theta^2-\theta_B^2}
    \int_{\theta_B}^\theta  u du \kappa(u)
  \labelprint{eqn:abx}
\end{equation}
is the mean surface density in the annulus $\theta_B < u < \theta$.
If we now solve the constraint Eqn.~\ref{eqn:abu} again, we find that
\begin{equation}
     b_B^2 = \theta_A\theta_B - \kbar_{AB} \theta_B (\theta_A-\theta_B)
  \labelprint{eqn:aby}
\end{equation}
where $\kbar_{AB}=\kbar(\theta_A,\theta_B)$ is the mean density in the 
annulus $\theta_B < \theta < \theta_A$ between the images.  
Thus, there is a degeneracy between the total mass interior to image B
and the mean surface density (mass) between the two
images.  There is no dependence on the distribution of the mass interior 
to $\theta_B$, the distribution of mass between the two images, or on
either the amount or distribution of mass exterior to $\theta_A$.
This is Gauss' law for gravitational lens models.  

If we normalize the mass scale at any point in the interior of the annulus
then the result will appear to depend on the distribution of the mass 
simply because the mass must be artificially
divided.  For example, suppose we model the surface density 
{\it locally} as a power law $\kappa \propto \theta^{1-n}$ with a mean
surface density $\kbar$ in the annulus $\theta_B<\theta<\theta_A$
between the images.  The mass inside the mean image radius $\rbar$ is
\begin{eqnarray}
   b_{\rbar}^2 &= &\theta_A\theta_B \left( 1-\kappa_0 \right) +\\
    && \dr^2 \kbar \left[ { n \over 4 } +
    \left( { \dr \over \rbar } \right)^2 {(4-n)(2-n)(1-n) \over 192 }+
     O\left(\left( { \dr \over \rbar } \right)^4 \right)  \right] \nonumber
\end{eqnarray}
where we have expanded the result in the ratio $\dr/\rbar$ (in fact, the
result as shown is exact for $n=2/3$, $1$, $2$, $4$ and $5$).  
We included in
this result an additional, global convergence $\kappa_0$ so that we
can contrast the local degeneracies due to the distribution of matter
between the images with the global degeneracies produced by a infinite
mass sheet.  The leading term $\theta_A\theta_B$ is the Einstein radius
expected for a point mass lens (Eqn.~\ref{eqn:abu}).  While the total enclosed 
mass ($\theta_A\theta_B$) is fixed, the mass associated with the lens
galaxy $b_{\rbar}^2$ must be modified in the presence of a global
convergence by the usual $1-\kappa_0$ factor created by the mass
sheet degeneracy (Falco, Gorenstein \& Shapiro~\cite{Falco1985p1}).  
The structure of the lens in the annulus leads
to fractional corrections to the mass of order $(\dr/\rbar)^2$ that
are proportional to $ n \kbar$ to lowest order.  

Only if you have additional images inside the annulus can you begin
to constrain the structure of the density in the annulus. The constraint
is not, unfortunately, a simple constraint on the density.   Suppose 
that we see an additional (pair) of images on the Einstein ring at
$\theta_0$, with $\theta_B<\theta_0<\theta_A$
This case is simpler than the general case because it divides our
annulus into two sub-annuli (from $\theta_B$ to $\theta_0$ and from
$\theta_0$ to $\theta_A$) rather than three.  Since we put the extra
image on the Einstein ring, we know that the mean surface density
interior to $\theta_0$ is unity (Eqn.~\ref{eqn:aah}).  The A and B images then constrain
a ratio
\begin{equation}
     { 1-\kbar_{B0} \over 1-\kbar_{A0} } = { \theta_B \over \theta_A }
         {\theta_A^2-\theta_0^2 \over \theta_0^2-\theta_B^2} 
      \simeq { \theta_A-\theta_0 \over \theta_0-\theta_B } \left[
                1 - {\theta_A-\theta_B \over 2\theta_0 } \cdots \right]
     \labelprint{eqn:ajl}
\end{equation}
of the average surface densities between the Einstein ring and image B
($\kbar_{B0}$) and the Einstein ring and image A ($\kbar_{A0}$). Since
a physical distribution must have $0 < \kbar_{A0} < \kbar_{B0}$, the
surface density in the inner sub-annulus must satisfy
\begin{equation}
     { \theta_A +\theta_B \over \theta_A } { \theta_0^2-\theta_A\theta_B
          \over \theta_0^2-\theta_B^2} < \kbar_{B0} < 1
\end{equation}
where the lower (upper) bound is found when the density in the outer sub-annulus
is zero (when $\kbar_{B0}=\kbar_{A0}$).  
The term $\theta_0^2-\theta_A\theta_B$ is the difference between
the measured critical radius $\theta_0$ and the critical radius implied
by the other two images for a lens with no density in the annulus 
(e.g. a point mass), $(\theta_A\theta_B)^{1/2}$.  Suppose we actually
have images formed by an SIS, so $\theta_A=\theta_0(1+x)$ and 
$\theta_B=\theta_0(1-x)$ with $0 < x=\beta/\theta_0 < 1$, then the
lower bound on the density in the inner sub-annulus is
\begin{equation}
     \kbar_{B0} > { 2 x \over (2-x)(1+x) } 
\end{equation}
and the fractional uncertainly in the surface density is unity for
images near the Einstein ring ($x\rightarrow 0$) and then steadily
diminishes as the A and B images are more asymmetric.  If you want
to constrain the monopole, the more asymmetric the configuration the better.
This rule becomes still more important with the introduction of angular structure. 
  

Fig.~\ref{fig:cusp} illustrates these issues. 
We arbitrarily picked a model consisting of an SIS lens with two sources.
One source is close to the origin and produces images at $\theta_A=1\farcs1$
and $\theta_B=0\farcs9$.  The other source is farther from the origin with
images at $\theta_A=1\farcs5$ and $\theta_B=0\farcs5$.  We then modeled
the lens with either a softened power law (Eqn.~\ref{eqn:abo}) or a
three-dimensional cusp (Eqn.~\ref{eqn:abp}).  We did not worry about 
the formation of additional images when the core radius becomes too large
or the central cusp is too shallow -- this would rule out models with 
very large core radii or shallow central cusps.  If there were only a single source,
either of these models can fit the data for any values of the parameters.  Once,
however, there are two sources,  most of parameter space is ruled out
except for degenerate tracks that look very different for the two mass
models.  Along these tracks, the models satisfy the additional constraint
on the surface density given by Eqn.~\ref{eqn:ajl}.  
The first point to make about Fig.~\ref{fig:cusp} is the importance of 
carefully defining parameters.  The input SIS model has very different
parameters for the two mass models -- while the exponent $n=2$ is the 
same in both cases, the SIS model is the limit $s\rightarrow0$ for 
the core radius in the softened power law, but it is the limit
$a\rightarrow\infty$ for the break radius in the cusp model.
Similarly, models with an inner cusp $n=0$ will closely resemble 
power law models whose exponent $n$ matches the outer exponent $m$
of the cuspy models.  Our frequent failure to explain these similarities is one
reason why lens modeling seems so confusing. 
The second point to make about Fig.~\ref{fig:cusp} is 
that the deflection profiles implied by these models are fairly similar
over the annulus bounded by the images.  Outside the annulus, particularly
at smaller radii, they start to show very large fractional differences.  
Only if we were to add a third set of multiple images or measure a time
delay with a known value of $H_0$ would the 
parameter degeneracy begin to be broken.

These general results show that studies of how lenses constrain the
monopole need the ability to simultaneously vary the mass scale, the
surface density of the annulus and possibly the slope of the density profile in 
the annulus to have the full range of freedom permitted by 
the data.  Most parametric studies constraining the monopole have had two parameters,
adjusting the mass scale and a correlated combination of the surface
density and slope (e.g. Kochanek~\cite{Kochanek1995p559},
Impey et al.~\cite{Impey1998p551}, Chae, Turnshek \& Khersonsky~\cite{Chae1998p609},
Barkana et al.~\cite{Barkana1999p54},
Chae~\cite{Chae1999p582}, Cohn et al.~\cite{Cohn2001p1216}, 
Mu\~noz et al.~\cite{Munoz2001p657}, Wucknitz et al.~\cite{Wucknitz2004p14}), 
although there are exceptions using models with additional degrees of freedom
(e.g. Bernstein \& Fischer~\cite{Bernstein1999p14},
Keeton et al.~\cite{Keeton2000p74},
Trott \& Webster~\cite{Trott2002p621}, Winn,
Rusin \& Kochanek~\cite{Winn2003p80}).  This
limitation is probably not a major handicap, because realistic density
profiles show a rather limited range of local logarithmic slopes.

\begin{figure}[ph]
\begin{center}
\centerline{\psfig{figure=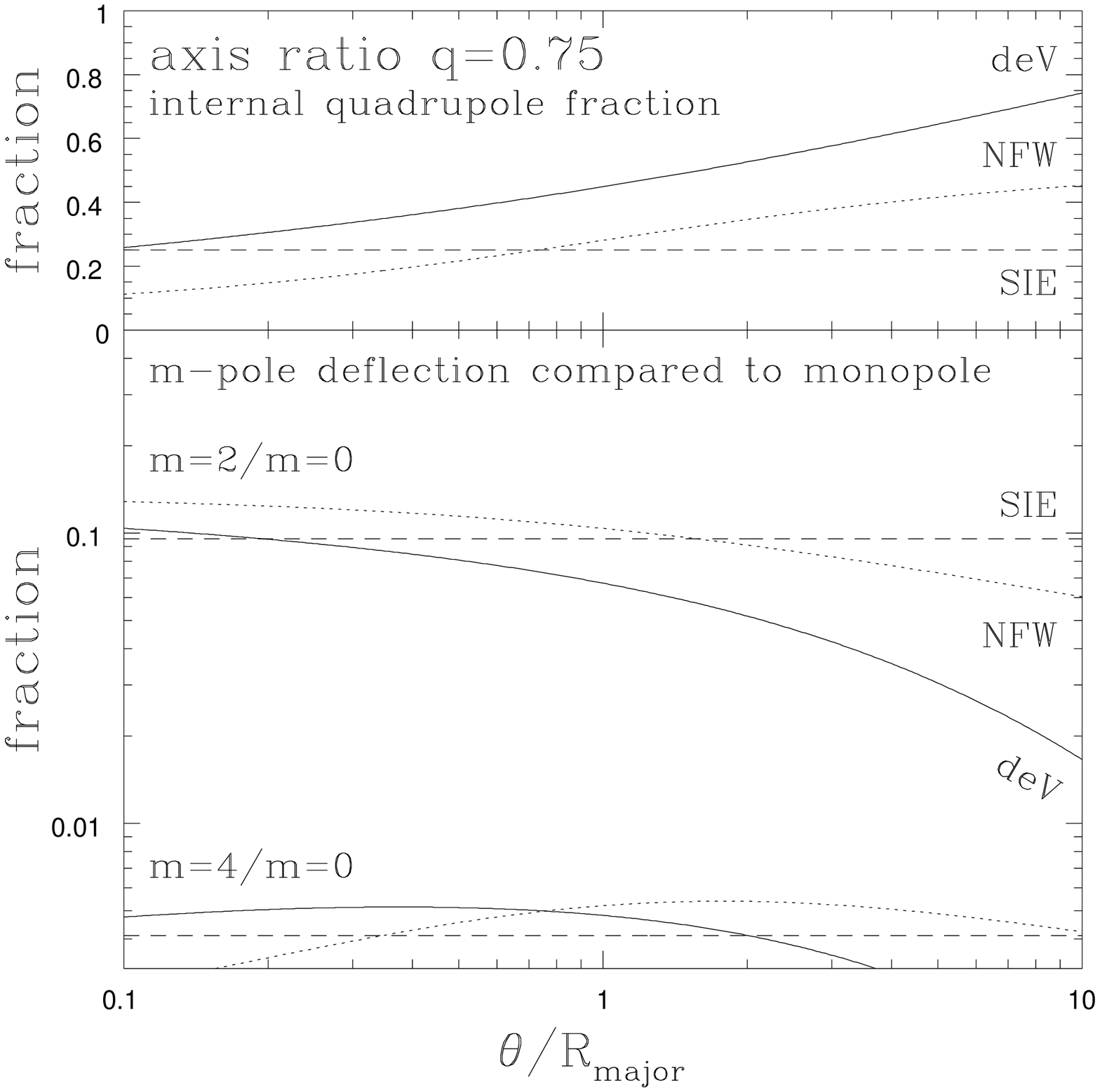,width=3.3in}}
\end{center}
\begin{center}
\centerline{\psfig{figure=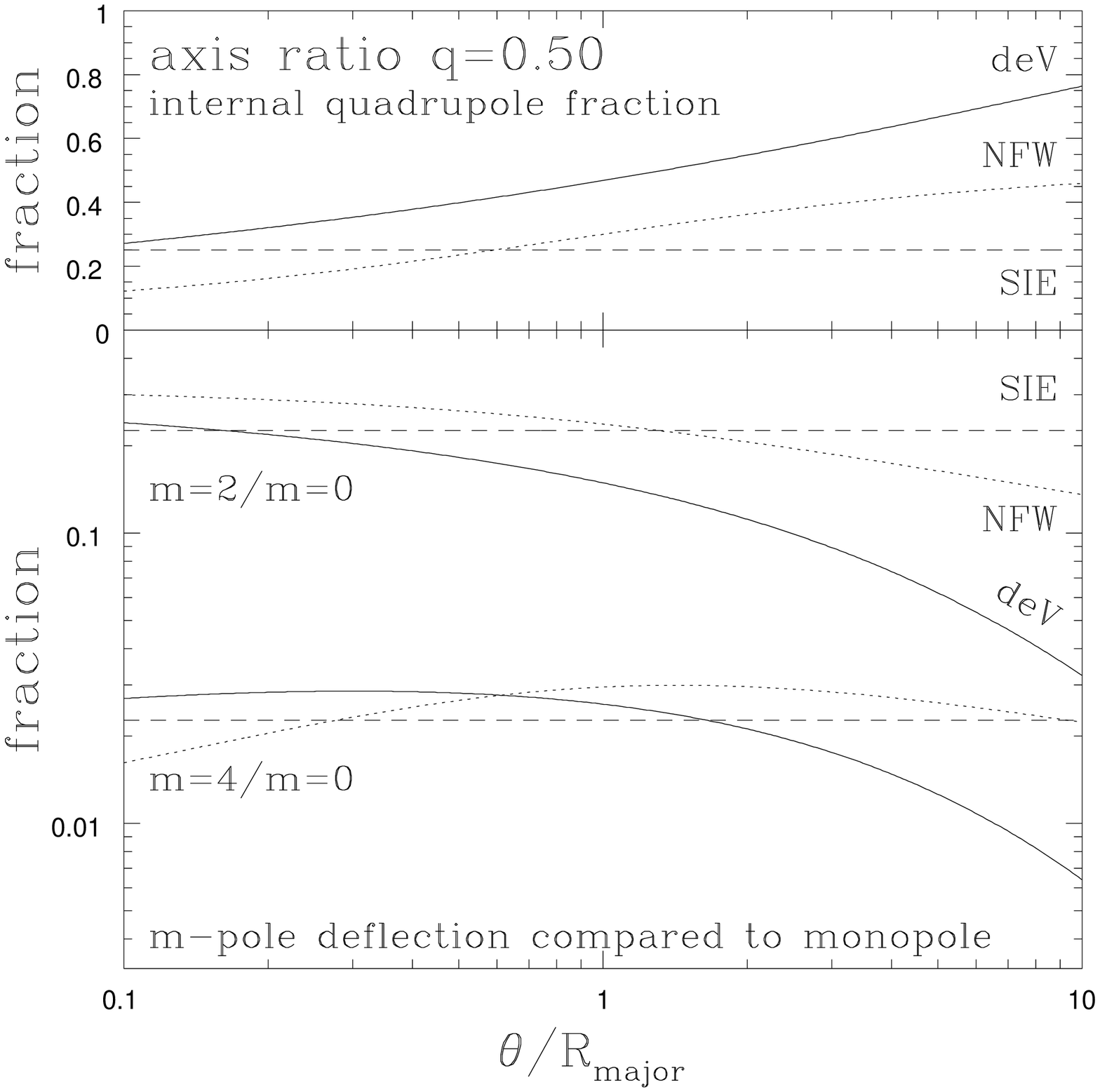,width=3.3in}}
\end{center}
\caption{
   Behavior of the angular multipoles for the de Vaucouleurs (solid), SIE (dashed) and
   NFW (dotted) models with axis ratios of either $q=0.75$ (Top) or $q=0.5$ (Bottom)
   as a function of radius from the lens center in units of the lens major axis scale
   $R_{major}$.  For each axis ratio, 
   the lower panel shows the ratio of the maximum angular deflections produced by the 
   quadrupole ($m=2$) and the $m=4$ pole relative to the deflection produced by the
   monopole ($m=0$).  The upper panel shows the fraction of the quadrupole generated by the
   mass interior to each radius.
   }
\labelprint{fig:quadmom}
\end{figure}

\subsection{The Angular Structure of Lenses \labelprint{sec:massquad}}

Assuming you have identified all the halos needed to model a particular lens,
there are three sources of angular structure in the potential.  The first 
source is the shape of the luminous lens galaxy,  the second source is
the dark matter in the halo of the lens, and the third source is perturbations 
from nearby objects or objects along the line of sight.  Of these, the 
only one which is easily normalized is the contribution from the stars
in the lens galaxy, since it must be tightly connected to the monopole
deflection of the stars.  The observed axis ratios of early-type galaxies
show a deficit of round galaxies, a plateau for axis ratios from $q\sim 0.9$
to $q\sim 0.5$ and then a sharp decline beyond $q\sim 0.5$ 
(e.g. Khairul \& Ryden~\cite{Khairul2002p610}).  Not surprisingly, the
true elliptical galaxies are rounder than the lenticular (S0) galaxies
even if both are grouped together as early-type galaxies.
In three dimensions, the stellar distributions are probably close to oblate
with very modest triaxialities (e.g. Franx et al.~\cite{Franx1991p112}).
Theoretical models of galaxy formation predict ellipticities and
triaxialities larger than observed for luminous galaxies and show
that the shapes of the dark matter halos are significantly modified
by the cooling baryons
(Dubinski~\cite{Dubinski1992p441},~\cite{Dubinski1994p617}, 
Warren et al.~\cite{Warren1992p405}, Kazantzidis et al.~\cite{Kazantzidis2004p1}).   
Local estimates of the shape
of dark matter halos are very limited (e.g. Olling \& Merrifield~\cite{Olling2001p361},
Buote et al.~\cite{Buote2002p183}).   Stellar isophotes also
show deviations from perfect ellipses (e.g. Bender et al.~\cite{Bender1989p35},
Rest et al.~\cite{Rest2001p2431}) and the deviations of simulated halos
from ellipses have a similar amplitude (Heyl et al.~\cite{Heyl1994p165},
Burkert \& Naab~\cite{Burkert2003}).

It is worth considering two examples to understand the relative
importance of the higher order multipoles of a lens.  The first is the
singular isothermal ellipsoid (SIE) introduced in \S\ref{sec:basics} 
(Eqns.~\ref{eqn:abg}-\ref{eqn:abi}).  Let
the major axis of the model lie on the $\theta_1$ axis, in which case only the
$\cos(m\chi)$ multipoles with $m=2,4,\cdots$ are non-zero.  All non-zero poles also have 
the same radial dependence, with $\kappa_{cm} = A_{m}/\theta$ and 
$\Psi_{cm} = -2 A_{m}\theta/(m^2-1)$.  The ratio of the internal
to the external multipole depends only on the index of the multipole,
$\Psi_{cm,int}/\Psi_{cm,ext}=(m-1)/(m+1)$.  Note, in particular,
that the quadrupole moment of an SIE is dominated by the matter
{\it outside} any given radius, with an internal quadrupole fraction of
\begin{equation}
   f_{int} = { \Psi_{c2,int} \over \Psi_{c2} } = { 1 \over 4  }.    
\end{equation}
{\it For lenses dominated by dark matter halos that have roughly flat global 
rotation curves, most
of the quadrupole moment is generated outside the Einstein ring
of the lens (i.e. by the halo!).}  
This will hold provided any halo truncation radius is large
compared to the Einstein ring radius.  The tangential deflection is larger
than the radial deflection, with $|\alpha_{cm,rad}/\alpha_{cm,tan}| = 1/m$.  
The final question is
the relative amplitudes between the poles. The ratio of the angular
deflection from the $m=2$ quadrupole to the radial deflection of the
monopole is 
\begin{equation}
   {\alpha_{c2,tan} \over \alpha_{0,rad}} \simeq { \epsilon \over 3 }
           \left[1+ {1 \over 2 }\epsilon + { 9 \over 32 }\epsilon^2 \cdots\right]
\end{equation}
while the ratio for the $m=4$ quadrupole is
\begin{equation}
   {\alpha_{c4,tan} \over \alpha_{0,rad}} \simeq { \epsilon^2 \over 20 }
           \left[1+ \epsilon + { 19 \over 24 }\epsilon^2 \cdots\right]
\end{equation}
where the axis ratio of the ellipsoid is $q=1-\epsilon$.  Each higher
order multipole has an amplitude $\Psi_{m} \propto \epsilon^{m/2}$ to
leading order.  

The relative importance
of the higher order poles can be assessed by computing the deflections
for a typical lens with the monopole deflection 
(essentially the Einstein radius) fixed to be one arc second.   Using the leading
order scaling of the power-series, but setting the numerical value to
be exact for an axis ratio $q=1/2$, the angular deflection from the
quadrupole is $0\farcs46\epsilon$ and that from the $m=4$ pole is
$0\farcs09\epsilon^2$, while the radial deflections will be 
smaller by a factors of 2 and 4 respectively.  Since typical astrometric
errors are of order $0\farcs005$, the quadrupole is quantitatively
important for essentially any ellipticity while the $m=4$ pole
becomes quantitatively  important only for $q \ltorder 0.75$ 
(and the $m=6$ pole becomes quantitatively important for $q \ltorder 0.50$).

In Fig.~\ref{fig:quadmom} we compare the SIE to ellipsoidal 
de Vaucouleurs and NFW models.  Unlike the SIE, these models are
not scale free, so the multipoles depend on the distance from the
lens center in units of the major axis scale length of the lens,
$R_{major}$.  The behavior of the de Vaucouleurs model will be
typical of any ellipsoidal mass distribution that is more centrally
concentrated than an SIE.  
Although the de Vaucouleurs model produces angular
deflections similar to those of an SIE on small scales (for the same axis
ratio), these are beginning to decay rapidly at the radii where
we see lensed images (1--2$R_{major}$) because most of the mass
is interior to the image positions and the amplitudes of the 
higher order multipoles decay faster with radius than the monopole (see Eqn.~\ref{eqn:ajk}).
Similarly, as more of the mass lies at smaller radii, the quadrupole
becomes dominated by the internal quadrupole.  The NFW model
has a somewhat different behavior because on small scales it
is less centrally concentrated than an SIE (a $\rho \propto 1/r$
central density cusp rather than $\propto 1/r^2$).   It 
produces a somewhat bigger quadrupole for a given axis ratio,
and an even larger fraction of that quadrupole is generated 
on large scales.  In a ``standard'' dark matter halo model, 
the region with $\theta < R_{major}$ is also where we see the 
lensed images.  On larger scales, the NFW profile is more centrally
concentrated than the SIE, so the quadrupole begins to decay
and becomes dominated by the internal component.

It is unlikely that mass distributions are true ellipsoids producing
only even poles ($m=2$, $4$, $\cdots$) with no twisting of the axes
with radius.  For model fits we need to consider the likely amplitude
of these deviations and the ability of standard terms to absorb and
mask their presence.  It is clear from Fig.~\ref{fig:quadmom} that the
amplitude of any additional terms must be of order the $m=4$ deflections
expected for an ellipsoid for them to be important.  
Here we illustrate the issues with the first
few possible terms.  

A dipole moment ($m=1$) corresponds to making the galaxy lopsided
with more mass on one side of the lens center than the other.  
Lopsidedness is not rare in disk galaxies ($\sim$30\% at
large radii,  Zaritsky \& Rix~\cite{Zaritsky1997p118}), 
but is little
discussed (and hence presumably small) for early-type galaxies.
Certainly in the CASTLES photometry of lens galaxies we never
see significant dipole residuals.
It is difficult (impossible) to have an equilibrium system 
supported by random stellar motions with a dipole moment 
because the resulting forces will tend to eliminate the dipole. Similar
considerations make it difficult to have a dark matter halo offset from
the luminous galaxy.  Only disks, which are supported by ordered rather than
random motion, permit relatively long-lived lopsided structures.  
Where a small dipole exists, it will have little effect on the lens
models unless the position of the lens galaxy is imposed as a stringent
constraint.  The reason is that a dipole adds terms to the 
effective potential of the form $\theta_1 G(\theta)$ whose
leading terms are degenerate with a change in the 
unknown source position.

Perturbations to the quadrupole (relative to an ellipsoid) arise from
variations in the ellipticity or axis ratio with radius.  Since realistic
lens models require an independent external shear simply to model the
local environment, it will generally be very difficult to detect these
types of perturbations or for these types of perturbations to significantly
modify any conclusions.  In essence, the amplitude and orientation of 
the external shear can capture most of their effects.  Their actual
amplitude is easily derived from perturbation theory. For example, if there
is an isophote twist of $\Delta\chi$ between the region inside the
Einstein ring and outside the Einstein ring, the fractional perturbations
to the quadrupole will be of order $\Delta\chi$, or approximately
$\epsilon\Delta\chi/3$ of the monopole -- independent of the ability
of the external shear to mimic the twist, the actual amplitude of the
perturbation is approaching the typical measurement precision unless
the twist is very large.  Only in Q0957+561 have models found reasonably
clear evidence for an effect arising from isophotal twists and ellipticity
gradients, but both distortions are unusually large in this system
(Keeton et al.~\cite{Keeton2000p74}).  In general, in the CASTLES photometry
of lens galaxies, deviations from simple ellipsoidal models are rare.

Locally we observe that the isophotes of elliptical galaxies are not perfect 
ellipses (e.g. Bender et al.~\cite{Bender1989p35}, Rest et al.~\cite{Rest2001p2431}) 
and simulated halos show deviations of similar amplitude 
(Heyl et al.~\cite{Heyl1994p165}, Burkert \& Naab~\cite{Burkert2003}).
For lensing calculations it is useful to characterize these perturbations
by a contribution to the lens potential and surface density of
\begin{equation}
    \Psi = { \epsilon_m \over m } \theta \cos m(\chi-\chi_m) 
    \quad\hbox{and}\quad
    \kappa_m = { \epsilon_m \over \theta } { 1-m^2 \over m }  \cos m(\chi-\chi_m) 
\end{equation}
respectively where the amplitude of the term is related to the usual 
isophote parameter $a_m = \epsilon_m |1-m^2|/mb$ for a lens with Einstein
radius $b$.  A typical early-type galaxy might have $|a_4| \sim 0.01$,
so their fractional effect on the deflections, $|\epsilon_4|/b\sim |a_4|/4 \sim 0.003$,
will be comparable to the astrometric measurement accuracy.

\subsection{Constraining Angular Structure \labelprint{sec:angstruc} }

The angular structure of lenses is usually simply viewed as an obstacle to
understanding the monopole.  This is a serious mistake.   The reason
angular structure is generally ignored is that the ability to accurately
constrain the angular structure of the gravitational field is nearly
unique to gravitational lensing.  Since we have not emphasized the ability
of lenses to measure angular structure and other methods cannot do so very
accurately, there has been little theoretical work on the angular structure
of galaxies with dark matter.  Both theoretical studies of halos and 
modelers of gravitational lenses need to pay more attention to the 
angular structure of the gravitational potential.

We start by analyzing a simple two-image lens using our non-parametric
model of the monopole (Eqn.~\ref{eqn:abw}) in an external shear (Eqn.~\ref{eqn:acc}).  The
two images are located at $\vec{\theta}_A=\theta_A(\cos\chi_A,\sin\chi_A)$,
and $\vec{\theta}_B=\theta_B(\cos\chi_B,\sin\chi_B)$ as illustrated
in Fig.~\ref{fig:geometry}.  To illustrate
the similarities and differences between shear and convergence, we will
also include a global convergence $\kappa_0$ in the model.  This corresponds
to adding a term to the lens potential of the form $(1/2)\kappa_0 \theta^2$.
The model now has five parameters -- two shear components, the mass
and surface density of the monopole model and the additional global
convergence.  We have only two astrometric constraints, and so can solve
for only two of the five parameters.  Since the enclosed mass is always
an interesting parameter, we can only solve for one of the two shear
components.  In general, we will find that the amplitude of $\gamma_c$
depends on the amplitude of $\gamma_s$.  There is, however, a special
choice of the shear axis, $\chi_\gamma=(\chi_A+\chi_B)/2+\pi/4$, such
that the shear parameters become independent of each other.  This allows
us to determine the ``invariant'' shear associated with the images, 
\begin{equation}
   \gamma_1 = { 
   \left(1-\kappa_0-\kbar_{AB}\right)(\theta_A^2-\theta_B^2)\sin(\chi_A-\chi_B) \over
        \Delta\theta^2 }
  \labelprint{eqn:aca}
\end{equation}
where $\Delta\theta=|\vec{\theta}_A-\vec{\theta}_B|$ is the image separation.
The monopole mass and the other shear component are degenerate,
\begin{eqnarray}
   b_B^2 +\gamma_2 \theta_A\theta_B =&&  \\
   \quad { \left( 1-\kappa_0 \right) 
     \left[ \Delta\theta^2(\theta_A^2+\theta_B^2)-(\theta_A^2-\theta_B^2)^2 \right]
      - \kbar_{AB} \left( \theta_A^2-\theta_B^2 \right) 
         \left(\Delta\theta^2-\theta_A^2+\theta_B^2\right) \over 2 \Delta\theta^2 }. &&
      \nonumber
  \labelprint{eqn:acb}
\end{eqnarray}
Several points are worth noting. First, the amplitude of the invariant shear
$\gamma_1$
has the same degeneracy with the (local) surface density between the images 
$\kbar_{AB}$ as it does with a global convergence $\kappa_0$.  More centrally
concentrated mass distributions with lower $\kbar_{AB}$ require higher external
shears to fit the same data.  Second, the other component $\gamma_2$ introduces
an uncertainty into the enclosed mass, with a series of somewhat messy trade offs
between $b_B^2$, $\gamma_2$, $\kbar_{AB}$ and $\kappa_0$.  As a practical matter,
the shear does not lead to an astronomically significant uncertainty in the mass,
since $\gamma_2\ltorder 0.1$ in all but the most extreme situations.

\begin{figure}[ph]
\begin{center}
\centerline{\psfig{figure=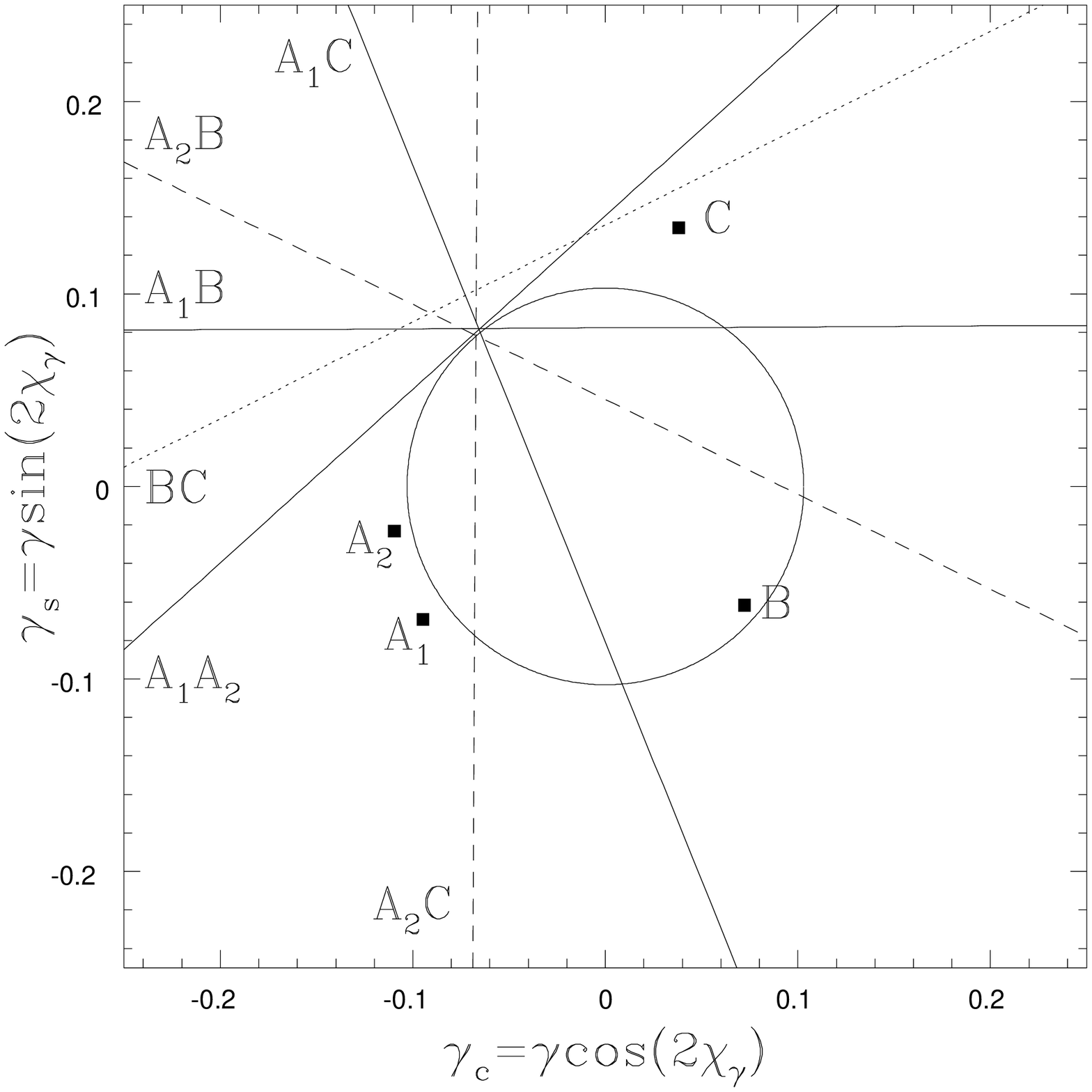,width=3.2in}}
\end{center}
\begin{center}
\centerline{\psfig{figure=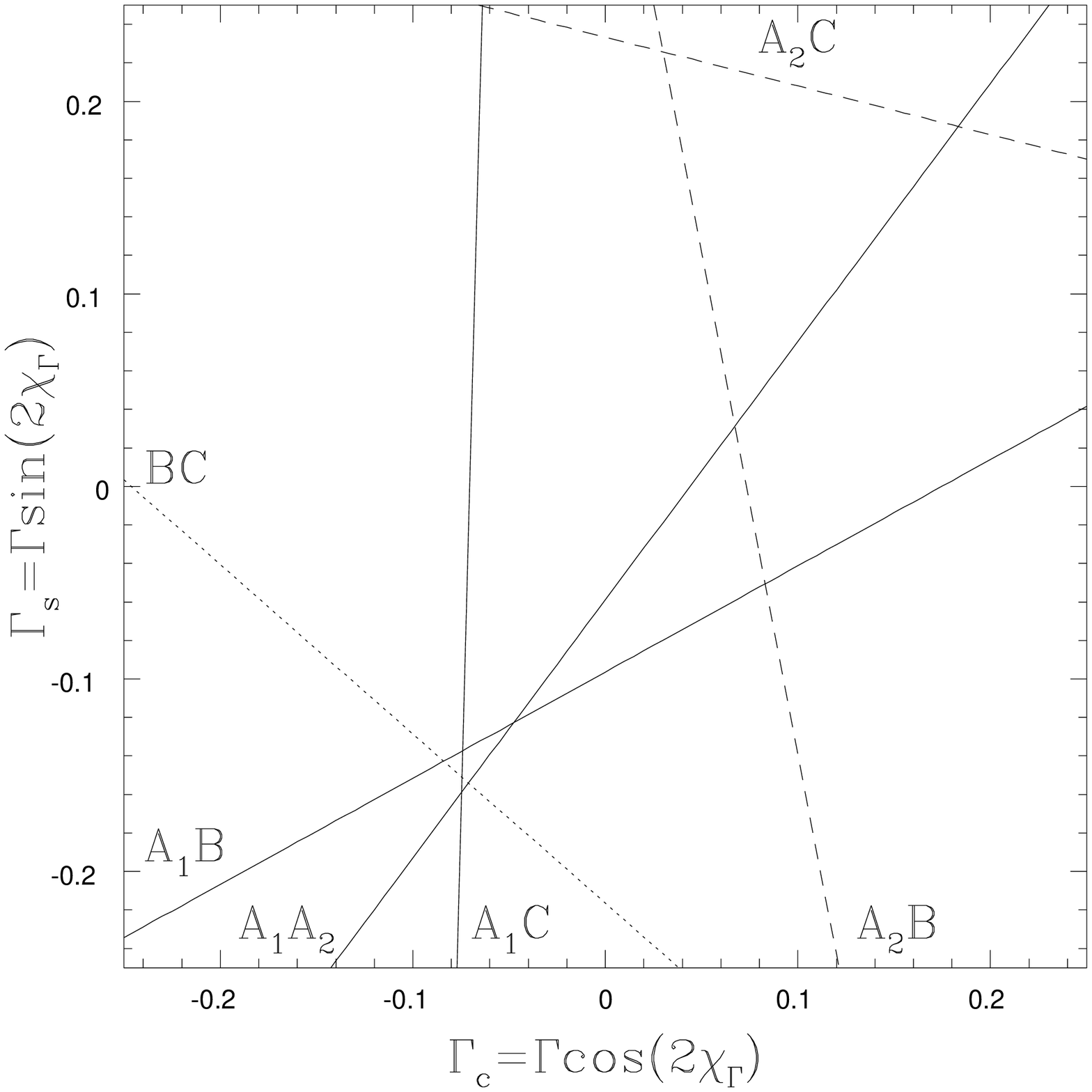,width=3.2in}}
\end{center}
\caption{
   The invariant shears for the lens PG1115+080 modeled using either an 
   external (top) or an internal (bottom) quadrupole and an SIS monopole.
   Each possible image pair among the A$_1$, A$_2$, B and C images,
   constrains the quadrupole to lie on the labeled line.  The amplitude
   and orientation of each invariant shear is given by the point where 
   the corresponding line passes closest to the origin.  Models of PG1115+080
   show that the quadrupole is dominated by external (tidal) shear.  Here
   we see that for the external quadrupole (top), the lines nearly cross
   at a point, so the data are consistent with an almost pure external shear.
   For an internal quadrupole (bottom), the A$_2$B and A$_2$C image pairs
   require shear parameters completely inconsistent with the other images.
   }
\labelprint{fig:invshear}
\end{figure}

The external shear is only one component of the quadrupole.  There is also an
internal shear due to the mass interior to the images (Eqn.~\ref{eqn:adl}).
The internal and external shears differ in their ``handedness''.  For the same
angular deflection ($d\Psi/d\chi$) they have opposite signs for the radial
deflection ($d\Psi/d\theta$).  The solution for two images is much the same as
for an external shear.  There is an invariant shear component, whose amplitude
scales with $1-\kappa_0-\kbar_{AB}$ but whose orientation differs from that
of the external shear solution.  The monopole mass $b_B^2$ is degenerate with
the $\gamma_2$ shear component and the $\kappa_0$ and $\kbar_{AB}$ surface
densities.  The actual expressions are too complex to be illuminating.

Fig.~\ref{fig:invshear} illustrates how the invariant shears combine to determine
the overall structure of the quadrupole for the lens PG1115+080.
For each image pair there is a line of permitted shears because of the 
degeneracy between the enclosed mass and the second shear component.  The
invariant shear component is the shear at the point where the line passes 
closest to the origin.  If the quadrupole model is correct, the lines for
all the image pairs will cross at a point, while if it is incorrect they
will not.  PG1115+080 is clearly going to be well modeled if the quadrupole
is dominated by an external shear and poorly modeled if it is dominated by
an internal shear.  This provides a simple geometric argument for why full
models of PG1115+080 are always dominated by an external shear (e.g.
Impey et al.~\cite{Impey1998p551}).  A
failure of the curves to cross in both cases is primarily evidence for a
mixture of external and internal quadrupoles or the presence of other multipoles
rather than for a problem in the monopole mass distribution.  In Fig.~\ref{fig:invshear}
we used an SIS for the monopole.  For a point mass monopole, the figure
looks almost the same provided we expand the scale -- the invariant shear
scales as $1-\kbar_{AB}$ so in going from a SIS with $1-\kbar_{AB}\simeq 1/2$
to a point mass with $1-\kbar_{AB}=1$ the shear will double.

This scaling of the quadrupole with the surface density of the monopole provides
an as yet unused approach to studying the monopole.  Since the mass enclosed by
the Einstein radius is nearly constant, the more centrally concentrated constant
mass-to-light ($M/L$) ratio models must have lower surface mass densities near 
the images than the SIE model.  As a result, they will require
quadrupole amplitudes that are nearly twice those of models
like the SIS with nearly flat rotation curves.  Since the typical SIE model
of a lens has an ellipticity that is comparable to the typical ellipticities
of the visible galaxies, the more centrally concentrated monopole of a constant
$M/L$ model requires an ellipticity much larger than the observed ellipticity
of the lens galaxy.  The need to include an external tidal shear to represent
the environment allows these models to produce acceptable fits, but the 
amplitudes of the required external shears are inconsistent with expectations
from weak lensing (\partweak).

\subsection{Model Fitting and the Mass Distribution of Lenses \labelprint{sec:modelfit} }

Having outlined (in perhaps excruciating detail) how lenses constrain the mass
distribution, we turn to the problem of actually fitting data.  These days the
simplest approach for a casual user is simply to down load a modeling package,
in particular the lensmodel package (Keeton~\cite{Keeton2001}) at 
http://cfa-www.harvard.edu/castles/, read the 
manual, try some experiments, and then apply it intelligently (i.e. read
the previous sections about what you can extract and what you cannot!).
Please publish results with a complete description of the models and the
constraints using standard astronomical nomenclature.

In most cases we are interested in the problem of fitting the positions 
$\vec{\theta}_i$ of $i=1 \cdots n$ images where the image positions have
been measured with accuracy $\sigma_i$.  We may also know the positions
and properties of one or more lens galaxies.  Time delay ratios also constrain
lens models but sufficiently accurate ratios are presently available for only one lens
(B1608+656, Fassnacht et al.~\cite{Fassnacht2002p823}), fitting them is 
already included in most packages, and 
they add no new conceptual difficulties.  Flux ratios constrain the lens
model, but we are so uncertain of their systematic uncertainties due to
extinction in the ISM of the lens galaxy, microlensing (\partmicro) and the
effects of substructure (see \S\ref{sec:substruc}) that we can never impose them with the
accuracy needed to add a significant constraint on the model.  

The basic issue with lens modeling is whether or not to invert the lens
equations (``source plane'' or ``image plane'' modeling).  
The lens equation supplies the source position
\begin{equation}
     \vec{\beta}_i = \vec{\theta}_i - \vec{\alpha}(\vec{\theta}_i,\vec{p})
\end{equation}
predicted by the observed image positions $\vec{\theta}_i$ and the current model 
parameters $\vec{p}$. Particularly
for parametric models it is easy to project the images on to the source
plane and then minimize the difference between the projected source 
positions.  This can be done with a $\chi^2$ fit statistic of the form
\begin{equation}
     \chi^2_{src} = \sum_i \left( 
   { \vec{\beta} -\vec{\beta}_i \over \sigma_i }\right)^2 
\end{equation} 
where we treat the source position $\vec{\beta}$ as a model parameter.  
The astrometric uncertainties $\sigma_i$ are typically a few milli-arcseconds.
Moreover, where VLBI observations give significantly smaller uncertainties, they
should be increased to approximately 0\farcs001--0\farcs005 because low mass 
substructures in the lens galaxy can produce systematic errors on this order
(see \S\ref{sec:substruc}). 
{\it You can impose astrometric constraints to no greater accuracy than
the largest deflection scales produced by lens components you are not
including in your models.}
The advantage of $\chi^2_{src}$ is that it is fast and has excellent 
convergence properties.  The disadvantages are that it is wrong, cannot
be used to compute parameter uncertainties, and may lead to a model producing
additional images that are not actually observed.  

\begin{figure}[ph]
\begin{center}
\centerline{\psfig{figure=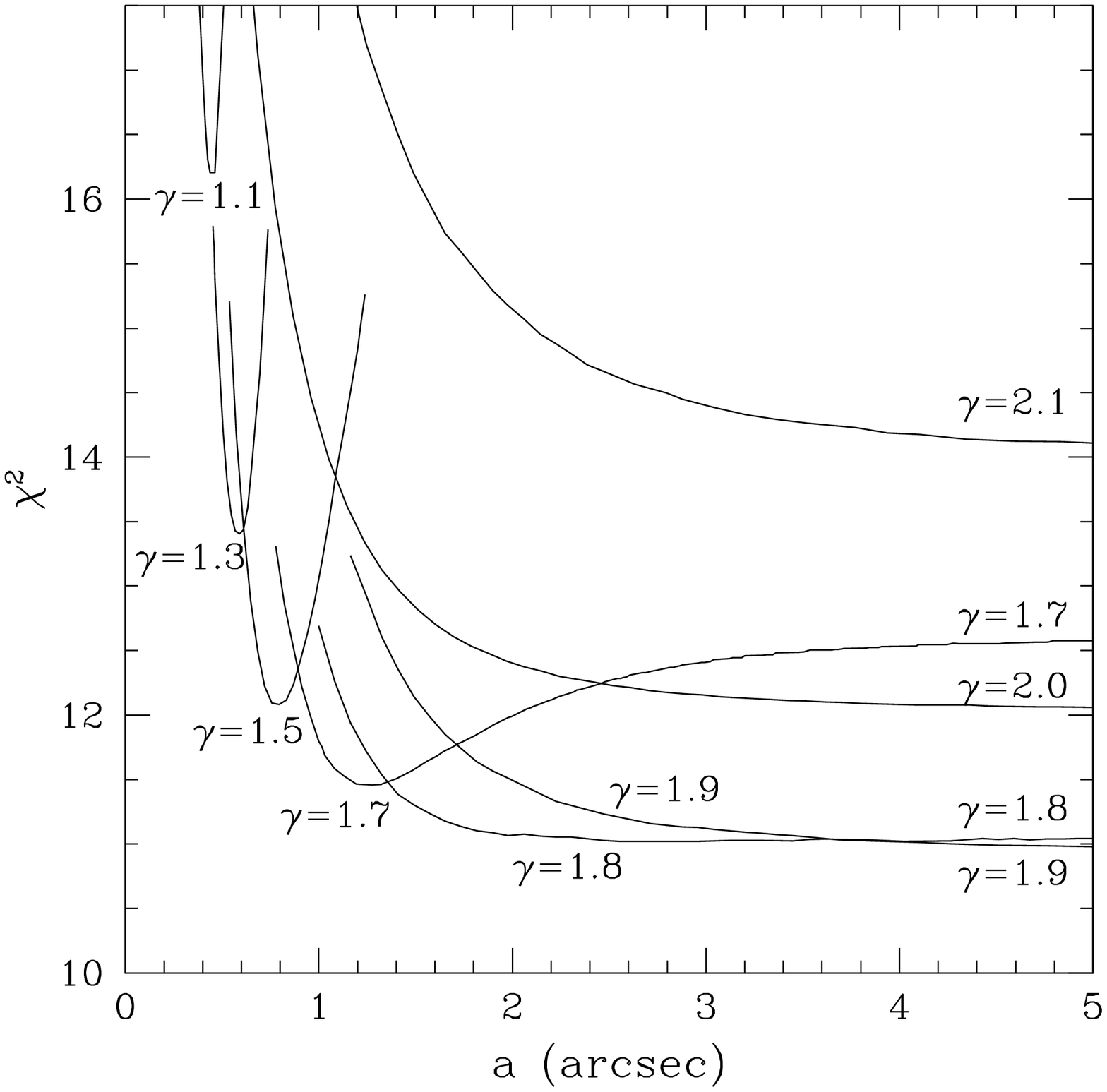,width=3.2in}}
\end{center}
\begin{center}
\centerline{\psfig{figure=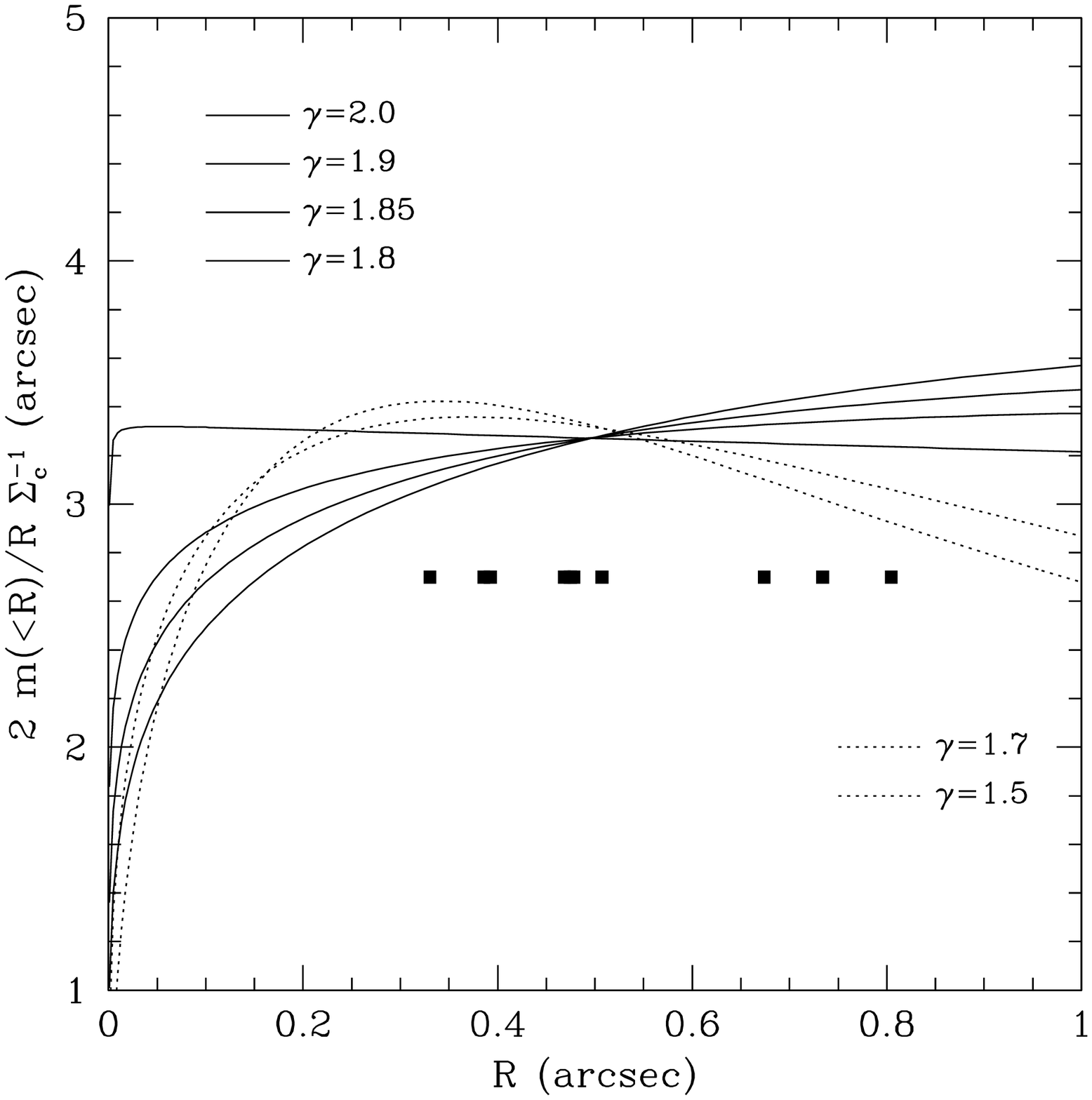,width=3.2in}}
\end{center}
\caption{ (Top)
   Goodness of fit $\chi^2$ for cuspy models of B1933+503 as a function of the
   inner density exponent $\gamma$ ($\rho\propto r^{-\gamma}$) and the profile
   break radius $a$.  Models with cusps significantly shallower or steeper
   than isothermal are ruled out, and acceptable models near isothermal must
   have break radii outside the region with the lensed images.
   }
\labelprint{fig:mod1933a}
\caption{ (Bottom)
   The monopole deflections of the B1933+503 models for the range of permitted
   cusp exponents $\gamma$.  The points show the radii of the lensed images,
   and the models only constrain the shape of the monopole in this region.
   The monopole deflection is closely related to the square of the rotation
   curve.  Note the similarity to Fig.~\ref{fig:cusp}.
   }
\labelprint{fig:mod1933b}
\end{figure}

The reason it is wrong and cannot be used to compute parameter errors is that
the uncertainty $\sigma_i$ in the image positions does not have any meaning
on the source plane.   This is easily understood if we Taylor expand the 
lens equation near the projected source point $\vec{\beta}_i$ corresponding
to an image
\begin{equation}
          \vec{\beta}-\vec{\beta}_i = M_i^{-1} (\vec{\theta}-\vec{\theta}_i)
\end{equation}
where $M_i^{-1}$ is the inverse magnification tensor at the observed location
of the image.  In the frame where the tensor is diagonal, we have that 
$\Delta\beta_\pm = \lambda_\pm \Delta\theta_\pm$ so a positional
error  $\Delta\beta_\pm$ on the source plane corresponds to a positional
error $\lambda_\pm^{-1}\Delta\beta_\pm$ on the image plane.  Since the 
observed lensed images are almost always magnified (usually $\lambda_+ =1+\kappa+\gamma\sim 1$
and $0.5 > |\lambda_-=1+\kappa-\gamma| < 0.05$)  there is always one direction in which 
small errors on the source plane are significantly magnified when projected
back onto the image plane.  Hence, if you find solutions with $\chi^2_{src}\sim N_{dof}$   
where $N_{dof}$ is the number of degrees of freedom, you will have source
plane uncertainties $\Delta\beta \ltorder \sigma_i$.  However, the actual
errors on the image plane are $\mu=|M|$ larger, so the $\chi^2$ on the image
plane is $\sim \mu^2 N_{dof}$ and you in fact have a terrible fit.

If you assume that in any interesting model
you are close to having a good solution, then this Taylor expansion provides
a means of using the easily computed source plane positions to still get
a quantitatively accurate fitting statistic,
\begin{equation}
     \chi^2_{int} = \sum_i 
   { \left(\vec{\beta} -\vec{\beta}_i\right)\cdot M_i^2 \cdot 
     \left(\vec{\beta} -\vec{\beta}_i\right) \over \sigma_i^2 }, 
     \label{eqn:fit1}
\end{equation} 
in which the magnification tensor $M_i$ is used to correct the error in the source
position to an error in the image position.  This procedure will be approximately
correct provided the observed and model image positions are close enough for the
Taylor expansion to be valid.  
Finally, there is the exact statistic where for the model source position $\vec{\beta}$
you numerically solve the lens equation to find the  exact image positions 
$\vec{\theta}_i(\vec{\beta})$ and then compute the goodness of fit on the image plane
\begin{equation}
     \chi^2_{img} = \sum_i \left( 
   { \vec{\theta}_i(\vec{\beta}) -\vec{\theta}_i \over \sigma_i } \right)^2. 
     \label{eqn:fit2}
\end{equation} 
This will be exact even if the Taylor expansion of $\chi^2_{int}$ is breaking down,
and if you find all solutions to the lens equations you can verify that the model
predicts no additional visible images.  Unfortunately, using the exact $\chi^2_{img}$
is also a much slower numerical procedure.

As we discussed earlier, even though lens models provide the most accurate mass
normalizations in astronomy, they can constrain the mass
distribution only if the source is more complex than a single compact 
component.  Here we only show examples where there are multiple point-like
components, deferring discussions of models with extended source structure
to \S\ref{sec:hosts}.  The most spectacular example of a multi-component source is
B1933+503 (Sykes et al.~\cite{Sykes1998p310}, see Fig.~\ref{fig:b1933merlin}) 
where a source consisting of a radio core and
two radio lobes has 10 lensed images because the core and one lobe are
quadruply imaged and the other lobe is doubly imaged.  Since we have 
many images spread over roughly a factor of two in radius, this lens should
constrain the radial mass distribution just as in our discussion for
\S\ref{sec:monofit}.  Mu\~noz et al.~(\cite{Munoz2001p657},
also see Cohn et al.~\cite{Cohn2001p1216} for softened power law models) fitted
this system with cuspy models (Eqn. 55 with $\alpha=2$ and $m=4$), varying the inner
density slope $n=\gamma$ ($\rho\propto r^{-n}$) and the break radius $a$.  
Fig.~\ref{fig:mod1933a} shows the resulting $\chi^2$ as a function of the
parameters and Fig.~\ref{fig:mod1933b} illustrates the range of the
acceptable monopole mass distributions -- both are very similar to 
Fig.~\ref{fig:cusp}. The best fit is for $\gamma=1.85$
with an allowed range of $1.6 < \gamma <2.0$ that completely excludes
the shallow $\gamma=1$ cusps of the Hernquist and NFW profiles and is
marginally consistent with the $\gamma=2$ cusp of the SIS model. 
A second example, which illustrates how the distribution of mass 
well outside the region with images has little effect on the models,
are the Winn et al.~(2003) models of the three-image lens PMNJ1632--0033
shown in Fig.~\ref{fig:mod1632}.  In these models the outer slope $\eta$, with 
$\rho \propto r^{-\eta}$ asymptotically, of the density was
also explored but has little effect on the results.  Unless the break
radius of the profile is interior to the B image, the mass profile
is required to be close to isothermal $1.89 < \beta < 1.93$.

\begin{figure}[t]
\begin{center}
\centerline{\psfig{figure=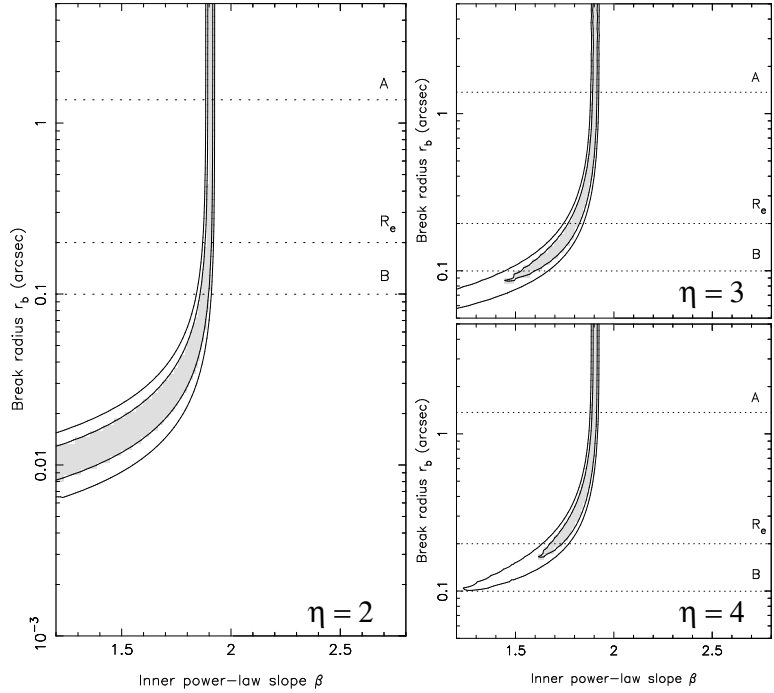,width=4.0in}}
\end{center}
\caption{ 
   Allowed parameters for cuspy models of PMNJ1632--0033 assuming that image
   C is a true third image.  Each panel shows the constraints on the inner
   density cusp $\beta$ ($\rho \propto r^{-\beta}$) and the break radius
   $r_b$ for three different asymptotic density slopes $\rho \propto r^{-\eta}$.
   A Hernquist model has $\beta=1$ and $\eta=4$, an NFW model has 
   $\beta=1$ and $\eta=3$, and a pseudo-Jaffe model has $\beta=2$ and
   $\eta=4$.  Unless the break radius is place interior to the B image,
   it is restricted to be close to isothermal ($\beta=2$).
   }
\labelprint{fig:mod1632}
\end{figure}

Unfortunately, systems like B1933+503 and PMNJ1632--0033 are a 
small minority of lens systems.  For most lenses, obtaining information
on the radial density profile requires some other information such
as a dynamical measurement (\S\ref{sec:dynamics}), a time delay measurement 
(\S\ref{sec:time})
or a lensed extended component of the source (\S\ref{sec:hosts}).  Even for these
systems, it is important to remember that the actual constraints on the
density structure really only apply over the range of radii spanned by
the lensed images -- the mass interior to the images is constrained but
its distribution is not, while the mass exterior to the images is completely
unconstrained.  This is not strictly true when we include the angular 
structure of the gravitational field and the mass distribution is 
quasi-ellipsoidal.

It is also important to keep some problems with parametric models in mind.
First, models that lack the degrees of freedom needed to describe the actual
mass distribution can be seriously in error.  Second, models with too many
degrees of freedom can be nonsense.  We can illustrate these two limiting
problems with the sad history of Q0957+561 for the first problem and attempts to
explain anomalous flux ratios (see \S\ref{sec:substruc}) with complex angular
structures in the density distribution for the dark matter for the second.

Q0957+561, the first lens discovered (Walsh, Carswell \& Weymann~\cite{Walsh1979p381})
and the first lens with a well measured time delay (see \S\ref{sec:time}, Schild \& 
Thomson~\cite{Schild1995p1970}, Kundi\'c et al.~\cite{Kundic1997p75} and references
therein), is an ideal lens for demonstrating the trouble you can get into using
parametric models without careful thought.  The lens consists of a cluster and its
brightest cluster galaxy with two lensed images of a radio source bracketing the
galaxy.  VLBI observations (e.g. Garrett et al.~\cite{Garrett1994p457}) resolve
the two images into thin, multi-component jets with very accurately measured 
positions (uncertainties as small as 0.1~mas, corresponding to deflections 
produced by a mass scale $\sim 10^{-8}$ of the primary lens!).   
Models developed along two lines.  One line focused on models in which the
cluster was represented as an external shear (e.g. Grogin \&
Narayan~\cite{Grogin1996p570}, Chartas et al.~\cite{Chartas1998p661},
Barkana et al.~\cite{Barkana1999p54}, Chae~\cite{Chae1999p582})
while the other explored more complex models for the cluster
(see Kochanek~\cite{Kochanek1991p517}, Bernstein, Tyson \&
Kochanek~\cite{Bernstein1993p816}, Bernstein \& Fischer~\cite{Bernstein1999p14})
and argued that external shear models had too few parameters to represent
the mass distribution given the accuracy of the constraints.  The latter
view was born out by the morphology of the lensed host galaxy 
(Keeton et al.~\cite{Keeton2000p74}) and direct X-ray observations of 
the cluster (Chartas et al.~\cite{Chartas2002p96}) which showed that the
lens galaxy was within about one Einstein radius of the cluster center
where a tidal shear approximation fails catastrophically.  The origin of 
the problem is that as a
two-image lens, Q0957+561 is critically short of constraints unless the
fine details of the VLBI jet structures are included in the models.  Many
studies imposed these constraints to the limit of the measurements
while not including all possible terms in the 
potential which could produce a deflection on that scale (i.e. the 
precision should have been restricted to milli-arcseconds rather than
micro-arcseconds).  Models would adjust the positions and masses of the cluster
and the lens galaxy in order to reproduce the small scale astrometric details 
of the VLBI jets without including less massive components of the mass 
distribution (e.g. the ellipticity gradient and isophote twist of the lens galaxy, 
Keeton et al.~\cite{Keeton2000p74}) that also affect the VLBI jet structure
on these angular scales.  Lens models must contain all 
reasonable structures producing deflections comparable to the scale of the
measurement errors. 

We are in the middle of an experiment exploring the second problem -- if
you include small scale structures but lack the constraints needed to measure
them, their masses easily become unreasonable unless constrained by common sense,
physical priors or additional data.  Lately this has become an issue in studies
(Evans \& Witt~\cite{Evans2003p1351},  Kochanek \& Dalal~\cite{Kochanek2004},
Quadri, Moller \& Natarajan~\cite{Quadri2003p659}, Kawano et al.~\cite{Kawano2004})
of whether the flux ratio anomalies in gravitational lenses could be due to
complex angular structure in the lens galaxy rather than CDM substructure or
satellites in the lens galaxy (see \S\ref{sec:substruc}). 
The problem, as we discuss in the next
section on non-parametric models (\S\ref{sec:nonparam}), is that lens modeling
with large numbers of parameters is closely related to solving linear equations
with more variables than constraints -- as the matrix inversion necessary to
finding a solution becomes singular, the parameters of the mass distribution
show wild, large amplitude fluctuations even as the fit to the constraints
becomes perfect.  Thus, a model including enough unconstrained parameters is 
guaranteed to ``solve'' the anomalous flux ratio problem even if it should not.  
For example, Evans \& Witt~(\cite{Evans2003p1351}) could match the flux ratios of 
Q2237+0305 even though for this lens we know from the time variability of the flux
ratios that the flux ratio anomalies are created by microlensing rather than
complex angular structures in the lens model (see \partmicro).   

\begin{figure}[ph]
\begin{center}
\centerline{\psfig{figure=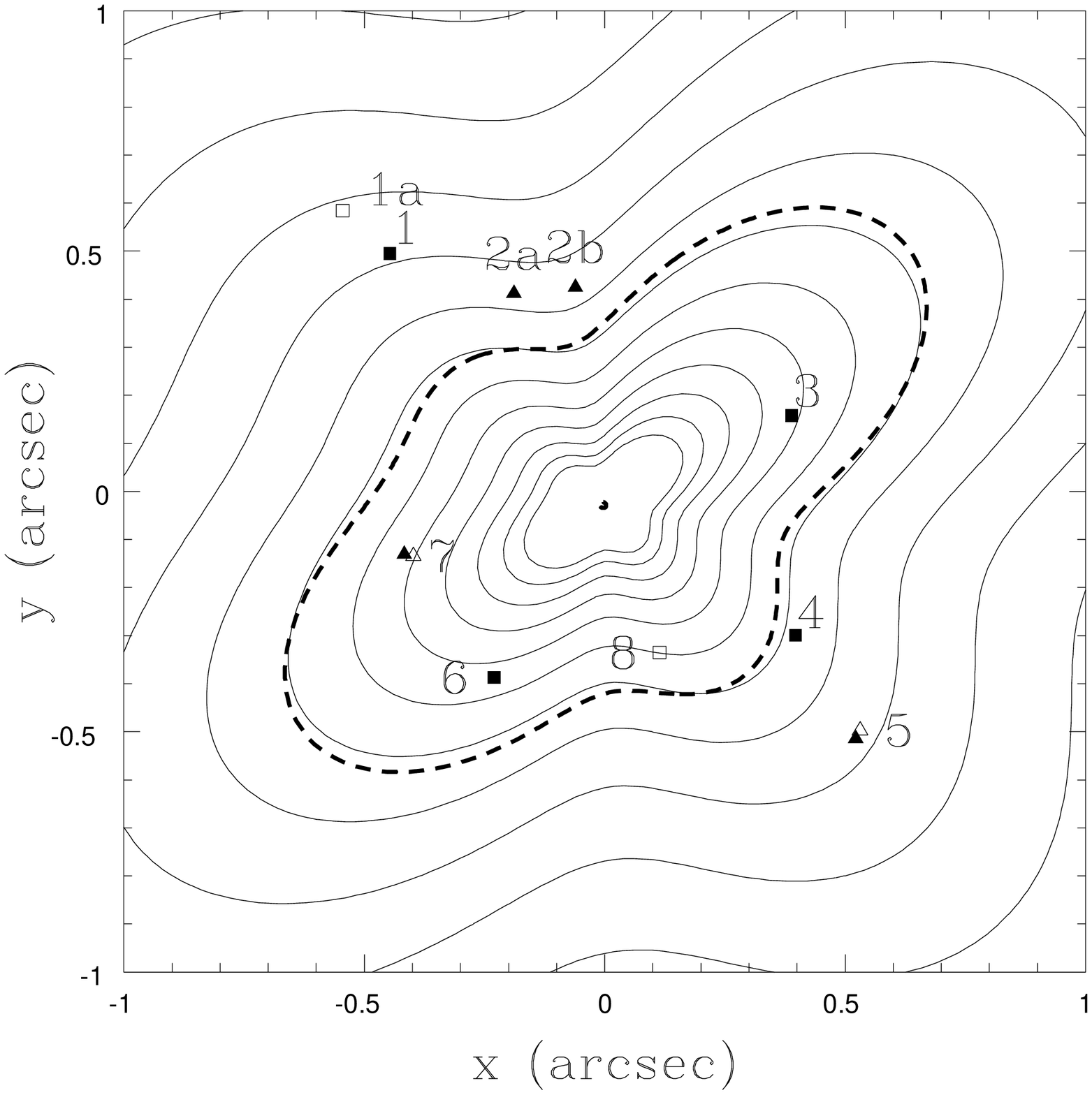,width=3.0in}}
\end{center}
\vspace{-0.25in}
\begin{center}
\centerline{\psfig{figure=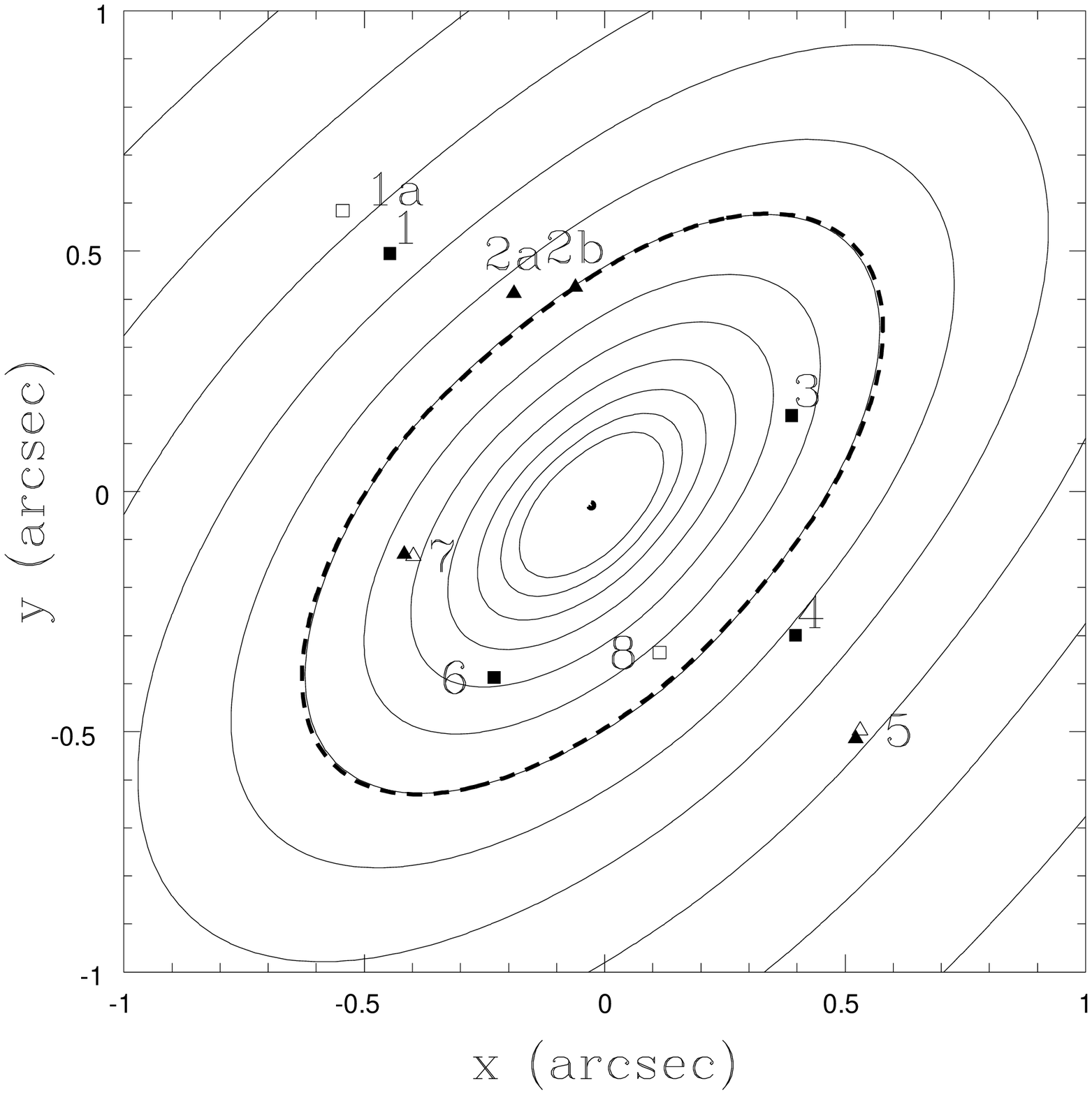,width=3.0in}}
\end{center}
\caption{Surface density contours for models of B1933+503 including misaligned
  $a_3$ and $a_4$ multipoles (thin lines).  The model in the top panel is
  constrained only by the 4 compact images (images 1, 3, 4 and 6, filled
  squares).  The model in the bottom panel is also constrained by the other
  images in the lens (the two-image system 1a/8, open squares; the four-image system
  2a/2b/5/7 filled triangles; and the two-image system comprising parts of 5/7,
  open triangles)   The tangential critical line of the model (heavy solid
  curve) must pass between the merging images 2a/2b, but fails to do so in
  the first model (top panel).
   }
\labelprint{fig:substruc4}
\end{figure}

If only the four compact images are modeled, then the flux ratio
anomalies can be greatly reduced or eliminated in almost all lenses at the
price of introducing deviations from an ellipsoidal density distribution far
larger than expected (see \S\ref{sec:massquad}).  In some cases, however, you
can test these solutions because the lens has extra constraints beyond the
four compact images.  We illustrate this in Fig.~\ref{fig:substruc4}. By adding 
large amplitude $\cos 3\theta$ and $\cos 4\theta$ perturbations to the surface 
density model for B1933+503,  Kochanek \& Dalal~(\cite{Kochanek2004})
could reproduce the observed image flux ratios if they fit only the four
compact sources. However, after adding the constraints from the other lensed 
components, the solution is driven back to being nearly ellipsoidal and the flux ratios 
cannot be fit.  In every case, Kochanek \& Dalal~(\cite{Kochanek2004})
found that the extra constraints drove the solution back toward an ellipsoidal
density distribution.  In short, a sufficiently complex model can fit 
underconstrained data, but that does not mean it makes any sense to do so.

\subsection{Non-Parametric Models \labelprint{sec:nonparam} }

The basic idea behind non-parametric mass models is that the effective
lens potential and the deflection equations are linear ``functions'' of the
surface density.  The surface density can be decomposed into multipoles 
(Kochanek~\cite{Kochanek1991p354}, Trotter, Winn \& Hewitt~\cite{Trotter2000p671},
Evans \& Witt~\cite{Evans2003p1351}), 
pixels (see Saha \& Williams~\cite{Saha1997p148}, \cite{Saha2004}, 
Williams \& Saha~\cite{Williams2000p439}), or any other form in which
the surface density is represented as  a linear combination of density 
functionals multiplied by unknown coefficients $\vec{\kappa}$.  In any
such model, the lens equation for image $i$ takes the form
\begin{equation}
     \vec{\beta} = \vec{\theta}_i - A_i \vec{\kappa}
\end{equation}
where $A_i$ is the matrix that gives the deflection at the position of 
image $i$ in terms of the coefficients of the surface density 
decomposition $\vec{\kappa}$.  For a lens with $i=1 \cdots n$ images
of the same source, such a system can be solved exactly if there are
enough degrees of freedom in the description of the surface density.  
For simplicity, consider a two-image lens so that we can eliminate
the source position by hand, leaving the system of equations
\begin{equation}
     \vec{\theta}_1- \vec{\theta}_2 =  (A_1-A_2) \vec{\kappa},
\end{equation}
which is easily solved by simply taking the inverse of the matrix
$A_1-A_2$ to find that 
\begin{equation}
     \vec{\kappa} = (A_1-A_2)^{-1}\left(\vec{\theta}_1- \vec{\theta}_2\right). 
\end{equation}
Sadly, life is not that simple because as soon as the density 
decomposition has more degrees of freedom than there are constraints,
the inverse $(A_1-A_2)^{-1}$ of the deflection operators is singular.

The solution to this problem is to instead consider the problem as a 
more general minimization problem with a $\chi^2$ statistic for the
constraints and some form of regularization to restrict the results to
plausible surface densities.
One possibility is linear regularization, in which
you minimize the function
\begin{equation}
    F = \chi^2 + \lambda \vec{\kappa} \cdot H \cdot \vec{\kappa}
\end{equation}
where the $\chi^2$ measures the goodness of fit to the lens constraints,
$H$ is a weight matrix and $\lambda$ is a Lagrange multiplier.  The
Lagrange multiplier controls the relative importance given to fitting the
lens constraints (minimizing the $\chi^2$) versus producing a smooth
density distribution (minimizing $\vec{\kappa} \cdot H \cdot \vec{\kappa}$).  
The simplest smoothing function is to minimize the variance of the
surface density ($H=I$, the identity matrix), or, equivalently, ignore $H$ and 
use the singular value 
decomposition for inverting a singular matrix. By using more complicated matrices you can
minimize derivatives of the density (gradients, curvature etc.).  Solutions
are found by adjusting the multiplier $\lambda$ until the goodness of fit
satisfies $\chi^2 \simeq N_{dof}$ where $N_{dof}$ is the number of degrees of freedom.
Another solution is to use linear programming methods to impose constraints
such as positive surface densities, negative density gradients from the lens
center or density symmetries (Saha \& Williams~\cite{Saha1997p148}, \cite{Saha2004},
Williams \& Saha~\cite{Williams2000p439}).  Time delays, which are also linear
functions of the surface density,  are easily included.  Flux ratios are 
more challenging because magnifications are quadratic rather than linear
functions of the surface density except for the special case of the generalized
singular isothermal models where $ \Psi = b\theta F(\chi)$ 
(Eqn.~\ref{eqn:wevans}, Witt, Mao \& Keeton~\cite{Witt2000p98},
Kochanek et al.~\cite{Kochanek2001p50}, Evans \& Witt~\cite{Evans2001p1260}).
The best developed, publicly available non-parametric models are those
by Saha \& Williams~(\cite{Saha2004}).  These are available at 
http://ankh-morpork.maths.qmc.ac.uk/$\sim$saha/astron/lens/.

Personally, I am not a fan of the non-parametric models, because
almost all the additional degrees of freedom they include are irrelevant to the 
problem. As I have tried to outline in the preceding sections, there is no
real ambiguity about the aspects of gravitational potentials either constrained
or unconstrained by lens models.  Provided the parametric models capture these
degrees of freedom and you do not get carried away with the precision of the
fits, you can ignore deviations of the $\cos(16\chi)$ term of the surface density
from that expected for an ellipsoidal model.  Similarly
for the monopole profile, the distribution of mass interior and exterior to the   
images is irrelevant and for the most part only the mean surface density between
the images has any physical effect. Nothing is gained by allowing arbitrary,
fine-grained distributions.

There are also specific physical and mathematical problems with non-parametric models
just as there are for parametric models.  First,
the trick of linearization only works if the lens equations are solved on the
source plane.  As we discussed when we introduced model fitting (\S\ref{sec:modelfit}), 
this makes it impossible to properly compute error bars on any parameters. The equations
become non-linear if they include either the magnification tensor 
(Eqn.~\ref{eqn:fit1}) or use the true image plane fit statistic (Eqn.~\ref{eqn:fit2}), 
and this greatly reduces the attractiveness of these methods.  Second, in
many cases the non-parametric models are not constrained to avoid creating 
extra images not seen in the observations -- the models reproduce the observed 
images exactly,
but come with no guarantee that they are not producing 3 other images somewhere
else.  Third, it is very difficult to guarantee that the resulting models are
physical.  For example, consider a simple spherical lens constrained to have
positive surface density.  For the implied three-dimensional density to also
be positive definite, the surface density must decline monotonically from the
center of the lens.  This constraint is usually applied by the 
Saha \& Williams~(\cite{Saha2004}) method.  
For the distribution function of the stars making up the galaxy to be positive
definite, the three dimensional density must also decline monotonically --
this implies a constraint on the second derivative of the surface density 
which is not imposed by any of these methods.   For the distribution to be 
dynamically stable it must satisfy a criterion on the derivative of the 
distribution function with respect to the orbital energy, and this implies a
criterion on the third derivative of the surface density which is also not 
imposed (see Binney \& Tremaine~\cite{Binney1987p747}).  
Worse yet, for a non-spherical system we cannot even write down the
constraints on the surface density required for the model to correspond to a
stable galaxy with a positive definite distribution function.  In short, most
non-parametric models will be unphysical -- they overestimate the degrees of 
freedom in the mass distribution.
The critique being made, parametric models have a role because they define the
outer limits of what is possible by avoiding the strong physical priors
implicit in parametric models of galaxies.

\begin{figure}[t]
\centerline{\psfig{figure=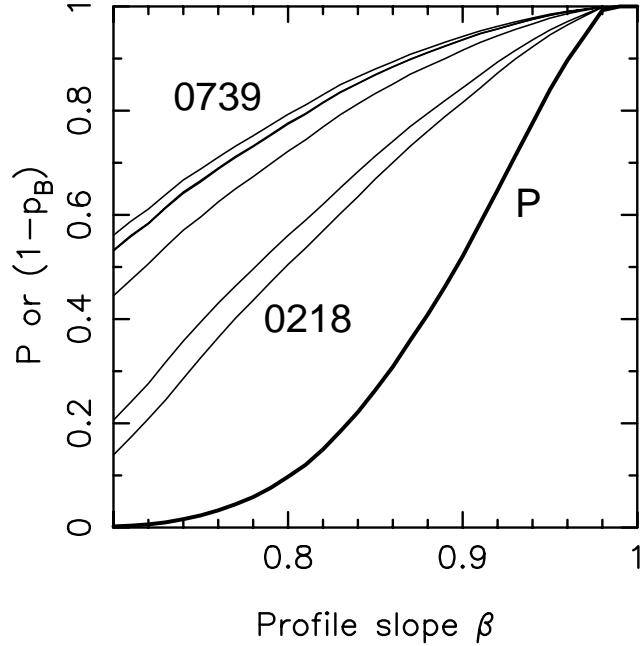,width=3.3in}}
\caption{Limits on the central density exponent for power-law density profiles
  $\rho \propto r^{-n} = r^{-1-\beta}$ from the absence of detectable 
  central images in
  a sample of 6 CLASS survey radio doubles (Rusin \& Ma~\cite{Rusin2001p33}).  
  The lighter curves
  show the limits for the individual lenses with the weakest constraint from
  B0739+366 and the strongest from B0218+357, and the heavy solid curve shows 
  the joint probability $P$.  }
\labelprint{fig:rusincusp}
\end{figure}

\begin{figure}[ph]
\begin{center}
\centerline{\psfig{figure=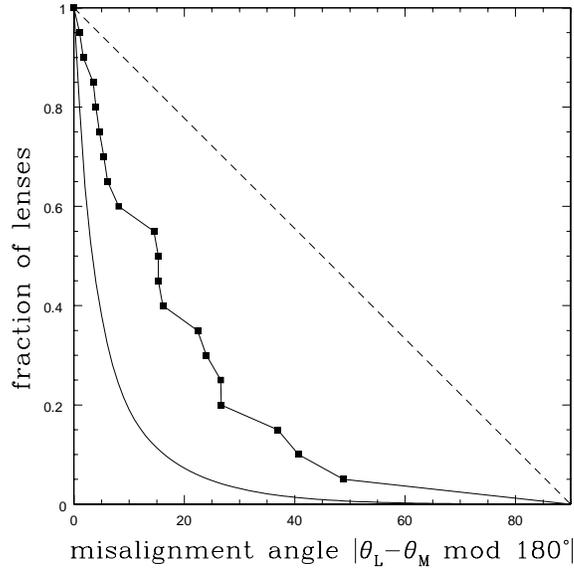,width=3.1in}}
\end{center}
\begin{center}
\centerline{\psfig{figure=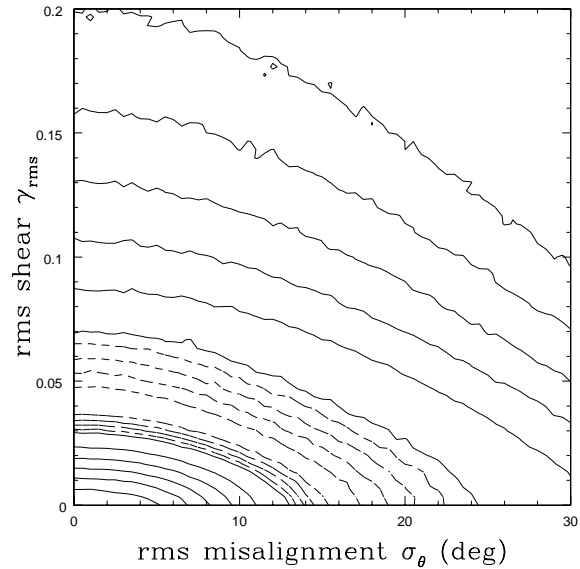,width=3.1in}}
\end{center}
\caption{ (Top)
   The integral distribution of misalignment angles $\Delta\chi_{LM}$ 
   between the major axes of the lens galaxy and an ellipsoidal lens 
   model (solid curve with points for each lens).  If the two angles
   were completely uncorrelated, the distribution would follow the dashed 
   line.  If the two angles were perfectly correlated they would follow the
   solid curve because of the measurement uncertainties in the two angles.  
   }
\vspace{-0.05in}
\labelprint{fig:statmod1a}
\caption{ (Bottom)
   Logarithmic contours of the probability for matching the distribution of
   misalignment angles as a function of the rms misalignment $\sigma_\theta$
   between the mass and the light and the typical tidal shear $\gamma_{rms}$.
   Theoretically we expect tidal shears $\gamma_{rms} \simeq 0.06$.  The solid
   contours are spaced by 0.5 dex and the dashed contours are spaced by 0.1 dex
   relative to the maximum likelihood contour.  The differences between dashed
   contours are not statistically significant, while those between solid contours
   are statistically significant.
   }
\labelprint{fig:statmod1b}
\end{figure}

\begin{figure}
\centerline{\psfig{figure=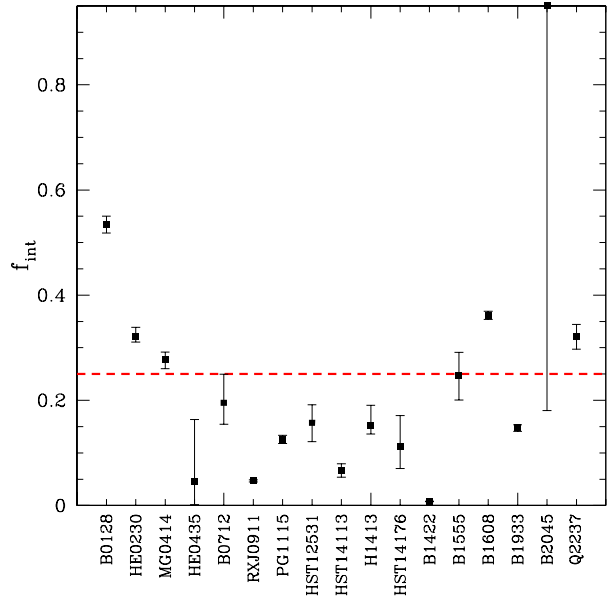,width=3.3in}}
\caption{ The internal shear fraction $f_{int}$ for the 
   four-image lenses.  Each system was fitted by an SIS combined with an internal shear 
   and an external shear and $f_{int}=|\Gamma|/(|\Gamma|+|\gamma|)$ is the fraction
   of the quadrupole amplitude due to the internal shear.  An SIE has $f_{int}=1/4$
   (see Fig~\ref{fig:quadmom}).  Most of the quads have $f_{int} \ltorder 1/4$ as
   expected for an SIE in an additional external (tidal) shear field.  
   Objects with very low $f_{int}$ (e.g. HE0435--1223, RXJ0911+0551, B1422+231) have nearby
   galaxies or clusters generating anomalously large external shears, while 
   objects with anomalously high $f_{int}$ (B1608+656, HE0230--2130, MG0414+0534) tend to have
   additional lens components like the second lens galaxy of B1608+656.  For some
   systems either the imaging data (e.g. B0128+437) or the models (e.g. B2045+265) 
   do not allow a clear qualitative explanation.
   }
\labelprint{fig:intfrac}
\end{figure}

\begin{figure}
\centerline{\psfig{figure=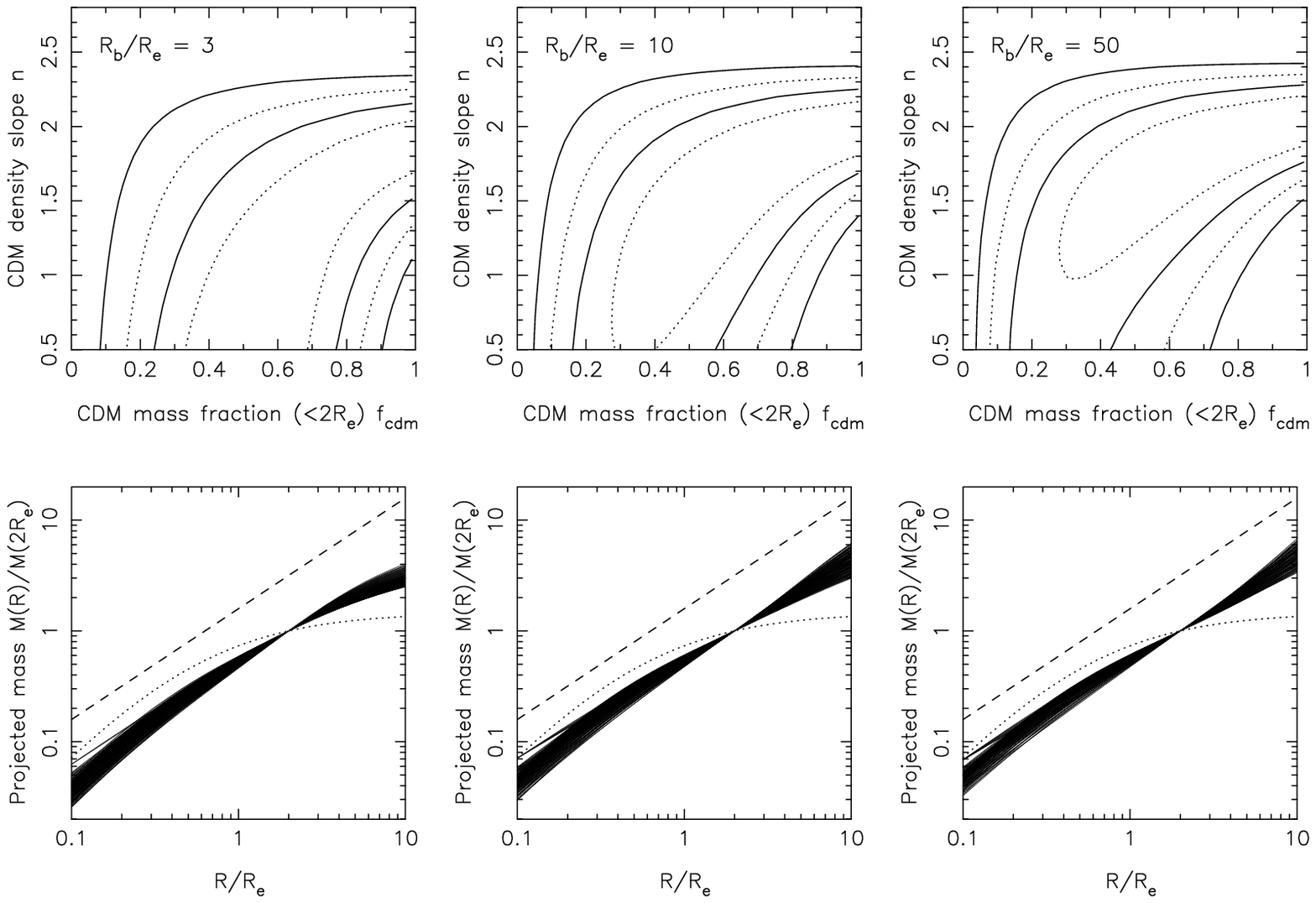,width=4.5in}}
\caption{  The structure of lens galaxies in self-similar models.  The top row
   shows the permitted region for the slope of the inner dark matter cusp
   ($\rho \propto r^{-n}$) and the projected fraction of the mass $f_{CDM}$ inside $2R_e$
   composed of dark matter.  The results are shown for three ratios $R_b/R_e$
   between the break radius $R_b$ of the dark matter profile and the effective
   radius $R_e$ of the luminous galaxy.  The solid (dashed) contours show the 68\% and 95\%
   confidence levels for two (one) parameter.  Note that the estimates of
   $n$ and $f_{CDM}$ depend little on the location of the break radius relative
   to the effective radius.  The bottom row shows all the mass profiles lying
   within the (two parameter) 68\% confidence region normalized to a fixed projected
   mass inside $2R_e$.  For comparison we show the mass enclosed by a de Vaucouleurs
   model (dotted line) and an SIS (offset dashed line).  While the allowed models
   exhibit a wide range of dark matter abundances, slopes and break radii, they
   all have roughly isothermal total mass profiles over the radial range spanned
   by the lensed images.
   }
\labelprint{fig:selfsim1}
\end{figure}

\subsection{Statistical Constraints on Mass Distributions \labelprint{sec:statfit} }

Where individual lenses may fail to constrain the mass distribution, ensembles
of lenses may succeed.  There are two basic ideas behind statistical constraints
on mass distributions.  The first idea is that models of individual lenses 
should be weighted by the likelihood of the observed configuration given the
model parameters.  The second idea is that the statistical properties of 
lens samples should be homogeneous.

An example of weighting models by the likelihood is the limit on the slopes
of central density cusps from the observed absence of central images.
Rusin \& Ma (\cite{Rusin2001p33}) considered 6 CLASS (see \S\ref{sec:stat})
survey radio
doubles and computed the probability $p_i(n)$ that lens $i$ would have
a detectable third image in the core of the lens assuming power law mass
densities $\Sigma \propto R^{1-n}$ and including a model for the
observational sensitivities and the magnification bias (see \S\ref{sec:magbias}) of the
survey.  They were only interested in the range $n<2$, because
as discussed in \S\ref{sec:basics}, density cusps with $n \geq 2$ never have central
images.  For most of the lenses they considered, it was possible to find
models of the 6 lenses that lacked detectable central images over a broad
range of density exponents.  However, the shallower the
cusp, the smaller the probability $p_i(n)$ of producing a lens without a visible 
central image.  For any single lens, $p_i(n)$ varies too little to set
a useful bound on the exponent, but the joint probability of the entire
sample having no central images, $P=\Pi_i (1-p_i(n))$, leads to a
strong (one-sided) limit that $n > 1.78$ at 95\% confidence 
(see Fig.~\ref{fig:rusincusp}).   
In practice, Keeton~(\cite{Keeton2003p17}) demonstrated that the central stellar densities
are sufficiently high to avoid the formation of visible central images in
almost all lenses given the dynamic ranges of existing radio observations
(i.e. stellar density distributions are sufficiently cuspy), and central
black holes can also assist in suppressing the central image (Mao, Witt \& 
Koopmans~\cite{Mao2001p301}).  However, the basic
idea behind the Rusin \& Ma (\cite{Rusin2001p33}) analysis is important and underutilized.

An example of requiring the lenses to be homogeneous is the estimate of 
the misalignment between the major axis of the luminous lens galaxy and
the overall mass distribution by Kochanek (\cite{Kochanek2002p62}).  
Fig.~\ref{fig:statmod1a} shows the misalignment angle $\Delta\chi_{LM}=|\chi_L-\chi_M|$
between the major axis $\chi_L$ of the lens galaxy and the major axis
$\chi_M$ of an ellipsoidal mass model for the lens.  The particular mass
model is unimportant because any single component model of a four-image lens
will give a nearly identical value for $\chi_M$ (e.g. Kochanek~\cite{Kochanek1991p354},
Wambsganss \& Paczynski~\cite{Wambsganss1994p1156}).  
The distribution of the 
misalignment angle $\Delta\chi_{LM}$ is not consistent with the mass and
the light being either perfectly correlated or uncorrelated.  This is not 
surprising, because a simple ellipsoidal model determines the position angle of the mean
quadrupole moment near the Einstein ring, which is a combination of the
quadrupole moment of the lens galaxy, the halo of the lens galaxy, and 
the local tidal shear (see \S\ref{sec:massquad}).  Even if the lens galaxy and the halo 
were perfectly aligned, we would still find that the orientation of the mean quadrupole
would differ from that of the light because of the effects of the tidal
shears.  We can model this by estimating the probability of reproducing
the observed misalignment distribution in terms of the strength of the
local tidal shear $\gamma_{rms}$ and the dispersion in $\sigma_\chi$ in the
angle between the major axis of the mass distribution and the light, as
shown in Fig.~\ref{fig:statmod1b}.   The observed mismatch can either 
be produced by having a typical tidal shear of $\gamma_{rms} \simeq 0.05$ 
or by having a typical misalignment between mass and light of 
$\sigma_\chi \simeq 20^\circ$.  We know, however, that the typical
tidal shear cannot be zero because it can be estimated from the statistics
of galaxies (e.g. Keeton, Kochanek \& Seljak~\cite{Keeton1997p604}, 
Holder \& Schechter~\cite{Holder2003p688}).
Keeton et al. (\cite{Keeton1997p604}) obtained $\gamma_{rms} \simeq 0.05$, 
in which case mass must align with light and we obtain
an upper limit of $\sigma_\chi \ltorder 10^\circ$.  
Holder \& Schechter~(\cite{Holder2003p688})
argue for a much higher rms shear of $\gamma_{rms} = 0.15 $ based on N-body
simulations, which is too high to be consistent with the observed alignment
of mass models and the luminous galaxy.  One possible explanation
(based on the results of White, Hernquist \& Springel~\cite{White2001p1}) is
that Holder \& Schechter~(\cite{Holder2003p688}) included parts 
of the lens galaxy's own halo in their estimate of the external shear.  
Alternatively, if lens galaxies are more compact than the SIE model
used by Kochanek (\cite{Kochanek2002p62}), then the lower surface 
density $\kbar$ raises the required shear (since $\gamma \propto (1-\kbar)$, 
Eqn.~\ref{eqn:aca}).  However, mass distributions similar to constant mass-to-light
ratio models of the lenses would be required, which would be inconsistent
with shear estimates from simulations in which galaxy masses are 
dominated by extended dark matter halos.

The trade-off between central concentration and shear leads to the 
the interesting question of where the quadrupole structure of
lenses originates.  As we discussed in \S\ref{sec:massquad}, we can break up the quadrupole
of the mass distribution into the internal quadrupole due to the matter 
interior to the Einstein ring (Eqn.~\ref{eqn:adl}) and the exterior quadrupole 
due to the matter outside the Einstein ring (Eqn.~\ref{eqn:acc}).  
While the internal quadrupole is due only to
the lens galaxy, the external quadrupole is a mixture of the quadrupole from
the parts of the galaxy outside the Einstein ring (i.e. the dark matter 
halo) and the tidal shear from the environment.  An important fact to 
remember is that for an isothermal ellipsoid, only $f_{int}=25\%$ of the quadrupole
is due to mass inside the Einstein ring (see Fig.~\ref{fig:quadmom}, \S\ref{sec:massquad})!  
Turner, Keeton \& Kochanek~(\cite{Turner2004}) explored this by fitting all
the available four-image lenses with an SIS monopole combined with an internal
and an external quadrupole.  They then computed the fraction of the quadrupole
$f_{int}$ associated with the mass interior to the Einstein ring to find the distribution
shown in Fig.~\ref{fig:intfrac}.  Most four-image lenses seem to be dominated by
the external quadrupole, with internal quadrupole fractions below the 
$f_{int}=0.25$ fraction expected for an isothermal ellipsoid.  Lenses clearly
in environments with very large tidal shears (e.g. RXJ0911+0551 which is 
near massive cluster, Bade et al.~\cite{Bade1997p13}, Kneib et al.~\cite{Kneib2000p35},
Morgan et al.~\cite{Morgan2001p1} or 
HE0435--1223 which is near a large galaxy, Wisotzki et al.~\cite{Wisotzki2002p17},
see Fig.~\ref{fig:basic4b}) show much smaller internal shear fractions.
B1608+656 (Myers et al.~\cite{Myers1995p5}, 
Fassnacht et al.~\cite{Fassnacht1999p498}), which has two lens galaxies inside
the Einstein ring, shows a significantly higher internal quadrupole fraction.
Combined with
the close correlation of mass model alignments with the luminous galaxies, this
seems to argue for significant dark matter halos aligned with the luminous
galaxy, but the final step of quantitatively assembling all the pieces has
yet to be done.

Statistical analyses can also be used to estimate the radial density distribution
from samples of lenses which individually cannot.
The existence of the fundamental plane (see \S\ref{sec:optical}) strongly suggests that the 
structure of early-type galaxies is fairly homogeneous -- in particular it is
consistent with galaxies having self-similar mass distributions 
in the sense that the halo structure can be scaled from the structure
of the visible galaxy.  As a particular example based on our theoretical
expectations, Rusin, Kochanek \& Keeton~(\cite{Rusin2003p29}) and Rusin \& 
Kochanek~(\cite{Rusin2004p1}) modeled the visible galaxy with a 
Hernquist (Eqn.~\ref{eqn:abn}) model scaled to match the observed effective radius 
of the lens galaxy, $R_e$, and then added a cuspy dark matter halo (Eqn.~\ref{eqn:abq} with 
a variable inner cusp $\eta$, $\alpha=2$ and $m=3$) where the inner density cusp 
($\rho \propto r^{-\eta}$), the halo break radius $r_b$ and the dark
matter fraction $f_{CDM}$ inside $2R_e$ were kept as variables.  The
assumption of self-similarity enters by keeping the ratio $r_b/R_e$ constant,
the dark matter fraction $f_{CDM}$ constant, and then scaling the mass-to-light
ratio of the stars $\Upsilon \propto L^x$ with the luminosity.\footnote{
They could also have allowed the CDM fraction to vary as $f_{CDM} \propto L^y$,
but these led to degenerate models where only the combination $x+y$ was
constrained.}  We recover the fundamental plane in this model when 
$x \simeq 0.25$.  Putting all the pieces together,
the projected mass inside radius $R$ is 
\begin{equation}
   M(<R) = \Upsilon_* L_* \left( { L(0) \over L_* } \right)^{1+x}
  \left[ g(R/R_e) + g(2) { f_{CDM} \over 1-f_{CDM} } m_{CDM}(R/R_e) \right]
   \labelprint{eqn:selfsimmass}
\end{equation}
where $\Upsilon_*$ is the mass-to-light ratio of the stars in an $L_*$ galaxy,
$\log L(0)= \log L(z) - e(z)$ is the luminosity of the lens galaxy evolved to
redshift zero (where we discuss estimates of the evolution rate $e(z)$ in
\S\ref{sec:optical}), $g(x)$ is the fraction of the light inside dimensionless radius $x=R/R_e$
($g(1)=1/2$) and $m_{CDM}(x)$ is the dimensionless dark matter mass inside radius $x$
with $m_{CDM}(2)=1$ so that the CDM mass fraction inside $x=2$ is $f_{CDM}$.

As we discussed earlier in \S\ref{sec:modelfit}, few lenses have sufficient constraints
to estimate all the parameters in such a complex model.  However, the assumption of 
self-similarity allows the average profile to be constrained statistically
(Rusin et al. 2003, 2004).  Suppose we saw lensed images generated by the 
same galaxy at a range of different source and lens redshifts.  Each observed
lens only reliably measures an aperture mass $M_{ap}(R<R_{Ein})=\pi \Sigma_c R_{Ein}^2$ where
$R_{Ein}$ is the Einstein radius.  But the physical scale $R_{Ein}$ varies
with redshift, so the ensemble of the lenses traces out the overall mass
profile.  Clearly we do not have ensembles of lenses generated by identical galaxies,
but the assumption of self-similarity allows us to use the same idea for lenses 
with a range of luminosities and scale lengths.  
For 22 lenses with redshifts and accurate photometry we compared
the measured aperture masses to the predicted aperture masses (the procedure
for two-image lenses is a little more complicated, see Rusin et al.~\cite{Rusin2003p29}) 
to estimate all the model parameters.  Fig.~\ref{fig:selfsim1} shows the
results for the parameters associated with the dark matter halo.  In the 
limit that $f_{CDM}\rightarrow 1$ we find that the mass distribution is
consistent with a simple SIS model (the limit $f_{CDM}\rightarrow 1$ and
$n\rightarrow 2$) almost independent of the break radius location.  There is
a slight trend with break radius because as the break to the steep $\rho \propto r^{-3}$
outer profile gets closer to the region with the lensed images the inner
cusp can be shallower while keeping the overall profile over the region with
images close to isothermal.  As we reduce $f_{CDM}$ and add mass to the stars,
the inner cusp becomes shallower, such that for a NFW ($n=1$) cusp the 
dark matter fraction inside $2R_e$ is $\sim 40$\%.  It is interesting to
note, however, that the total mass distribution (light + dark) changes little
over the full range of allowed parameters (bottom panels of Fig.~\ref{fig:selfsim1}) --
lensing constrains the global mass distribution not how it is divided into
luminous and dark subcomponents.  Note the resemblance of the statistical
results to the results for detailed models of B1933+503 in Fig.~\ref{fig:mod1933b}.  

\subsection{Stellar Dynamics and Lensing \labelprint{sec:dynamics} }

Stellar dynamical analyses of gravitational lenses have reached the level of
studies of local galaxies approximately 15--20 years ago.  The analyses are based
on the spherical Jeans equations (see Binney \& Tremaine~\cite{Binney1987p747}) 
with simple models of the orbital
anisotropy and generally ignore  both deviations from sphericity and higher order
moments of the velocity distributions.  The spherical Jeans equation
\begin{equation}
    { 1 \over \nu } { d \nu \sigma_r^2 \over d r} + { 2 \beta(r) \over r }\sigma_r^2
    = - { G M(r) \over r^2}
  \labelprint{eqn:acd}
\end{equation}
relates the radial velocity dispersion $\sigma_r=\langle v_r^2\rangle^{1/2}$, 
the isotropy parameter $\beta(r)=1-\sigma_\theta^2/\sigma_r^2$ characterizing the
ratio of the tangential dispersion to the radial dispersion, the luminosity density
of the stars $\nu(r)$ and the mass distribution $M(r)$.  A well known
result from dynamics is that you cannot infer the mass distribution 
$M(r)$ without constraining the isotropy $\beta(r)$ (e.g. Binney \& Mamon~\cite{Binney1982p361}). 
Models with $\beta=0$ are called
isotropic models (i.e. $\sigma_r=\sigma_\theta$), while models with
$\beta \rightarrow 1$ are dominated by radial orbits ($\sigma_\theta \rightarrow 0$) and models
with $\beta \rightarrow -\infty$ are dominated by tangential orbits ($\sigma_r \rightarrow 0$) .
These 3D components of the velocity dispersion must then be projected
to measure the line-of-sight velocity dispersion $\langle v_{los}^2\rangle^{1/2}$,
\begin{equation}
    \Sigma(R) \langle v_{los}^2\rangle (R) = 2 \int_0^\infty dz \nu 
          \left[  {z^2 \over r^2} \sigma_r^2 + {R^2 \over r^2} \sigma_\theta^2 \right]
    = 2 \int_0^\infty dz \nu \sigma_r^2 
          \left[  1 - {R^2 \over r^2} \beta(r) \right] \labelprint{eqn:ace}
\end{equation}
where $\Sigma(R) = 2 \int_0^\infty dz \nu(r)$ is the projected surface
brightness and $z$ is the coordinate along the line of sight.  

Modern
observations of local galaxies break the degeneracy between mass and
isotropy by measuring higher order moments ($\langle v_{los}^n\rangle$)
of the line-of-sight velocity distribution (LOSVD) because the shape
of the LOSVD is affected by the isotropy of the orbits.  Because the
velocity dispersions are measured starting from a Gaussian fit to the
LOSVD, the higher order moments are described by the amplitudes $h_n$
of a decomposition of the LOSVD into Gauss-Hermite polynomials
(e.g. van der Marel \& Franx~\cite{van_der_Marel1993p525}).  In general, the
rms velocity (i.e. combining dispersion and rotation) and higher order
moment profiles of early-type galaxies are fairly self-similar, with 
nearly flat rms velocity profiles, modest values of $h_4 \simeq 0.01 \pm 0.03$
and slightly radial orbits $\langle\beta\rangle \simeq 0.1$--$0.2$
(e.g. Romanowsky \& Kochanek~\cite{Romanowsky1999p18}, Gerhard et al.~\cite{Gerhard2001p1936}).

Stellar dynamics is used for two purposes in lensing studies.  The first is to
provide a mass normalization for lens models used in studies of lens statistics.
We will discuss this in \S\ref{sec:stat}.  The second is to use comparisons
between a mass estimated from the geometry of a lens and the velocity dispersion
of the lens galaxy to constrain the mass distribution (e.g. Romanowsky \& 
Kochanek~\cite{Romanowsky1999p18}, Trott \& Webster~\cite{Trott2002p621},
Koopmans \& Treu~\cite{Koopmans2002p5}, \cite{Koopmans2003p606},
Treu \& Koopmans~\cite{Treu2002p6}, \cite{Treu2002p87}, 
Koopmans et al.~\cite{Koopmans2003p70}).  It is important to understand that
the systematic uncertainties in combining lensing and stellar dynamics to 
determine mass distributions are different from using either in isolation.
For local galaxies we measure a velocity dispersion profile. The 
normalization of the profile sets the mass scale and the changes in
the profile (and any higher order moments) with radius constrains the mass 
distribution.  To lowest order, a simple scaling error in the velocity
measurements will lead to errors in the mass scale rather than in 
the mass distribution.  For lens galaxies, it is the comparison between
the velocity dispersion and the mass determined by the geometry of the
images that constrains the mass distribution.  Thus, estimates of
the mass distribution are directly affected by any calibration errors
in the velocity dispersions.  

We can understand the differences with a simple thought experiment.  Suppose
we have a mass distribution $M = M_0 (R/R_0)^x$ in projection and we have
mass estimates $M_1$ at $R_1$ and $M_2$ at $R_2$.  Combining them we can
solve for the exponent describing the mass distribution, 
$x=\ln(M_1/M_2)/\ln(R_1/R_2)$.  In a dynamical observation the mass
estimate is some sort of virial estimator $M \propto \sigma_v^2 R/G$
while in a lensing measurement it is a direct measurement of $M$.  Standard
velocity dispersion measurements start from the best fit Gaussian
line width $\hat{\sigma}$ and then subtract an intrinsic line width
$\sigma_c$ due to the instrument and the intrinsic line width of the
star in quadrature to estimate the portion of the line width due to
the motions of the stars.  Thus $\sigma_v^2=f^2(\hat{\sigma}^2-\sigma_c^2)$
where $f \simeq 1 $ is a scale factor to account for deviations from 
spherical symmetry and non-Gaussian line of sight velocity distributions (LOSVDs).
In a purely dynamical study, uncertainties in $f$ and $\sigma_c$ produce
bigger fractional errors in the absolute mass scale $M_0$ than in the exponent
$x$. For example, given measurements $\sigma_1$ and $\sigma_2$ at radii $R_1$ and $R_2$,
the exponent, $x=1+\ln(\sigma_1^2/\sigma_2^2)/\ln(R_1/R_2)$, depends only
on velocity dispersion ratios in which calibration errors tend to cancel. 
This is obvious for the scale factor $f$, which cancels exactly if it does
not vary with radius.  Since studies of lens dynamics use a comparison
between a dynamical mass and a lensing mass to estimate the mass distribution, 
the results are more sensitive to calibration problems because these 
cancellations no longer occur.  If we combine a velocity dispersion 
measurement $\sigma_1$ with a lensing mass measurement $M_2$ our 
estimate of the exponent becomes $x=\ln(\sigma_1^2 R_1/GM_2)/\ln(R_1/R_2)$
and the uncertainties are linear in the scale factor $f$ rather than
canceling.  An error analysis for the effects of $\sigma_c$ is messier,
but you again find that the sensitivity in the mixed lensing and dynamics
constraint to errors in $\sigma_c$ is greater than in a purely dynamical
study.

Velocity dispersions have now been measured for 10 lenses 
(0047--2808 Koopmans \& Treu~\cite{Koopmans2003p606}; 
CFRS03.1077 Treu \& Koopmans~\cite{Treu2004p1};
Q0957+561 Falco et al.~\cite{Falco1997p70}, Tonry \& Franx~\cite{Tonry1999p512}; 
PG1115+080~Tonry~\cite{Tonry1998p1});
HST14176+5226 Ohyama et al.~\cite{Ohyama2002p2903}, Gebhardt et al.~\cite{Gebhardt2003p1},
   Treu \& Koopmans~\cite{Treu2004p1};
HST15433+5352  Treu \& Koopmans~\cite{Treu2004p1};
MG1549+3047 Leh\'ar et al.~\cite{Lehar1996p1812};
B1608+656 Koopmans et al.~\cite{Koopmans2003p70};
MG2016+112 Koopmans \& Treu~\cite{Koopmans2002p5};
Q2237+0305 Foltz et al.~\cite{Foltz1992p43}).  With the exception
of Romanowsky \& Kochanek~(\cite{Romanowsky1999p18}), who fitted for the
distribution function of the lens,
the analyses of the data have used the spherical
Jeans equations with parameterized models for the isotropy $\beta(r)$.
They include the uncertainties in $\sigma_c$ about as
well as any other dynamical study, although it is worth bearing in mind that
this is tricky because we lack nearby stars with the appropriate metallicity
and the problem of matching the spectral resolution for the galaxy and the
template stars lacks direct checks of the success of the procedure.  A useful
rule of thumb to remember is that repeat measurements of velocity dispersions
by different groups almost always show larger scatter than is consistent with
the reported uncertainties.  For example, the three velocity dispersion 
measurements for the lens HST14176+5226 
($224\pm15$~km/s by Ohyama et al.~\cite{Ohyama2002p2903}, 
$202\pm9$~km/s by Gebhardt et al.~\cite{Gebhardt2003p1}, and $230\pm14$~km/s
by Treu \& Koopmans~\cite{Treu2004p1}) are mutually consistent only if the
uncertainties are broadened by 30\%.

\begin{figure}[p]
\centerline{ \psfig{figure=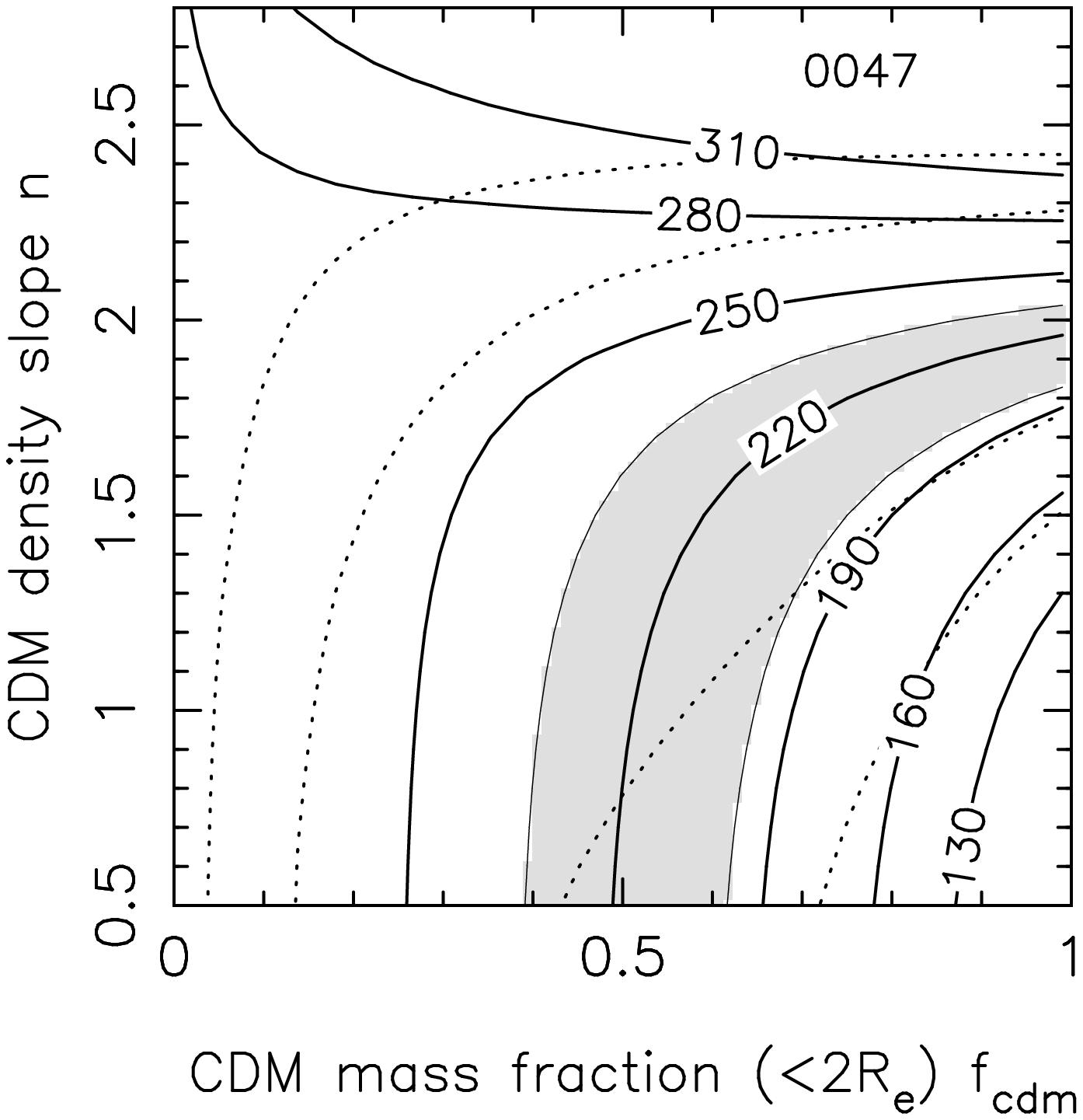  ,width=1.5in}
             \psfig{figure=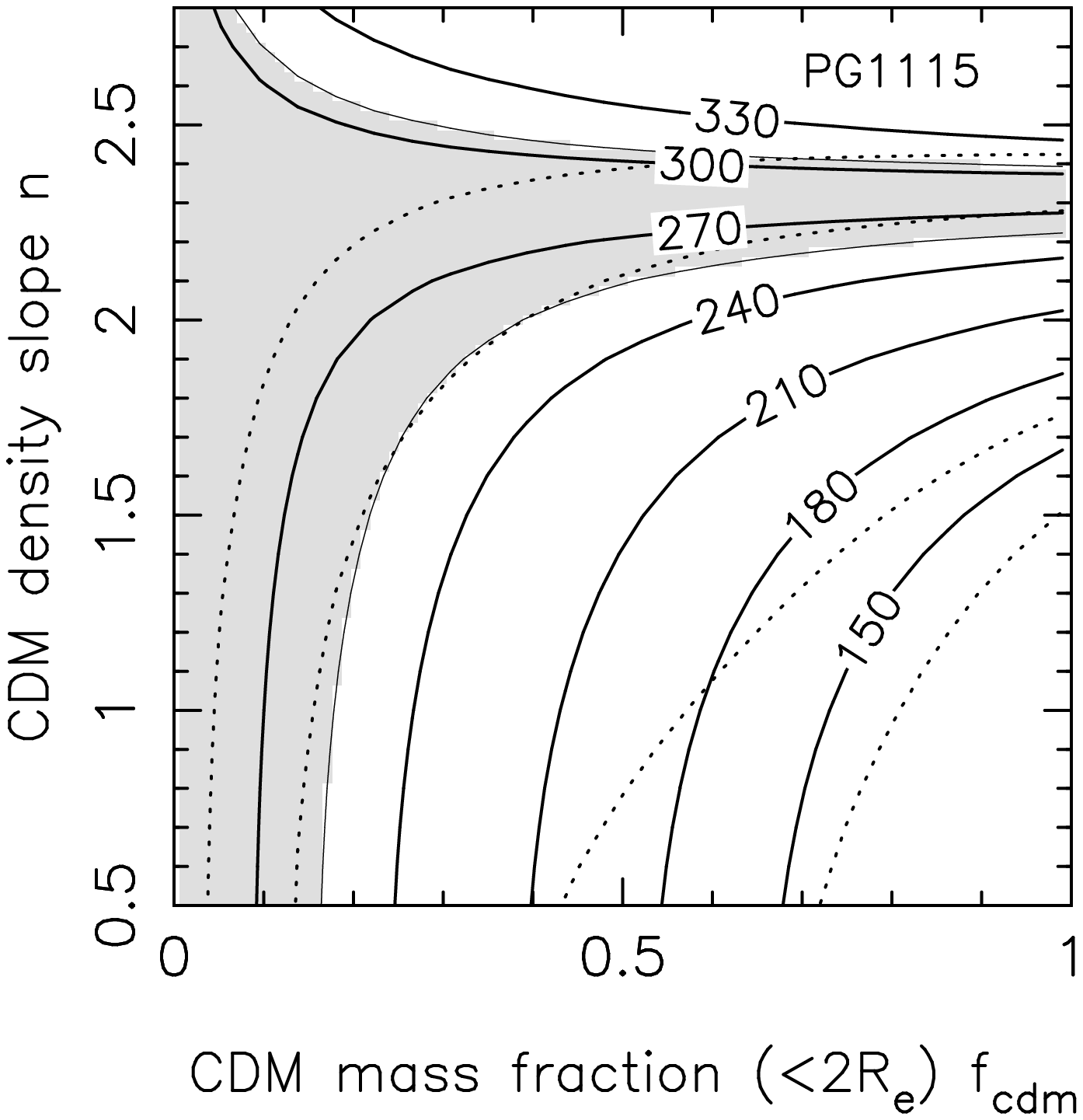  ,width=1.5in}
             \psfig{figure=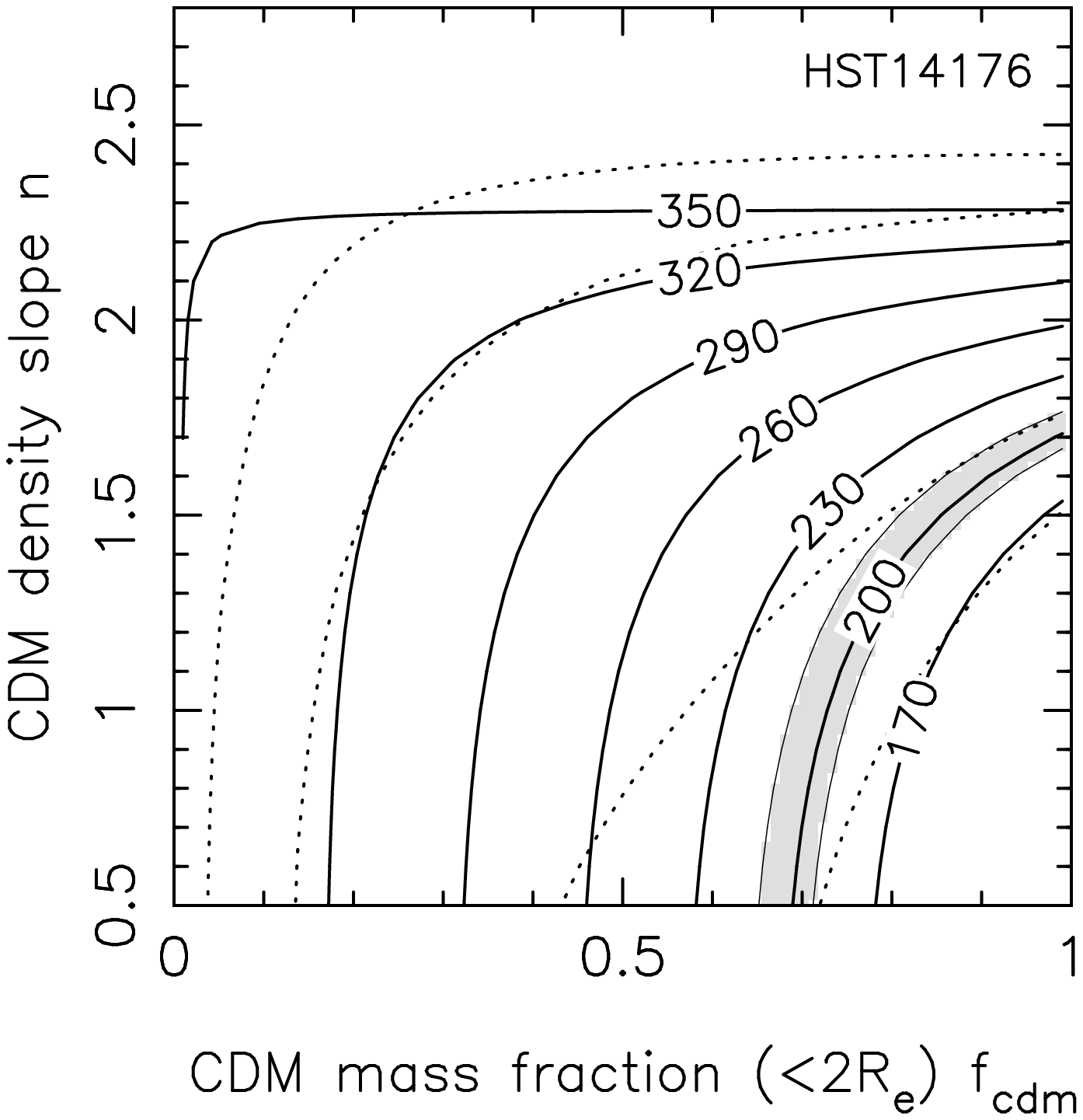 ,width=1.5in} }
\centerline{ \psfig{figure=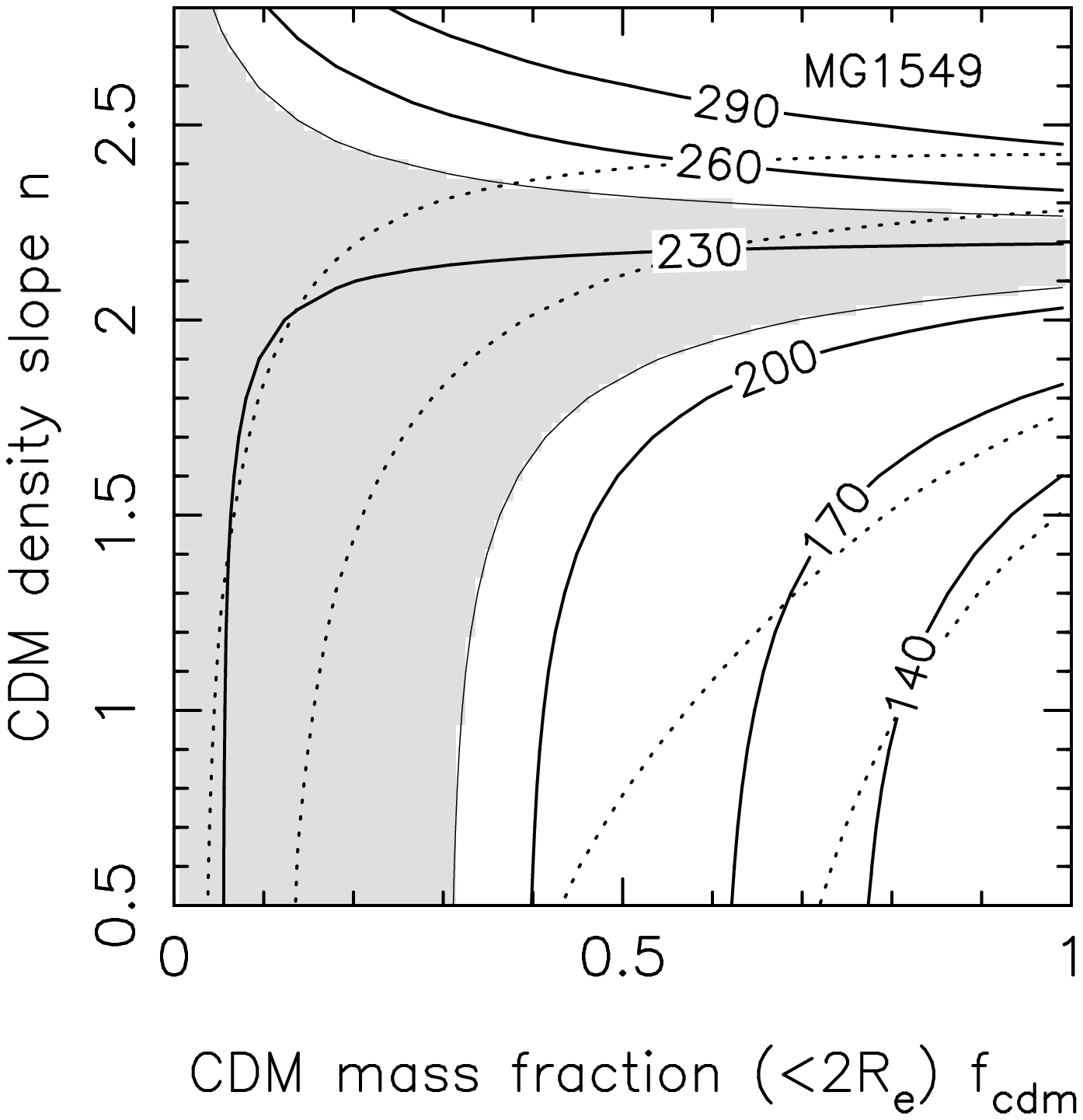  ,width=1.5in}
             \psfig{figure=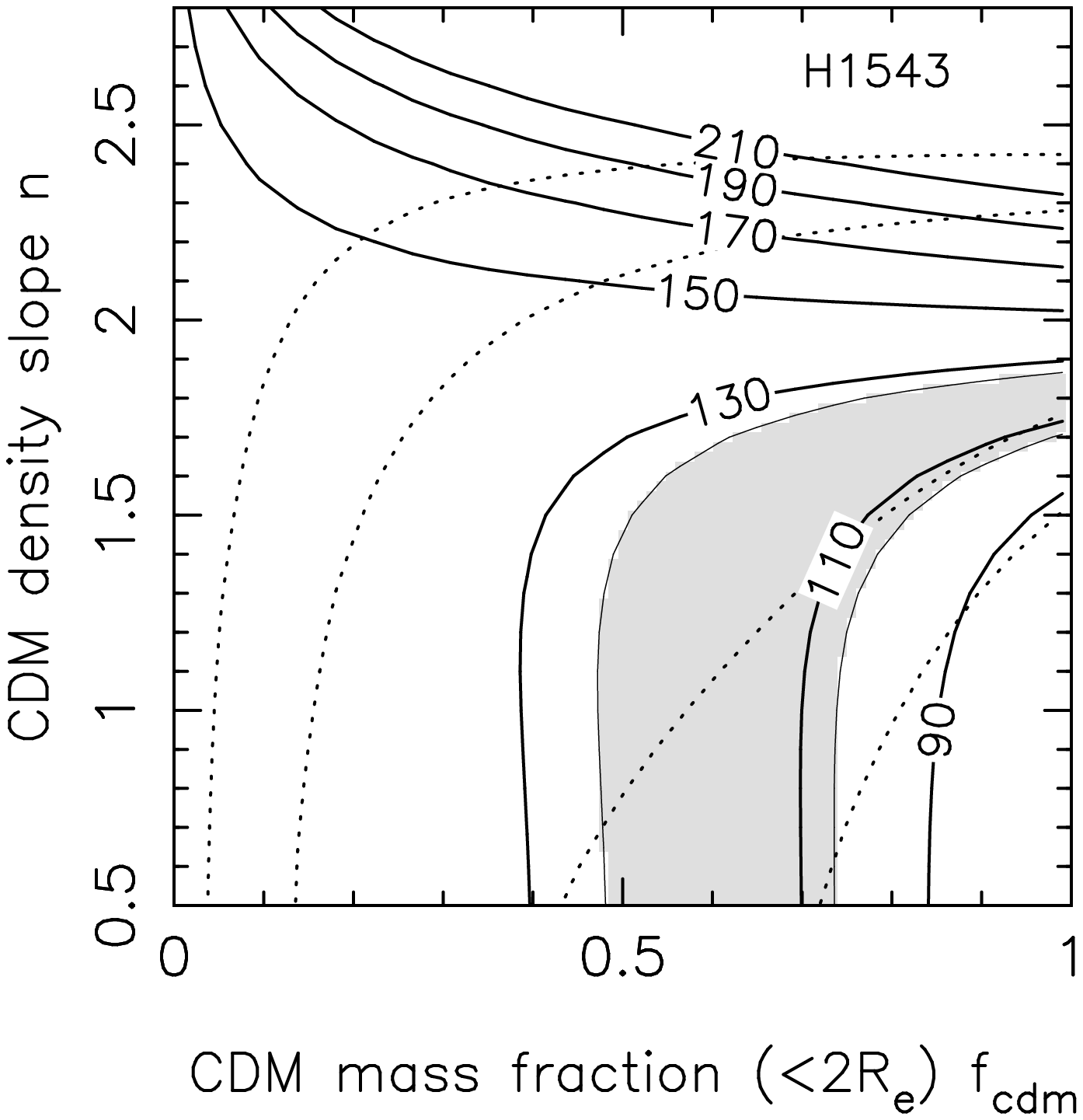 ,width=1.5in}
             \psfig{figure=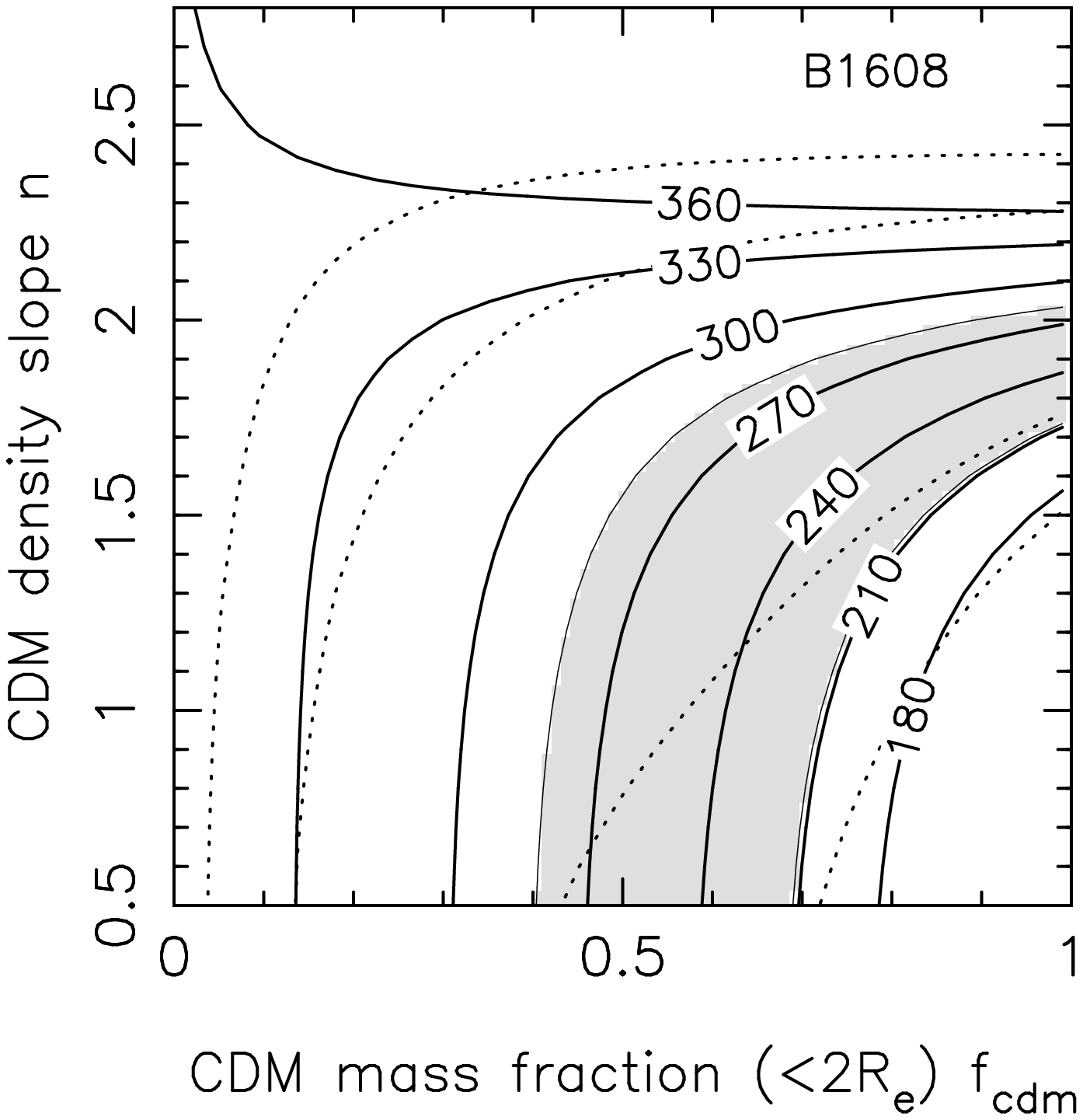  ,width=1.5in} }
\centerline{ \psfig{figure=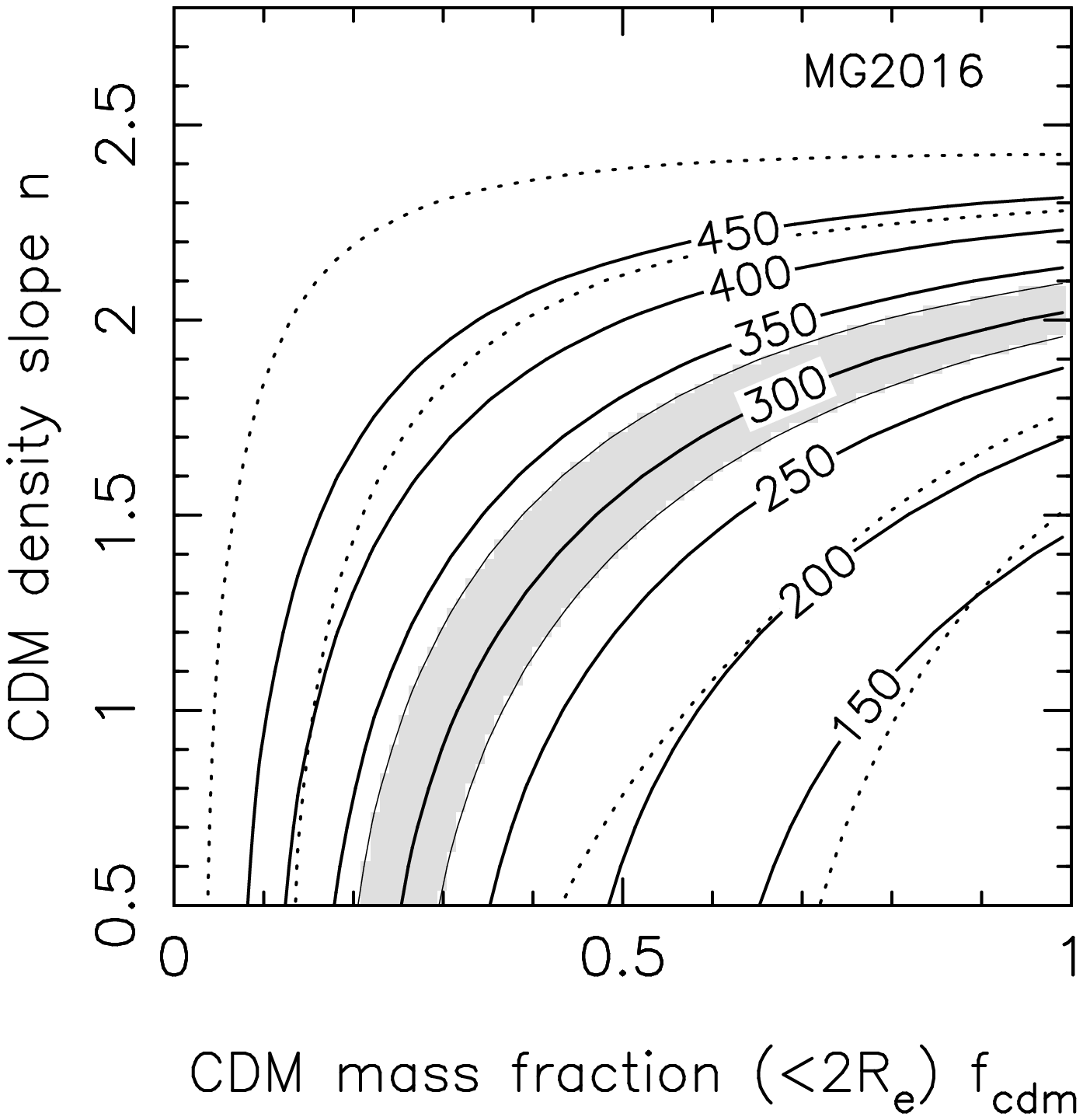  ,width=1.5in}
             \psfig{figure=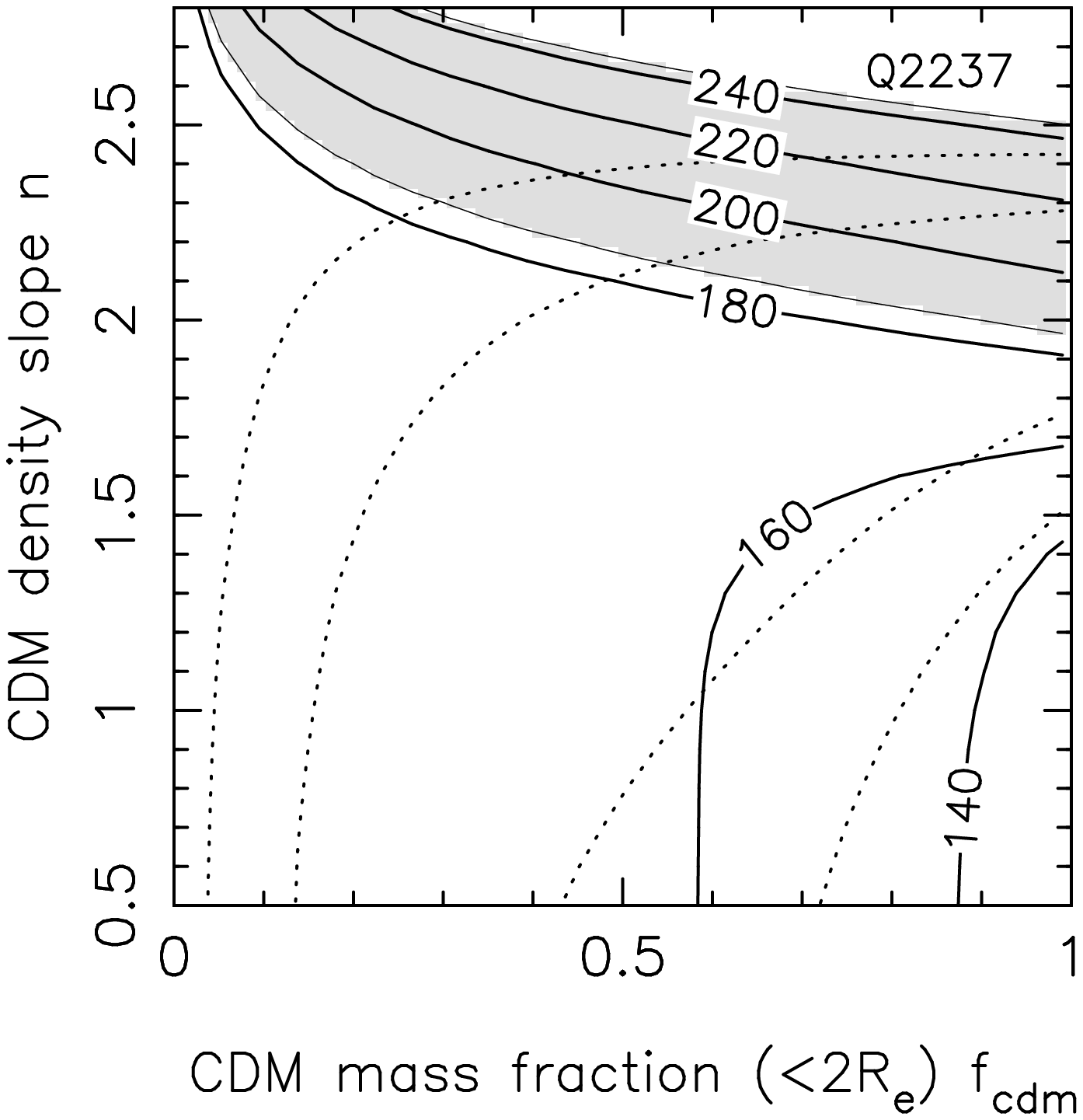  ,width=1.5in}
             \psfig{figure=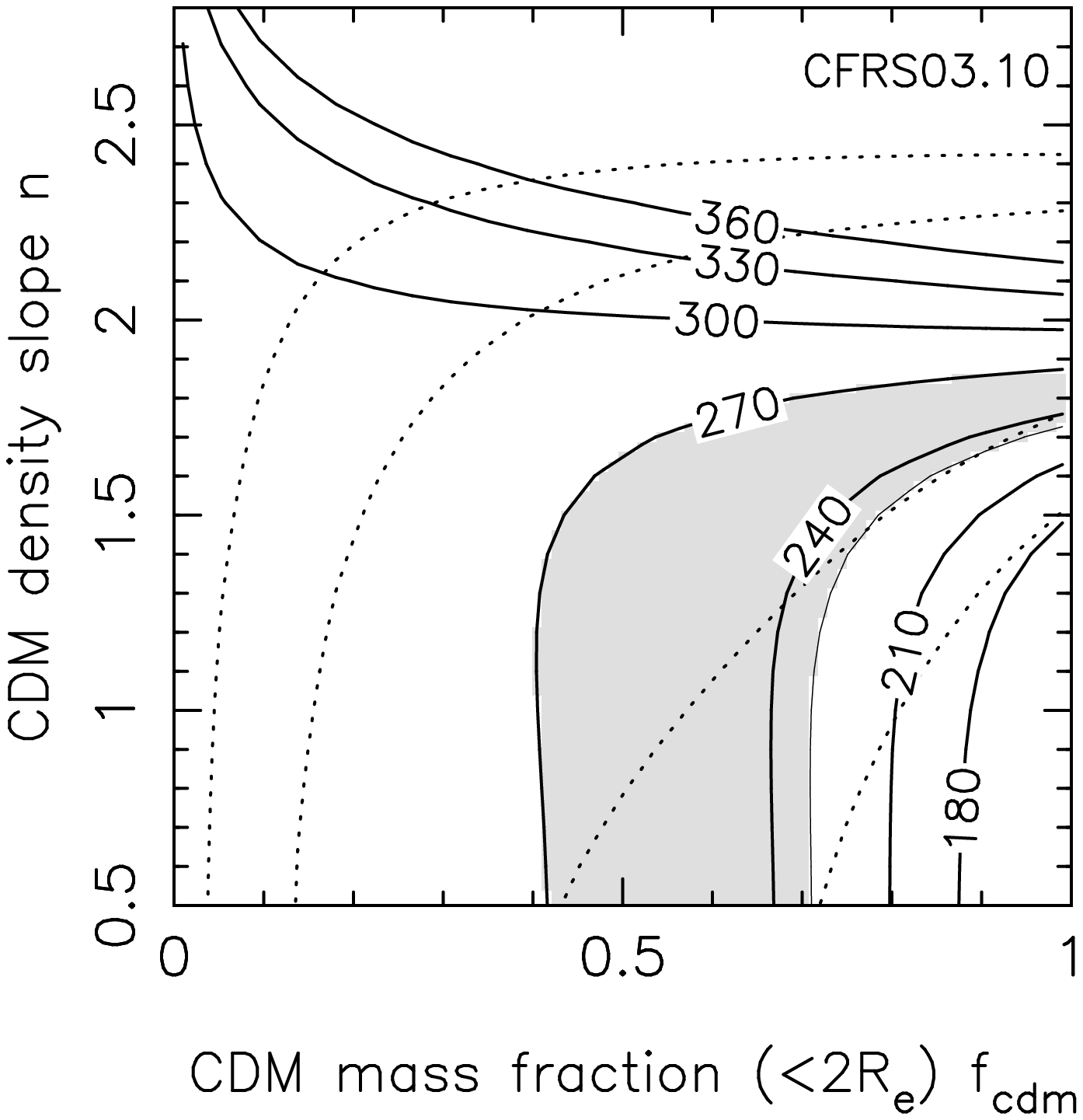,width=1.5in} }
\caption{  Constraints from lens velocity dispersion measurements on the
   self-similar mass distributions of Eqn.~\ref{eqn:selfsimmass} and 
   Fig.~\ref{fig:selfsim1}. The dotted contours show the 68\% and 95\%
   confidence limits from the self-similar models for $R_b/R_e=50$.
   The shaded regions show the models allowed (68\% confidence)
   by the formal velocity dispersion measurement errors, and the heavy solid
   lines show contours of the velocity dispersion in km/s.  We used
   the low Gebhardt et al.~(\cite{Gebhardt2003p1}) velocity dispersion
   for HST14176+5226 because it has the smallest formal error.  These
   models assumed isotropic orbits, thereby underestimating the full
   uncertainties in the stellar dynamical models.  
   }
\labelprint{fig:vdisp}
\end{figure}

In Fig.~\ref{fig:vdisp} we summarize the dynamical constraints for 9 of these systems 
using the self-similar mass distribution from Rusin \& Kochanek~(\cite{Rusin2004p1},
Eqn.~\ref{eqn:selfsimmass}).  This model is very similar to that used by
Treu \& Koopmans~(\cite{Treu2004p1}).  For most of the lenses, the
region producing a good fit to the combined lensing and dynamical data 
overlaps the same region preferred by the Rusin \& Kochanek~(\cite{Rusin2004p1})
self-similar models, shows the same general parameter degeneracy and is 
consistent with a simple SIS mass distribution ($f_{cdm}\rightarrow 1$ and $n=2$).  
This is particularly true of 0047--2808, HST15433+5352, B1608+656, MG2016+112 and 
CFRS03.1077.  Only Q2237+0305, where the lens is the bulge of a nearby spiral 
and we might not expect this mass model to be applicable, shows a very 
different trend (e.g. see the models of Trott \& Webster~\cite{Trott2002p621}).
PG1115+080 and to a lesser extent MG1549+3047 might have steeper than isothermal 
mass distributions (falling rotation curves) and the possibility of being consistent 
with a constant mass-to-light ratio model (Treu \& Koopmans~\cite{Treu2002p6}).  
HST14176+5226 and to a lesser extent HST15433+5352 could have shallower than 
isothermal mass distribution (rising rotation curves).  Along the degeneracy
direction for each lens we will find similar mass distributions with very
different decompositions into luminous and dark matter, just as in Fig.~\ref{fig:selfsim1}. 
The problem raised by this panorama is whether it shows that the halo structure
is largely homogeneous with some measurement outliers, or that the structure
of early-type is heterogeneous with important implications for understanding
time delays (\S\ref{sec:time}) and galaxy evolution (\S\ref{sec:optical}).

My own view tends toward the first interpretation -- that the dynamical data
supports the homogeneity of early-type galaxy structure.  The permitted bands
in Fig.~\ref{fig:vdisp} show the 68\% confidence regions given the formal 
measurement errors and the simple, spherical, isotropic Jeans equation models
 -- this means that the true 68\% confidence
regions are significantly larger. We have already argued that the formal errors 
on dynamical measurements tend to be underestimates.  For example, the need
for HST14176+5226 to have a rising rotation curve would be considerably reduced 
if we used the higher velocity dispersion measurements from Ohyama et 
al.~(\cite{Ohyama2002p2903}) or Treu \& Koopmans~(\cite{Treu2004p1})
or if we broadened the uncertainties by the 30\% needed to make the three
estimates statistically consistent.  Moreover, the existing analyses have also neglected
the systematic uncertainties arising from the scaling factor $f$.  There
are two important issues that make $f \neq 1$.  The first issue is that
standard velocity dispersion measurements are the width of the best fit Gaussian
model for the LOSVD, and this is not the same as the mean square velocity
($\langle v_{los}^2\rangle^{1/2}$)
 appearing
in the Jeans equations used to analyze the data unless the LOSVD is also a
Gaussian.  Stellar dynamics has adopted the dimensionless coefficients $h_n$
of a Gauss-Hermite polynomial series to model the deviations of the LOSVD
from Gaussian, and a typical early-type galaxy has $|h_4|\ltorder 0.03$
(e.g. Romanowsky \& Kochanek~\cite{Romanowsky1999p18}).  This
leads to a systematic difference between the measured dispersions and the mean square
velocity of $\langle v_{los}^2 \rangle^{1/2} \simeq \sigma (1+\sqrt{6}h_4)$
(e.g. van der Marel \& Franx~\cite{van_der_Marel1993p525})), so $|f-1|\sim 7\%$
for $|h_4|\simeq 0.03$.  Only the Romanowsky \& Kochanek~\cite{Romanowsky1999p18})
models of Q0957+561 and PG1115+080 have properly included this uncertainty.
In fact, Romanowsky \& Kochanek~(\cite{Romanowsky1999p18}) demonstrated that
there were stellar distribution functions in which the mass distribution of
PG1115+080 is both isothermal and agrees with the measured velocity dispersion. 
While it is debatable whether these models allowed too much freedom, it is 
certainly true that models using the Jeans equations and ignoring the LOSVD
have too little freedom and will overestimate the constraints.
 
The second issue is that lens galaxies are not spheres.  Unfortunately
there are few simple analytic results for oblate or triaxial systems like
early-type galaxies in which the ellipticity is largely due to anisotropies
in the velocity dispersion tensor rather than rotation.  For the system as a whole,
the tensor virial theorem provides a simple global relationship between the
major and minor axis velocity dispersions
\begin{equation}
   {  \sigma_{major} \over \sigma_{minor} } \simeq 1 + { 1 \over 5}e^2
         + { 9 \over 70} e^4 + \cdots
\end{equation}
for an oblate ellipsoid of axis ratio $q$ and eccentricity $e=(1-q^2)^{1/2}$
(e.g. Binney \& Tremaine~\cite{Binney1987p747}).
The velocity dispersion viewed along the major
axis is larger than that on the minor, and the correction can be quite
large since a typical galaxy with $q=0.7$ will have a ratio
$\sigma_{major}/\sigma_{minor}\simeq 1.16$ that is much larger than
typical measurement uncertainties.  If galaxies are oblate, this provides no
help for the case of PG1115+080 because making the line-of-sight dispersion 
too high requires a prolate galaxy.  However, it is a very simple means of
shifting HST14176+5226.  Crudely, if we start with the low 209~km/s 
velocity dispersion and assume that the lens is an $q=0.7$ galaxy viewed
pole on, then $\sigma_{major}/\sigma_{minor}\simeq 1.14$ and the corrections
for the shape are large enough to make HST14176+5226 consistent with the
other systems.  

A final caveat is that neglecting necessary degrees of freedom in your lens
model can bias inferences from the stellar dynamics of lenses just as it can
in pure lens modeling.  For example, Sand et al.~(\cite{Sand2002p129}, \cite{Sand2003p1})
used a comparison of lensed
arcs in clusters to velocity dispersion measurements of the central cluster
galaxy to argue that the cluster dark matter distribution could not have
the $\rho \propto 1/r$ cusp of the NFW model for CDM halos.  However,
Bartelmann \& Meneghetti~(\cite{Bartelmann2003p1}) and
Dalal \& Keeton~(\cite{Dalal2003p1}) show that the data are consistent with
an NFW cusp if the lens models include a proper treatment of the non-spherical
nature of the clusters.  This has not been an issue in the stellar dynamics
of strong lenses where the lens models used to determine the mass scale have
always included the effects of ellipticity and shear, but it is well worth
remembering.

\section{Time Delays \labelprint{sec:time} }

Nothing compares to the measurement of the Hubble constant in bringing out the
worst in astronomers.  As we discussed in the previous section on lens modeling,
many discussions of lens models seem obfuscatory rather than illuminating, and the
tendency in this direction increases when the models are used to 
estimate $H_0$.  In this section we discuss the relationship between time
delay measurements, lens models and $H_0$.  All results in the literature are
consistent with this discussion, although it may take you several days and a
series of e-mails to confirm it for any particular paper.  The basic idea is
simple.  We see images at extrema of the virtual time delay surface
(e.g. Blandford \& Narayan~\cite{Blandford1986p568}, \partintro) so the propagation
time from the source to the observer differs for each image.   The differences
in propagation times, known as time delays,  are proportional to $H_0^{-1}$ because 
the distances between the observer, the lens and the source depend on the
Hubble constant (Refsdal~\cite{Refsdal1964p307}).  When the source
varies, the variations appear in the images separated by the time delays
and the delays are measured by cross-correlating the light curves.
There are recent reviews of time delays and the Hubble constant by 
Courbin, Saha \& Schechter~(\cite{Courbin2002p1}) and Kochanek \& Schechter~(\cite{Kochanek2004p117}).
Portions of this section are adapted from Kochanek \& Schechter~(\cite{Kochanek2004p117})
since we were completing that review at about the same time as we presented
these lectures.

To begin the discussion we start with our standard simple model, the circular
power law lens from \S\ref{sec:basics}.  As a circular lens, we see two images at radii
$\theta_A$ and $\theta_B$ from the lens center and we will assume that
$\theta_A > \theta_B$ (Fig.~\ref{fig:geometry}).  
Image A is a minimum, so source variability will
appear in image A first and then with a time delay $\Delta t$ in the 
saddle point image B.  We can easily fit this data with an SIS lens model
since (see Eqn.~\ref{eqn:aap} and \ref{eqn:aaq}) 
to find that $\theta_A=\beta+b$ and $\theta_B=b-\beta$ 
where $b=(\theta_A+\theta_B)/2$ is the critical (Einstein) radius of the lens 
and $\beta=(\theta_A-\theta_B)/2$ is the source 
position.  The light travel time for each image relative to a fiducial
unperturbed ray is (see \partintro) 
\begin{equation}
    \tau(\vec{\theta}) = { D_d D_s \over c D_{ds} } \left[
     { 1 \over 2 } \left( \vec{\theta}-\vec{\beta} \right)^2
     - \Psi(\vec{\theta}) \right]
  \labelprint{eqn:acf}
\end{equation}
where the effective potential $\Psi=b\theta$ for the SIS lens.  Remember
that the distances are comoving angular diameter distances rather than the more familiar
angular diameter distances and this leads to the vanishing of the extra
$1+z_l$ factor that appears in the numerator if you insist on using angular
diameter distances.  The propagation time scales as $H_0^{-1}\simeq 10h^{-1}$~Gyr
because of the $H_0^{-1}$ scalings of the distances.  After substituting
our lens model, and differencing the delays for the two images, we find that 
\begin{equation}
  \Delta t_{SIS} = \tau_B - \tau_A = { 1 \over 2 } { D_d D_s \over c D_{ds} }
             \left( \theta_A^2-\theta_B^2 \right).
  \labelprint{eqn:acg}
\end{equation} 
The typical deflection angle $b \sim 3 \times 10^{-6}$~radians (so $R_A^2 \sim 10^{-11}$)
converts the $10h^{-1}$~Gyr propagation time into a time delay of months or years
that can be measured by a graduate student.
Naively, this result suggests that the problem of interpreting time delays to
measure $H_0$ is a trivial problem in astrometry.  

We can check this assumption by 
using our general power-law models from \S\ref{sec:basics} instead of an
SIS.  The power-law models correspond to density 
distributions $\rho \propto r^{-n}$, surface densities $\kappa \propto R^{1-n}$
and circular velocities $v_c \propto r^{(2-n)/2}$ of which the SIS model is
the special case with $n=2$.  These models have effective potentials
\begin{equation}
    \Psi(\vec{\theta}) = { b^2 \over 3-n} \left( { \theta \over b } \right)^{3-n}.
  \labelprint{eqn:ach}
\end{equation}
As we discussed in the \S\ref{sec:massmono} we can fit our simple, circular two-image lens
with any of these models to determine $b(n)$ and $\beta(n)$ (Eqn.~\ref{eqn:abv}),
which we can then
substitute into the expression for the propagation time to find that the time
delay between the images is
\begin{equation}
   \Delta t(n) = (n-1) \Delta t_{SIS} \left[ 1 - { (2-n)^2 \over 12 }
           \left( { \dr \over \rbar } \right)^2 + \cdots \right]
  \labelprint{eqn:aci}
\end{equation}
where we have expanded the result as a series in the ratio between the mean
radius of the images $\rbar=(\theta_A+\theta_B)/2$ and the
thickness of the radial annulus separating them $\dr=\theta_A-\theta_B$.  While
the expansion assumes that $\dr/\rbar \sim \beta/b$ is small, we can usually
ignore higher order terms even when $\dr/\rbar$ is of order unity.  We now
see that the time delay depends critically on the density profile, with more
centrally concentrated mass distributions (larger values of $n$) producing  
longer time delays or implying larger Hubble constants for a fixed time
delay.  

The other idealization of the SIS model, the assumption of a circular lens, turns out
to be less critical.  A very nice analytic example is to consider a 
singular isothermal model with arbitrary angular structure in which 
$\kappa = b F(\chi)/2 \theta$ where $F(\chi)$ is an arbitrary function
of the azimuthal angle.  The singular isothermal ellipsoid (Eqn.~\ref{eqn:abf}) is
an example of this class of potential.  For this model family, 
$\Delta t = \Delta t_{SIS}$ independent of the actual angular
structure $F(\chi)$ (Witt, Mao \& Keeton~\cite{Witt2000p98}).

\subsection{A General Theory of Time Delays \labelprint{sec:timetheory} }

Just as for estimating mass distributions (\S\ref{sec:mass}), the aspect of modeling time
delays that creates the greatest suspicion is the need to model the gravitational
potential of the lens.   Just as for mass distributions, this problem is 
largely of our own making, arising from poor communication, understanding
and competition between groups.  Here we will use simple mathematical
expansions to show exactly what properties of the potential determine 
time delays.  Any models which have these generic properties have all the
degrees of freedom needed to properly interpret time delays.  This does not,
unfortunately, avoid the problem of degeneracies between the mass models
and the Hubble constant.
 
\begin{figure}[ph]
\begin{center}
\centerline{\psfig{figure=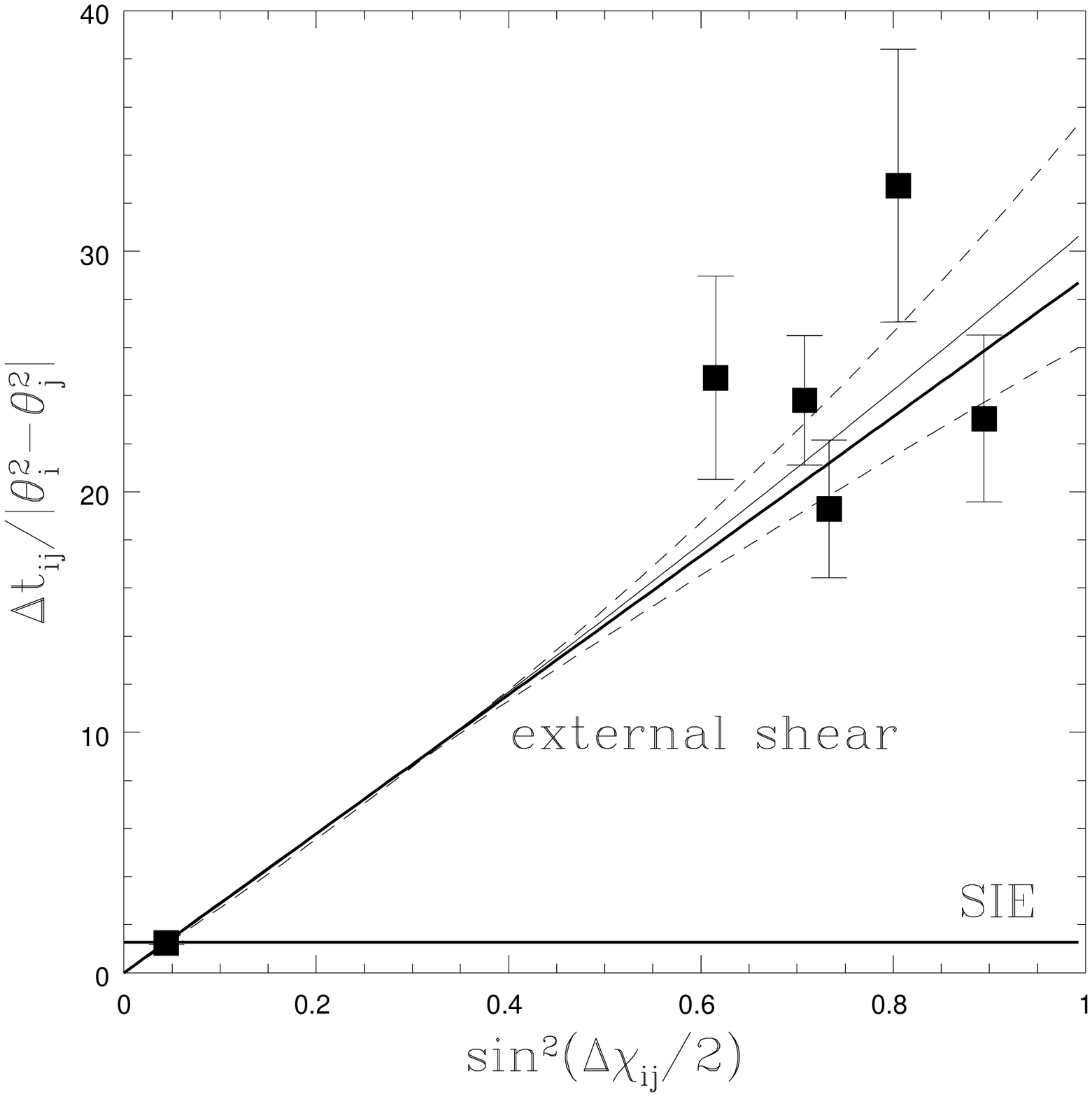,width=3.2in}}
\end{center}
\begin{center}
\centerline{\psfig{figure=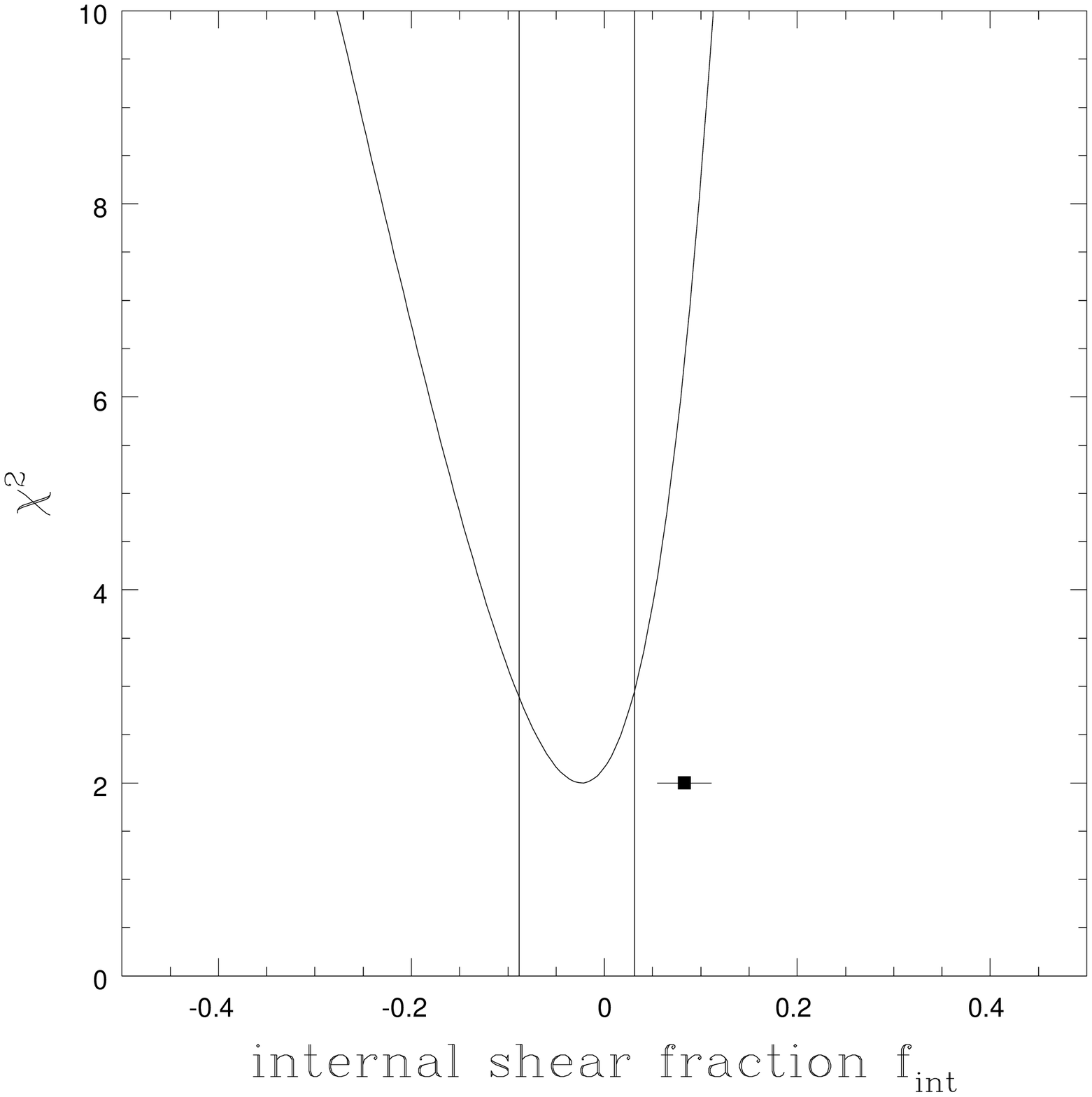,width=3.2in}}
\end{center}
\caption{  (Top) The PG1115+080 time delays scaled by the astrometric
   factor $\theta_i^2-\theta_j^2$ appearing in $\Delta t_{SIS}$
   (Eqn.~\ref{eqn:acg}) as a function of
   the leading angular dependence of the time delay $(\sin^2\Delta\chi_{ij}/2)$
   (Eqn.~\ref{eqn:timequad}).  The light solid curve and the dashed curves
   show the dependence for the best fit internal shear fraction $f_{int}$
   and its 68\% confidence limits.  A true external shear $f_{int}=0$ is shown by the
   heavy solid curve inside the confidence limits, and the scaling for an
   SIE ($f_{int}=1/4$) is shown by the horizontal line.
   (Bottom) The $\chi^2$ goodness of fit for the internal shear fraction
    $f_{int}$ from the time delay ratios is shown by the curve with the
   68\% confidence region bracketed by the vertical lines.  The estimate
   of $f_{int}$ from the image astrometry is shown by the point with an
   error bar.
   }
\labelprint{fig:timequad}
\end{figure}

The key to understanding time delays comes from Gorenstein, Falco \& Shapiro
(\cite{Gorenstein1988p693}, Kochanek~\cite{Kochanek2002p25}, 
see also Saha~\cite{Saha2000p1654}) who showed that the time 
delay in a circular
lens depends only on the image positions and {\it the surface density 
$\kappa(\theta)$ in the annulus between the images}.  The two lensed images
at radii $\theta_A > \theta_B$ define an annulus bounded by their radii,
with an interior region for $\theta < \theta_B$ and an exterior region
for $\theta>\theta_A$ (Fig.~\ref{fig:geometry}).  As we discussed in
\S\ref{sec:massmono}, the mass in the interior region is implicit in 
the image positions and constrained by the astrometry.  From
Gauss' law we know that the distribution of the mass in the interior
and the amount or distribution of mass in the exterior region is 
irrelevant (see \S\ref{sec:monofit}).  A useful approximation is to assume that the surface density
in the annulus can be {\it locally} approximated by a power law,
$\kappa(\theta)\propto \theta^{1-n}$ for $\theta_B<\theta<\theta_A$,
with a mean surface density in the annulus of $\kbar=\langle\Sigma\rangle/\Sigma_c$.
The time delay between the images is then (Kochanek~\cite{Kochanek2002p25})
\begin{equation}
    \Delta t = 2 \Delta t_{SIS} \left[ 1-\kbar - { 1- n\kbar \over 12 }
           \left( { \dr \over \rbar } \right)^2 +
         O\left(\left( { \dr \over \rbar } \right)^4 \right) \right]
  \labelprint{eqn:adj}
\end{equation}
where $\rbar=(\theta_A+\theta_B)/2$ and $\dr=\theta_A-\theta_B$ as before.
Thus, the time delay is largely determined by the average surface density 
$\kbar$ in the annulus with only modest corrections from the local shape of the surface
density distribution even when $\dr/\rbar \sim 1$.  This second order
expansion is exact for an SIS lens ($\kbar=1/2$, $n=2$), and it reproduces
the time delay of a point mass lens ($\kbar=0$) to better than 1\%
even when $\dr/\rbar=1$.  The local model also explains the scalings of
the global power-law models.  A $\kappa \propto \theta^{1-n}$ global power
law has surface density $\kbar=(3-n)/2$ near the Einstein ring, so the
leading term of the time delay is 
$\Delta t =2 \Delta_{SIS}(1-\kbar)=(n-1)\Delta t_{SIS}$
just as in Eqn.~\ref{eqn:aci}.  

The role of the angular structure of the lens is easily incorporated into
the expansion through the multipole expansion of \S\ref{sec:mass}.  A
quadrupole term in the potential with dimensionless amplitude $\epsilon_\Psi$
produces ray deflections of order $O(\epsilon_\Psi b)$ at the Einstein radius
$b$ of the lens.  In a four-image lens, the quadrupole deflections are
comparable to the fractional thickness of the annulus, $\epsilon_\Psi \simeq \dr/\rbar$,
while in a two-image lens they are smaller.  For an ellipsoidal density
distribution, the $\cos(2m\chi)$ multipole amplitude is smaller than 
the quadrupole amplitude by $\epsilon_{2m} \sim \epsilon_\Psi^m \ltorder (\dr/\rbar)^m$.
Hence, to lowest order in the expansion we only need to include the internal
and external quadrupoles of the potential but not the changes of the
quadrupoles in the annulus or any higher order multipoles.  Remember that
what counts is the angular structure of the potential rather than of the
density, and that potentials are always much rounder than densities with
a typical scaling of $m^{-2}$:$m^{-1}$:$1$ between the potential, deflections
and surface density for the $\cos m\chi$ multipoles (see \S\ref{sec:massquad})  

While the full expansion for independent internal and external quadrupoles
is too complex to be informative, the leading term for the case when the
internal and external quadrupoles are aligned is informative.  We have
an internal shear of amplitude $\Gamma$ and an external shear of amplitude
$\gamma$ with $\chi_\gamma=\chi_\Gamma$ as defined in Eqns.~\ref{eqn:acc} 
and \ref{eqn:adl}.  The leading term of the time delay is
\begin{equation}
    \Delta t \simeq 2 \Delta t_{SIS} \left( 1 -\kbar \right)
    { \sin^2 \left( \Delta\chi_{AB}/2 \right)
      \over 1 - 4 f_{int} \cos^2 \left( \Delta\chi_{AB}/2 \right) }
     \labelprint{eqn:timequad}
\end{equation}
where $\Delta\chi_{AB}$ is the angle between the images (Fig.~\ref{fig:geometry})
and $f_{int}=\Gamma/(\Gamma+\gamma)$ is the internal quadrupole fraction we
explored in Fig.~\ref{fig:intfrac}.  We need not worry about a singular
denominator -- successful models of the image positions do not allow such
configurations.  

A two-image lens has too few astrometric constraints to fully constrain a model
with independent, misaligned internal and external quadrupoles.  Fortunately,
when the lensed images lie on opposite sides of the lens galaxy 
($\Delta\chi_{AB} \simeq \pi+\delta$ with $|\delta| \ll 1$), the time
delay becomes insensitive to the quadrupole structure.  Provided the
angular deflections are smaller than the radial deflections 
($|\delta|\rbar \ltorder \dr$), the leading term of the time delay reduces
to the result for a circular lens, $\Delta t = 2\Delta t_{SIS}(1-\kbar)$
if we minimize the total shear of the lens.  In the minimum shear solution
the shear converges to the invariant shear ($\gamma_1$) and the other
shear component $\gamma_2=0$ (see \S\ref{sec:angstruc}). 
 If, however, you allow the other shear
component to be non-zero, then you find that $\Delta t = 2\Delta t_{SIS}(1-\kbar-\gamma_2)$
to lowest order -- the second shear component acts like a contribution to
the convergence.   In the absence of any other constraints, this adds a modest
additional uncertainty (5--10\%) to interpretations of time delays in two-image lenses.
To first order its effects should average out in an ensemble of lenses because 
the extra shear has no preferred sign.

A four image lens has more astrometric constraints and can constrain a model
with independent, misaligned internal and external quadrupoles -- this was 
the basis of the Turner et al.~(\cite{Turner2004}) summary of the internal to total 
quadrupole ratios shown in Fig.~\ref{fig:intfrac}.  If the external shear
dominates, then $f_{int} \simeq 0$ and the leading term of the delay 
becomes $\Delta t= 2\Delta t_{SIS} (1-\kbar) \sin^2 \Delta\chi_{AB}/2$.
If the model is isothermal, like the $\Psi=\theta F(\chi)$ model we 
introduced in Eqn.~\ref{eqn:wevans}, then $f_{int}=1/4$ and we obtain the
Witt et al.~(\cite{Witt2000p98}) result that the time delay is 
independent of the angle between the images 
$\Delta t \simeq 2\Delta t_{SIS} (1-\kbar)$.  Thus, delay ratios
in a four-image lens are largely determined by the angular structure and
provide a check on the potential model.  Unfortunately, the only lens
with precisely measured delay ratios, B1608+656 
(Fassnacht et al.~\cite{Fassnacht2002p823}), also has two galaxies inside
the Einstein ring and is a poor candidate for a simple multipole treatment
(although it is dominated by an internal quadrupole as expected, see
Fig.~\ref{fig:intfrac}).
The delay ratios for PG1115+080 are less well measured (Schechter et 
al.~\cite{Schechter1997p85}, Barkana~\cite{Barkana1997p21}, Chartas~\cite{Chartas2003p1}),
but should be dominated by external shear since the estimate from the
image astrometry is that $f_{int}=0.083$ ($0.055 < f_{int} < 0.111$ at
95\% confidence).  Fig.~\ref{fig:timequad} shows the dependence of the
PG1115+080 delays on the leading angular dependence of the time delay
(Eqn.~\ref{eqn:timequad}) after scaling out the 
standard astrometry factor for the different radii of the images 
(Eqn.~\ref{eqn:acg}).  Formally, the estimate from the time delays
that $f_{int}=-0.02$ ($-0.09 < f_{int} < 0.03$ at 68\% confidence) is
a little discrepant, but the two estimates agree at the 95\% confidence
level and there are still some systematic uncertainties in the shorter
optical delays of PG1115+080.  Changes in $f_{int}$ between lenses is
the reason Saha~(\cite{Saha2004p425}) found significant scatter between
time delays scaled only by $\Delta t_{SIS}$, since the time delay lenses
range from external shear dominated systems like PG1115+080 to internal
shear dominated systems like B1608+656.

\subsection{Time Delay Lenses in Groups or Clusters \labelprint{sec:timecluster}}

Most galaxies are not isolated, and many early-type lens galaxies are
members of groups or clusters, so we need to consider the effects of
the local environment on the time delays.  Weak perturbations are
easily understood since they will simply be additional contributions
to the surface density ($\kappa_c$) and the external shear/quadrupole 
($\gamma_c$) we discussed in \S\ref{sec:mass}.  In general the
effects of the external shear $\gamma_c$ are minimal because they
either have little effect on the delays (two-image lenses) or
are tightly constrained by either the astrometry or delay ratios
(four-image lenses or systems with lensed host galaxies see \S\ref{sec:hosts}).
The problems arise from either the degeneracies associated with the
surface density $\kappa_c$ or the need for a complete, complicated 
cluster model.

The problem with $\kappa_c$ is the infamous {\it mass-sheet degeneracy}
(\partintro, Falco, Gorenstein \& Shapiro~\cite{Falco1985p1}).  
If we have a model predicting a time delay $\Delta t_0$ 
and add a sheet of constant surface density $\kappa_c$, then the time delay 
is changed to $(1-\kappa_c)\Delta t_0$ without changing the image positions,
flux ratios, or time delay ratios.  Its effects can be understood from
\S\ref{sec:timetheory} as a contribution to the annular surface density with
$\kbar=\kappa_c$ and $\eta=1$.  Its only observable effect other than that
on the time delays is a reduction in the mass of the lens galaxy that
could be detected given an independent estimate of the lens galaxy's mass
such as a velocity dispersion
(e.g. \S\ref{sec:dynamics} see Romanowsky \& Kochanek~\cite{Romanowsky1998p64}
for an attempt to to this for Q0957+561).  It can also be done given
an independent estimate of the properties of the group or cluster
using weak lensing (e.g.  Fischer et al.~\cite{Fischer1997p521}
  in Q0957+561), cluster galaxy velocity dispersions 
  (e.g., Angonin-Willaime, Soucail, \& Vanderriest~\cite{Angonin1994p411}
  for Q0957+561, Hjorth et al.~\cite{Hjorth2002p11}
  for RXJ0911+0551) or X-ray temperatures/luminosities (e.g., Morgan et al.~\cite{Morgan2001p1}
  for RXJ0911+0551 or Chartas et al.~\cite{Chartas2002p96} for Q0957+561).  
  The accuracy of these methods is uncertain at present because each
  suffers from its own systematic uncertainties, and they probably
 cannot supply the 10\% or higher precision measurements of $\kappa_c$
 needed to strongly constrain models. 
When the convergence is due to an object like a cluster, there is a strong 
correlation between the convergence $\kappa_c$ and the shear $\gamma_c$ that is
controlled by the density distribution of the cluster (for an 
isothermal model $\kappa_c=\gamma_c$).  When the lens is in
  the outskirts of a cluster, as in RXJ0911+0551, it is probably 
  reasonable to assume that $\kappa_c \leq \gamma_c$, as most mass 
  distributions are more centrally concentrated than isothermal
 (see Eqn.~\ref{eqn:aae2}).
Neglecting the extra surface density coming
from nearby objects (galaxies, groups, clusters) leads to
an overestimate of the Hubble constant, because these objects
all have $\kappa_c >0$.  For most time delay systems this correction
is probably $\ltorder 10\%$.

If the cluster or any member galaxies are sufficiently close, then we cannot 
ignore the higher-order 
perturbations in the expansion of Eqn.~(\ref{eqn:aau}).  This 
is certainly true for Q0957+561 (as discussed in \S\ref{sec:modelfit})
where the lens galaxy is the brightest cluster galaxy and located very
close to the center of the cluster.  It is easy to gauge when they 
become important by simply comparing the deflections produced by any
higher order moments of the cluster beyond the quadrupole with the
uncertainties being used for the image positions.  For a cluster
of critical radius $b_c$ at distance $\theta_c$ from a lens of
Einstein radius $b$, these perturbations are of order 
$b_c(b/\theta_c)^2 \sim b \gamma_c (b/\theta_c)$.  Because the
astrometric precision of the measurements is so high, these higher
order terms can be relatively easy to detect.  For example, models of
PG1115+080 (e.g. Impey et al.~\cite{Impey1998p551}) find that using
a group potential near the optical centroid of the nearby galaxies
produces a better fit than simply using an external shear.  In this
case the higher order terms are fairly small and affect the results 
little, but results become very misleading if they are important 
but ignored. 

\begin{table}[t]
  \begin{center}
  \caption{Time Delay Measurements}
  \begin{tabular}{lccccc}
\hline               
System       &$N_{im}$&$\Delta t$ (days)  &Astrometry  &Model      &Ref. \\
\hline
HE1104--1805 &2       &$161\pm 7$         &$+$    &``simple''         & 1 \\
PG1115+080   &4       &$ 25\pm 2$         &$+$    &``simple''         & 2 \\
SBS1520+530  &2       &$130\pm 3$         &$+$    &``simple''         & 3 \\
B1600+434    &2       &$ 51\pm 2$         &$+/-$  &``simple''         & 4 \\
HE2149--2745 &2       &$103\pm12$         &$+$    &``simple''         & 5 \\
\hline
RXJ0911+0551 &4       &$146\pm 4$         &$+$    & cluster/satellite & 6 \\
Q0957+561    &2       &$417\pm 3$         &$+$    & cluster           & 7 \\ 
B1608+656    &4       &$ 77\pm 2$         &$+/-$  & satellite         & 8 \\
\hline
B0218+357    &2       &$10.5\pm0.2$       &$-$    &``simple''         & 9 \\
PKS1830--211 &2       &$26 \pm 4$         &$-$    &``simple''         & 10 \\
\hline               
B1422+231    &4       &($  8 \pm  3)$     &$+$    &``simple''         & 11 \\
\hline               
  \end{tabular}
  \end{center}
  \labelprint{tab:delays}
  \noindent
   $N_{im}$ is the number of images. 
   $\Delta t$ is the longest of the measured delays and its 1$\sigma$ 
     error; delays in parenthesis require further confirmation.  
   The ``Astrometry'' column indicates the quality of the astrometric data for 
   the system: $+$ (good), $+/-$ (some problems), $-$ (serious problems). 
   The ``Model'' column indicates the type of model needed to interpret the
      delays.  ``Simple'' lenses can be modeled as a single primary lens
      galaxy in a perturbing tidal field.  More complex models are needed 
      if there is a satellite galaxy inside the Einstein ring (``satellite'') 
      of the primary lens galaxy, or if the primary lens belongs to a
      cluster.
    References: 
    (1) Ofek \& Maoz~\cite{Ofek2003p101}, Wyrzykowski et al.~\cite{Wyrzykowski2003p229};
    (2) Barkana~\cite{Barkana1997p21}, based on Schechter et al.~\cite{Schechter1997p85}; 
    (3) Burud et al.~\cite{Burud2002p481};       
    (4) Burud et al.~\cite{Burud2000p117}, also Koopmans et al.~\cite{Koopmans2000p391}; 
    (5) Burud et al.~\cite{Burud2002p71}; 
    (6) Hjorth et al.~\cite{Hjorth2002p11}; 
    (7) Kundi\'c et al.~\cite{Kundic1997p75}, also Schild \& 
        Thomson~\cite{Schild1997p130} and 
        Haarsma et al.~\cite{Haarsma1999p64}; 
    (8) Fassnacht et al.~\cite{Fassnacht2002p823}; 
    (9) Biggs et al.~\cite{Biggs1999p349}, also Cohen et al.~\cite{Cohen2000p578};
    (10) Lovell et al.~\cite{Lovell1998p51};
    (11) Patnaik \& Narasimha~\cite{Patnaik2001p1403}.
\end{table}

\begin{figure}[ph]
\begin{center}
\centerline{\psfig{figure=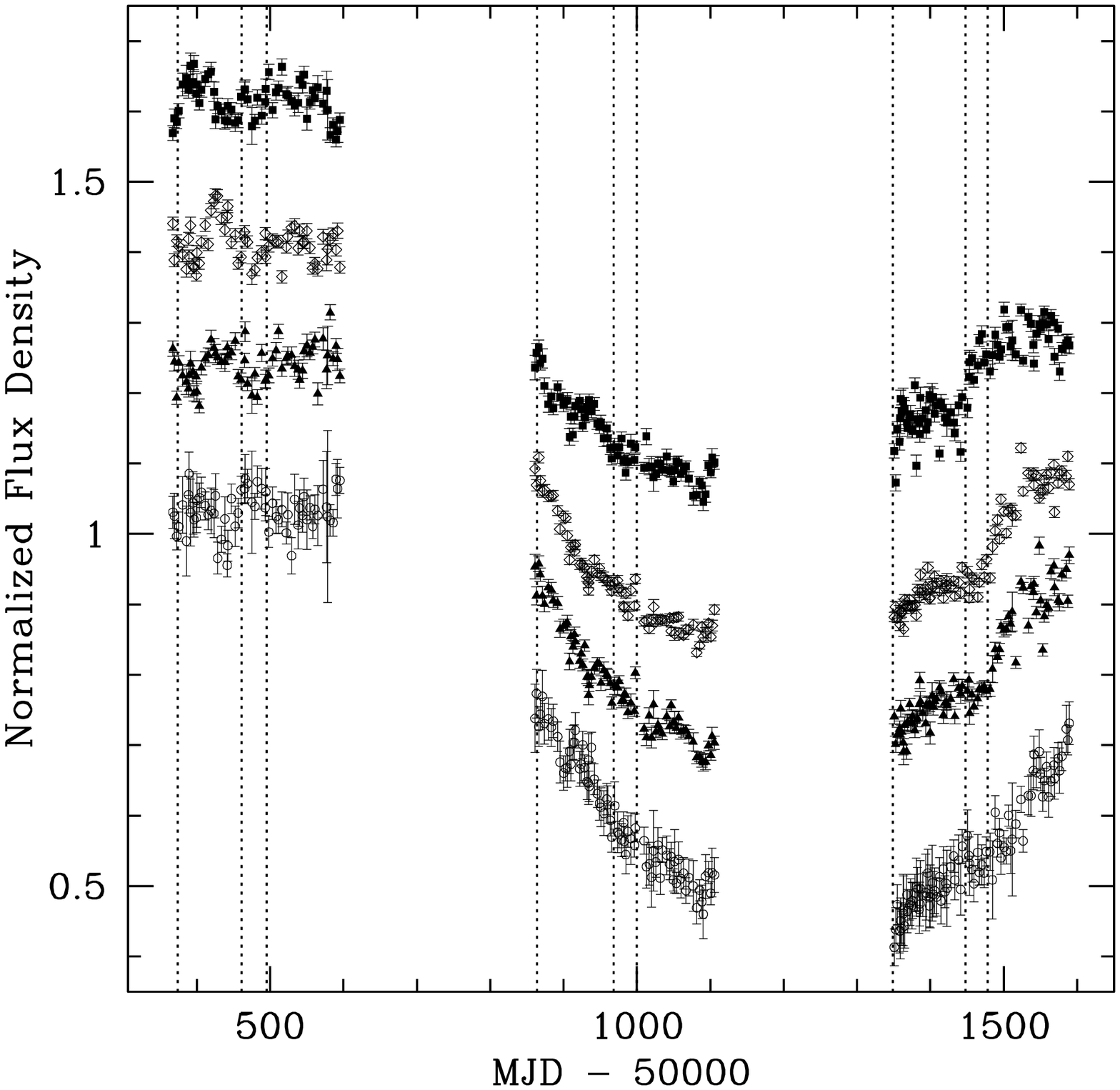,width=3.2in}}
\end{center}
\begin{center}
\centerline{\psfig{figure=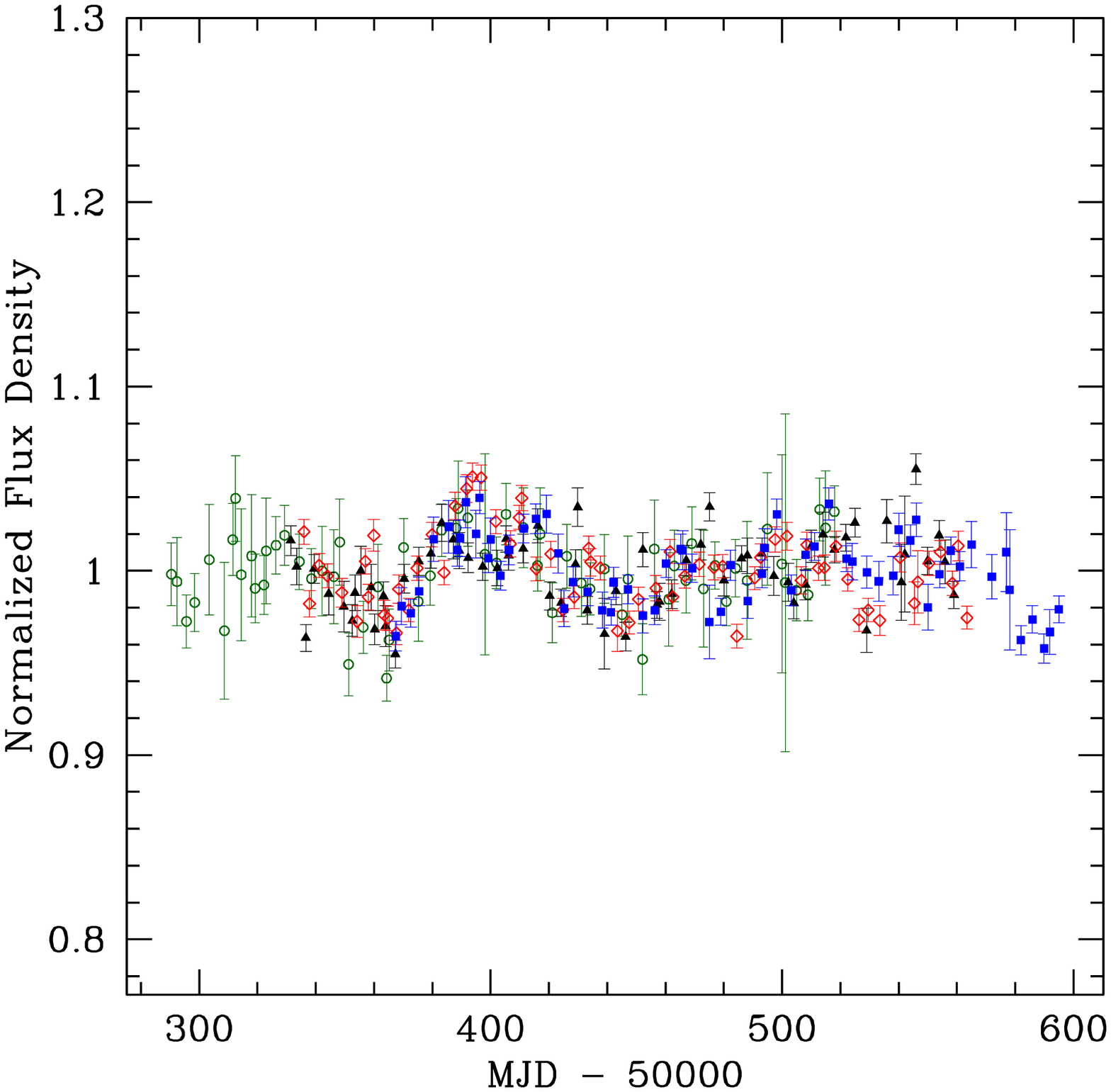,width=3.2in}}
\end{center}
\caption{ 
VLA monitoring data for the four-image lens B1608+656.  The top panel shows 
(from top to bottom) the normalized light curves for the B (filled squares), A 
(open diamonds), C (filled triangles) and D (open circles) images as a 
function of the mean Julian day.  The bottom panel shows the composite light 
curve for the first monitoring season after cross correlating the light curves 
to determine the time delays ($\Delta t_{AB}=31.5\pm1.5$, 
$\Delta t_{CB}=36.0\pm1.5$ and $\Delta t_{DB}=77.0\pm1.5$~days) and the flux 
ratios.  (From Fassnacht et al.~\cite{Fassnacht2002p823}.)
   }
\labelprint{fig:lcurve}
\end{figure}

\subsection{Observing Time Delays and Time Delay Lenses \label{sec:timedata}}

The first time delay measurement, for the gravitational lens Q0957+561,
was reported in 1984 (Florentin-Nielsen~\cite{Florentin1984p19}). 
Unfortunately, a controversy then developed between a short 
delay ($\simeq 1.1$~years, Schild \& Cholfin~\cite{Schild1986p209}; Vanderriest et
al.~\cite{Vanderriest1989p215}) and a long delay ($\simeq 1.5$~years,
Press, Rybicki, \& Hewitt~\cite{Press1992p404},~\cite{Press1992p416}), which was finally
settled in favor of the short delay only after 5 more years of 
effort (Kundi\'c et al.~\cite{Kundic1997p75}; also Schild \& 
Thomson~\cite{Schild1997p130}
and Haarsma et al.~~\cite{Haarsma1999p64}).  Factors contributing to
the intervening difficulties included the small amplitude of the
variations, systematic effects, which, with hindsight, appear to be due to
microlensing and scheduling difficulties (both technical and
sociological).  

While the long-running controversy over Q0957+561 led to poor publicity for
the measurement of time delays, it allowed the community to 
come to an understanding of the systematic problems in measuring time delays, 
and to develop a broad range of methods for reliably determining time delays 
from typical data.  Only
the sociological problem of conducting large monitoring projects remains
as an impediment to the measurement of time delays in large numbers. 
Even these are slowly being overcome, with the result that the last
five years have seen the publication of time delays in 11 systems (see
Table~\ref{tab:delays}). 
 
The basic procedures for measuring a time delay are simple.  A monitoring
campaign must produce light curves for the individual lensed images that 
are well sampled compared to the time delays.  During this period, the
source quasar in the lens must have measurable brightness fluctuations
on time scales shorter than the monitoring period.
The resulting light curves are cross correlated by one or more methods
to measure the delays and their uncertainties  
(e.g., Press et al.~~\cite{Press1992p404}~\cite{Press1992p416}; Beskin \& 
Oknyanskij~\cite{Beskin1995p341};
 Pelt et al.~\cite{Pelt1996p97}; references in Table~ 1.1).
Care must be taken because there can be sources of 
uncorrelated variability between the images due to systematic errors
in the photometry and real effects such as microlensing of the individual
images (e.g., Koopmans et al.~\cite{Koopmans2000p391}; Burud et al.\cite{Burud2002p481}; 
Schechter et al.~\cite{Schechter2003p657}).  Figure~\ref{fig:lcurve} shows an example, the
beautiful light curves from the radio lens B1608+656 by Fassnacht
et al.~(\cite{Fassnacht2002p823}), where the variations of all four lensed images
have been traced for over three years.  One of the 11 systems,
B1422+231, is limited by systematic uncertainties in the delay
measurements.

We want to have uncertainties in the time delay measurements that 
are unimportant for the estimates of $H_0$.  For the present, 
uncertainties of order 3\%--5\% are adequate (so improved delays
are still needed for PG1115+080, HE2149--2745, and PKS1830--211).
In a four-image lens we can measure three independent time delays,
and the dimensionless ratios of these delays provide
additional constraints on the lens models (see \S\ref{sec:timetheory}).
These ratios are well measured in B1608+656 (Fassnacht et al.~\cite{Fassnacht2002p823}),
poorly measured in PG1115+080 (Barkana~\cite{Barkana1997p21}; Schechter et 
al.~\cite{Schechter1997p85}; Chartas~\cite{Chartas2003p1}) and unmeasured 
in either RXJ0911+0551
or B1422+231.  Using the time delay lenses as very precise 
probes of $H_0$, dark matter and cosmology will eventually require
still smaller delay uncertainties ($\sim 1\%$).  Once a delay is
known to 5\%, it is relatively easy to reduce the uncertainties
further because we can accurately predict when flux variations 
will appear in the other images and the lens will need to be monitored 
more intensively.

The expression for the time delay in an SIS lens (Eqn.~\ref{eqn:acg}) 
reveals the other data that are necessary to interpret time delays.  
First, the source and lens redshifts
are needed to compute the distance factors that set the scale of the
time delays.  Fortunately, we know both redshifts for all 11 systems 
in Table~\ref{tab:delays} even though missing redshifts are a problem
for the lens sample as a whole (see \S\ref{sec:data}).  
The dependence of the distances $D_d$, $D_s$ and $D_{ds}$ 
on the cosmological model is unimportant until our total uncertainties 
approach 5\%.
Second, we require accurate relative positions for the images and the
lens galaxy.  These uncertainties are always dominated by the position
of the lens galaxy relative to the images.
For most of the lenses in Table~\ref{tab:delays}, observations with radio 
interferometers (VLA, Merlin, VLBA) and HST have measured the
relative positions of the images and lenses to accuracies
$\ltorder 0\farcs005$.  Sufficiently deep HST images can obtain
the necessary data for almost any lens, but dust in the lens
galaxy (as seen in B1600+434 and B1608+656) can limit the accuracy 
of the measurement even in a very deep image.  For 
B0218+357 and PKS1830--211, however, the position of the lens galaxy
relative to the images is not known to sufficient precision or 
determined only from models (see Biggs et al.~\cite{Biggs1999p349},
L\'ehar et al.~\cite{Lehar2000p584}; 
Courbin et al.~\cite{Courbin2002p95}; 
Winn et al.~\cite{Winn2002p103}, Wucknitz, Biggs \& Browne~\cite{Wucknitz2004p14},
York et al.~\cite{York2004p1}).

We can also divide the systems by the complexity of the
required lens model.  We define eight of the lenses as ``simple,'' in the
sense that the available data suggests that a model consisting of a
single primary lens in a perturbing shear (tidal gravity) field should
be an adequate representation of the gravitational potential.  In some
of these cases, an external potential representing a nearby galaxy or
parent group will improve the fits, but the differences between the
tidal model and the more complicated perturbing potential are small
(see \S\ref{sec:timecluster}).  If we neglect the convergence produced
by the group, then $H_0$ may be overestimated.  
We include the quotation marks because the classification is based 
on an impression of the systems from the available data and models. 
Remember also that there are convergence fluctuations along the line
of sight that add a low level of cosmic variance to the time delays
of individual lenses (\S\ref{sec:singlescreen} and Fig.~\ref{fig:los}).
While we cannot guarantee that a system is simple, we can easily
recognize two complications that will require more complex models.

The first complication is
that some primary lenses have less massive satellite galaxies inside
or near their Einstein rings.  This includes two of the time delay
lenses, RXJ0911+0551 and B1608+656.  RXJ0911+0551 could simply be 
a projection effect, since neither lens galaxy shows irregular 
isophotes.  Here the implication for models may simply be the need
to include all the parameters (mass, position, ellipticity $\cdots$)
required to describe the second lens galaxy, and with more parameters
we would expect greater uncertainties in $H_0$.
In B1608+656, however, the lens galaxies show the 
disturbed isophotes of dusty galaxies possibly undergoing a disruptive 
interaction.  How one
should model such a system is unclear. If there was once dark matter 
associated with each of the galaxies, how is it distributed now?  Is it still 
associated with the individual galaxies?  Has it settled into an equilibrium
configuration?  While B1608+656 can be well fit with standard lens
models (Fassnacht et al.~\cite{Fassnacht2002p823}, Koopmans et al.~\cite{Koopmans2003p70}), 
these complications have yet to be explored in detail. 

The second complication occurs when the primary lens is a member of a 
more massive (X-ray) cluster, as in the time delay lenses RXJ0911+0551 
(Morgan et al.~\cite{Morgan2001p1}) and Q0957+561 (Chartas et al.~\cite{Chartas2002p96}).  The cluster
model is critical to interpreting these systems (see \S\ref{sec:timecluster}).  
The cluster surface density at the position of the lens ($\kappa_c \gtorder 0.2$)
leads to large corrections to the time delay estimates and the higher-order 
perturbations are crucial to obtaining a good model.  For example, models
in which the Q0957+561 cluster was treated simply as an external 
shear were grossly incorrect (see \S\ref{sec:modelfit}, 
Keeton et al.~\cite{Keeton2000p74}).  In addition to the uncertainties in
the cluster model itself, we must also decide how to include and
model the other cluster galaxies near the primary lens.  Thus,
lenses in clusters have many reasonable degrees of freedom beyond those
of the ``simple'' lenses.

\subsection{Results: The Hubble Constant and Dark Matter \labelprint{sec:timeresults}}

With our understanding of the theory and observations of the lenses
we will now explore their implications for $H_0$.  We focus on
the ``simple'' lenses PG1115+080, SBS1520+530, B1600+434, and 
HE2149--2745.  We only comment on the interpretation of the HE1104--1805
delay because the measurement is too recent to have been interpreted
carefully.  We will briefly discuss the more complicated systems
B0218+357, RXJ0911+0551, Q0957+561, and B1608+656, and we will not discuss 
the systems with problematic time delays or astrometry.  

The most common, simple, realistic model of a lens consists of a singular
isothermal ellipsoid (SIE) in an external (tidal) shear field (see \S\ref{sec:mass}).
The model has 7 parameters (the lens position, mass, ellipticity, major axis
orientation for the SIE, and the shear amplitude and orientation).
It has many degrees of freedom associated with the angular
structure of the potential, but the radial structure is fixed with
$\kbar\simeq 1/2$.  For comparison, a two-image (four-image) lens
supplies 5 (13) constraints on any model of the potential: 2 (6) from the
relative positions of the images, 1 (3) from the flux ratios of the images,
0 (2) from the inter-image time delay ratios, and 2 from the lens position.
With the addition of extra components (satellites/clusters) for the
more complex lenses, this basic model provides a good fit to all the
time delay lenses except Q0957+561.  Although a naive counting of the
degrees of freedom ($N_{dof}=-2$ and $6$, respectively) suggests that
estimates of $H_0$ would be under constrained for two-image lenses and
over constrained for four-image lenses, the uncertainties are actually
dominated by those of the time delay measurements and the astrometry in
both cases.  This is what we expect from \S\ref{sec:timetheory} --- the
model has no degrees of freedom that change $\kbar$ or $\eta$, so there
will be little contribution to the uncertainties in $H_0$ from the
model for the potential.

If we use a model that includes parameters to control 
the radial density profile (i.e., $\kbar$), 
for example by adding a halo truncation radius $a$ to the SIS profile 
[the pseudo-Jaffe model, $\rho \propto r^{-2} (r^2+a^2)^{-1}$; e.g., Impey
et al.~1998; Burud et al.~2002a],\footnote{This is simply an example.
  The same behavior would be seen for any other parametric model in which
  the radial density profile can be adjusted. } 
then we
find the expected correlation between $a$ and $H_0$ --- as we make the
halo more concentrated (smaller $a$), the estimate of $H_0$ rises
from the value for the SIS profile ($\kbar=1/2$ as $a\rightarrow\infty$)
to the value for a  point mass ($\kbar=0$ as $a\rightarrow 0$),
with the fastest changes occurring when $a$ is similar
to the Einstein radius of the lens. We show an example of such
a model for PG1115+080 in Figure~\ref{fig:scaling}.  This case is somewhat
more complicated than a pure pseudo-Jaffe model because there is an 
additional contribution to the surface
density from the group to which the lens galaxy belongs. 
 As long as the structure of the radial density profile is 
fixed (constant $a$), the uncertainties are again dominated by the
uncertainties in the time delay.  Unfortunately, 
the goodness
of fit, $\chi^2(a)$, shows too little dependence on $a$ to determine
$H_0$ uniquely.  In general, two-image lenses have too
few constraints, and the extra constraints supplied by a four-image
lens constrain the angular structure rather than the
radial structure of the potential.
This basic problem holds for all
existing models of the current sample of time delay lenses.

\begin{figure}[ph]
\centerline{\psfig{figure=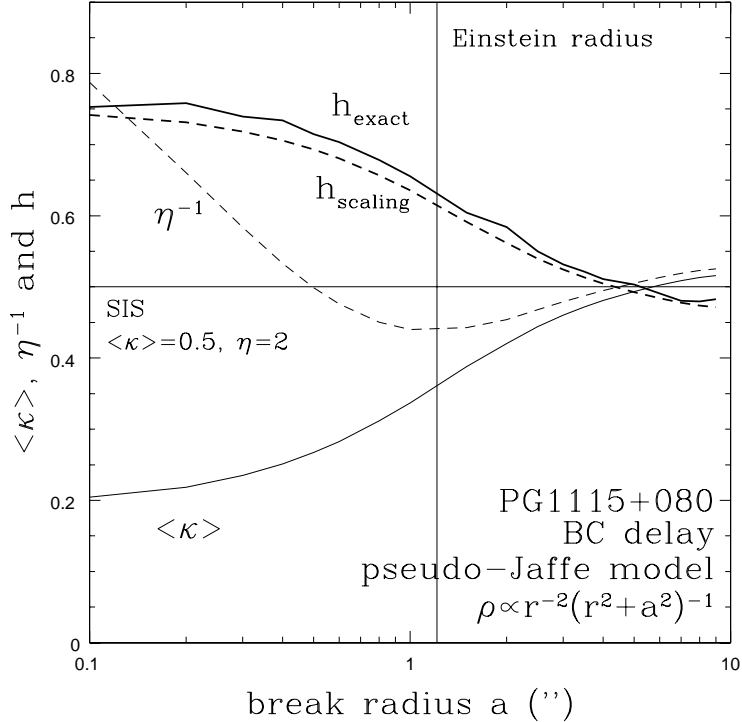,width=4.0in}}
\caption{ 
$H_0$ estimates for PG1115+080.  The lens galaxy is modeled as an ellipsoidal
pseudo-Jaffe model, $\rho \propto r^{-2}(r^2+a^2)^{-1}$, and the nearby group
is modeled as an SIS.  As the break radius $a \rightarrow \infty$ the
pseudo-Jaffe model becomes an SIS model, and as the break radius
$a\rightarrow 0$ it becomes a point mass.  The heavy solid curve ($h_{exact}$)
shows the dependence of $H_0$ on the break radius for the exact, nonlinear
fits of the model to the PG1115+080 data.  The heavy dashed curve
($h_{scaling}$) is the value found using our simple theory
(\S\protect{\ref{sec:timetheory}}) of time delays.  The agreement of the exact and
scaling solutions is typical. The light solid line shows the average surface
density $\langle\kappa\rangle$ in the annulus between the images, and the
light dashed line shows the {\it inverse} of the logarithmic slope $\eta$ in
the annulus ($\kappa\propto \theta^{1-\eta}$).  
For an SIS model we would have $\langle\kappa\rangle=1/2$ and
$\eta^{-1}=1/2$, as shown by the horizontal line.  When the break radius is
large compared to the Einstein radius (indicated by the vertical line), the
surface density is slightly higher and the slope is slightly shallower than
for the SIS model because of the added surface density from the group.  As we
make the lens galaxy more compact by reducing the break radius, the surface
density decreases and the slope becomes steeper, leading to a rise in $H_0$.
As the galaxy becomes very compact, the surface density near the Einstein ring
is dominated by the group rather than the galaxy, so the surface density
approaches a constant and the logarithmic slope approaches the value
corresponding to a constant density sheet ($\eta=1$).
\labelprint{fig:scaling}}
\end{figure}

The inability of the present time delay lenses to directly constrain the
radial density profile is the major problem for using them to determine
$H_0$. Fortunately, it is a consequence of the available data on the current 
sample rather than a fundamental limitation.
It is, however, a simple trade-off -- models with less dark matter 
(lower $\kbar$, more centrally concentrated densities) produce higher 
Hubble constants than those with more dark matter.  
We do have some theoretical limits on the value of $\kbar$. In particular, we
can be confident that the surface density is bounded by 
two limiting models.  The mass distribution should not be more compact
than the luminosity distribution, so a constant mass-to-light ratio ($M/L$)
model should set a lower limit on $\kbar \gtorder \kbar_{M/L} \simeq 0.2$, and
an upper limit on estimates of $H_0$.   We are also confident 
that the typical 
lens should not have a rising rotation curve at 1--2 optical effective radii from 
the center of the lens galaxy.  
Thus, the SIS model is probably the least concentrated reasonable 
model, setting an upper bound on $\kbar \ltorder \kbar_{SIS}=1/2$,  and a 
lower limit on estimates of $H_0$.  Figure~\ref{fig:timeproblem} 
shows joint estimates of $H_0$ from the four simple lenses for these two 
limiting mass distributions (Kochanek~2003b).  The results for the 
individual lenses are mutually consistent and are unchanged by the
new $0.149\pm0.004$~day delay for the A$_1$-A$_2$ images in PG1115+080
(Chartas~\cite{Chartas2003p1}).  For galaxies with isothermal profiles we
find $H_0=\vsis $~\kmsmpc, and for galaxies with constant $M/L$ 
we find $H_0=\vml $~\kmsmpc.  While our best prior estimate for the
mass distribution is the isothermal profile (see \S\ref{sec:modelfit}),
the lens galaxies would have to have constant $M/L$ to match 
Key Project estimate of $H_0=72 \pm 8$~\kmsmpc\  (Freedman et al.~\cite{Freedman2001p47})
or the WMAP estimate of $H_0=72 \pm 5$~\kmsmpc\ for a flat universe with a
cosmological constant (Spergel et al.~\cite{Spergel2003p175}). 

\begin{figure}[t]
\centerline{\psfig{figure=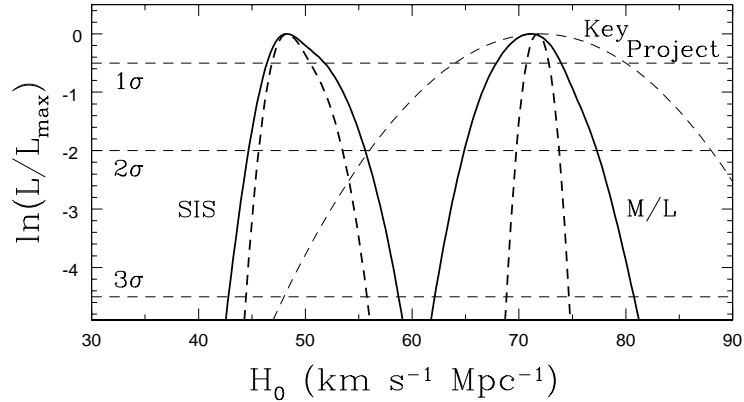,width=4.0in}}
\caption{
$H_0$ likelihood distributions.  The curves show the joint likelihood
functions for $H_0$ using the four simple lenses PG1115+080, SBS1520+530,
B1600+434, and HE2149--2745 and assuming either an SIS model (high
$\protect{\kbar}$, flat rotation curve) or a constant $M/L$ model (low
$\protect{\kbar}$, declining rotation curve).  The heavy dashed curves show
the consequence of including the X-ray time delay for PG1115+080 from Chartas
(2003) in the models. The light dashed curve shows a Gaussian model for the
Key Project result that $H_0=72\pm8$~\kmsmpc.
\labelprint{fig:timeproblem}}
\end{figure}

The difference between these two limits is entirely explained by the
differences in $\kbar$ and $\eta$ between the SIS and constant
$M/L$ models.  In fact, it is possible to reduce the $H_0$ estimates
for each simple lens to an approximation formula,
$H_0 = A(1-\kbar) + B \kbar (\eta -1)$. The coefficients, 
$A$ and $|B| \approx A/10$, are derived from the image positions and the time
delay using 
the simple theory from \S\ref{sec:timetheory}.
These approximations reproduce numerical results using ellipsoidal
lens models to accuracies of $3$~\kmsmpc\  (Kochanek~\cite{Kochanek2002p25}). 
For example, in Figure~\ref{fig:scaling} we also show the estimate of
$H_0$ computed based on the simple theory of \S\ref{sec:timetheory} and the
annular surface density ($\kbar$) and slope ($\eta$) of the numerical models.  
The agreement
with the full numerical solutions is excellent, even though the
numerical models include both the ellipsoidal lens galaxy and
a group.  No matter what the mass distribution is, the five lenses PG1115+080, 
SBS1520+530, B1600+434, PKS1830--211,\footnote{
PKS1830--211 is included based on the Winn et al.~(\cite{Winn2002p103}) model of the
{\it HST}\ imaging data as a single lens galaxy.  Courbin et al.~(\cite{Courbin2002p95}) prefer 
an interpretation with multiple lens galaxies which would invalidate the 
analysis.}
and HE2149--2745 have very similar dark matter halos.  For a fixed slope 
$\eta$, the five systems are consistent with a common value for the surface 
density of
\begin{equation}
   \kbar = 1 - 1.07 h + 0.14 (\eta-1)(1-h) \pm 0.04
\end{equation}
where $H_0=100h$~\kmsmpc\  and there is an upper limit of 
$\sigma_\kappa \ltorder 0.07$ on the intrinsic scatter of $\kbar$.  Thus,
time delay lenses provide a new window into the structure and homogeneity
of dark matter halos, regardless of the actual value of $H_0$.

There is an enormous range of parametric models that can illustrate
how the extent of the halo affects $\kbar$ and hence $H_0$ ---  the 
pseudo-Jaffe model we used above is only one example. It is
useful, however, to use a physically motivated model where the lens
galaxy is embedded in a standard NFW (Navarro, Frenk, \& White 1996) profile 
halo as we discussed at the end of \S\ref{sec:massmono}.  
The lens galaxy consists of the baryons that have cooled to form stars, 
so the mass of the visible galaxy can be parameterized using the cold baryon
fraction $f_{b,cold}$ of the halo, and for these CDM halo models the value
of $\kbar$ is controlled by the cold baryon fraction (Kochanek~\cite{Kochanek2003p49}).
A constant $M/L$ model is the limit 
$f_{b,cold} \rightarrow 1$ (with $\kbar \simeq 0.2$, $\eta \simeq 3$).
Since the baryonic mass fraction of a CDM halo should not exceed
the global fraction of $f_b \simeq 0.17 \pm0.03$ (e.g.,
Spergel et al.~\cite{Spergel2003p175}),
we cannot use constant $M/L$ models without also abandoning CDM.
As we reduce $f_{b,cold}$, we are adding mass to an extended 
halo around the lens, leading to an increase in $\kbar$ and a decrease
in $\eta$.  For $f_{b,cold} \simeq 0.02$ the model closely resembles
the SIS model ($\kbar\simeq 1/2$, $\eta \simeq 2$).  If we reduce $f_{b,cold}$
further, the mass distribution begins to approach that of the NFW halo
without any cold baryons. Figure~\ref{fig:timecdm} shows how $\kbar$ and
$H_0$ depend on $f_{b,cold}$ for PG1115+080, SBS1520+530, B1600+434 and
HE2149--2745.  When $f_{b,cold} \simeq 0.02$, the CDM models have parameters
very similar to the SIS model, and we obtain a very similar estimate
of $H_0=52\pm6$~\kmsmpc\  (95\% confidence).  If all baryons cool, and 
$f_{b,cold}=f_b$, then we obtain $H_0=65\pm6$~\kmsmpc\  (95\% confidence), 
which is still lower than the Key Project estimates.

\begin{figure}[p]
\begin{center}
\centerline{\psfig{figure=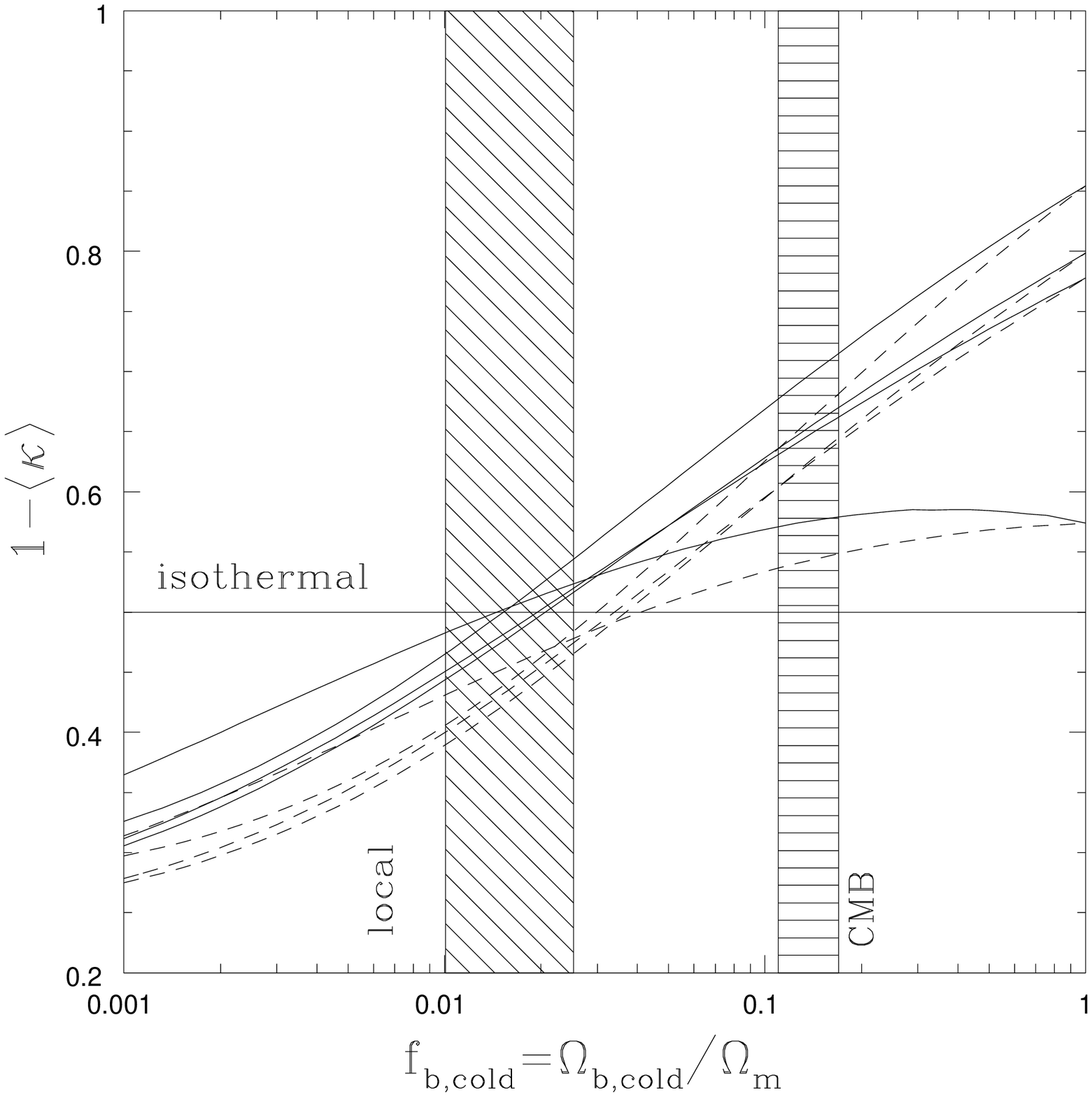,width=3.1in}}
\end{center}
\begin{center}
\centerline{\psfig{figure=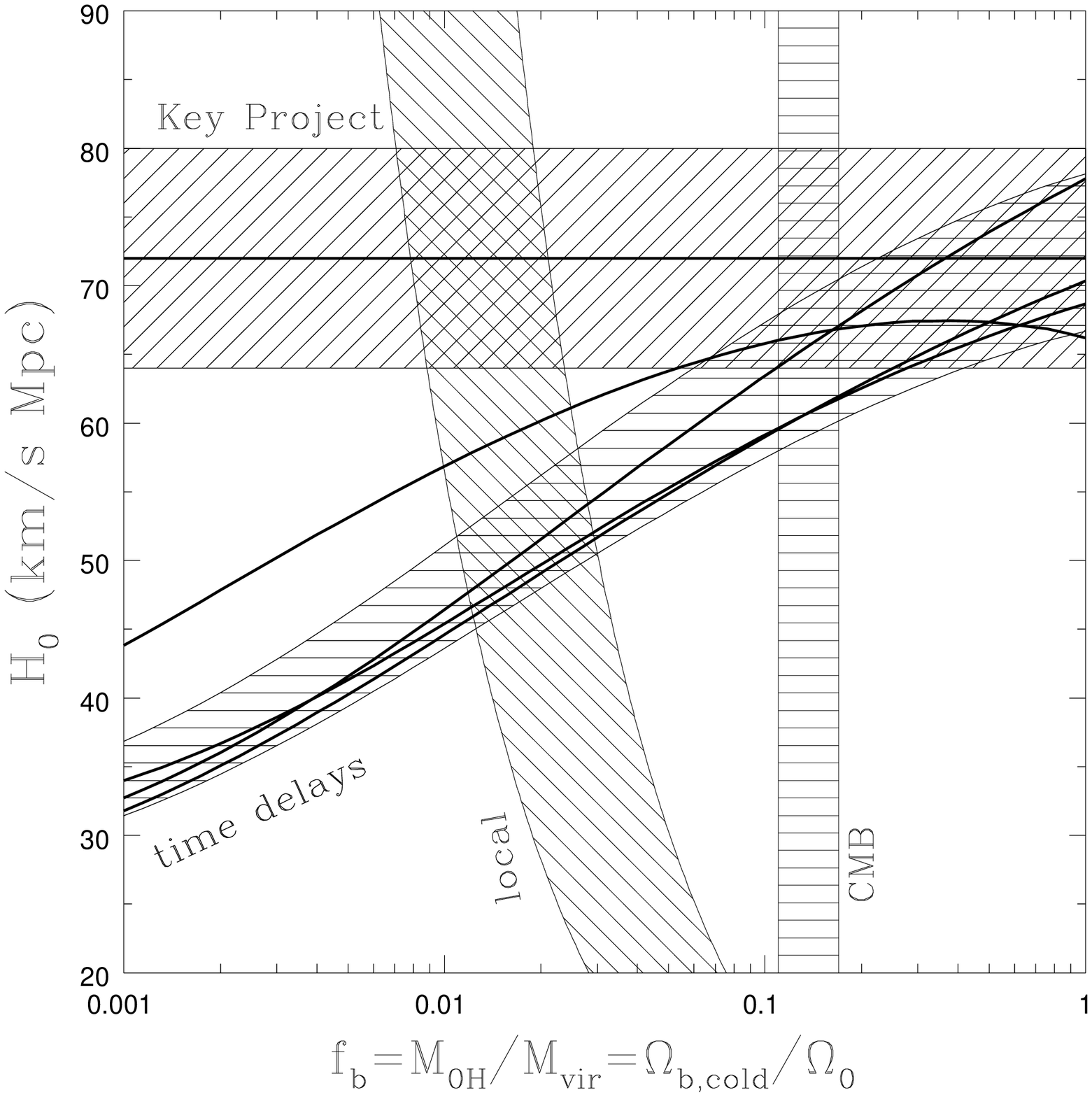,width=3.1in}}
\end{center}
\caption{
$H_0$ in CDM halo models.  The top panel shows $1-\kbar$ for the ``simple''
lenses (PG1115+080, SBS1520+530, B1600+434, and HE2149--2745) as a function of
the cold baryon fraction $f_{b,cold}$.  The solid (dashed) curves include
(exclude) the adiabatic compression of the dark matter by the baryons. The
horizontal line shows the value for an SIS potential.  The bottom panel shows
the resulting estimates of $H_0$, where the shaded envelope bracketing the
curves is the 95\% confidence region for the combined lens sample.  The
horizontal band shows the Key Project estimate.  For larger $f_{b,cold}$, the
density $\kbar$ decreases and the local slope $\eta$ steepens, leading to
larger values of $H_0$.  The vertical bands in the two panels show the lower
bound on $f_b$ from local inventories and the upper bound from the CMB.
\labelprint{fig:timecdm}}
\end{figure}

We excluded the lenses requiring significantly
more complicated models with multiple lens galaxies or very strong
perturbations from clusters.  If we have yet
to reach a consensus on the mass distribution of relatively isolated
lenses, it seems premature to extend the discussion to still more
complicated systems.  We can, however, show that the clusters lenses
require significant contributions to $\kbar$ from the cluster in 
order to produce the same $H_0$ as the more isolated systems.
As we discussed in \S\ref{sec:data} the three more complex systems
are RXJ0911+0551, Q0957+561 and B1608+656.

RXJ0911+0551 is very strongly perturbed by the nearby X-ray cluster
(Morgan et al.~\cite{Morgan2001p1}; Hjorth et al.~\cite{Hjorth2002p11}).  
Kochanek (\cite{Kochanek2003p1}) found
$H_0=49 \pm 5$~\kmsmpc\  if the primary lens and its satellite were 
isothermal and $H_0=67 \pm 5$~\kmsmpc\  if they had constant mass-to-light
ratios.  The higher value of   
$H_0=71\pm4$~\kmsmpc\  obtained by Hjorth et al.~(\cite{Hjorth2002p11}) can be
understood by combining \S\ref{sec:timetheory} and \S\ref{sec:timecluster}
with the differences in the models.  In particular, Hjorth et al.~(\cite{Hjorth2002p11})
truncated the halo of the primary lens near the Einstein radius
and used a lower mass cluster, both of which lower $\kbar$ and 
raise $H_0$.  The Hjorth et al.~(\cite{Hjorth2002p11}) models also included many
more cluster galaxies assuming fixed masses and halo sizes.

Q0957+561 is a special case because the primary lens galaxy is the
brightest cluster galaxy and it lies nearly at the cluster center
(Keeton et al.~\cite{Keeton2000p74}; Chartas et al.~\cite{Chartas2002p96}).  As a result, the 
lens modeling problems are particularly severe, and Keeton 
et al.~(\cite{Keeton2000p74}) found that all previous models (most recently,
Barkana et al.~\cite{Barkana1999p54}; Bernstein \& Fischer~\cite{Bernstein1999p14}; 
and Chae~\cite{Chae1999p582}, see \S\ref{sec:modelfit}) 
were incompatible with the observed geometry of the lensed host 
galaxy.  While Keeton et al.~(\cite{Keeton2000p74}) found models consistent with
the structure of the lensed host, they covered a range of almost 
$\pm25\%$ in their estimates of $H_0$.  A satisfactory
treatment of this lens remains elusive.    

HE1104--1805 has the most recently measured time delay(Ofek \& Maoz~\cite{Ofek2003p101}, 
Wyrzykowski et al.~\cite{Wyrzykowski2003p229}).
Given the $\Delta t=161\pm7$~day delay,
a standard SIE model of this system predicts a very high 
$H_0 \simeq 90$~\kmsmpc.  The geometry of this system and the
fact that the inner image is brighter than the outer image both
suggest that HE1104--1805 lies in an anomalously high tidal shear
field, while the standard model includes a prior to keep the 
external shear small.  A prior is needed because a two-image lens 
supplies too few constraints to determine both the
ellipticity of the main lens and the external shear simultaneously.
Since the images and the lens in HE1104--1805 are nearly collinear,
the anomalous $H_0$ estimate for the standard model may be an
example of the shear degeneracy we briefly mentioned in \S\ref{sec:timetheory}.
At present the model surveys needed to understand the new delay
have not been made. Observations of the geometry of the host 
galaxy Einstein ring will resolve any ambiguities due to the shear
in the near future (see \S\ref{sec:hosts}). 

The lens B1608+656 consists of two interacting galaxies, and, as
we discussed in \S\ref{sec:data}, this leads to a 
greatly increased parameter space.  Fassnacht et al.~(\cite{Fassnacht2002p823})
used SIE models for the two galaxies to find $H_0=61-65$~\kmsmpc,
depending on whether the lens galaxy positions are taken from the $H$-band or 
$I$-band lens {\it HST}\ images (the statistical errors are 
negligible).  The position differences are probably created by extinction
effects from the dust in the lens galaxies.
Like isothermal models of the ``simple'' lenses,
the $H_0$ estimate is below local values, but the disagreement is
smaller.  These models correctly match the observed time delay
ratios.    Koopmans et al.~(\cite{Koopmans2003p70}) obtain a still higher
estimate of $H_0=75\pm7$~\kmsmpc\, largely because the lens galaxy positions shift after they
include extinction corrections.  They use a foreground screen model
to make the extinction corrections, which is a better approximation
than no extinction corrections, but is unlikely allow precise correction
in a system like B1608+656 where the dust and stars are mixed and there 
is no simple relation between color excess and optical depth (e.g. Witt, 
Thronson \& Capuano~\cite{Witt1992p393}).

Despite recent progress both in modeling the VLBI structure (Wucknitz et al.~\cite{Wucknitz2004p14})
and obtaining deep images (York et al.~\cite{York2004p1}) it is unclear whether 
B0218+357 has escaped its problems with astrometry and models.  While York
et al.~(\cite{York2004p1}) have clearly measured the position of the lens galaxy,
the dependence of the position on the choice of the PSF model remains a significant
source of uncertainty for estimates of $H_0$.  Models of the system using
power law models find a slope very close to isothermal $\eta = 2.04 \pm 0.02$
($\rho \propto r^{-\eta}$).  Unfortunately, these models have too few degrees
of freedom given the small astrometric uncertainties in the VLBI structures providing the
constraints (because the only angular structure in the model is the ellipsoidal
potential used for the main lens galaxy), and this makes
the limits on the power slope suspect (see \S\ref{sec:modelfit}).  For example, while
it is true that Leh\'ar et al.~(\cite{Lehar2000p584}) estimated that the environmental
shear near B0218+357 was small, even a $\gamma=0.01$ external tidal shear produces
deflections (3 milli-arcseconds) that are large compared to the accuracy of the
constraints used for the models and so must be included for the models to be 
reliable.  
With these caveats, B0218+357 (like the models of B1608+656 with
significant extinction corrections) support a nearly isothermal mass distribution
with $H_0 =73\pm8$~\kmsmpc.

\subsection{The Future of Time Delay Measurements \label{sec:timefuture} }

We understand the theory of time delays very well -- the only important variable
in the lens structure is the average surface density $\kbar$ of the lens near 
the images for which the delay is measured.  The angular structure of the potential
has an effect on the delays, but it is either small or well-constrained by the
observed image positions.  Provided a lens does not lie in a cluster where the
cluster potential cannot be described by a simple expansion, any lens model that
includes the parameters needed to vary the average surface density of the lens near
the images and to change the ratio between the quadrupole moment of the lens and
the environment includes all the variables needed to model time delays, to estimate
the Hubble constant, and to understand the systematic uncertainties in the results.
Unfortunately, there is a tendency in the literature to confuse rather than to
illuminate this understanding, even though all differences between estimates of
the Hubble constant for the simple time delay lenses can be understood on this
basis.

The problem with time delays lies with the confusing state of the data.  The
four simplest time delay lenses, PG1115+080, SBS1520+530, B1600+434 and HE2149--2745,
can only match the currently preferred estimate of $H_0 \simeq 72\pm 8$~\kmsmpc\
(Freedman et al.~\cite{Freedman2001p47}, Spergel et al.~\cite{Spergel2003p175})
if they have nearly constant $M/L$ mass distributions.  If they have the favored
quasi-isothermal mass distributions, then $H_0 \simeq 48\pm 3$~\kmsmpc.  This leads to
a conundrum: why do simple lenses with time delay measurements have falling rotation
curves, while simple lenses with direct estimates of the mass profile do not?
This is further
confused by B1608+656 and B0218+357, which due to their observational complexity
would be the last systems I would attempt to understand, but in current analyses
can be both isothermal and have high $H_0$.    In resolving this problem it is
not enough to search for a ``Golden Lens.'' {\it There is no such thing!}
While chanting ``My lens is better than your lens!'' may be satisfying, it
contributes little to understanding the basic problem.

The difficulty at the moment is that systems I would view as problematic
(B0218+357 due to problems in astrometry or B1608+656 due to the interacting
lens galaxies) allow both mass distributions with flat rotation curves and
$H_0=72$~\kmsmpc, while systems that should be simpler to interpret
(the simple lenses in Table~\ref{tab:delays}) do not.   Yet the preponderance
of evidence on the mass distributions of lens galaxies suggests that they
are fairly homogeneous in structure and have roughly flat rotation curves
(\S\ref{sec:mass}).  The simplest way to clarify this problem is to measure
accurate time delays for many more systems.  At a fixed value of the Hubble
constant we will either find significant scatter in the surface densities
near the images of simple lenses or we will not.

\section{Gravitational Lens Statistics \labelprint{sec:stat} }

It is the opinion of the author that the statistics of lenses as a method for
determining the cosmological model has largely 
ceased to be interesting.  However, it is important to understand the underlying
physics because it determines the types of lenses we detect.  
While most recent analyses have found cosmological results
consistent with the concordance model (Chae et al.~\cite{Chae2002p1301}, Chae~\cite{Chae2003p746},
Davis, Huterer \& Krauss~\cite{Davis2003p1029}, Mitchell et al.~\cite{Mitchell2004p1})
there are still large statistical uncertainties
and some dangerous systematic assumptions.  More importantly, there is little prospect
at present of lens statistics becoming competitive with other methods.   
Gravitational lenses statistics arguably begins with Press \& Gunn~(\cite{Press1973p397}),
although the ``modern'' era begins with the introduction of magnification bias
(Turner~\cite{Turner1980p135}), the basic statistics of normal galaxy lenses 
(Turner, Ostriker \& Gott~\cite{Turner1984p1}), cross sections and optical depths
for more general lenses (Blandford \& Kochanek~\cite{Blandford1987p133},
Kochanek \& Blandford~\cite{Kochanek1987p676}), explorations of the effects
of general cosmologies (Fukugita et al.~\cite{Fukugita1990p24}, 
Fukugita \& Turner~\cite{Fukugita1991p99}) and lens structure (Maoz \& Rix~\cite{Maoz1993p425},
Kochanek~\cite{Kochanek1996p638}) and the development of the general
methodology of interpreting observations (Kochanek~\cite{Kochanek1993p12},
\cite{Kochanek1996p638}).

\subsection{The Mechanics of Surveys \labelprint{sec:statmech}}

There are two basic approaches to searching for gravitational lenses
depending on whether you start with a list of potentially lensed 
sources or a list of potential lens galaxies.  Of the two, only a 
search of sources for lensed sources has a significant
cosmological sensitivity -- for a non-evolving population of lenses
in a flat cosmological model we will find in \S\ref{sec:statcross} that
the number of lensed sources scales with the volume between the observer 
and the source $D_s^3$.  If you search potential lens galaxies for those
which have actually lensed a source, 
then the cosmological dependence enters only through
distance ratios, $D_{ds}/D_s$, and you
require a precise knowledge of the source redshift distribution. Thus,
while lenses found in this manner are very useful for many projects
(mass distributions, galaxy evolution etc.), they are not very useful
for determining the cosmological model.  This changes for the case of
cluster lenses where you may find multiple lensed sources at different
redshifts behind the same lens (e.g. Soucail, Kneib \& Gorse~\cite{Soucail2004p33}).

Most lenses have been found by searching for lensed sources because the
number of targets which must be surveyed is considerably smaller.  This
is basically a statement about the relative surface densities of candidate
sources and lenses.  The typical lens is a galaxy with an Einstein radius
of approximately $b\simeq 1\farcs0$ so it has a cross section of order
$\pi b^2$.  If you search $N$ lenses with such a cross section for 
signs of a lensed source, you would expect to find $N \pi b^2 \Sigma_{source}$
lenses where $\Sigma_{source}$ is the surface density of detectable
sources.   If you search $N$ sources for a lens galaxy in front of 
them, you would expect to find $N \pi b^2 \Sigma_{lens}$ lenses,
where $\Sigma_{lens}$ is the surface density of lens galaxies.  Since
the surface density of massive galaxies is significantly higher than
the surface density of easily detectable higher redshift sources 
($\Sigma_{lens} \gg \Sigma_{source}$), you need examine fewer sources
than lens galaxies to find the same number of lensed systems.
This is somewhat mitigated by the fact the surface density of 
potential lens galaxies is high enough to allow you to examine
many potential lenses in a single observation, while the surface
density of sources is usually so low that they can be examined only
one at a time.
  
For these reasons, we present a short synopsis of searches for sources
behind lenses and devote most of this section to the search for lenses
in front of sources.  The first method for finding sources behind lenses
is a simple byproduct of redshift surveys.  Redshift surveys take 
spectra of  the central regions of low redshift galaxies
allowing the detection of spectral features from any lensed images inside
the aperture used for the spectrum.  Thus, the lens Q2237+0305 was found
in the CfA redshift survey (Huchra et al.~\cite{Huchra1985p691}) and 
SDSS0903+5028 (Johnston et al.~\cite{Johnston2003p2281}) was found in the 
SDSS survey.  Theoretical estimates
(Kochanek~\cite{Kochanek1992p381},
Mortlock \& Webster~\cite{Mortlock2000p879}) 
suggest that the discovery rate should one per $10^4$--$10^5$ redshift
measurements, but this does not seem to be borne out by the number
of systems discovered in this age of massive redshift surveys (the
origin of the lower rate in the 2dF survey is discussed by 
Mortlock \& Webster~\cite{Mortlock2001p629}).
  Miralda-Escude \& Leh\'ar~(\cite{MiraldaEscude1992p31})
proposed searching for lensed optical (emission line) rings, a strategy
successfully used by Warren et al.~(\cite{Warren1996p139}) to find 0047--2808
and by Ratnatunga, Griffiths \& Ostrander~(\cite{Ratnatunga1999p86}) to 
find lenses in the HST Medium Deep Survey (MDS).
There is also a hybrid approach whose main
objective is simply to find lenses with minimal follow up observations
by looking for high redshift radio lobes that have non-stellar optical
counterparts (Leh\'ar et al.~\cite{Lehar2001p60}).  
Since radio lobes have no intrinsic optical emission,
a lobe superposed on a galaxy is an excellent lens candidate.  The 
present limitation on this method is the low angular resolution of
the available all sky radio surveys (FIRST, NVSS) and the magnitude
limits and star/galaxy separation problems of the current all-sky
optical catalogs.  Nonetheless, several systems have been 
discovered by this technique.

\begin{figure}[t]
\begin{center}
\centerline{\psfig{figure=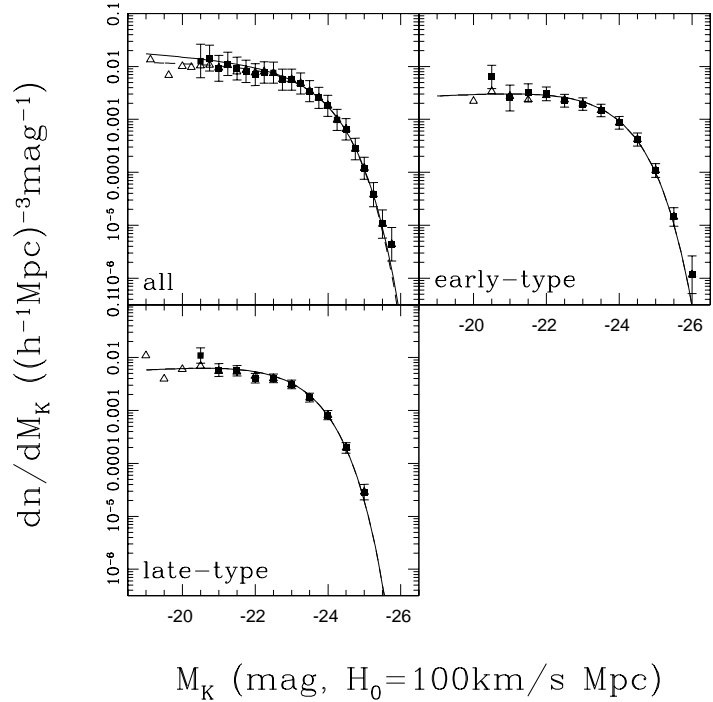,width=4.0in}}
\end{center}
\caption{
  Example of a local galaxy luminosity function.  These are the K-band 
  luminosity functions for either all galaxies or by morphological type
  from Kochanek et al.~(\cite{Kochanek2001p566}).  The curves show the
  best fit Schechter models for the luminosity functions while the
  points with error bars show a non-parametric reconstruction.
  }
\labelprint{fig:lfunc}
\end{figure}

The majority of lens surveys, however, have focused on either
optical quasars or radio sources because they are source populations
known to lie at relatively high redshift ($z_s \gtorder 1$) and that are
easily detected even when there is an intervening lens galaxy.
Surveys of optical quasars (Crampton, McClure \& Fletcher~\cite{Crampton1992p23},
Yee, Fillipenko \& Tang~\cite{Yee1993p7}, Maoz et al.~\cite{Maoz1993p28},
Surdej et al.~\cite{Surdej1993p2064}, Kochanek, Falco \& Schild~\cite{Kochanek1995p109}) 
have the advantage that the 
sources are bright, and the disadvantages that the bright sources
can mask the lens galaxy and that the selection process is modified
by dust in the lens galaxy and emission from the lens galaxy.  We
will discuss these effects in \S\ref{sec:optical}.  While many more lensed quasars have
been discovered since these efforts, none of the recent results have been presented
as a survey.  
Surveys of all radio sources 
(the MIT/Greenbank survey, Burke, Leh\'ar \& Conner~\cite{Burke1992p237}) 
have the advantage that most lensed radio sources are 
produced by extended, steep spectrum sources 
(see Kochanek \& Lawrence~\cite{Kochanek1990p1700}), and the 
disadvantage that the complex intrinsic structures of extended
radio sources make the follow up observations difficult.  Surveys of
flat spectrum radio sources (the CLASS survey, Browne et al.~\cite{Browne2003p341},
the PANELS survey, Winn, Hewitt \& Schechter~\cite{Winn2001p61}) have the advantage that the follow
up observations are relatively simple because most unlensed
flat spectrum sources are (nearly) point sources.
There are disadvantages
as well -- because the source structure is so simple, flat spectrum
lenses tend to provide fewer constraints on mass models than steep
spectrum lenses. The radio sources tend to be optically faint, making
it difficult to determine their redshifts in many cases. 

\begin{figure}[t]
\centerline{\psfig{figure=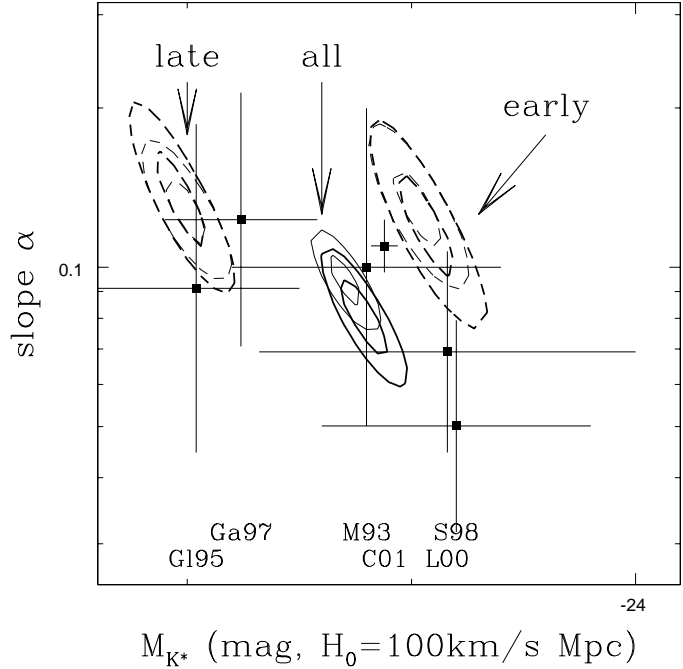,width=4.0in}}
\caption{
  Schechter parameters $\alpha$ and $M_*$ for the 2MASS luminosity functions
  shown in Fig.~\ref{fig:lfunc}.  Note there is a significant correlations
  not only between $\alpha$ and $M_*$ but also with the comoving density
  scale $n_*$ that should be included
  in lens statistical calculations but generally are not.
  }
\labelprint{fig:lfunc2}
\end{figure}

The second issue for any survey is to understand the method by which the sources
were originally identified.  For example, it is important to know whether
the flux of a lensed source in the input catalog is the total flux of all 
the images or only a part of the flux (e.g. the flux of the brightest
image).  This will have a significant effect on the statistical 
corrections for using a flux-limited catalog, a correction known
in gravitational lensing as the ``magnification bias'' (see \S\ref{sec:magbias}).
All large, published surveys were essentially drawn from samples
which would include the total flux of a lensed system.  It is also 
important to know whether the survey imposed any criterion for the
sources being point-like, since lensed sources are not, or any color 
criterion that might be violated by lensed sources with bright lens
galaxies or significant extinction.

The third issue for any survey is to consider the desired selection
function of the observations.  This is some combination of resolution,
dynamic range and field of view.  These determine the range of lens 
separations that are detectable, the nature of any background sources,
and the cost of any follow up observations.  Any survey is a trade-off
between completeness (what fraction of all lenses in sample that can
be discovered), false positives (how many objects selected as lenses
candidates that are not), and the cost of follow-up observations.  The
exact strategy is not critical provided it is well-understood. The 
primary advantages of the surveys of flat spectrum radio sources are
the relatively low false positive rates and follow up costs produced by using
a source population consisting almost entirely of point sources with
no contaminating background population.  This does not mean that the
flat spectrum surveys are free of false positives -- core-jet sources
can initially look like asymmetric two-image lenses.
On small angular scales
($\Delta\theta \ltorder 3\farcs0$) the quasar surveys share this
advantage, but for wider separations there is contamination from
binary quasars (see \S\ref{sec:binaryqsos}) 
and Galactic stars (see Kochanek~\cite{Kochanek1993p438}).

\begin{figure}[t]
\centerline{\psfig{figure=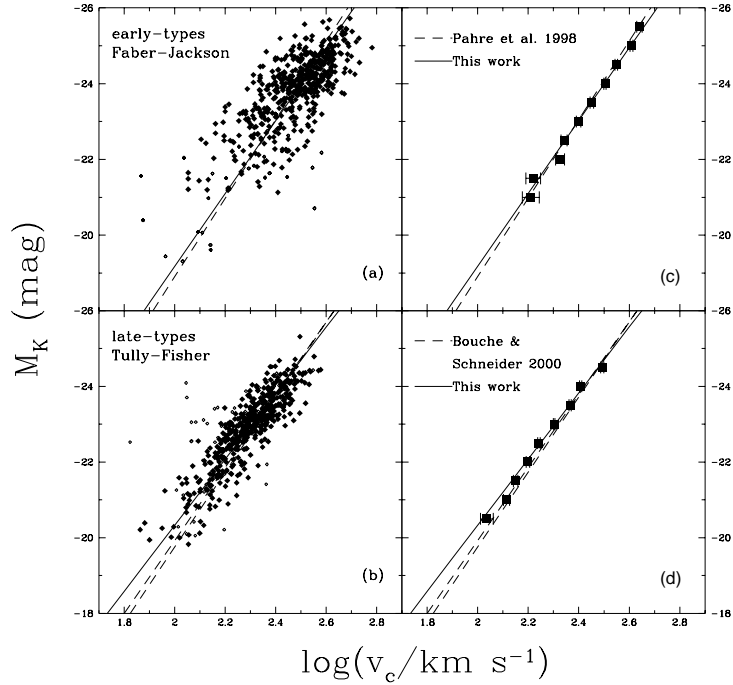,width=4.0in}}
\caption{
  K-band kinematic relations for 2MASS galaxies.  The top panels show the
  Faber-Jackson relation and the bottom panels show the Tully-Fisher relations
  for 2MASS galaxies with velocity dispersions and circular velocities drawn
  from the literature. The left hand panels show the individual galaxies,
  while the right hand galaxies show the mean relations.  Note the
  far larger scatter of the Faber-Jackson relation compared to the
  Tully-Fisher relation.
  }
\labelprint{fig:kinematic}
\end{figure}

\subsection{The Lens Population \labelprint{sec:statgals}}

The probability that a source has an intervening lens requires a model for the
distribution of the lens galaxies.  In almost all cases these are based on
the luminosity function of local galaxies combined with the assumption that
the comoving density of galaxies does not evolve with redshift.  Of course
luminosity is not mass, so a model for converting the luminosity of a local
galaxy into its deflection scale as a lens is a critical part of the process.
For our purposes, the distributions of galaxies in luminosity are well-described
by a Schechter~(\cite{Schechter1976p297}) function,
\begin{equation}
     { dn \over dL } = { n_* \over L_*} \left( { L \over L_* } \right)^\alpha
       \exp\left(-L/L_*\right).
\end{equation}
The Schechter function has three parameters:  a characteristic luminosity $L_*$
(or absolute magnitude $M_*$), an exponent $\alpha$ describing the rise at
low luminosity, and a comoving density scale $n_*$.   All these parameters
depend on the type of galaxy being described and the wavelength of the
observations.  In general, lens calculations have divided the galaxy population
into two broad classes:  late-type (spiral) galaxies and early-type (E/S0)
galaxies.   Over the period lens statistics developed, most luminosity 
functions were measured in the blue, where early and late-type galaxies 
showed similar characteristic luminosities.  The definition of a galaxy type
is a slippery problem -- it may be defined by the morphology of the surface
brightness (the traditional method), spectral classifications (the modern 
method since it is easy to do in redshift surveys), colors (closely
related to spectra but not identical), and stellar kinematics (ordered rotational
motions versus random motions).  Each approach has advantages and disadvantages,
but it is important to realize that the kinematic definition is the one most
closely related to gravitational lensing and the one never supplied by local
surveys.  Fig.~\ref{fig:lfunc} shows an example of a luminosity function, in this case
K-band infrared luminosity function by Kochanek et al.~(\cite{Kochanek2001p566}, also Cole et 
al.~\cite{Cole2001p255}) where
$M_{K*e}=-23.53\pm0.06$~mag, $n_{*e}=(0.45\pm0.06)\times 10^{-2}h^3$~Mpc$^{-3}$,
and $\alpha_e=-0.87\pm0.09$ for galaxies which were morphologically early-type galaxies
and $M_{K*l}=-22.98\pm0.06$~mag, $n_{*l}=(1.01\pm0.13)\times 10^{-2}h^3$~Mpc$^{-3}$,
and $\alpha_l=-0.92\pm0.10$ for galaxies which were morphologically late-type galaxies.
Early-type galaxies are less common but brighter than late-type galaxies at K-band.
It is important to realize that the parameter estimates
of the Schechter function are correlated, as shown in Fig.~\ref{fig:lfunc2}, and
that it is dangerous to simply extrapolate them to fainter luminosities than
were actually included in the survey 

\begin{figure}[t]
\centerline{\psfig{figure=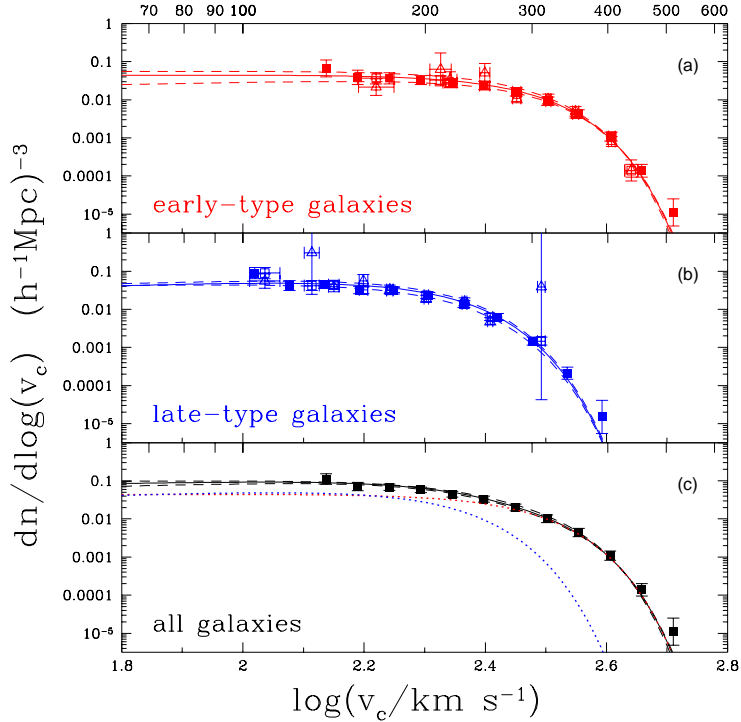,width=4.0in}}
\caption{
  The resulting velocity functions from combining the K-band luminosity
  functions (Fig.~\ref{fig:lfunc}) and kinematic relations (Fig.~\ref{fig:kinematic})
  for early-type (top), late-type (middle) and all (bottom) galaxies. The points
  show partially non-parametric estimates of the velocity function based 
  on the binned estimates in 
  the right hand panels of Fig.~\ref{fig:kinematic} rather than power-law
  fits.  Note that early-type galaxies dominate for high circular velocity.  }
\labelprint{fig:vfunc}
\end{figure}

However, light is not mass, and it is mass which determines lensing properties.
One approach would simply be to assign a mass-to-light ratio to the galaxies
and to the expected properties of the lenses.  This was attempted only
in Maoz \& Rix~(\cite{Maoz1993p425}) who found that for normal stellar mass
to light ratios it was impossible to reproduce the data (although it is 
possible if you adjust the mass-to-light ratio to fit the data, also see 
Kochanek~\cite{Kochanek1996p638}).  Instead, most studies convert the 
luminosity functions $dn/dL$ into a velocity functions $dn/dv$ using 
the local kinematic properties of galaxies and then relate the stellar
kinematics to the properties of the lens model.
As Fig.~\ref{fig:kinematic} 
shows (for the same K-band magnitudes of our luminosity function example),
both early-type and late-type galaxies show correlations between luminosity
and velocity.  For late-type galaxies there is a tight correlation known
as the Tully-Fisher~(\cite{Tully1977p661}) relation between luminosity $L$ and
circular velocity $v_c$ and for early-type galaxies
there is a loose correlation known as the Faber-Jackson~(\cite{Faber1976p668})
relation between luminosity and central velocity dispersion $\sigma_v$. 
 Early-type galaxies do show a much tighter correlation known
as the fundamental plane (Dressler et al.~\cite{Dressler1987p42},
Djorgovski \& Davis~\cite{Djorgovski1987p59}) but it is a three-variable correlation
between the velocity dispersion, effective radius and surface brightness (or luminosity)
that we will discuss in \S\ref{sec:optical}.  While there is probably some effect
of the FP correlation on lens statistics, it has yet to be found.  For lens calculations,
the circular velocity of late-type galaxies is usually converted into an equivalent
(isotropic) velocity dispersion using $v_c = \sqrt{2} \sigma_v$.  We can derive
the kinematic relations for the same K-band-selected galaxies used in the
Kochanek et al.~(\cite{Kochanek2001p566}) luminosity function, finding the 
Faber-Jackson relation
\begin{equation}
   M_k - 5\log h = (-23.83 \pm 0.03) - 2.5(4.04\pm 0.18)(\log v_c -2.5)
   \labelprint{eqn:kbandtf}
\end{equation}
and the Tully-Fisher relation
\begin{equation}
   M_k - 5\log h = (-22.92 \pm 0.02) - 2.5(3.96\pm 0.08)(\log v_c -2.3).
   \labelprint{eqn:kbandfj}
\end{equation}
These correlations, when combined with the K-band luminosity function have the
advantage that the magnitude systems for the luminosity function and the kinematic
relations are identical, since magnitude conversions have caused problems for
a number of lens statistical studies using older photographic luminosity 
functions and kinematic relations.  For these relations, the characteristic
velocity dispersion of an $L_*$ early-type galaxy is $\sigma_{*e} \simeq 209$~km/s
while that of an $L_*$ late-type galaxy is $\sigma_{*l} \simeq 143$~km/s.  These
are fairly typical values even if derived from a completely independent set 
of photometric data.

\begin{figure}[t]
\centerline{\psfig{figure=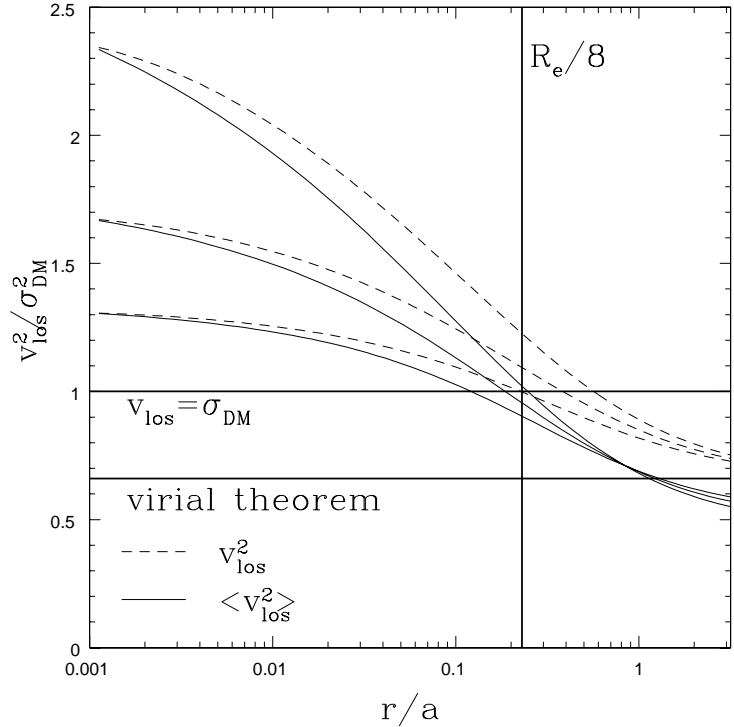,width=4.0in}}
\caption{
  Stellar velocity dispersions $v_{los}$ for a Hernquist distribution of stars in an
  isothermal halo of dispersion $\sigma_{DM}$.  The solid curves show the 
  local value $v_{los}$ and the dashed curves show the mean interior to the radius
   $\langle v_{los}^2\rangle$.   Local velocity dispersions are typically
  measured on scales similar to $R_e/8$ where the stellar and dark matter
  dispersions are nearly equal rather than matching the viral theorem limit
  which would be reached in an infinite aperture.  The upper, lower and
  middle curves are for stars with isotropies of $\beta=0.2$ (somewhat
  radial), $\beta=0$ (isotropic) and $\beta=-0.2$ (somewhat tangential).
  }
\labelprint{fig:dispersion}
\end{figure}

Both the Faber-Jackson and Tully-Fisher relations are power-law relations between
luminosity and velocity, $L/L_* \propto (\sigma_v/\sigma_*)^{\gamma_{FJ}}$.  This allows
a simple variable transformation to convert the luminosity function into a 
velocity function,
\begin{equation}
    { d n \over d \sigma_v } = { d n \over dL } \left| { d L \over d\sigma_v }\right| =
          \gamma_{FJ} { n_* \over \sigma_* } 
         \left( { \sigma_v \over \sigma_* } \right)^{(1+\alpha)\gamma_{FJ}-1}
          \exp\left(-(\sigma_v/\sigma_*)^\gamma_{FJ} \right).
       \label{eqn:vfunc}
\end{equation}
There are three caveats to keep in mind about this variable change.  First, we have
converted to the distribution in stellar velocities, not some underlying velocity
characterizing the dark matter distribution.  
Many early studies assumed a fixed transformation
between the characteristic velocity of the stars and the lens model.  In
particular, Turner, Ostriker \& Gott~(\cite{Turner1984p1}) 
introduced the assumption $\sigma_{dark} = (3/2)^{1/2} \sigma_{stars}$ for
an isothermal mass model based on the stellar dynamics 
(Jeans equation, Eqn.~\ref{eqn:acd} and \S\ref{sec:dynamics}) 
of a $r^{-3}$ stellar density distribution in a 
$r^{-2}$ isothermal mass distribution.  Kochanek (\cite{Kochanek1993p12},
\cite{Kochanek1994p56}) showed that this 
oversimplified the dynamics and that if you embed a real stellar luminosity
distribution in an isothermal mass distribution you actually find that the central
stellar velocity dispersion is close to the velocity dispersion characterizing
the dark matter halo.  Fig.~\ref{fig:dispersion} compares the stellar velocity
dispersion to the dark matter halo dispersion for a Hernquist distribution of
stars in an isothermal mass distribution.  Such a normalization calculation is
required for any attempt to match the observed velocity functions with a particular
mass model for the lenses.     
Second, in an ideal world, the luminosity function and the kinematic relations should 
be derived from a consistent set of photometric data, while in practice they
rarely are.  As we will see shortly, the cross
section for lensing scales roughly as $\sigma_*^4$, so small errors in
estimates of the characteristic velocity have enormous impacts on the
resulting cosmological results -- a 5\% velocity calibration error leads
to a 20\% error in the lens cross section.  Since luminosity functions and
kinematic relations are rarely derived consistently (the exception is
Sheth et al.~\cite{Sheth2003p225}), the resulting systematic errors creep 
into cosmological estimates.  Finally, for the early-type galaxies
where the Faber-Jackson kinematic relation has significant scatter, 
transforming the luminosity function using the mean relation as we
did in Eqn.~\ref{eqn:vfunc} while ignoring the scatter underestimates the number 
of high velocity dispersion galaxies (Kochanek~\cite{Kochanek1994p56},
Sheth et al.~\cite{Sheth2003p225}).  This leads to underestimates of both
the image separations and the cross sections.  The fundamental lesson of
all these issues is that the mass scale of the
lenses should be ``self-calibrated'' from the observed separation distribution
of the lenses rather than imposed using local observations  
(as we discuss below in \S\ref{sec:cosmo}). 

\begin{figure}[t]
\centerline{\psfig{figure=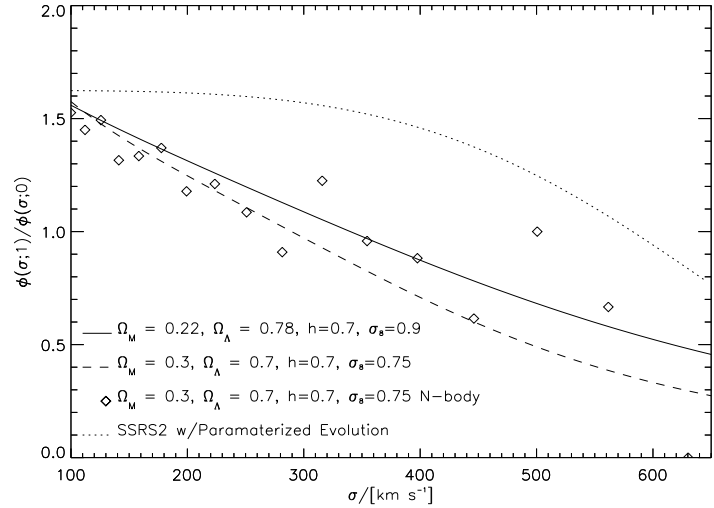,width=4.0in}}
\caption{
  The ratio of the velocity function of halos at $z=1$ to that at $z=0$ from
  Mitchell et al.~(\cite{Mitchell2004p1}).  The
  solid curve shows the expectation for an $\Omega_\Lambda\simeq0.78$ flat cosmological
  model. The points show results from an N-body simulation with $\Omega_\Lambda\simeq 0.7$
  and the dashed curve shows the theoretical expectation.   For comparison, the
  dotted curve shows the evolution model used by Chae \& Mao~(\cite{Chae2003p61}).
  }
\labelprint{fig:dispevolve}
\end{figure}

Most lens calculations have assumed that the comoving density of the lenses
does not evolve with redshift.  For moderate redshift sources this only requires
little evolution for $z_l < 1$ (mostly $z_l < 0.5$), but for higher redshift
sources it is important to think about evolution as well.  
The exact degree of evolution is the subject of some debate, but a standard
theoretical prediction for the change between now and redshift unity is 
shown in Fig.~\ref{fig:dispevolve} (see Mitchell et al.~\cite{Mitchell2004p1} and
references therein).  Because lower mass systems merge to
form higher mass systems as the universe evolves, low mass systems are
expected to be more abundant at higher redshifts while higher mass systems
become less abundant.  For the $\sigma_v \sim \sigma_* \sim 200$~km/s galaxies
which dominate lens statistics, the evolution in the number of galaxies is
actually quite modest out to redshift unity, so we would expect galaxy evolution
to have little effect on lens statistics.  Higher mass systems evolve rapidly 
and are far less abundant at redshift unity, but these systems will tend to
be group and cluster halos rather than galaxies and the failure of the baryons
to cool in these systems is of greater importance to their lensing effects
than their number evolution (see \S\ref{sec:cluster}).  There have been
a number of studies examining lens statistics with number evolution 
(e.g. Mao~\cite{Mao1991p9}, Mao \& Kochanek~\cite{Mao1994p569}, Rix et al.~\cite{Rix1994p49})   
and several attempts to use the lens data to constrain the evolution
(Ofek, Rix \& Maoz~\cite{Ofek2003p101}, Chae \& Mao~\cite{Chae2003p61},
Davis, Huterer \& Krauss~\cite{Davis2003p1029}).

\subsection{ Cross Sections \labelprint{sec:statcross}}

The basic quantity we need for any statistical analysis is the
cross section of the lens for producing the desired lensing effect (e.g. multiple
images, two images, bright images...).  The simplest cross section is the
multiple imaging cross section of the SIS lens -- 
the angular area on the source plane in which a source will produce two lensed 
images.  We know from Eqns.~\ref{eqn:aap} and \ref{eqn:aaq} that the source 
must lie within Einstein radius
$b$ of the lens center to produce multiple images, so the cross section
is simply $\sigma_{SIS} = \pi b^2$.  Since the Einstein radius 
$b = 4\pi(\sigma_v/c)^2 D_{ls}/D_s$ depends on the velocity dispersion and
redshift of the lens galaxy, we will need a model for the distribution of
lenses in redshift and velocity dispersion to estimate the 
optical depth for lensing.   If we are normalizing directly to stellar
dynamical measurements of lenses, then we will also need a dynamical
model (e.g. the Jeans equations of \S\ref{sec:dynamics}) to relate the
observed stellar velocity dispersions to the characteristic dark matter
velocity dispersion $\sigma_v$ appearing as a parameter of the SIS model.
We can also compute cross sections for obtaining different image morphologies.
For example, in Eqn.~\ref{eqn:aba} 
we calculated the caustic boundaries for the four-image
region of an SIS in an external shear $\gamma$.  If we integrate to find the
area inside the caustic we obtain the four-image cross section
\begin{equation}
     \sigma_4 = { 3 \pi \over 2 } { \gamma^2 b^2 \over 1-\gamma^2 },
\end{equation}
while (provided $|\gamma|<1/3$) the two-image cross section is
$\sigma_2=\sigma_{SIS}-\sigma_4 \simeq \sigma_{SIS}$. If the shear
is larger, then the tips of astroid caustic extend beyond the 
radial (pseudo-)caustic and the lens has regions producing two images,
three images in the disk geometry (Fig.~\ref{fig:cusplens}), and
four images with no simple expression for the cross sections.
There are no analytic results for the singular isothermal ellipsoid
(Eqn.~\ref{eqn:abf} with $s=0$), but we can power expand the cross
section as a series in the ellipticity to find at lowest order that 
\begin{equation}
     \sigma_4 = {  \pi \over 6 } b^2 \epsilon^2
\end{equation}
for a lens with axis ratio $q=1-\epsilon$, while the total cross section
is $\sigma_{SIS}=\pi b^2$ (e.g. Kochanek~\cite{Kochanek1996p595},
Finch et al.~\cite{Finch2002p51}).  As a general rule,
a lens of ellipticity $\epsilon$ is roughly equivalent to a spherical lens
in an external shear of $\gamma \simeq \epsilon/3$.
According to the cross sections, the fraction of four-image lenses should 
be of order $\sigma_4/\sigma_{SIS} \sim \gamma^2 \sim (\epsilon/3)^2 \sim 0.01$ 
rather than the observed 30\%.  Most of this difference is a consequence 
of the different magnification biases of the two image multiplicities.  

There is an important subtlety when studying lens statistics with models
covering a range of axis ratios, namely that the definition of the 
critical radius $b$ in (say) the SIE model (Eqn.~\ref{eqn:abf}) depends
on the axis ratio and exactly what quantity you are holding fixed in
your calculation (see Keeton, Kochanek \& Seljak~\cite{Keeton1997p604},
Keeton \& Kochanek~\cite{Keeton1998p157}, Rusin \& Tegmark~\cite{Rusin2001p709},
Chae~\cite{Chae2003p746}).  For example, if we compare a singular
isothermal sphere to a  face on Mestel disk with the same equatorial
circular velocity, the Einstein radius of the disk is $2/\pi$ smaller
than the isothermal sphere because for the same circular velocity a
disk requires less mass than a sphere.  Since we usually count galaxies
locally and translate these counts into a dynamical variable, this means
that lens models covering a range of ellipticities must be normalized in
terms of the same dynamical variables as were used to count the galaxies.

Much early effort focused of the effects of adding a finite core radius
to these standard models (e.g.  Blandford \& Kochanek~\cite{Blandford1987p658}, 
Kochanek \& Blandford~\cite{Kochanek1987p676}, Kovner~\cite{Kovner1987p22},
Hinshaw \& Krauss~\cite{Hinshaw1987p468}, Krauss \& White~\cite{Krauss1992p385},
Wallington \& Narayan~\cite{Wallington1993p517}, Kochanek~\cite{Kochanek1996p638}).
The core radius $s$ leads to an evolution of the
caustic structures (see \partintro, Blandford \& Narayan~\cite{Blandford1986p568})
with the ratio between the core radius and the
critical radius  $s/b$.  Strong lenses with $s/b \ll 1$ act like 
singular models.  Weak, or marginal, lenses with $s/b \sim 1$ have
significantly reduced cross sections but higher average magnifications
such that the rising magnification bias roughly balances the diminishing
cross section to create a weaker than expected effect of core
radii on the probability of finding a lens (see Kochanek~\cite{Kochanek1996p638}).  
As the evidence that lenses are effectively singular
has mounted, interest in these models has waned, and we will not discuss
them further here.  There is some interest in these models as a probe
of large separation lenses due to groups and clusters where a finite
core radius is replaced by effects of the shallow $\rho \propto r^{-1}$
NFW density cusp,  and we will consider this problem in
\S\ref{sec:cluster} where we discuss large separation lenses.

\subsection{Optical Depth \labelprint{stat:opdepth}}

The optical depth associated with a cross section is the fraction of the sky 
in which you can place a source and see the effect.  This simply requires 
adding up the contributions from all the lens galaxies between the observer
and the redshift of the source.  For the SIS lens we simply need to know
the comoving density of lenses per unit dark matter velocity dispersion $dn/d\sigma$
(which may be a function of redshift)
\begin{equation}
   \tau_{SIS} = \int_0^{z_s} { dV \over dz_l} dz_l \int_0^\infty { d n \over d\sigma_v} 
          { \sigma_{SIS} \over 4 \pi } d\sigma_v
    \labelprint{eqn:opdepthdef}
\end{equation}
where $dV/dz_l$ is the comoving volume element per unit redshift
(e.g. Turner, Ostriker \& Gott~\cite{Turner1984p1}).  For a flat cosmology, which 
we adopt from here on, the comoving volume element is simply 
$dV = 4\pi D_d^2 dD_d$ where $D_d$ is the comoving distance to the lens
redshift (Eqn.~\ref{eqn:volume}).
As with most lens calculations, this means that the
expression simplifies if expressed in terms of the comoving angular
diameter distances,
\begin{equation}
   \tau_{SIS} = \int_0^{D_s} dD_d D_d^2 \left( { D_{ds} \over D_s }\right)^2 
          \int_0^\infty { d n \over d\sigma_v} 16 \pi^2 \left( { \sigma_v \over c } \right)^4
\end{equation}
(Gott, Park \& Lee~\cite{Gott1989p1}, Fukugita, Futamase \& Kasai~\cite{Fukugita1990p24}).
If the comoving density of the lenses does not depend on redshift, the 
integrals separate to give
\begin{equation}
    \tau_{SIS} = { 8\pi^2 \over 15 } D_s^3 \int_0^\infty d\sigma_v 
         { d n \over d\sigma}_v \left( { \sigma_v \over c } \right)^4 
\end{equation}
(Fukugita \& Turner~\cite{Fukugita1991p99}).
If we now assume that the galaxies can be described by the combination of 
Schechter luminosity functions and kinematic relations described in 
\S\ref{sec:statgals}, then we can do the remaining integral to find that
\begin{equation}
    \tau_{SIS} = { 8\pi^2 \over 15 } n_* \left( { \sigma_* \over c } \right)^4
             D_s^3 \Gamma[1+\alpha+6/\gamma]
           = { 1 \over 30 } \tau_* r_H^{-3} D_s^3  \Gamma[1+\alpha+6/\gamma]
\end{equation}
where $\Gamma[x]$ is a Gamma function, $r_H=c/H_0$ is the Hubble radius 
and the optical depth scale is
\begin{equation}
   \tau_*=16\pi^3 n_* r_H^3 \left( { \sigma_*\over c} \right)^4  
      = 0.026 \left( { n_* \over 10^{-2} h^3\hbox{Mpc}^{-3} } \right)
              \left( { \sigma_* \over 200\hbox{km s}^{-1} } \right)^4.
\end{equation}
 Thus, lens statistics are essentially
a volume test of the cosmology (the $D_s^3$), predicated on knowing the comoving
density of the lenses ($n_*$) and their average mass ($\sigma_*$).  The 
result does not depend on the Hubble constant -- all determinations of $n_*$
scale with the Hubble constant such that $n_* D_s^3$ is independent of $H_0$.

Two other distributions, those in image separation and in lens redshift at
fixed image separation, are easily 
calculated for the SIS model and useful if numerical for any other lens.  The
SIS image separation is $\Delta\theta=8\pi(\sigma_v/c)^2 D_{ds}/D_s$, so 
\begin{eqnarray}
  { d \tau_{SIS} \over d \Delta\theta} &= { 1 \over 2 } D_s^3 \hat{\Delta\theta}^2
   &\left( \Gamma\left[1+\alpha-2/\gamma_{FJ}, \xi \right] \right. \\
   && \qquad \left.  -  2\hat{\Delta\theta} 
     \Gamma\left[1+\alpha-4/\gamma_{FJ}, \xi\right] 
    +\hat{\Delta\theta}^2 
    \Gamma\left[1+\alpha-6/\gamma_{FJ}, \xi\right]
   \right) \nonumber
    \labelprint{eqn:sepdist}
\end{eqnarray}  
where $\xi=(\Delta\theta/\Delta\theta_*)^{\gamma_{FJ}/2}$ and
\begin{equation}
   \Delta\theta_* = 8\pi \left( { \sigma_* \over c }\right)^2 =
          2\farcs3 \left( { \sigma_* \over 200\hbox{km s}^{-1} } \right)^2
\end{equation}
is the maximum separation produced by an $L_*$ galaxy. The mean image separation,
\begin{equation}
     \langle \Delta\theta\rangle = { \Delta\theta_* \over 2 }
            {  \Gamma[1+\alpha+8/\gamma] \over 
             \Gamma[1+\alpha+6/\gamma]^{1/2}},
\end{equation}
depends only on the properties of the lens galaxy and not on cosmology.  
If the cosmological model is not flat, a very weak dependence on cosmology
is introduced (Kochanek~\cite{Kochanek1993p453}).  For a known separation
$\Delta\theta$, the probability distribution for the lens redshift becomes
\begin{equation}
   { d P \over d z_l } \propto { D_d^2 \over D_s } { dD_d \over dz_l} 
         \exp\left[ - \left( 
        { \Delta \theta \over \Delta\theta_* } { D_s \over D_{ds}} \right)^{1/2} \right] 
    \labelprint{eqn:zpdist}
\end{equation}  
(we present the result only for Schechter function $\alpha=-1$ and 
Faber-Jackson $\gamma_{FJ}=4$).  The location of the exponential cut off
introduced by the luminosity function has a strong cosmological dependence,
so the presence or absence of lens galaxies at higher redshifts dominates
the cosmological limits.  The structure of this function is quite different
from the total optical depth, which in a flat cosmology is a slowly varying
function with a mean lens distance equal to one-half the distance to the 
source.  The mean redshift changes with cosmology because of the changes
in the distance-redshift relations, but the effect is not as dramatic as
the redshift distributions for lenses of known separation.   

We end this section by discussing the Keeton~(\cite{Keeton2002p1}) ``heresy''.  
Keeton~(\cite{Keeton2002p1})
pointed out that if you used a luminosity function derived at intermediate
redshift rather than locally, then the cosmological sensitivity of the 
optical depth effectively vanishes when the median redshift of the lenses
matches the median redshift of the galaxies used to derive the 
luminosity function.  The following simple thought experiment shows that
this is true at one level. Suppose there was only one kind of galaxy
and we make a redshift survey and count all the galaxies in a thin 
shell at redshift $z$, finding $N$ galaxies between $z$ and $z+\Delta z$.
The implied comoving density of the galaxies, $n = N/(\Delta z dV/dz)$, 
depends on the cosmological model with the same volume factor appearing in the 
optical depth calculation (Eqn.~\ref{eqn:opdepthdef}).  To the extent that
the redshift ranges and weightings of the galaxy survey and a lens survey
are similar, the cosmological sensitivity of the optical depth vanishes
because the volume factor cancels and the optical depth depends only on
the number of observed galaxies $N$.  This does not occur when we use a
local luminosity function because changes in cosmology have no effect on 
the local volume element.  The problem with the  Keeton~(\cite{Keeton2002p1})
argument is that it basically says that
if we could use galaxy number counts to determine the cosmological model
then we would not need lensing to do so because the two are redundant.
To continue our thought experiment, we also have local estimates $n_{local}$
for the density of galaxies, and as we vary the cosmology we would find
that $n$ and $n_{local}$ agree only for a limited range of cosmological
models and this would restore the cosmological sensitivity. The problem 
is that the comparison of 
near and distant measurements of the numbers of galaxies is tricky because
it depends on correctly matching the galaxies in the presence of galaxy
evolution and selection effects -- in essence, you cannot use this
argument to eliminate the cosmological sensitivity of lens surveys unless
you think you understand galaxy evolution so well that you can use galaxy
number counts to determine the cosmological model, a program of research
that has basically been abandoned.  

\subsection{ Spiral Galaxy Lenses \labelprint{sec:spirals}}

Discussions of lens statistics, or even lenses in general, focus on early-type
galaxies (E/S0).  The reason is that spiral lenses are relatively rare.
The only morphologically obvious spirals are
B0218+357 (Sc, York et al.~\cite{York2004p1}),
B1600+434 (S0/Sa, Jaunsen \& Hjorth~\cite{Jaunsen1997p39}),
PKS1830--211  (Sb/Sc, Winn et al.~\cite{Winn2002p103}),
PMNJ2004--1349 (Sb/Sc, Winn, Hall \& Schechter \cite{Winn2003p672}),
and Q2237--0305 (Sa, Huchra et al.~\cite{Huchra1985p691}).
Other small separation systems may well be spiral galaxies, but we
do not have direct evidence from imaging.
There are studies of individual spiral lenses or the statistics of
spiral lenses by 
Maller, Flores \& Primack~(\cite{Maller1997p681}),
Keeton \& Kochanek~(\cite{Keeton1998p157}),
Koopmans et al.~(\cite{Koopmans1998p534}),
Maller et al.~(\cite{Maller2000p194}), Trott \& Webster~(\cite{Trott2002p621}),
and Winn, Hall \& Schechter~(\cite{Winn2003p672}).

The reason lens samples are dominated by early-type galaxies is that the
early-type galaxies are more massive even if slightly less numerous 
(e.g. Fukugita \& Turner~\cite{Fukugita1991p99}, see \S\ref{sec:statgals}).  The relative numbers
of early-type and late-type lenses should be the ratio of their 
optical depths, $(n_l/n_e)(\sigma_l/\sigma_e)^4$, based on the 
comoving densities and characteristic velocity dispersions of the
early and late-type galaxies.  For example, in the Kochanek et al.~(\cite{Kochanek2001p566})
K-band luminosity function $n_l/n_e \simeq 2.2$ while the ratio of the
characteristic velocity dispersions is $\sigma_{*l}/\sigma_{*e}=0.68$ 
giving an expected fraction of 32\% spiral.  This is modestly higher
than the values using other luminosity functions (usually closer to
20\%) or the observed fraction.  
Because the typical separation of the spiral lenses will also be smaller
by a factor of $(\sigma_{*l}/\sigma_{*e})^2=0.46$, they will be much harder
to resolve given the finite resolution of lens surveys.  Thus, survey 
selections functions discriminate more strongly against late-type lenses
than against early-type lenses.  The higher prevalence of dust in late-type
lenses adds a further bias against them in optical surveys.

\subsection{Magnification Bias \labelprint{sec:magbias}}

The optical depth calculation suggests that the likelihood of finding that
a $z_s\simeq 2$ quasar is lensed is very small ($\tau \sim 10^{-4}$) , 
while observational surveys of bright quasars typically find that of
order 1\% of bright quasars are lensed.  The origin of the discrepancy
is the effect known as ``magnification bias'' (Turner~\cite{Turner1980p135}),
which is really the correction
needed to account for the selection of survey targets from flux limited
samples.  Multiple imaging 
always magnifies the source, so lensed sources are brighter than the population
from which they are drawn.  For example, the mean magnification of all multiply
imaged systems is simply the area over which we observe the lensed images divided
by the area inside the caustic producing multiple images because the
magnification is the Jacobean relating area on the image and source
planes, $d^2\vec{\beta} = |\mu|^{-1} d^2\vec{\theta}$.  For example, an
SIS lens with Einstein radius $b$ produces multiple images over a region
of radius $b$ on the source plane (i.e. the cross section is $\pi b^2$),
and these images are observed over a region of radius $2b$ on the image
plane, so the mean multiple-image magnification is 
$\langle\mu\rangle = (4\pi b^2)/(\pi b^2) = 4$.

Since fainter sources are almost always more 
numerous than brighter sources, magnification bias almost always increases your
chances of finding a lens.  The simplest example is to imagine a lens which
always produces the same magnification $\mu$ applied to a population with
number counts $N(F)$ with flux $F$.  The number counts of the lensed 
population are then $N_{lens}(F)=\tau \mu^{-1} N(F/\mu)$, so the fraction 
lensed objects (at flux $F$) is larger than the number expected from the
optical depth if fainter objects are more numerous than the magnification times 
the density of brighter objects.  Where did the extra factor of magnification come 
from?  It has to be there to conserve the total number of sources or 
equivalently the area on the source and lens planes -- you
can always check your expression for the magnification bias by
computing the number counts of lenses and checking to make sure that the 
total number of lenses equals the total number of sources if the optical
depth is unity.  

Real lenses do not produce unique magnifications, so it is necessary to work
out the magnification probability distribution $P(>\mu)$ (the probability of a
magnification larger than $\mu$) or its differential $dP/d\mu$ and then convolve
it with the source counts.  Equivalently we can
define a magnification dependent cross section, $d\sigma/d\mu=\sigma dP/d\mu$ 
where $\sigma$ is the total cross section.  We can do this
easily only for the SIS lens, where a source at $\beta$ produces two images
with a total magnification of $\mu=2/\beta$ with $\mu > 2$ in the
multiple image region (Eqns~\ref{eqn:aap}, \ref{eqn:aaq}), 
to find that $P(>\mu)=(2/\mu)^2$ and $dP/d\mu=8/\mu^3$.  The
structure at low magnification depends on the lens model, but all sensible
lens models have $P(>\mu) \propto \mu^{-2}$ at high magnification because this
is generic to the statistics of fold caustics (\partintro,
Blandford \& Narayan~\cite{Blandford1986p568}). 

Usually people have defined a magnification bias factor $B(F)$ for sources of
flux $F$ so that the probability $p(F)$ of finding a lens with flux $F$
is related to the
optical depth by $p(F) = \tau B(F)$.  The magnification bias factor is
\begin{equation}
    B(F) = N(F)^{-1} \int { d\mu \over \mu } { dP \over d\mu } N \left( { F \over \mu } \right)
     \labelprint{eqn:bias1}
\end{equation}
for a source with flux $F$, or
\begin{equation}
    B(m) = N(m)^{-1} \int d\mu { dP \over d\mu } N \left(m+2.5\log \mu\right)
     \labelprint{eqn:bias2}
\end{equation}
for a source of magnitude $m$.  Note the vanishing of the extra $1/\mu$ factor 
when using logarithmic number counts $N(m)$ for the sources rather than
the flux counts $N(F)$.  Most standard models have magnification probability 
distributions similar to the SIS model, with $P(>\mu)\simeq (\mu_0/\mu)^2$ for
$\mu >\mu_0$, in which case the magnification bias factor for sources with 
power law number counts $N(F)=dN/dF\propto F^{-\alpha}$ is
\begin{equation}
    B(F) = { 2 \mu_0^{\alpha-1} \over 3-\alpha } 
\end{equation}
provided the number counts are sufficiently shallow ($\alpha < 3$).  For
number counts as a function of magnitude $N(m)=dN/dm \propto 10^{am}$
(where $a=0.4(\alpha-1)$) the bias factor is
\begin{equation}
    B(F) = { 2 \mu_0^{2.5a} \over 2.5a - 2 }. 
\end{equation}
The steeper the number counts and the brighter the source is relative to
any break between a steep slope and a shallow slope, the greater the
magnification bias.  For radio sources a simple power law model
suffices, with $\alpha \simeq 2.07\pm 0.11$ for the CLASS survey
(Rusin \& Tegmark~\cite{Rusin2001p709}), leading to a magnification
bias factor of $B\simeq 5$.  For quasars, 
however, the bright quasars have number counts steeper than this critical
slope, so the location of the break from the steep slope of the bright 
quasars to the shallower
slope for fainter quasars near $B\sim 19$~mag is critical to determining the 
magnification bias.   Fig.~\ref{fig:seleff} shows an example of a typical
quasar number counts distribution as compared to several (old) models for
the distribution of lensed quasars.  The changes in the magnification bias
with magnitude are visible as the varying ratio between the lensed and 
unlensed counts, with a much smaller ratio for bright quasars (high 
magnification bias) than for faint quasars (low magnification bias) and
a smooth shift between the two limits as you approach the break in the
slope of the counts at $B \sim 19$~mag.

For optically-selected lenses, magnification bias is ``undone'' by extinction
in the lens galaxy because extinction provides an effect that makes lensed
quasars dimmer than their unlensed counterparts.  Since the
quasar samples were typically selected at blue wavelengths, the rest
wavelength corresponding to the quasar selection band at the redshift
of the lens galaxy where it encounters the dust is similar to the U-band.
If we use a standard color excess $E(B-V)$ for the amount of dust, then
the images become fainter by of order $A_U E(B-V)$ magnitudes where
$A_U \simeq 4.9$.  Thus, if lenses had an average extinction of only
$E(B-V)\simeq 0.05$ mag, the net magnification of the lensed images
would be reduced by about 25\%.  If all lenses had the same
demagnification factor $f<1$ then the modifications to the magnification
bias would be straight forward.  For power-law number counts
$N(F) \propto F^{-\alpha}$, the magnification bias is reduced
by the factor $f^\alpha$ and a $E(B-V)=0.05$ extinction leads to
a 50\% reduction in the magnification bias for objects with a 
slope $\alpha \simeq 2$ (faint quasars) and to still larger reductions
for bright quasars.  Some examples of the changes with the addition 
of a simple mean extinction
are shown in the right panel of Fig.~\ref{fig:seleff}, although the
levels of extinction shown there are larger than observed in typical
lenses as we discuss in \S\ref{sec:ism}.   Comparisons between the
statistics of optically-selected and radio-selected samples can be
used to estimate the magnitude of the correction.  The only
such comparison found estimated extinctions consistent with the
direct measurements of \S\ref{sec:ism} 
(Falco, Kochanek \& Mu\~noz~\cite{Falco1998p47}).  However, the ISM
of real lenses is presumably far more complicated, with a distribution
of extinctions and different extinctions for different images which
may be a function of orientation and impact parameter relative to the
lens galaxy, for which we have no good theoretical model.  

The flux of the lens galaxy also can modify the magnification bias for 
faint quasars, although the actual sense of the effect is complex. 
The left panel Fig.~\ref{fig:seleff} shows the effect of dropping lenses
in which the lens galaxy represents some fraction of the total flux of 
the lensed images.  The correction is unimportant for bright quasars 
because lens galaxies with $B < 19$~mag are rare.  In this picture, the
flux from the lens galaxy leads to the loss of lenses because the 
added flux from the lens galaxy makes the colors of faint lensed quasars
differ from those of unlensed quasars so they are never selected as quasars to
begin with.  Alternatively, if one need not worry about color contamination,
then the lens galaxy increases the magnification bias by supplying extra
flux that makes lensed quasars brighter.  

Any other selection effect, such as the dynamic range allowed for flux
ratios between images as a function of their separation will also have
an effect on the magnification bias.  Exactly how the effect enters
depends on the particular class of images being considered.  For example,
in the SIS lens (or more generally for two-image lenses), a limitation on
the detectable flux ratio $0 < f_{min} < 1$ sets a minimum detectable 
magnification $\mu_{min}=2(1+f_{min})/(1-f_{min})>\mu_0=2$.  Since most lens
samples have significant magnification bias, which means that most lenses
are significantly magnified, such flux limits have only modest effects. 
The other limit, which cannot be captured in the SIS model, is that 
almost all bright images are merging pairs on folds (or triplets on cusps)
so the image separation decreases as the magnification increases.  The
contrast between the merging images and any other images also increases
with increasing magnification -- combined with limits on the detectability
of images, these lead to selection effects against highly magnified images.
This is also usually a modest effect -- while magnification bias is important,
the statistics are dominated by modestly magnified systems rather than 
very highly magnified images.
In fact, there are have been few attempts at complete
studies of the complicated interactions between finding
quasars, finding lenses, selection effects and magnification bias.
There is an early general study by Kochanek~(\cite{Kochanek1991p517})
and a detailed practical application of many of these issues to the
SDSS survey by  Pindor et al.~(\cite{Pindor2003p2325}).  Unfortunately,
Pindor et al.~(\cite{Pindor2003p2325}) 
seem to arrive at a completeness estimate from their selection model that is
too high given the number of lenses they found in practice.  Some of this
may be due to underestimating the luminosity of lens galaxies, the effects
of the lens galaxy or extinction on the selection of quasars or the 
treatment of extended, multicomponent lenses compared to normal quasars
in the photometric pipeline.  These difficulties, as well as the larger
size of the present radio-selected lens samples, are the reason that almost
all recent statistical studies have focused exclusively on radio lenses.  

\begin{figure}[t]
\begin{center}
\centerline{\psfig{figure=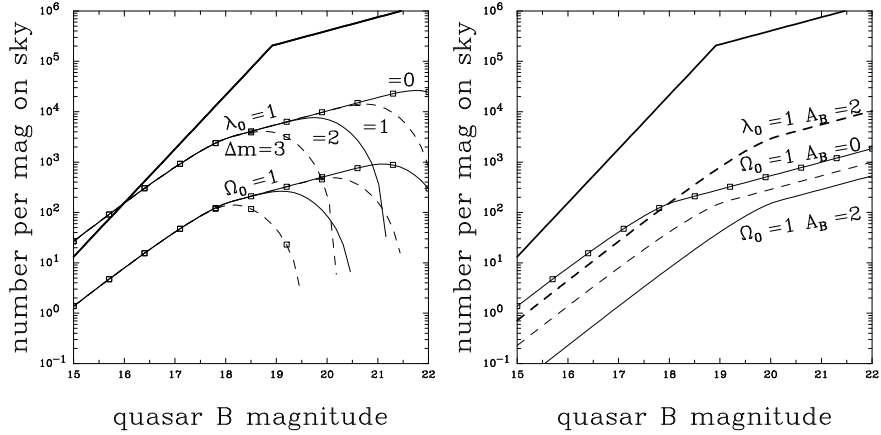,width=4.5in}}
\end{center}
\caption{
   Examples of selection effects on optically selected lens samples.  The
   heavy solid curves in the two panels shows a model for the magnitude
   distribution of optically-selected quasars.  The light curves labeled
   $\Omega_M=1$ and $\lambda_0=1$ show the distribution of lensed quasars
   for flat cosmologies that are either pure matter or pure cosmological
   constant.  The change in the ratio between the lensed curves and the
   unlensed curves illustrates the higher magnification bias for bright
   quasars where the number count distribution is steeper than for faint
   quasars.  In the left panel the truncated curves show the effect of
   losing the lensed systems where the lens galaxy is $\Delta m=1$, $2$
   or $3$ magnitudes fainter than the quasars.  Once surveys are searching
   for lensed quasars with $B \gtorder 20$~mag, the light from the
   lens galaxy becomes an increasing problem, particularly since the
   systems with the brightest lens galaxies will also have the largest
   image separations that would otherwise make them easily detected.
   In the left panel we illustrate the effect of adding a net extinction
   of $A_B=1$ or $2$~mag from dust in the lens galaxies. These correspond
   to larger than expected color excesses of $E(B-V)\simeq0.2$ and
   $0.4$~mag respectively.  Note how the extinction ``undoes'' the
   magnification bias by shifting the lensed distributions to fainter
   magnitudes.
   }
\labelprint{fig:seleff}
\end{figure}

The standard magnification bias expressions (Eqns.~\ref{eqn:bias1} and \ref{eqn:bias2})
are not always appropriate.
They are correct for the statistics of lenses selected from source populations
for which the total flux of the source (including all images of a lensed source)
is defining $F$ (or $m$).  This is true of most existing surveys -- for example
the CLASS radio survey sources were originally selected from single dish observations
with very poor resolution compared to typical image separations (see Browne et al.
\cite{Browne2003p341}).  If, however,
the separation of the images is large compared to the resolution of the observations
and the fluxes of the images are considered separately, then the bias must be
computed in terms of the bright image used to select sources to search for 
additional images.  This typically reduces the bias.  More subtle effects can
also appear.  For example, the SDSS survey selects quasar candidates based on 
the best fit point-source magnitudes, which will tend to be an underestimate of
the flux of a resolved lens.  Hence the magnification bias for lenses found 
in the SDSS survey will be less than in the standard theory.  Samples selected
based on more than one frequency can have more complicated magnification biases
depending on the structure of the multidimensional number counts  
(Borgeest, von Linde \& Refsdal~\cite{Borgeest1991p35},
Wyithe, Winn \& Rusin~\cite{Wyithe2003p58}).
The exact behavior is complex, but
the magnification bias can be tremendously increased if the fluxes in the 
bands are completely uncorrelated or tightly but nonlinearly correlated. 
For example, if the luminosities in bands A and B are related by 
tight, nonlinear correlation of the form $L_A \propto L_B^{1/2}$, 
then the lensed examples of these objects will
lie off the correlation.  At present, there are too few deep, wide-area
multiwavelength catalogs to make good use of this idea, but this is changing
rapidly.

\begin{figure}[t]
\centerline{\psfig{figure=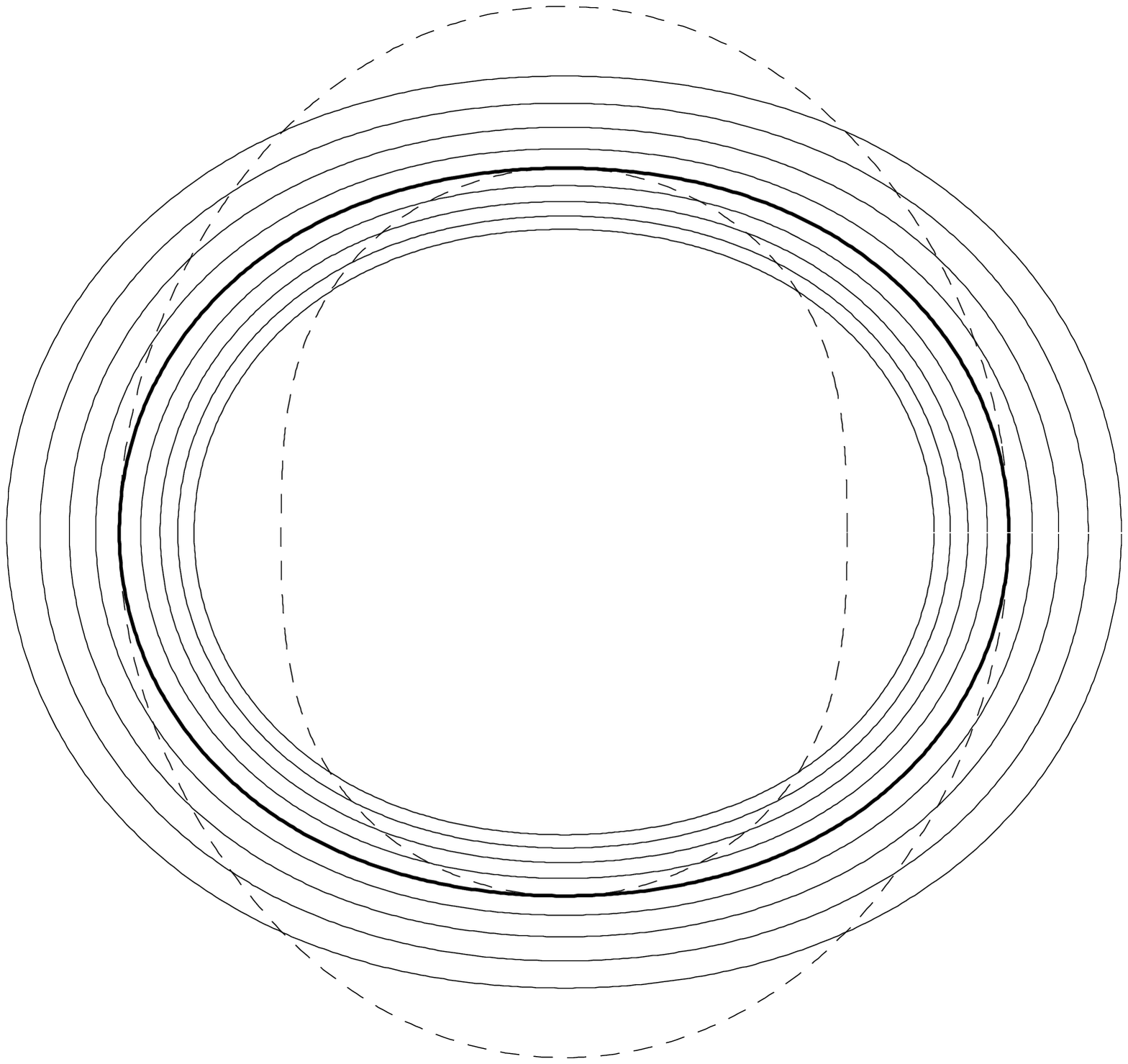,width=2.5in}
            \psfig{figure=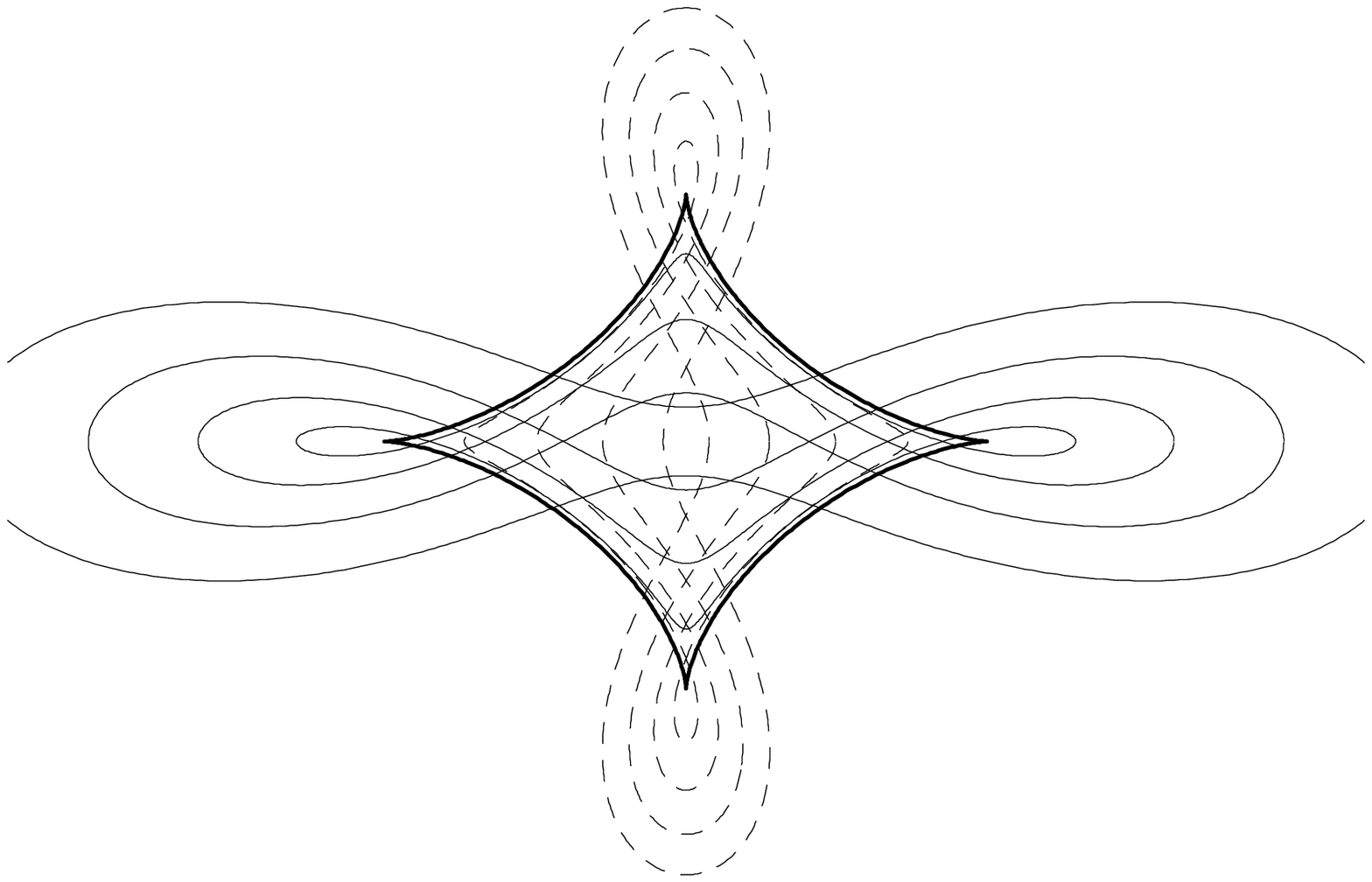,width=2.5in}}
\caption{
   Magnification contours on the image (left) and source (right) planes for
   an SIS in an external shear.  The heavy solid contours show the tangential
   critical line (left) and its corresponding caustic (right).  On the image
   plane (left), the light curves are magnification contours. These are 
   positive outside the critical curve and negative inside the critical
   curve.  The images found in a four-image lens are all found in the 
   region between the two dashed contours -- when two images are merging
   on the critical line, the other two images lie on these curves.  On 
   the source plane the solid (dashed) curves show the projections of the
   positive (negative) magnification contours onto the source plane.  
   Note that the high magnification regions are dominated by the four-image
   systems with the exception of the small high magnification regions 
   found just outside the tip of each cusp.  
   }
\labelprint{fig:magcont}
\end{figure}

In general, the ellipticity of the
lenses has little effect on the expected number of lenses, allowing the use
of circular lens models for statistical studies that are uninterested in the
morphologies of the images (e.g. Keeton, Kochanek \& Seljak~\cite{Keeton1997p604},
Rusin \& Tegmark~\cite{Rusin2001p709}, Chae~\cite{Chae2003p746}).
However, the effects of ellipticity are trivially observable in the
relative numbers of two-image and four-image lenses.
We noted earlier that the expectation from the cross section
is that four-image lenses should represent order 
$\epsilon_\Psi^2\sim \gamma^2 \sim 0.01$ of lenses
where $\epsilon_\Psi$ is the ellipticity of the lens potential.  Yet in
\S\ref{sec:data} we saw that four-image lenses represent roughly one third
of the observed population.  The high abundance of four-image lenses
is a  consequence of the different magnification biases of the two-image 
multiplicities -- the four-image lenses are more highly magnified than
the two-image lenses so they have a larger magnification bias factor.

\begin{figure}[ph]
\begin{center}
\centerline{\psfig{figure=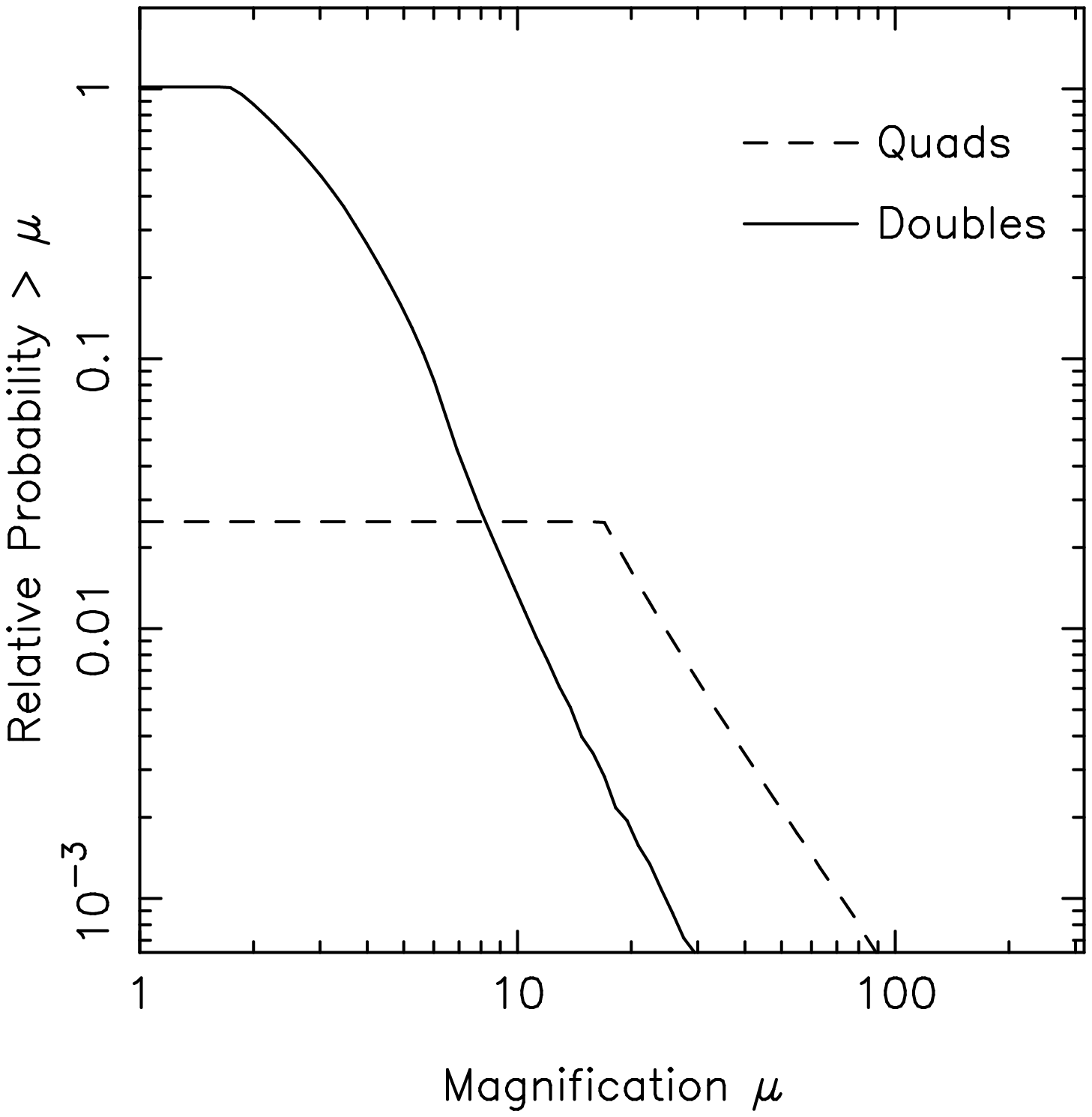,width=2.8in}}
\end{center}
\caption{
   The integral magnification probability distributions for a singular isothermal ellipsoid
   with an axis ratio of $q=0.7$ normalized by the total cross section for finding
   two images. Note that the total four-image cross section is only of order 
   $\epsilon_\Psi^2 \sim (\epsilon/3)^2\sim 0.01$ of the total, but that the 
   minimum magnification for the four-image systems ($\mu_{min}\sim 1/\epsilon \sim 10$)
   is much larger than that for the two-image systems ($\mu_{min}\sim 2$ just as
   for an SIS).  The entire four-image probability distribution is well approximated
   by the $P(>\mu) \propto \mu^{-2}$ power law expected for fold caustics, while
   the two-image probability distribution is steeper since highly magnified images
   can only be created by the cusps.  Figure courtesy of D. Rusin.
   }
\labelprint{fig:magdist}
\begin{center}
\centerline{\psfig{figure=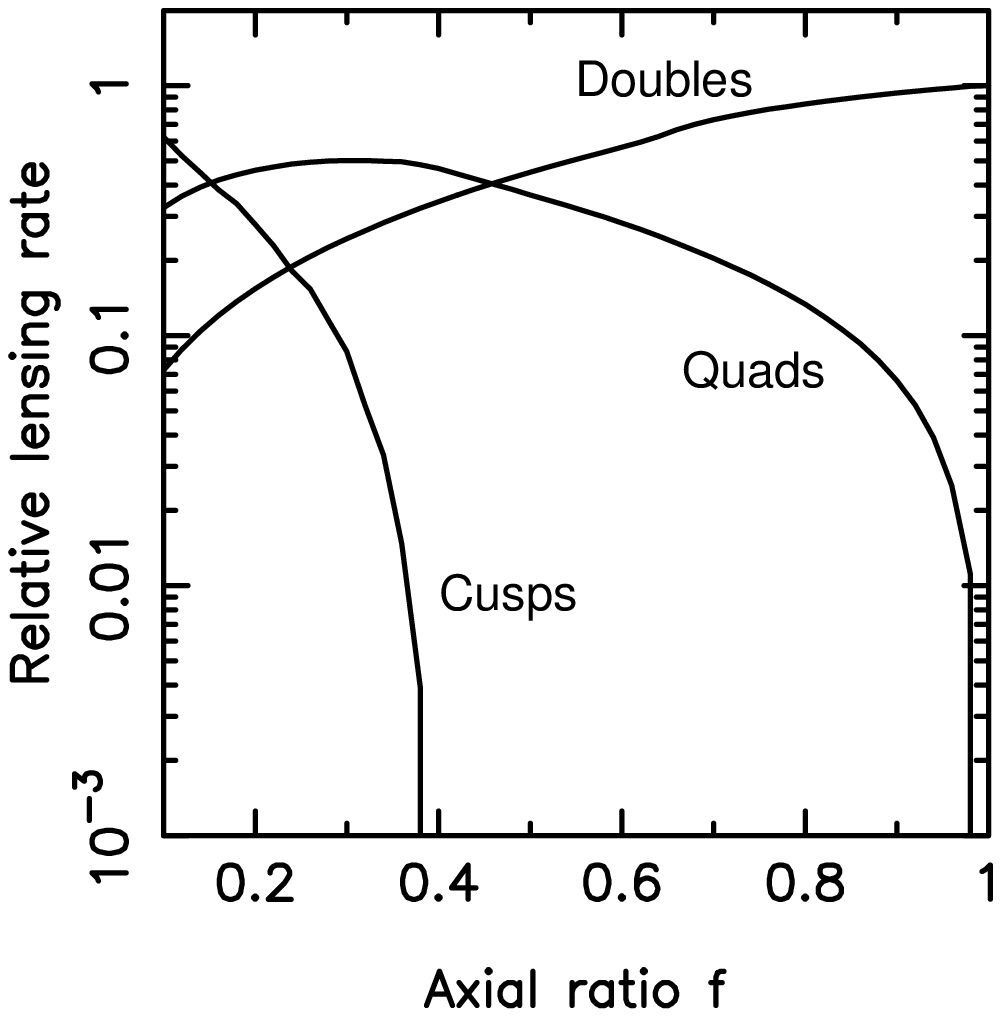,width=2.8in}}
\end{center}
\caption{
  The expected number of two-image, four-image and three-image (disk or cusp)
  lenses as a function of axis ratio $f$ for the CLASS sample. From
  Rusin \& Tegmark~(\cite{Rusin2001p709}).
   }
\labelprint{fig:quadfrac}
\end{figure}

Fig.~\ref{fig:magcont}
shows the image magnification contours for an SIS lens in an external shear
on both the image and source planes.  The highly magnified regions are confined
to lie near the critical line.  If we Taylor expand the inverse magnification
radially, then $\mu^{-1} = \Delta x |d\mu^{-1}/dx|$ where $\Delta x$ is the distance
from the critical line, so
the magnification drops inversely with the distance from the critical line.
If we Taylor expand the lens equations, then we find that the change in 
source plane coordinates is related to the change in image plane coordinates
by $\Delta \beta = \mu^{-1} \Delta x \propto \mu^{-2}$.  Thus, if $L$ is the
length of the astroid curve, the probability of a magnification larger than
$\mu$ scales as $P(>\mu)\propto \mu^{-2} L / |d\mu^{-1}/dx|$.  This applies only to the
four image region, because the only way to get a high magnification in the
two-image region is for the source to lie just outside the tip of a cusp.
The algebra is overly complex to present, but the generic result is that
the region producing magnification $\mu$ extends $\mu^{-2}$ from the 
cusp tip but has a width that scales as $\mu^{-1/2}$, leading to an 
overall scaling that the asymptotic cross section declines as $P(>\mu)\propto \mu^{-7/2}$
rather than $P(>\mu)\propto \mu^{-2}$.  This can all be done formally (see 
Blandford \& Narayan~\cite{Blandford1986p568}) so that asymptotic
cross sections can be derived for any model (e.g. Kochanek \& 
Blandford~\cite{Kochanek1987p676}, Finch et al.~\cite{Finch2002p51}), but
a reasonable approximation for the four-image region is to compute the
magnification, $\mu_0$, for the cruciform lens formed when the source is directly
behind the lens and then use the estimate that $P(>\mu)=(\mu_0/\mu)^2$.  Unfortunately,
such simple estimates are not feasible for the two-image region.  These
distributions are relatively easy to compute numerically, as in the
example shown in Fig.~\ref{fig:magdist}.

Because the minimum magnification of a four-image lens 
increases $\mu_0 \propto \gamma^{-1}$ even as
the cross section decreases as $\sigma_4 \propto \gamma^{2}$, the expected number
of four-image lenses in a sample varies much more slowly with ellipticity
than expected from the cross section.  The product 
$\sigma_4 B(F) \propto \gamma^2 \mu_0^{\alpha-1}$,
of the four-image cross section, $\sigma_4$, 
and the magnification bias, $B(F)$,
scales as $\gamma^{3-\alpha} \propto \gamma $ for the CLASS survey
($\alpha \simeq 2$), which is a much more gentle dependence on ellipticity
than the quadratic variation expected from the cross section.  There is
a limit, however, to the fraction of four-image lenses.  If the
potential becomes too flat, the astroid caustic extends outside the
radial caustic (Fig.~\ref{fig:cusplens}), to produce three-image 
systems in the ``disk'' geometry rather than additional four-image lenses.
In the limit that the axis ratio goes to zero (the lens becomes a line),
only the disk geometry is produced.  The existence of a maximum four-image
lens fraction, and its location at an axis ratio inconsistent with the
observed axis ratios of the dominant early-type lenses has made it 
difficult to explain the observed fraction of four image lenses
(King \& Browne~\cite{King1996p67}, Kochanek~\cite{Kochanek1996p595},
Keeton, Kochanek \& Seljak~\cite{Keeton1997p604}, Keeton \& Kochanek~\cite{Keeton1998p157},Rusin \& Tegmark~\cite{Rusin2001p709}).  Recently, Cohn \& Kochanek~(\cite{Cohn2001p1216})
argued that satellite galaxies of the lenses provide the explanation by
somewhat boosting the fraction of four-image lenses while at the same
time explaining the existence of the more complex lenses like B1359+154
(Myers et al.~\cite{Myers1999p2565}, Rusin et al.~\cite{Rusin2001p594}) 
and PMNJ0134--0931 (Winn et al.~\cite{Winn2002p143}, Keeton \& Winn~\cite{Keeton2003p39}) 
formed by having multiple lens galaxies with more complex caustic structures.
It is not, however, clear in the existing
data that four-image systems are more likely to have satellites to the lens
galaxy than two-image systems as one would expect for this explanation.

Gravitational lenses can produce highly magnified images without multiple
images only if they are highly elliptical or have a low central density.
The SIS lens has a single-image magnification probability distribution of
$\tau dP/d\mu = 2\pi b^2/(\mu-1)^3$ with $\mu < 2$ compared to 
$\tau dP/d\mu = 2\pi b^2/\mu^3$ with $\mu \geq 2$ for the multiply imaged region,
so single images are never magnified by more than a factor of 2.  For
galaxies, where we always expect high central densities, the only way
to get highly magnified single images is when the astroid caustic extends outside
the radial caustic (Fig.~\ref{fig:cusplens}).  A source just outside an exposed
cusp tip can be highly magnified with a magnification probability
distribution $dP/d\mu \propto \mu^{-7/2}$.   Such single image magnifications
have recently been a concern for the luminosity function of high redshift
quasars (e.g. Wyithe~\cite{Wyithe2004p49}, Keeton, Kuhlen \& Haiman~\cite{Keeton2004p1}) 
and will be the high magnification
tail of any magnification perturbations to supernova fluxes (e.g. Dalal et al.
\cite{Dalal2003p11}).  As a general rule for galaxies, the probability 
of a single image being magnified by more than a factor of two is comparable
to the probability of being multiply imaged.

\subsection{Cosmology With Lens Statistics \labelprint{sec:cosmo} }

The statistics of lenses, in the sense of the number of lenses expected in a sample
of sources as a function of cosmology, is a volume test of the cosmological model
because the optical depth (at least for flat cosmologies) is proportional to 
$D_s^3$.  However, the number of lenses also depends on the comoving density
and mass of the lenses ($n_*$, $\sigma_*$ and $\alpha$ in the simple SIS 
model).  While $n_*$ could plausibly be estimated locally, the $\sigma_*^4$ 
dependence on the mass scale makes it very difficult to use local estimates of
galaxy kinematics or masses to normalize the optical depth.
The key step to eliminating this problem is to note that there is an intimate 
relation between the cross section, the observed image separations and the 
mass scale.  While this will hold for any mass model, the SIS model is the
only simple analytic example.  The mean image separation for the lenses
should be
independent of the cosmological model for flat cosmologies (and only weakly
dependent on it otherwise).  Thus, in any lens sample you can eliminate the
dependence on the mass scale by replacing it with the observed mean image
separation, $\tau_{SIS} \propto n_* \langle\Delta\theta\rangle^2 D_s^3$.
Full calculations must include corrections for angular selection effects.
Most odd results in lens cosmology arise in calculations that ignore the
close coupling between the image separations and the cross section.

In practice, real calculations are based on variations of the maximum likelihood
method introduced by Kochanek (\cite{Kochanek1993p12}, \cite{Kochanek1996p638}).  
For each lens $i$ you compute the 
probability $p_i$ that it is lensed including magnification bias and selection
effects.  The likelihood of the observations is then
\begin{equation}
    \ln L_0 = \sum_{lenses} \ln p_i + \sum_{unlensed} \ln(1-p_i)
            \simeq \sum_{lenses} \ln p_i - \sum_{unlensed} p_i
\end{equation}
where $\ln(1-p_i) \simeq -p_i $ provided $p_i \ll 1$.  This simply encodes the
likelihood of finding the observed  number of lenses given the individual
probabilities that the objects are lensed.  Without further information, this 
likelihood could determine the limits on the cosmological model only to the
extent we had accurate prior estimates for $n_*$ and $\sigma_*$.  

If we add, however, a term for the probability that each detected lens has
its observed separation (Eqn.~\ref{eqn:sepdist} plus any selection effects)
\begin{equation}
    \ln L = \ln L_0 + \sum_{lensed} \ln \left( { p_i(\Delta\theta_i) \over p_i } \right),  
\end{equation}
then the lens sample itself can normalize the typical mass scale of the
lenses (Kochanek~\cite{Kochanek1993p12}).  
This has two advantages.  First, it eliminates any systematic
problems arising from the dynamical normalization of the lens model and its
relation to the luminosity function.  Second, it forces the cosmological
estimates from the lenses to be consistent with the observed image separations --
it makes no sense to produce cosmological limits that imply image separations
inconsistent with the observations.  In theory the precision exceeds that
of any local calibration very rapidly.  The fractional spread of the 
separations about the mean is $\sim 0.7$, so the fractional uncertainty in the mean separation
scales as $0.7/N^{1/2}$ for a sample of $N$ lenses.  Since the cross section
goes as the square of the mean separation, the uncertainty in the mean cross
section $1.4/N^{1/2}$ exceeds any plausible accuracy of a local normalization
for $\sigma_*$ (10\% in $\sigma_*$, or 20\% in $\langle\theta\rangle\propto \sigma_*^2$,
or 40\% in $\tau \propto \sigma_*^4$) with only $N\simeq 10$ lenses.

Any other measurable property of the lenses can be added to the likelihood, but
the only other term that has been seriously investigated is the probability of the
observed lens redshift given the image separations and the source redshift
(Kochanek~\cite{Kochanek1992p1}, \cite{Kochanek1996p638}, 
Helbig \& Kayser~\cite{Helbig1996p359}, Ofek, Rix \& Maoz~\cite{Ofek2003p639}).  In general, cosmologies with
a large cosmological constant predict significantly higher lens redshifts
than those without, and in theory this is a very powerful test because of
the exponential cutoff in Eqn.~\ref{eqn:zpdist}.  The biggest
problem in actually using the redshift test, in fact so big that it probably cannot be used
at present, is the high incompleteness of the lens redshift measurements 
(\S\ref{sec:data}).  There will be a general tendency, even at fixed separation,
for the redshifts of the higher redshift lens galaxies to be the ones that
are unmeasured.  Complete samples could be defined for a separation range,
usually by excluding small separation systems, but a complete analysis needs
to include the effects of groups and cluster boosting image separations beyond
the splitting produced by an isolated galaxy.  For example, how do we include
Q0957+561 with its separation of 6\farcs2 that is largely due to the lens 
galaxy but has significant contributions from the surrounding cluster? 

\begin{figure}[p]
\begin{center}
\centerline{\psfig{figure=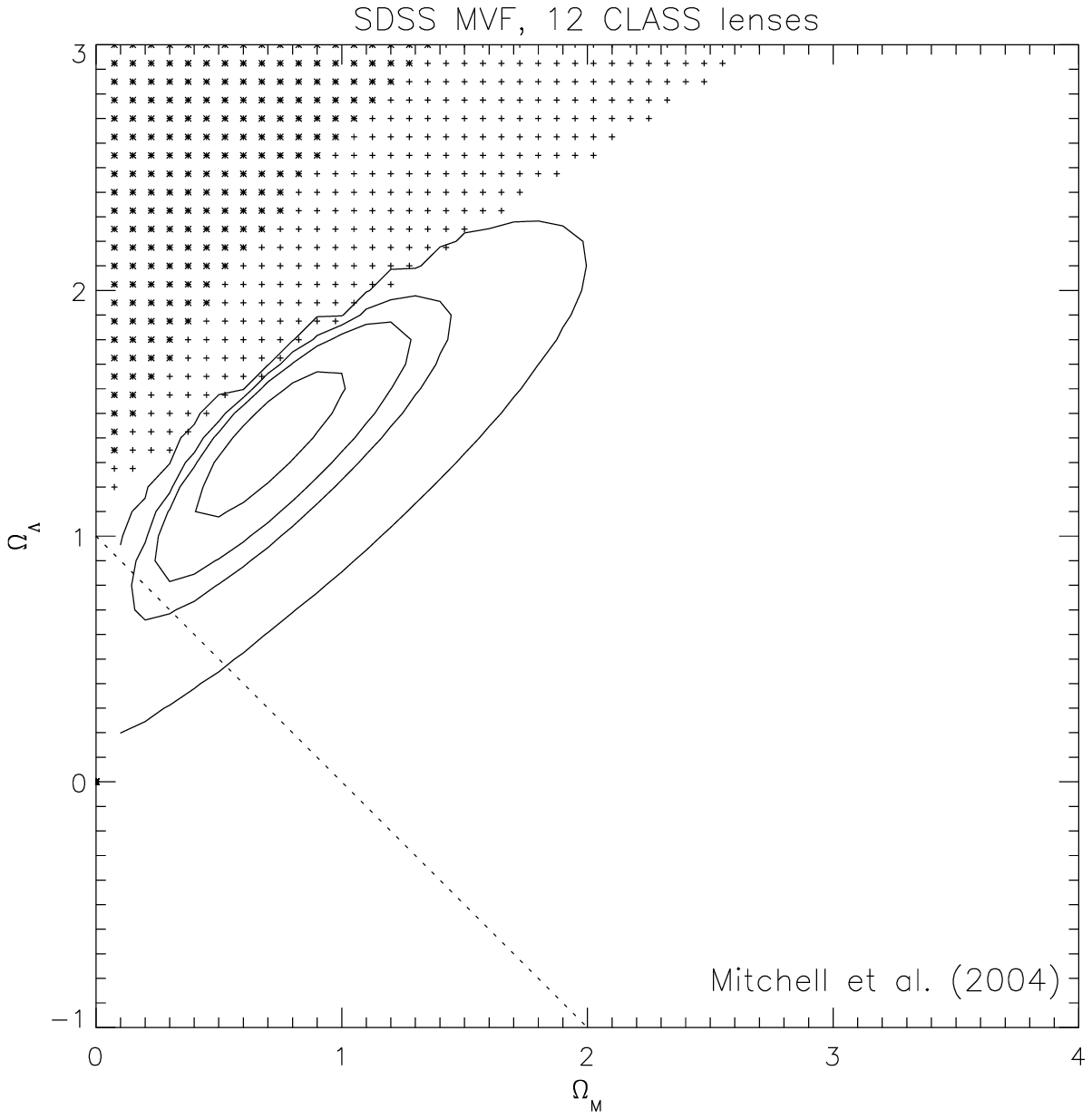,width=4.0in}}
\end{center}
\begin{center}
\centerline{\psfig{figure=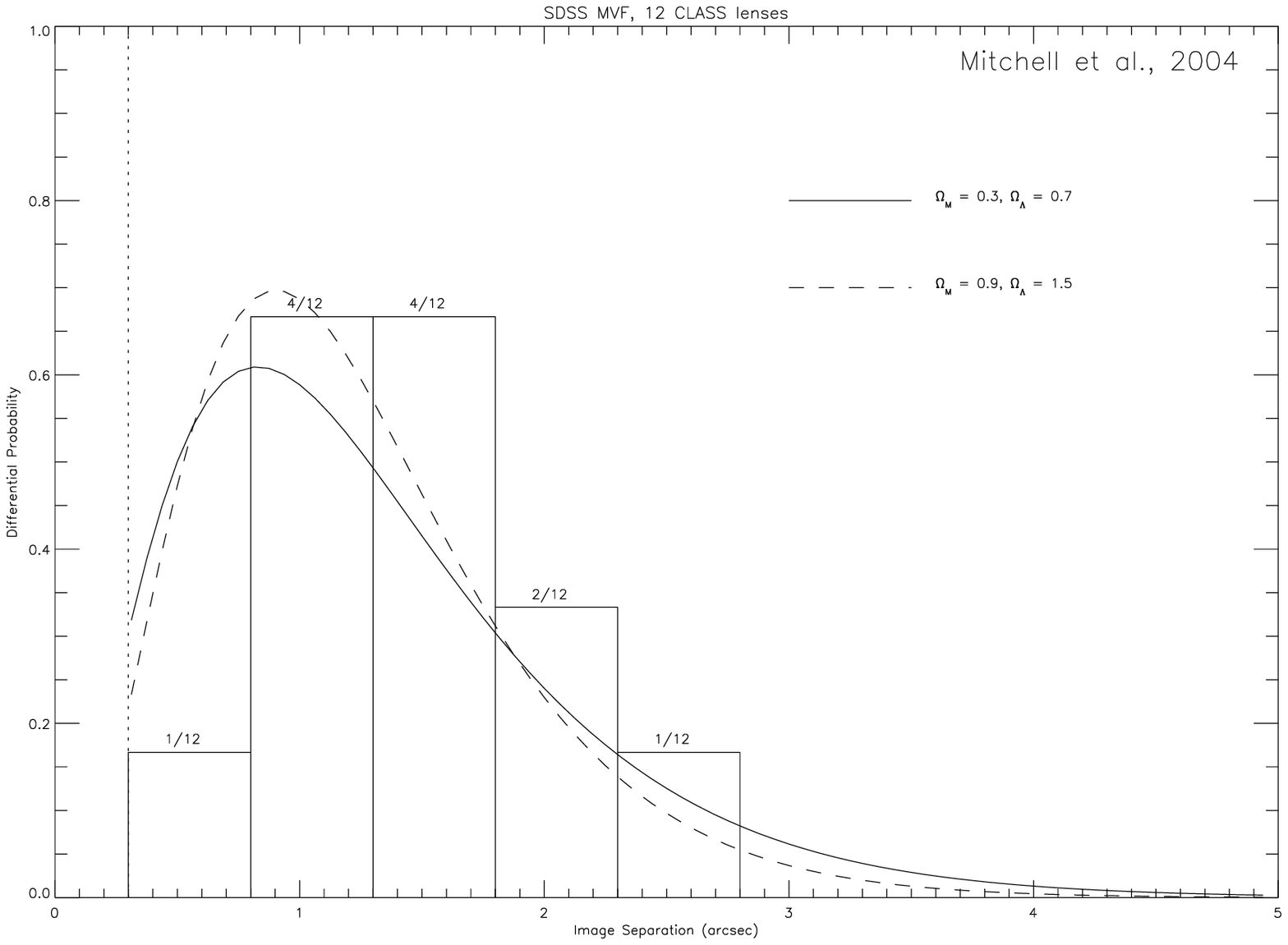,width=4.0in}}
\end{center}
\caption{
   (Top) Likelihood functions for the cosmological model 
   from Mitchell et al.~(\cite{Mitchell2004p1}) using the velocity
   function of galaxies measured from the SDSS survey and a sample of 12 CLASS lenses.
   The contours show the 68, 90, 95 and 99\% confidence intervals on
   the cosmological model.  In the shaded regions the cosmological distances
   either become imaginary or there is no big bang.
   (Bottom) The histogram shows the separation distribution of the 12 CLASS lenses
   used in the analysis and the curve shows the distribution predicted by the 
   maximum likelihood model including selection effects.
   }
\labelprint{fig:coslim}
\end{figure}

\subsection{The Current State}

Recent analyses of lens statistics have focused exclusively on the CLASS
flat spectrum radio survey (Browne et al.~\cite{Browne2003p341}).  
Chae et al.~(\cite{Chae2002p1301}),  Chae~(\cite{Chae2003p746}) 
and Mitchell et al.~(\cite{Mitchell2004p1}) focus on estimating the
cosmological model and find results in general agreement with estimates
from Type Ia supernovae (e.g. Riess et al.~\cite{Riess2004p665}).  The general approach of both groups is
to use variants of the maximum likelihood methods described above in
\S\ref{sec:cosmo}.   Chae~(\cite{Chae2003p746}) uses an obsolete
estimate of the galaxy luminosity function combined with a Faber-Jackson
relation and the variable transformation of Eqn.~\ref{eqn:vfunc} but
normalized the velocity scale using the observed distribution of lens
separations.  Mitchell et al.~(\cite{Mitchell2004p1}) use the true
velocity dispersion function from the SDSS survey (Sheth et al.~\cite{Sheth2003p225})
and incorporate a Press-Schechter~(\cite{Press1974p425}) model for the
evolution of the velocity function.   Chae~(\cite{Chae2003p746}) 
used ellipsoidal galaxies, although this has little 
cosmological effect, while Mitchell et al.~(\cite{Mitchell2004p1}) 
considered only SIS models.  Fig.~\ref{fig:coslim} shows the 
cosmological limits from Mitchell et al.~(\cite{Mitchell2004p1}), which
are typical of the recent results .  There are also attempts to use
lens statistics to constrain dark energy (e.g. Chae et al.~\cite{Chae2004p71},
Kuhlen, Keeton \& Madau~\cite{Kuhlen2004p104}), but far larger, well-defined
samples are needed before the resulting constraints will become 
interesting.

Chae \& Mao~(\cite{Chae2003p61}), Davis, Huterer \& Krauss~(\cite{Davis2003p1029})
and Ofek, Rix \& Maoz~(\cite{Ofek2003p639}) focused on galaxy properties
and evolution in a fixed, concordance cosmology rather than on determining
the cosmological models.  Mitchell et al.~(\cite{Mitchell2004p1}) compared
models where the lenses evolved following the predictions of CDM models in
comparison to non-evolving models. Because lens statistical estimates are unlikely
to complete with other means of estimating the cosmological models, these 
are more promising applications of gravitational lens statistics for
the future.  Attempts to estimate the evolution of the lens population
usually allow the $n_*$ and $\sigma_*$ parameters of the velocity function
(Eqn.~\ref{eqn:vfunc}) to evolve as power laws with redshift.  Mitchell 
et al.~(\cite{Mitchell2004p1}, Fig.~\ref{fig:dispevolve})  point out that
CDM halo models make specific predictions for the evolution of the velocity
function that have a different structure from simple power laws in 
redshift, but with the present data the differences are probably 
unimportant.  All these evolution studies came to the conclusion that
the number density of the $\sigma_v \sim \sigma_*$ galaxies which dominate
lens statistics has changed little ($\ltorder \pm 50\%$) between the present day 
and redshift unity.    

I have three concerns about these analyses and their focus on the ``complete''
CLASS lens samples.  First, a basic problem with the CLASS
survey is that we lack direct measurements of the redshift distribution
of the source population forming the lenses (e.g Marlow et al.~\cite{Marlow2000p2629},
Mu\~noz et al.~\cite{Munoz2003p684}).  In particular, Mu\~noz et al.~(\cite{Munoz2003p684})  
note that the radio source population is changing radically from nearly all
quasars to mostly galaxies as you approach the fluxes of the CLASS source
population.  This makes it dangerous to extrapolate the source population
redshifts from the brighter radio fluxes where the redshift samples are nearly
complete to the fainter samples where they are not.  The second problem
is that no study has a satisfactory treatment of the lenses with satellites
or associated with clusters.  All the analyses use isolated lens models
and then either include lenses with satellites but ignore the satellites
or drop lenses with satellites and ignore the fact that they have been
dropped.  The analysis by  Cohn \& Kochanek~(\cite{Cohn2004p1}) of lens
statistics with satellites shows that neither approach is satisfactory --
dropping the satellites biases the results to underestimate cross sections
while including them does the reverse.  Cohn \& Kochanek~(\cite{Cohn2004p1}) 
concluded that including they systems with satellites probably has fewer
biases than dropping them.   A similar problem probably arises from the
effects of the group halos to which many of the lenses belong (e.g. Keeton et al.
\cite{Keeton2000p129}, Fassnacht \& Lubin~\cite{Fassnacht2002p627}). 
My third concern is that the separations of
the radio lenses seen to be systematically smaller than the optically
selected lenses even though the optical HST Snapshot survey 
(Maoz et al.~\cite{Maoz1993p28}) 
had the greatest sensitivity to small separation systems.  It is possible
that this is simply due to selection effects in the optical samples, but
I have seen no convincing scenario for producing such a selection effect.
We see no clear correlation of extinction with image separation (see \S\ref{sec:ism}),
emission from the lens galaxy is less important for small separation systems
than for large separation systems, and the selection function due to the
resolution of the observations is fairly simple to model.

\begin{figure}[t]
\begin{center}
\centerline{\psfig{figure=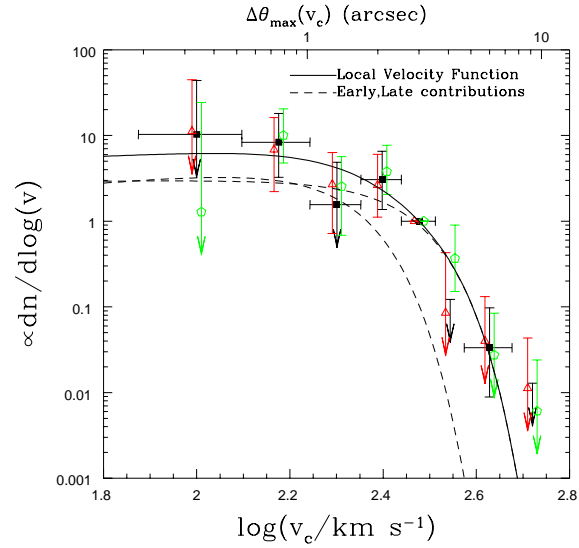,width=3.2in}}
\end{center}
\caption{
   Non-parametric reconstructions of the velocity function from the observed
   separations of gravitational lenses assuming an SIS lens model.  The
   velocity functions are all normalized to the bin centered at $300$~km/s.
   The filled squares use only the lenses in the flat spectrum radio
   surveys, the triangles use all radio-selected lenses and the pentagons
   include all radio lenses and all quasar lenses.  The horizontal error
   bars on the filled squares show the bin widths.  The triangles and
   pentagons are horizontally offset from the squares to make them more
   visible.  The curves show the velocity function estimated from the 2MASS
   sample from Fig.~\ref{fig:vfunc}.   The horizontal scale at the top
   of the figure shows the maximum separation produced by a lens of the
   corresponding circular velocity.  The mean separation produced by such
   a lens will be one-half the maximum.
   }
\labelprint{fig:nonpar}
\end{figure}

\begin{figure}[t]
\begin{center}
\centerline{\psfig{figure=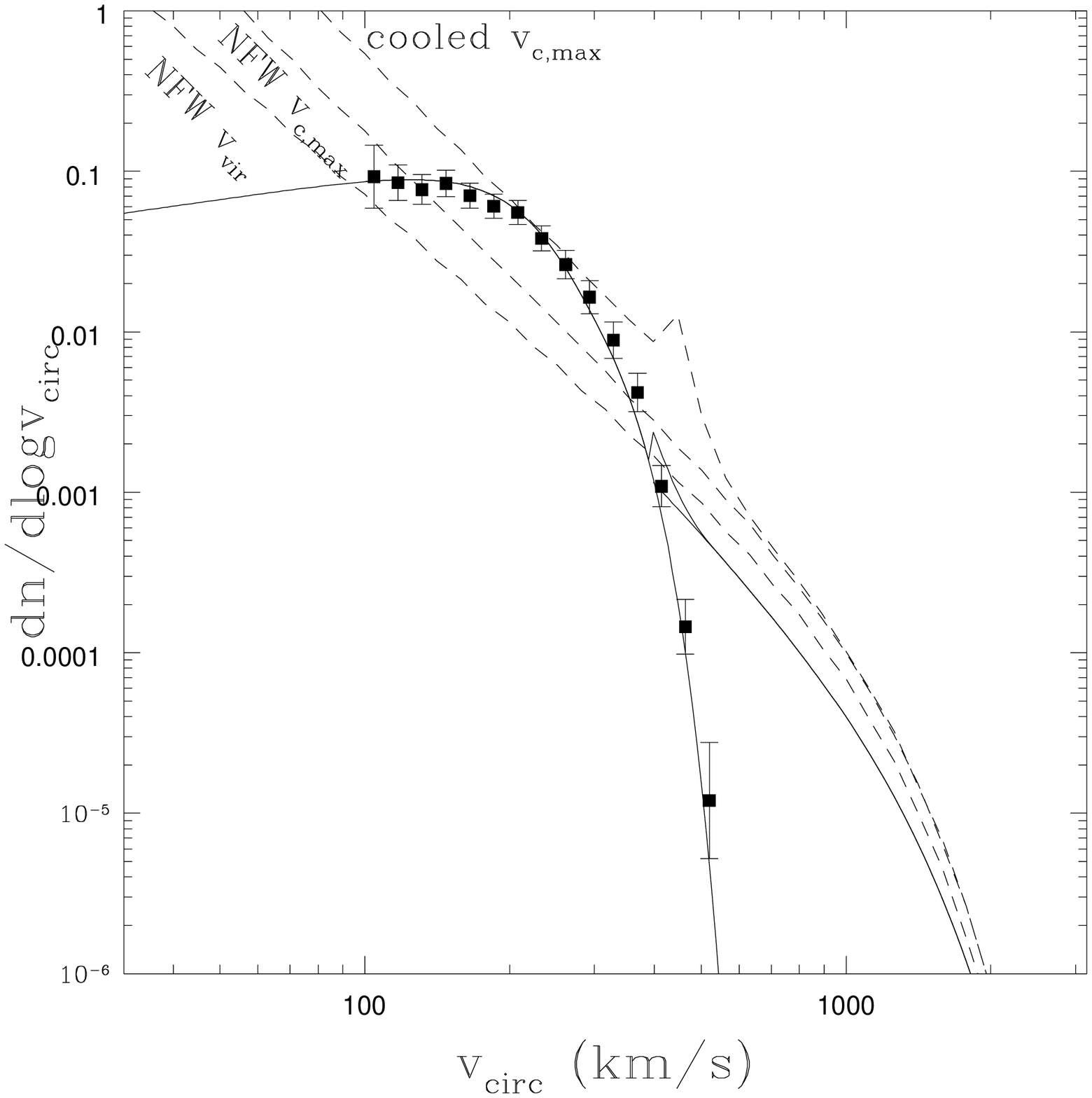,width=3.2in}}
\end{center}
\caption{The expected circular velocity function $dn/d\log v_c$ of CDM halos
  The lowest 
  dashed curve labeled NFW $v_{vir}$ shows the velocity function using the NFW halo virial 
  velocity $v_{vir}$ for the circular velocity (see \S\ref{sec:massmono}).  The middle
  dashed curve labeled NFW $v_{c,max}$ shows the velocity function if the peak circular
  velocity of the halo is used rather than the virial velocity.  The upper dashed curve
  is a model in which the baryons of halos with $M \ltorder 10^{13}M_\odot$ cool, raising 
  the central density and circular velocity.  The solid curve with the points shows the
  estimate of the local velocity function of galaxies (Fig.~\ref{fig:vfunc}) and the
  solid curve extending to higher velocities is an estimate of the local velocity 
  function of groups and clusters.
  }
\labelprint{fig:cdmvfunc}
\end{figure}

On the other hand, the various lens samples may all consistent.  One
way to compare the different data sets is to non-parametrically construct
the velocity function from the observed image separations of the samples.
To do this we assume an SIS lens model for the conversion from
image separations to circular velocities, and then adopt the standard 
non-parametric methods used to construct luminosity functions from redshift
surveys to construct the velocity function from the image separations
(Kochanek~\cite{Kochanek2003p139}).  The results for the flat-spectrum
lens surveys (CLASS, JVAS, PANELS), all radio surveys and all radio surveys
plus the quasar lenses are shown in Fig.~\ref{fig:nonpar}.  We
normalized the estimates to the density at $v_c=300$~km/s to eliminate
any dependence on the cosmological model.  The lens data can estimate
the velocity function from roughly $v_c \sim 100$~km/s to $500$~km/s.
At lower velocities the finite resolution of the observations makes 
the uncertainties in the density explode, and at higher velocities the
surveys have not searched large enough angular regions around the lens
galaxies.  The shape of the velocity function is consistent with local
estimates (Fig.~\ref{fig:vfunc}) except in the highest circular velocity
bin where we begin to see the contribution from clusters we will consider
in \S\ref{sec:cluster}.  Fig.~\ref{fig:nonpar} also makes it clear why
constraints on the evolution of the lenses are so weak -- evolution 
estimates basically try to compare the low-redshift separation distribution
to the high redshift separation distribution, and we simply do not have
large enough lens samples to begin subdividing them in redshift (to say
nothing of dealing with unmeasured redshifts) and still have small
statistical uncertainties.

\section{What Happened to The Cluster Lenses? \labelprint{sec:cluster} }

One would think from the number of conference proceeding covers featuring HST
images of cluster arcs that these are by far the most common type of lens. In
fact, this is an optical delusion created by the ease of finding the rich 
clusters even though they are exponentially rare.  The most common kind of
lens is the one produced by a typical massive galaxy -- as we saw in 
in Fig.~\ref{fig:nonpar}.  For a comparison, Fig.~\ref{fig:cdmvfunc}
shows several estimates of the velocity function based on standard CDM mass
functions and halo models (from Kochanek \& White~\cite{Kochanek2001p531}
and Kochanek~\cite{Kochanek2003p139}, using the Sheth \& Tormen~\cite{Sheth1999p119} 
mass function combined with the NFW halo model from \S\ref{sec:massmono}).
We see for high masses or circular velocities that the predicted distribution
of halos agrees with the observed distribution of clusters.  At the velocities
typical of galaxies, the observed density of galaxies is nearly an order
of magnitude higher than expected for a CDM halo mass function.  
At very low velocities we expect
many more halos than we observe galaxies.  The velocity function estimated
from the observed image separations matches that of galaxies with the beginnings of 
a tail extending onto the distribution of clusters at the high velocity 
end (Fig.~\ref{fig:cdmvfunc}). At low velocities the limited resolution of
the present surveys means that the current lens data does not probe the low
velocity end very well.  In this section we
discuss the difference between cluster and galaxy lenses and explain
the origin of the break between galaxies and clusters.  In
\S\ref{sec:substruc} on CDM substructure we will discuss the divergence
at low circular velocities. 

\begin{figure}[t]
\centerline{\psfig{figure=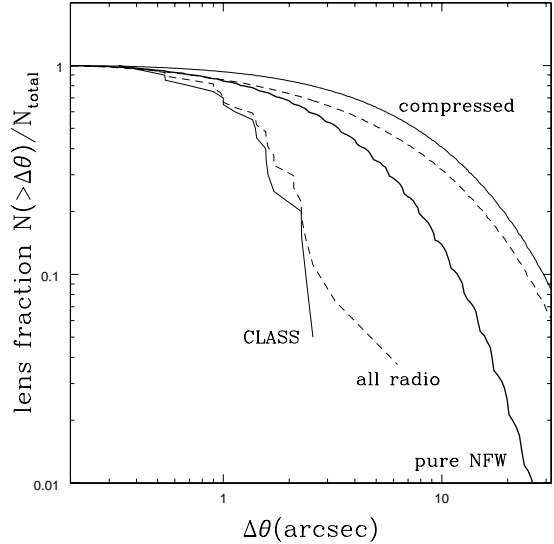,width=3.0in}}
\caption{
   Predicted image separation distributions assuming the structure of halos
   does not change with halo mass.  The heavy solid line shows the prediction
   for pure NFW models while the light solid (dashed) curves shows the 
   predictions after 5\% of the baryons have cooled into a disk (a disk
   plus a bulge with 10\% of the baryonic mass in the bulge).  The curves
   labeled CLASS (for the CLASS survey lenses) and all radio (for all
   radio selected lenses) show the observed distributions.  
   }
\labelprint{fig:clust0}
\end{figure}

\begin{figure}[p]
\begin{center}
\centerline{\psfig{figure=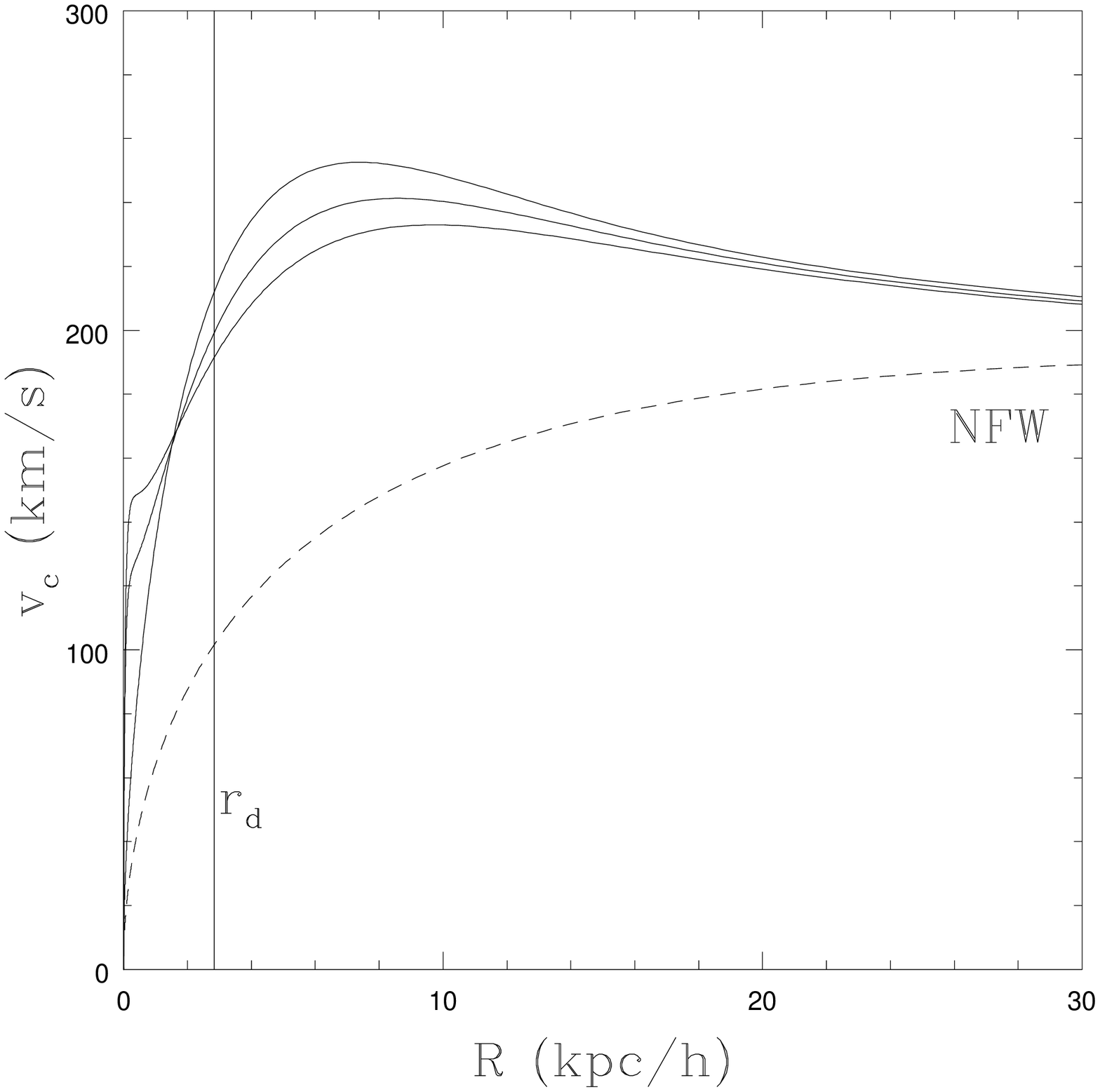,width=3.0in}}
\end{center}
\begin{center}
\centerline{ \psfig{figure=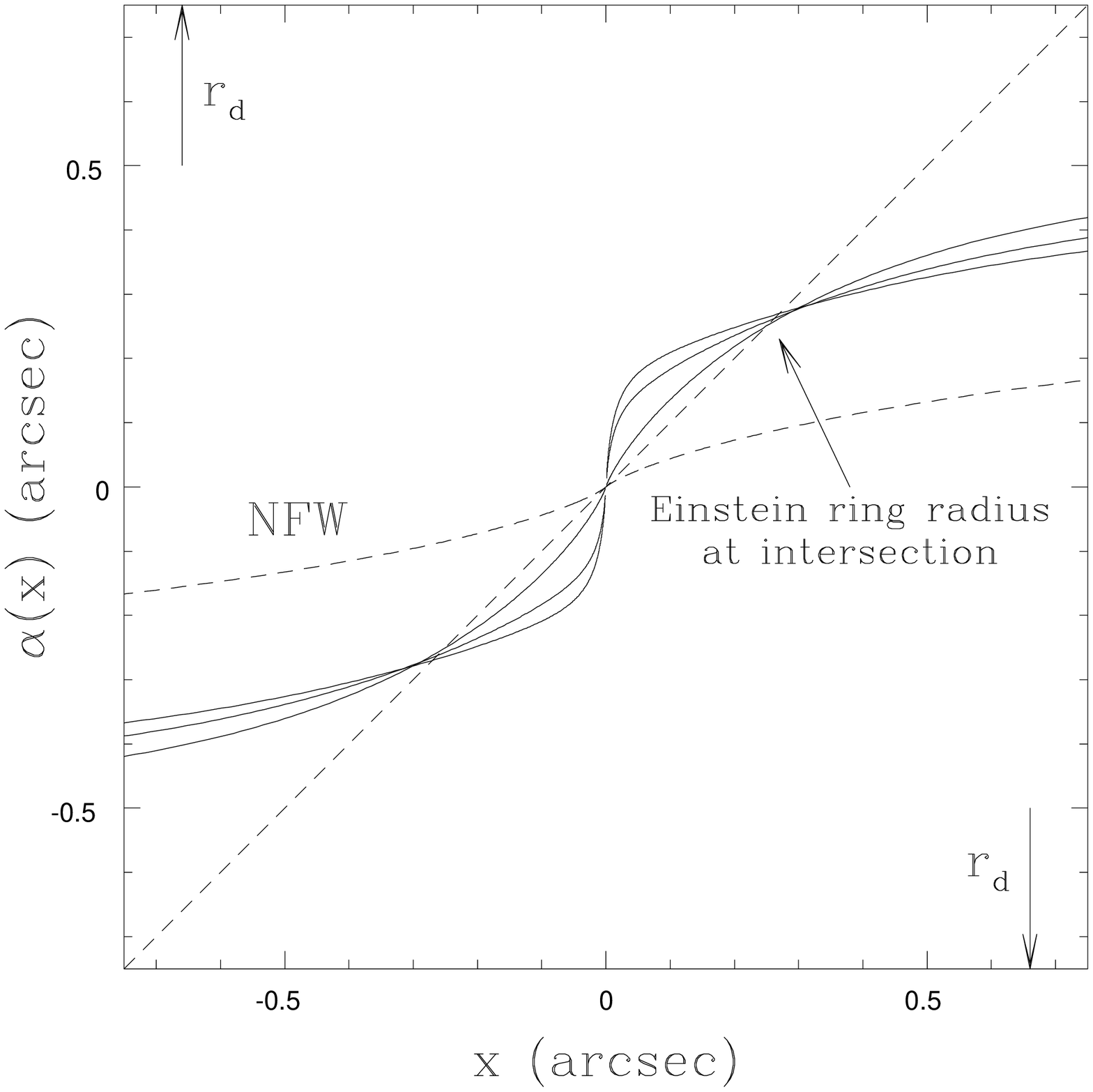,width=3.0in}}
\end{center}
\caption{(Top) The rotation curve and (Bottom) the bending angle $\alpha(x)$
  for a $10^{12}M_\odot$ halo at $z_l=0.5$ with a concentration of $c=8$
  lensing a source at $z_s=2.0$.  The dashed curves show the results for
  the initial NFW halo, while the solid curves show the results after 
  allowing 5\% of the mass to cool conserving angular momentum (spin
  parameter $\lambda=0.04$) and adiabatically compressing the dark matter.
  The three solid curves show the effect of putting 0\%, 10\% or 20\%
  of the baryonic mass into a central bulge.  Higher bulge masses raise
  the central circular velocity and steepen the central deflection profile.
  The final disk scale length is $r_d$.  Compare these to the bending angles
  of our simple models in Figs.~\ref{fig:pntmass}--\ref{fig:moore2}.
   }
\labelprint{fig:clust1}
\end{figure}

The standard halo mass function is roughly a power law with $dn/dM \sim M^{-1.8}$
combined with an exponential cutoff at the mass scale corresponding to the largest
clusters that could have formed at any epoch (e.g. the Sheth \& Tormen \cite{Sheth1999p119}
halo mass function).  Typically these rich clusters 
have internal velocity dispersions above $1000$~km/s and can produce image 
splittings of $\sim 30$~arcsec.  If halo structure was independent of mass,
then we would expect the separation distribution of gravitational lenses to
show a similar structure -- a power law out to the mass scale of rich clusters
followed by an exponential cutoff.  In Fig.~\ref{fig:clust0} we compare
the observed distribution of radio lenses to that expected from the halo mass
function assuming either NFW halos or NFW halos in which the baryons, representing
5\% of the halo mass, have cooled and condensed into the centers of the halos
(Kochanek \& White~\cite{Kochanek2001p531}).  We would
find similar curves if we used simple SIS models rather than these more 
complex CDM-based models (Keeton 1998, Porciani \& Madau~\cite{Porciani2000p679}).
In practice, the most complete survey for
multiply imaged sources, the CLASS survey, found a largest separation of 
$4\farcs5$ (B2108+213) despite checking candidates out to 
separations of $15\farcs0$ (Phillips et al.~\cite{Phillips2001p1001}).
The largest lens found in a search for multiply
imaged sources has an image separation of roughly 15~arcsec (SDSS1004+4112, Inada 
et al.~\cite{Inada2003p810}).
The overall separation distribution (see Fig.~\ref{fig:clust0}) has a sharp cutoff
on scales of 3~arcsec corresponding to galaxies with velocity dispersions
of $\sim 250$~km/s.    The principal searches for wide separation lenses
are Maoz et al.~(\cite{Maoz1997p75}), Ofek et al.~(\cite{Ofek2001p463}) and 
Phillips et al.~(\cite{Phillips2001p1001}),
although most surveys searched for image separations of at least 6\farcs0.
A large number of studies focused only on the properties
of lenses produced by CDM mass functions 
(e.g. Narayan \& White~\cite{Narayan1988p97}, Wambsganss et al.~\cite{Wambsganss1995p274}, \cite{Wambsganss1998p29}, Kochanek~\cite{Kochanek1995p545},
Maoz et al.~\cite{Maoz1997p75}, Flores \& Primack~\cite{Flores1996p5},
Mortlock \& Webster~\cite{Mortlock2000p872}, Li \& Ostriker~\cite{Li2002p652},
Keeton \& Madau~\cite{Keeton2001p25}, Wyithe, Turner \& Spergel~\cite{Wyithe2001p504}).
We will not discuss these in detail because such models cannot reproduce the
observed separation distributions of lenses.  Most recent analyses allow for
changes in the density distributions between galaxies and clusters.

Physically the important difference between galaxies and clusters is that the
baryons in the galaxies have cooled and condensed into the center of the halo
to form the visible galaxy.  As the baryons cool, they also drag some of the
dark matter inward through a process known as adiabatic compression (Blumenthal et al.~\cite{Blumenthal1986p27}), 
although this is less important than the cooling.  As we show in 
Fig.~\ref{fig:clust1}, standard dark matter halos are terrible lenses
because their central cusps ($\rho \propto r^{-\gamma}$ and 
$1.5 \geq \gamma \geq 1$) are too shallow.  In this case, a standard
NFW halo with a total mass of $10^{12}M_\odot$ and a concentration of
$c=8$ (see Eqns.~\ref{eqn:abr}--\ref{eqn:abt})  at a redshift of 
$z_l=0.5$ is unable to produce
multiple images of a source at redshift $z_s=2$ despite having an
asymptotic circular velocity of nearly $200$~km/s.  If we now assume that
5\% of the mass is in baryons starting with a typical halo angular
momentum and then cooling into a disk of radius $r_d$ while conserving
angular momentum we see that the rotation curve becomes flatter and the
galaxy is now able to produce multiple images.  Putting some fraction of
the mass into a still more compact, central bulge make the lens even 
more supercritical and the bending angle diagram begins to resemble that
of an SIS lens (see Fig.~\ref{fig:sis1}).  Thus, the cooling of the baryons converts
a sub-critical dark matter halo into one capable of producing multiple 
images.

\begin{figure}[ph]
\begin{center}
\centerline{\psfig{figure=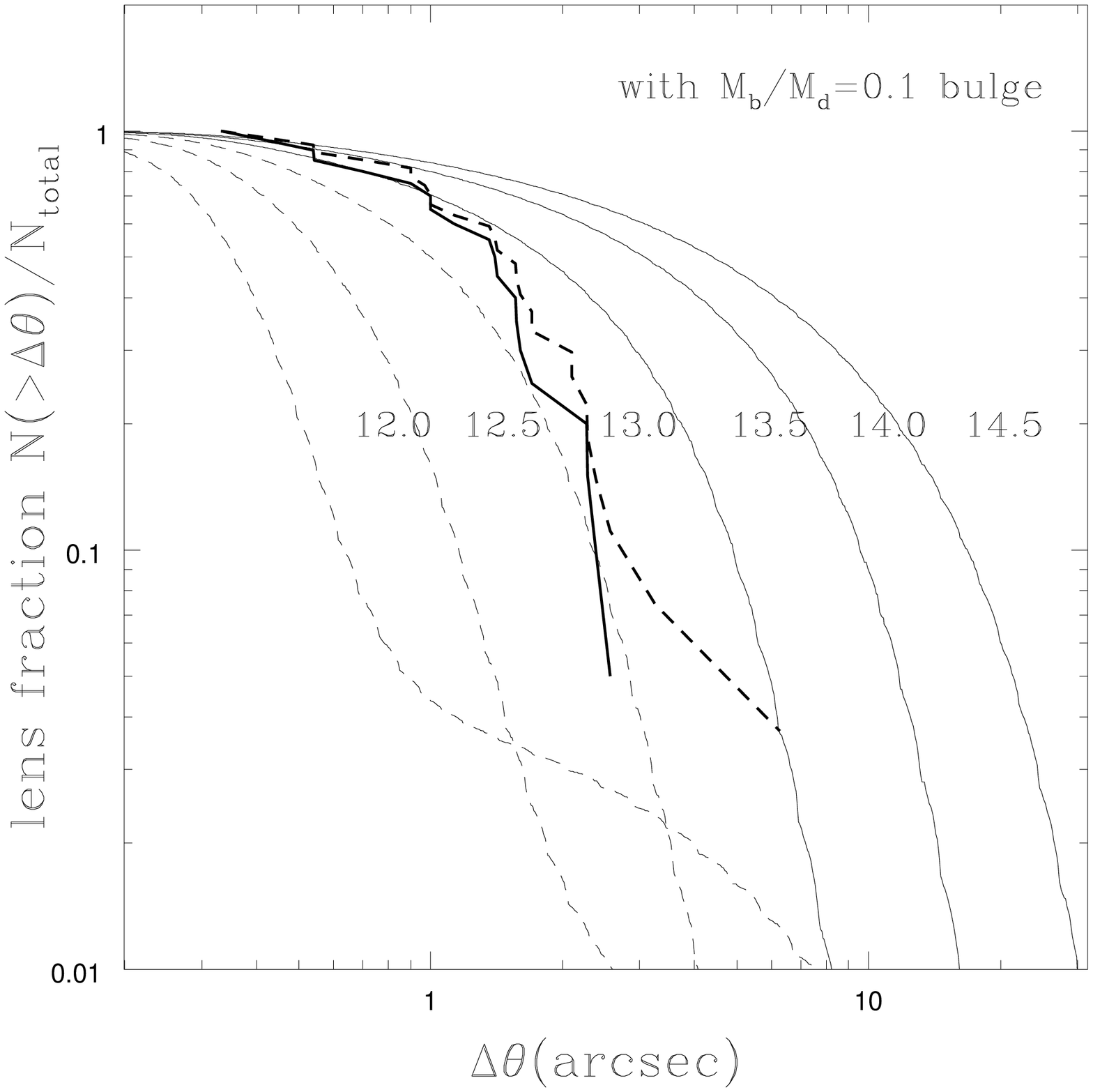,width=3.0in}}
\end{center}
\vspace{-0.25in}
\begin{center}
\centerline{\psfig{figure=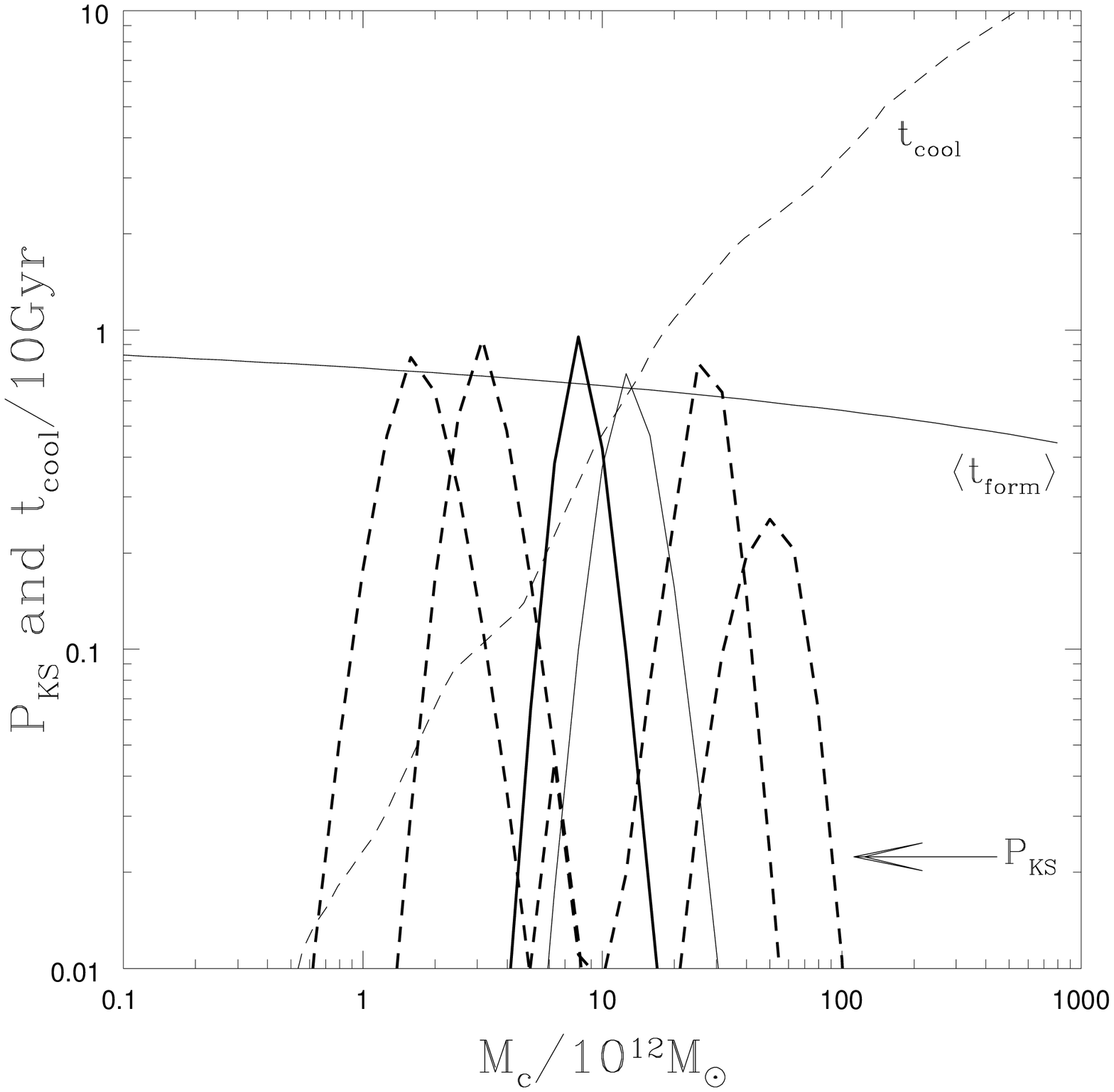,width=3.0in}}
\end{center}
\vspace{-0.25in}
\caption{(Top) Predicted separation distributions as a function of the
  cooling mass scale $M_c$ in which 5\% of the mass cools with 90\% of the
  cooled material in a disk and 10\% in a bulge.  The dashed curves show
  the distributions for $M_c=10^{12}M_\odot$, $3\times 10^{12}M_\odot$
  and $10^{13}M_\odot$, while the solid curves show the distributions for
  $M_c=3\times 10^{13}M_\odot$, $10^{14}M_\odot$ and $3\times 10^{14}M_\odot$.
  The heavy solid (dashed) curves shows the observed distribution of the
  CLASS (all radio-selected) lenses.  
   }
\labelprint{fig:clust2a}
\caption{(Bottom) The Kolmogorov-Smirnov probability, $P_{KS}$, of fitting the
  observed distribution of CLASS lenses as a function of the cooling mass
  scale $M_c$.  The heavy solid curves show the results when 5\% of the 
  mass cools without (with) 10\% of that mass in a bulge.  The heavy
  dashed curves show the results for models where lower (1\% and 2\%)
  or higher (10\% and 20\%) halo mass fractions cool, where the optimal
  cooling mass scale $M_c$ decreases as the cold baryon fraction increases.
  For comparison, the light dashed line shows the cooling time 
  $t_{cool}$ in units of 10~Gyr for the radius enclosing 50\% of the
  baryonic mass in the standard model.  The light solid line shows the
  average formation epoch, $\langle t_{form} \rangle$, also in units
  of $10$~Gyr.  
   }
\labelprint{fig:clust2b}
\end{figure}

The key point is that only intermediate mass halos contain baryons which
have cooled.  High mass halos (groups and clusters) have cooling times
longer than the Hubble time so they have not had time too cool (e.g. Rees \& 
Ostriker~\cite{Rees1977p541}).  
Most low mass halos also probably resemble dark matter halos more than
galaxies with large quantities of cold baryons because they lost 
their baryons due to heating from the UV background
during the initial period of star formation (e.g. Klypin et al.~\cite{Klypin1999p82}
Bullock, Kravtsov \& Weinberg~\cite{Bullock2000p517}, see \S\ref{sec:substruc}). 
Here we ignore the very low mass halos and consider
only the distinction between galaxies and groups/clusters.  The
fundamental realization in recent studies (e.g. Porciani \& Madau~\cite{Porciani2000p679}, 
Kochanek \& White~\cite{Kochanek2001p531}, Kuhlen, Keeton \& Madau~\cite{Kuhlen2004p104},
Li \& Ostriker~\cite{Li2003p603})
 is that introducing a cooling mass scale $M_c$ below which
the baryons cool to form galaxies and above which they do not supplies
the explanation for the difference between the observed separation distribution
of lenses and naive estimates from halo mass functions.  

Once we recognize the necessity of introducing a distinction between cluster
and galaxy mass halos, we can use the observed distribution of lens separations
to constrain the mass scale of the break and the physics of cooling.
Fig.~\ref{fig:clust2a} shows the most common version of these studies,
where separation distributions are computed as a function of the cooling
mass scale $M_c$.  We show the separation distributions for various
cooling mass scales assuming that 5\% of the mass cools into a disk
plus a bulge with 10\% of the baryonic mass in the bulge for all halos
with $M<M_c$.  If the 
cooling mass is either too low or too high we return to the models
of Fig.~\ref{fig:clust0}, while at some intermediate mass scale we
get the break in the separation distribution to match the observed
angular scale.  For these parameters, the optimal cooling mass scale
is $M_c \simeq 10^{13}M_\odot$ (Fig.~\ref{fig:clust2b}).  This agrees
reasonably well with Porciani \& Madau (\cite{Porciani2000p679}) and
Kuhlen, Keeton \& Madau~(\cite{Kuhlen2004p104}) who found a somewhat
higher mass scale $M_c\simeq 3\times 10^{13}M_\odot$ using SIS models
for galaxies.  Cosmological hydrodynamic simulations by Pearce et al. 
(\cite{Pearce1999p99}) also found that approximately 50\% of the baryons had cooled
on mass scales near $10^{13}M_\odot$.  Note, however, that the mass
scale needed to fit the data depends on the assumed fraction of the mass in cold baryons.
With fewer cold baryons a halo becomes a less efficient lens producing
smaller image separations so $M_c$ must increase to keep the break
at the observed scale.  If the cold baryon fraction is too low ($\ltorder 1\%)$, 
it becomes impossible to explain the data at all.  Crudely, the
cooling mass scale depends exponentially on the cold baryon fraction with
$\log M_c/M_\odot \simeq 13.6 - (\hbox{cold fraction})/0.15$.  

\begin{figure}[ph]
\begin{center}
\centerline{\psfig{figure=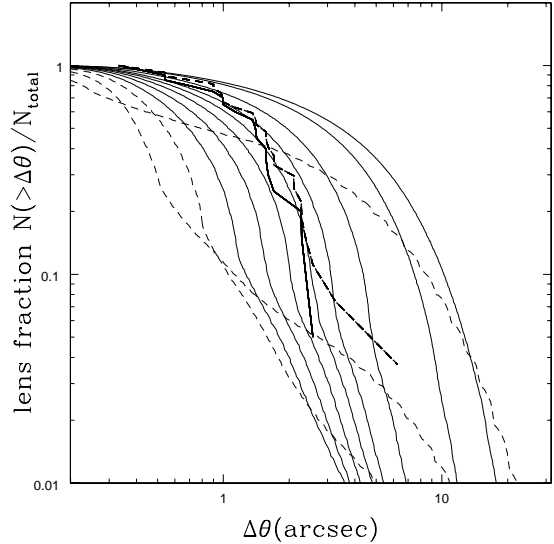,width=3.0in}}
\end{center}
\vspace{-0.25in}
\begin{center}
\centerline{\psfig{figure=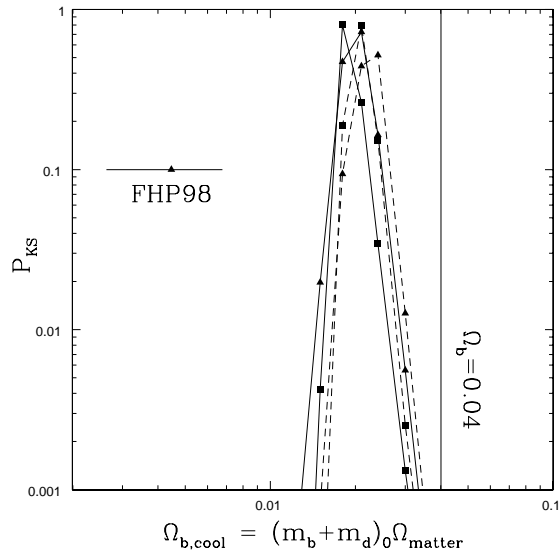,width=3.0in}}
\end{center}
\vspace{-0.25in}
\caption{(Top) Predicted separation distributions as a function of the
  cosmological cold baryon density $\Omega_{b,cool}$.  The dashed curves
  show the results for $\Omega_{b,cool}=0.003$, $0.006$ and $0.009$ 
  (right to left at large separation) and the solid curves show the
  results for $\Omega_{b,cool}=0.0012$, $0.015$, $0.018$, $0.021$,
  $0.024$, $0.030$, $0.045$ and $0.060$ (from left too right at large
  separation).  The models have 10\% of the cold baryons in a bulge.
  The heavy solid (dashed) curves show the observed distribution of
  CLASS (all radio) lenses.
   }
\labelprint{fig:clust3a}
\caption{(Bottom) The Kolmogorov-Smirnov probability, $P_{KS}$, of fitting the
  observed distribution of lenses as a function of the cold baryon
  density $\Omega_{b,cool}$.  The squares (triangles) indicate models
  with no bulge (10\% of the cooled material in a bulge), and the
  solid (dashed) lines correspond to fitting the CLASS (all radio) lenses.  
  For comparison, the horizontal error bar is the estimate by Fukugita,
  Hogan \& Peebles (\cite{Fukugita1998p518}) for the cold baryon (stars, remnants, cold gas)
  content of local galaxies. The vertical line marks the total baryon
  content of the concordance model.
   }
\labelprint{fig:clust3b}
\end{figure}

The mass scale of the break and the cold baryon fraction are not independent
parameters and should be derivable from the physics of the cooling gas.  In
its full details this must include not only the cooling of the gas but also
reheating of the gas in galaxies due to feedback from star formation.
Fig.~\ref{fig:clust2b} also shows the dependence of the cooling time
scale and the formation time scale for halos of mass $M_c$.  For this
model (based on the semi-analytic models of Cole et al.~\cite{Cole2000p168}), the 
cooling time 
becomes shorter than the age of the halo very close to the mass scale
required to explain the distribution of image separations.  These
semi-analytic models suggest an alternate approach in where the 
cooling mass scale need not be added as an {\it ad hoc} parameter.  
We could instead follow the semi-analytic models and use the cooling
function to determine the relative cooling rates of halos with different
masses.  We leave as the free parameter, the final cosmological density
in cold baryons $\Omega_{b,cool} \leq \Omega_b\simeq 0.04$ (i.e. some
baryons may never cool or cool and are reheated by feedback).  Low
$\Omega_{b,cool}$ models have difficulty cooling, making them equivalent
to models with a high cooling mass scale.  High $\Omega_{b,cool}$ models
cool easily, making them equivalent to models with a high cooling mass
scale.  Models with $0.015 \ltorder \Omega_{b,cool} \ltorder 0.025$ 
agree with the observations.  The result depends little on whether
we add a bulge, fit the CLASS sample or all radio lenses or adjust
the cooling curve by a factor of two.  Thus, the characteristic scale
of the gravitational lens separation distribution is a probe of the
cosmological baryon density $\Omega_b$ and the fraction of those
baryons that cool in the typical massive galaxy.  While it would be
premature to use this as a method for determining $\Omega_b$, it
is interesting to note that our estimate is significantly below 
current cosmological estimates that $\Omega_b\simeq 0.04$ which
would be consistent with feedback from star formation and other processes
preventing all baryons from cooling, but well above the estimates
of the cold baryon fraction in local galaxies 
($0.0045 \ltorder \Omega_{b,cool}\ltorder 0.0068$, Fukugita, Hogan
\& Peebles~\cite{Fukugita1998p518}).  These are also the models 
generating the velocity function estimate with baryonic cooling
in Fig.~\ref{fig:cdmvfunc}.  The cooling of the baryons shifts the
more numerous low velocity halos to higher circular velocities so
that the models match the observed density of $\sigma_v \simeq \sigma_*$ galaxies.
The models do not correctly treat the break region because they allow
``over-cooled'' massive groups, but then merge back onto the peak
circular velocity distribution of the CDM halos at higher velocities.
Since the models allow all low mass halos to cool, there is still
a divergence at low circular velocities which is closely related to
the problem of CDM substructure we discuss in \S\ref{sec:substruc}.
  
\subsection{The Effects of Halo Structure and the Power Spectrum}

Estimating the structure of clusters using gravitational lensing is
primarily a topic for \partweak, so we include only an abbreviated
discussion of lensing by clusters here.  For a fixed cosmological
model, two parameters largely control the abundance of cluster 
lenses.  First, the abundance of clusters varies nearly exponentially
with the standard normalization $\sigma_8 \simeq 1$ of the power
spectrum on $8h^{-1}$~Mpc scales.  Second, the cross sections of
the individual clusters depend strongly on the exponent of the
central density cusp of the cluster.  There are recent studies of
these issues by Li \& Ostriker~(\cite{Li2002p652}, \cite{Li2003p603}),
Huterer \& Ma~(\cite{Huterer2004p7}), Kuhlen, Keeton \& Madau~(\cite{Kuhlen2004p104}),
Oguri et al.~(\cite{Oguri2004p78}), and Oguri \& Keeton~(\cite{Oguri2004p1}).   

We can understand the general effects of halo structure very easily from
our simple power law model in Eqn.~\ref{eqn:aaf}.  In \S\ref{sec:basics}
we normalized the models to have the same Einstein radius, but we now
want to normalize them to all have the same total mass interior to 
some much larger radius $R_0$.  This is roughly what happens when we
keep the virial mass and break radius of the halo constant but vary
the central density exponent $\rho \propto r^{-n}$.  The deflection
profile becomes 
\begin{equation}
     \alpha(\theta) = { b_0^2 \over R_0 } \left( { \theta \over R_0 } \right)^{2-n}
\end{equation}
where $b_0 \ll R_0 $ sets the mass interior to $R_0$ and we recover our old
example if we let $b=b_0=R_0$.  The typical image separation is determined
by the tangential critical line at $\theta_t=R_0(b_0/R_0)^{2/(n-1)}$, so 
more centrally concentrated lenses (larger $n$) produce larger image separations
when $b_0/R_0 \ll 1$.
The radial caustic lies at $\beta_r = f(n) \theta_t$ where $f(n)$ is a
not very interesting function of the index $n$, so the cross section
for multiple imaging $\sigma \propto \beta_r^2 \propto R_0^2 (b_0/R_0)^{4/(n-1)}$ --
for an SIS profile $\sigma \propto b^4/R_0^2$, while the cross section for a 
Moore profile ($n=3/2$) $\sigma \propto b^8/16 R_0^6$ is significantly smaller.  
We cannot go to the limit of an NFW profile
($n=1$) because our power law model has a constant surface density rather than
a logarithmically divergent surface density in the limit as $n\rightarrow 1$, 
but we can see that as the density profile becomes shallower the multiple image
cross section drops rapidly when the models have constant mass inside a radius
which is much larger than their Einstein radius.  As a result, the numbers of
group or cluster lenses depends strongly on the central exponent of the density
distribution even when the mass function of halos is fixed.  Magnification bias
will weaken the dependence on the density slope because the models with shallower
slopes and smaller cross sections will generally have higher average magnifications.
The one caveat to
these calculations is that many groups or clusters will have central galaxies,
and the higher surface density of the galaxy can make the central density profile
effectively steeper than the CDM halo in isolation.        

\subsection{Binary Quasars \labelprint{sec:binaryqsos} }

Weedman et al.~(\cite{Weedman1982p5}) reported the discovery of the third
``gravitational lens'', Q2345+007, a pair of $z=2.15$ quasars separated by
7\farcs3.  The optical spectra of the two images are impressively similar 
(e.g. Small et al.~\cite{Small1997p2254}), but repeated attempts to find a 
lens have failed in both the optical (e.g. Pello et al.~\cite{Pello1996p73})
and with X-rays (Green et al.~\cite{Green2002p721}).  Q2345+007 is 
the founding member of a class of objects seen in the optical as a pair
of quasars with very similar spectra, small velocity differences and
separations $3\farcs0 \ltorder \Delta\theta \ltorder 15\farcs0$.  The
most recent compilation contained 15 examples (Mortlock, Webster \& 
Francis~\cite{Mortlock1999p836}).     The incidence of these quasar pairs
in surveys is roughly 2 per 1000 LBQS quasars (see Hewett et al.~\cite{Hewett1998p383}) and 
1 per 14000 CLASS radio sources (Koopmans et al.~\cite{Koopmans2000p815}).
The separations of these objects correspond to either very massive galaxies
or groups/clusters.  Obvious lenses on these scales, in the sense that
we see the lens, are rare but have an incidence consistent with 
theoretical expectations (see Fig.~\ref{fig:clust0}).  If, however, even
a small fraction of the objects like Q2345--007 are actually gravitational
lenses, then dark lenses outnumber normal groups and clusters and 
dominate the halo population on mass scales above
$M \gtorder 10^{13}M_\odot$. 

If the criterion of possessing a visible lens is dropped, so as to allow
for dark lenses, proving objects are lenses becomes difficult.  There 
are two unambiguous tests -- measuring a time delay between the images,
which is very difficult given the the long time delays expected for 
lenses with such large separations, or using deep imaging to find that
the host galaxies of the quasars show the characteristic arcs or 
Einstein rings of lensed hosts (Figs.~\ref{fig:basic4a}, \ref{fig:basic4b}).
The latter test is feasible with HST\footnote{We detected the host galaxies
  of the Q2345--007 quasars in the CASTLES H-band image.  Their 
  morphology is probably inconsistent with the lens hypothesis, but
  we viewed the data as too marginal to publish the result}
and will be trivial with JWST.  Spectral comparisons have been the main
area of debate.  In the optical, many of the pairs have alarmingly 
similar spectra if they are actually binary quasars (e.g. Q2345+007 or
Q1634+267, see Small et al.~\cite{Small1997p2254})  -- indeed, some of
these dark lens candidates have more similar spectra than genuinely
lensed quasars (see Mortlock, Webster \& Francis~\cite{Mortlock1999p836}).
The clearest examples of dark lens candidates that have to be binary
quasars are the cases in which only one quasar is radio loud.  These
objects, such as PKS1145--071 (Djorgovski et al.~\cite{Djorgovski1987p17})
or MGC2214+3550 (Mu\~noz et al.~\cite{Munoz1998p9}), represent 4 of the 
15 candidates.  Similarly, the dramatic difference in the flux ratio 
between optical and X-ray wavelengths of Q2345+007 is the strongest
direct argument for this object being a binary quasar (Green et al.~\cite{Green2002p721}).

Two statistical arguments provide the strongest evidence that these 
objects must be binary quasars independent of any weighting of spectral
similarities.  The first argument, due to Kochanek, Mu\~noz \& 
Falco~(\cite{Kochanek1999p590}), is that the existence of binary
quasars like MGC2214+3550 in which only one of the quasars is radio
loud predicts the incidence of pairs in which both are radio quiet.
We can label the quasar pairs as either $O^2R^2$, where both quasars
are seen in the optical (O) and the radio (R), $O^2R$, where only
one quasar is seen in the radio, or $O^2$ where neither quasar is
seen in the radio.  Lenses must be either $O^2R^2$ or $O^2$ pairs.
Surveys of quasars find that only $P_R\simeq 10\%$ of quasars are 
radio sources with 3.6~cm fluxes above 1~mJy (e.g. Bischof \&
Becker~\cite{Bischof1997p2000}).  If all the quasar pairs were binary
quasars and the probability of being radio loud is independent of 
whether a quasar is in a binary, then the relative number of $O^2$, 
$O^2R$ and $O^2R^2$ binaries should be 1 to $2 P_R=0.2$ to $P_R^2=0.01$.
Given that we observed 4 $O^2R$ binaries we should observe 
20 $O^2$ binaries and $0.2$ $O^2R^2$ binaries.  This statistical
pattern matches the data, and Kochanek, Mu\~noz \& 
Falco~(\cite{Kochanek1999p590}) found that the most probable 
solution was that all quasar pairs were binary quasars with an 
upper limit of only 8\% (68\% confidence) on the fraction that 
could be dark lenses.  With the subsequent expansion of the quasar
pair sample and the discovery of the first $O^2R^2$ binary
(B0827+525, Koopmans et al.~\cite{Koopmans2000p815}), these
limits could be improved.

The second statistical argument is that the dark lens candidates
do not have the statistical properties expected for lenses.  Three
aspects of the quasar pairs make them unlikely to be lenses simply
given the properties of gravitational lensing.  First, there are no 
four-image dark lens candidates even though a third of the normal 
lenses are quads. Second, many of the dark lens candidates have
very high flux ratios between the images -- 4 of the 9 ambiguous
quasar pairs considered by Rusin~(\cite{Rusin2002p705}) have flux
ratios of greater than 10:1.  Magnification bias makes such large
flux ratios very improbable for true gravitational lenses
(\S\ref{sec:magbias}, Kochanek~\cite{Kochanek1995p545}).  Third, 
the suppression of central/third/odd images in the lens population
is a consequence of baryonic cooling and the resulting increase 
of the central surface density.  Standard dark matter halos with
their shallow central cusps, $\rho \propto r^{-1}$, generally 
produce detectable third images.  Since it is probably a requirement
for a lens to remain dark that the baryons in the halo cannot 
cool (or they would form stars), you would expect the typical 
dark lens to resemble APM08279+5255 and have an easily detectable
third image (Rusin~\cite{Rusin2002p705}).  Thus, in the context of
CDM we would expect dark lenses to be standard cuspy density distributions
like the NFW model (Eqn.~\ref{eqn:abr}).  Rusin~(\cite{Rusin2002p705})
evaluated the likelihood of the quasar pairs assuming that dark lenses
have the structure of CDM halos and found that the observed flux ratios
and the lack of three-image dark lenses were extremely unlikely.  Only
the real lens APM08279+5255 had a significant probability of being 
produced by a dark CDM halo, although for this case I think the exposed
cusp/disk lens explanation for the morphology is more likely.

The evidence overwhelmingly favors interpreting the quasar pairs
as binary quasars.  However, as originally pointed out by 
Djorgovski~(\cite{Djorgovski1991p349}), the one problem with the
binary hypothesis is that the incidence of the quasar pairs is
two orders of magnitude above that expected from an extrapolation
of the quasar-quasar correlation function on scales of Mpc. As
discussed in  Kochanek, Mu\~noz \& Falco~(\cite{Kochanek1999p590})
and Mortlock, Webster \& Francis~(\cite{Mortlock1999p836}) the
incidence can be increased if the incipient merger of the two
host galaxies is triggering the quasar activity. The separation
distribution of the binary quasars is crudely compatible with 
tidally triggered activity when the merger starts followed by
a coalescence of the host galaxies driven by tidal friction. 
Small separation binary quasars ($\Delta\theta<3\farcs0$) are
rare because the decay of the host galaxy orbits accelerates as
their separation diminishes.  Well-measured angular distributions of
binary quasars, potentially obtainable from SDSS, might allow detailed
explorations of the triggering and merging physics.

\begin{figure}[t]
\centerline{\psfig{figure=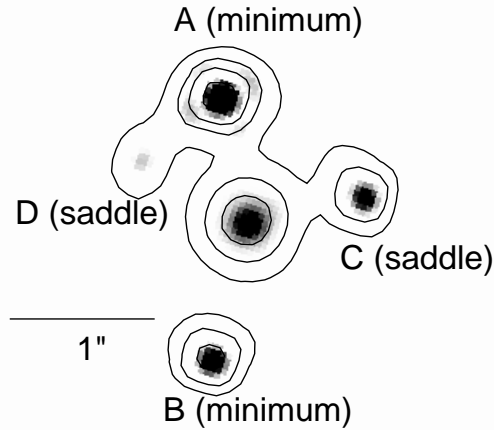,width=3.0in}}
\vspace{0.2in}
\caption{
   The most spectacular example of an anomalous flux ratio, SDSS0924+0219
  (Inada et al.~2003).  In this CASTLES infrared HST image, the D image should 
  be comparable in brightness to the
  A image, but is actually an order of magnitude dimmer.  The A and B images
  are minima, while C and D are saddle points.  The contours are spaced by
  factors of two from the peak of the A image. The lens galaxy is seen at the
  center.  At present we do not know whether the suppression of the saddle point
  in this lens is due to microlensing or substructure.  If it is microlensing,
  ongoing monitoring programs should see it return to its expected flux within
  approximately 10 years.
   }
\labelprint{fig:substruc0}
\end{figure}

\section{The Role of Substructure \labelprint{sec:substruc} }

Simulations of CDM halos predicted many more
small satellites than were actually observed in the Milky Way (e.g. Kauffmann et 
al.~\cite{Kauffmann1993p201}, Moore et al.~\cite{Moore1999p19},  
Klypin et al.~\cite{Klypin1999p82}).  Crudely
5-10\% of the mass was left in satellites with perhaps 1-2\% at the projected
separations of 1--2$R_e$ where we see most lensed images
(e.g. Zentner \& Bullock~\cite{Zentner2003p49}, Mao et al.~\cite{Mao2004p5}).  This is far
larger than the observed fraction of 0.01--0.1\% in observed satellites (e.g.
Chiba~\cite{Chiba2002p17}).
Solutions were proposed in three broad classes: hide the satellites by preventing
star formation so they are present but dark (e.g. Klypin et al.~\cite{Klypin1999p82},
Bullock et al.~\cite{Bullock2000p517}), destroy them using self-interacting dark 
matter (e.g. Spergel \& Steinhardt~\cite{Spergel2000p3760}), or avoid
forming them by changing the power spectrum to something similar to warm dark matter
with significantly less power on the relevant mass scales 
(e.g. Bode et al.~\cite{Bode2001p93}).  These hypotheses left the major observational 
challenge of distinguishing dark satellites from non-existent ones.
This became known as the CDM substructure problem.  

It was well known in the lensing community that the fluxes of lensed images 
were usually poorly fit by lens models.  There was a long litany of reasons for 
ignoring them arising from possible systematic errors which can corrupt image fluxes.
Differential effects between the images from the interstellar medium of the lens can
corrupt the fluxes (dust in the optical/IR, scatter broadening in the radio,
see \S\ref{sec:ism}). Time
delays combined with source variability can corrupt any single-epoch measurement. 
Microlensing by the stars in the lens galaxy can modify the fluxes of any sufficiently
compact component of the source (at a minimum the quasar accretion disk, see \partmicro).  
The most peculiar problem was the of anomalous
flux ratios in radio lenses.  Radio sources are essentially unaffected by the
ISM of the lens galaxy in low resolution observations that minimize the effects
of scatter broadening (VLA rather than VLBI), true absorption appears to
be rare, radio sources generally show little variability even when monitored,
and most of the flux should come from regions too large to be affected by microlensing.
Yet in B1422+231, for example, the three cusp images violated the cusp relation for
their fluxes (that the sum of the signed magnifications of the three images should be
zero, see Metcalf \& Zhao~\cite{Metcalf2002p5},
Keeton, Gaudi \& Petters~\cite{Keeton2003p138}, or
Schneider, Ehlers \& Falco~\cite{Schneider92}).\footnote{In specific models
there can also be global invariants relating image positions and magnifications
(e.g. Witt \& Mao~\cite{Witt2000p689},
Hunter \& Evans~\cite{Hunter2001p1227}, Evans \& Hunter~\cite{Evans2002p68}).
These results are usually for simple softened power law models using either
ellipsoidal potentials or an external shear rather than ellipsoidal
cuspy density distributions with an external shear, so their applicability to
the observed lenses is unclear.
  }

\begin{figure}
\begin{center}
\centerline{\psfig{figure=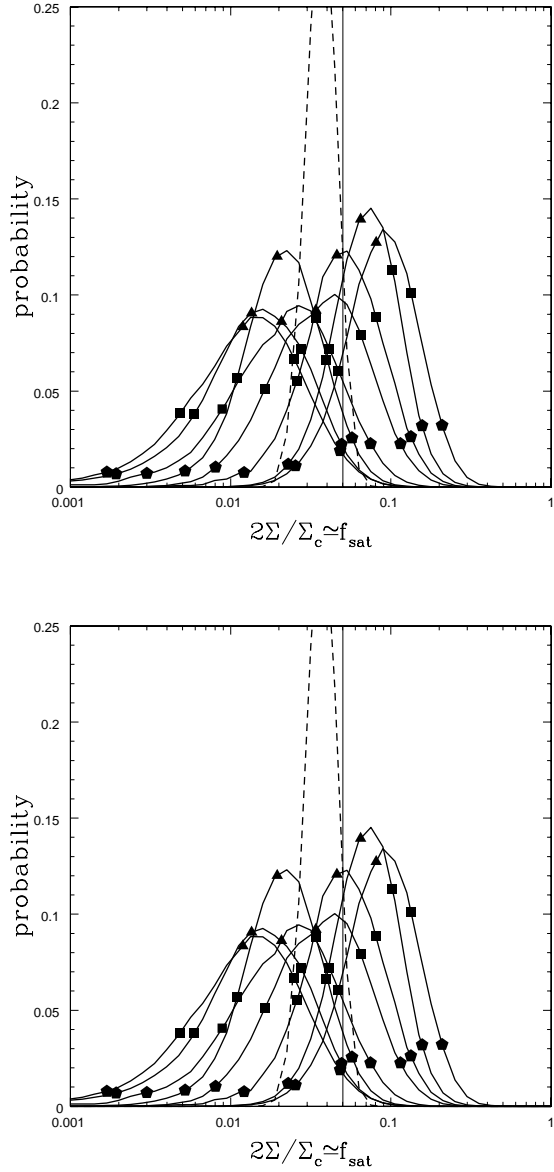,width=3.0in}}
\end{center}
\vspace{-0.2in}
\begin{center}
\centerline{\psfig{figure=substruc1.eps,width=3.0in}}
\end{center}
\vspace{-0.2in}
\caption{
   (Top) A Monte Carlo test for estimating substructure surface densities.  The heavy
   curves show the estimated probability distribution for the substructure 
   surface density fraction in a sample of 7 four-image lenses in which the
   input fraction was 5\% (marked by the vertical line).  The points on the
   curve show the median, $1\sigma$ and $2\sigma$ confidence limits.  The output
   distributions are consistent with the true input fraction.  The dashed line
   shows how the accuracy would improve given a sample of 56 lenses (i.e. 
   multiplying the 8 trials of 7 images each).
   }
\labelprint{fig:substruc1}
\caption{
   (Bottom) The same method applied to the real data.  The three distributions show the
   effects of changing assumptions on the actual flux measurement errors -- the
   greater the measurement uncertainties the less substructure surface density is
   required to explain the flux ratio anomalies.  The middle case (10\%) is
   probably slightly too conservative (20\% is ridiculously conservative and
   5\% is probably too optimistic).  
   }
\labelprint{fig:substruc2}
\end{figure}

It is easier to outline the problem of anomalous flux ratios near a fold caustic
(such as images A and D in SDSS0924+0219, see Fig.~\ref{fig:substruc0}), 
than a cusp caustic.  Near a fold, the lens equations
can be reduced to a one-dimensional model with
\begin{equation}
        \beta = \theta\left( 1-\Psi'' \right) - { 1 \over 2 } \Psi''' \theta^2
             \rightarrow  - { 1 \over 2 } \Psi''' \theta^2
\end{equation}
and inverse magnification
\begin{equation}
        \mu^{-1} = \left( 1-\Psi'' \right) - \Psi''' \theta 
             \rightarrow - \Psi'''\theta
\end{equation}
where we choose our coordinates such that there is a critical line at $\theta=0$
(i.e. $1-\Psi''=0$) and the primes denote derivatives of the potential.  These equations
are easily solved to find that you have images at $\theta_{\pm}=\pm (-2\beta/\Psi''')^{1/2}$
if the argument of the square root is positive and no solutions otherwise -- as you cross the fold
caustic ($\beta=0$) two images are created or destroyed on the critical line
at $\theta=0$.  Their inverse
magnifications of $\mu_{\pm}^{-1} = \mp (-2\beta\Psi''')^{1/2}$ are equal in 
magnitude but reversed in sign.  Hence, if the assumptions of the Taylor expansion
hold, the images merging at a fold should have identical fluxes.  Either by
guessing or by tedious algebra you can determine that the fractional correction
to the magnification from the next order term is of order $\theta_{\pm}\Psi^{(4)}/\Psi'''$.
For any reasonable central potential where the images are at radius $\theta_0$ from
the lens center, the fractional correction will be of order 
$\theta_\pm/\theta_0 \sim 0.1$ for the typical pair of anomalous images.  Hence, using
gravity to produce the anomalous flux ratios requires terms in the potential
with a length scale comparable to the separation of the images to 
significantly violate the rule that they should have similar fluxes.  
Mao \& Schneider (\cite{Mao1998p587}) pointed out that a very simple way of achieving this
was to put a satellite near the images, and they found that this could explain
the anomaly in B1422+231.  Metcalf \& Madau (\cite{Metcalf2001p9}, also see
Bradac et al.~\cite{Bradac2002p373} for images of the magnification patterns
expected from a CDM halo) put these two 
pieces together, pointing out that if normal satellite galaxies were too rare to 
make anomalous flux ratios common, the missing CDM substructure was not.  
They predicted that in CDM, anomalous flux ratios should be common. 

If we add a population of satellites with surface density $\kappa_{sat}=\Sigma_{sat}/\Sigma$
near the images we can estimate the nature of the perturbations.  If we model them as
pseudo-Jaffe potentials with critical radius $b$ and break 
radius\footnote{ This is the tidal truncation radius for an SIS of
  critical radius $b$ orbiting in an SIS of critical radius $b_0 > b$. The
  total satellite mass is $\simeq \pi a b \Sigma_c$.}
$a=(b b_0)^{1/2}$,
then the satellites produce a deflection perturbation of order
\begin{equation}
    \langle \delta\theta^2 \rangle^{1/2}
          \sim 10^{-3} b_0 \left( { 10 \Sigma_{sat} \over \Sigma_c } \right)^{1/2}
               \left( { 10^3 b \over b_0 }\right)^{3/4}.
\end{equation}
Only massive satellites will be able to produce deflection perturbations
large enough to be detected given typical astrometric errors.  Because the astrometric
constraints for lenses are so accurate, generally better than 0\farcs005, satellites with deflection
scales larger than $b \gtorder 10^{-2} b_0$ will usually have observable effects on model
fits and must be included in the basic lens model. 
The shear perturbation
\begin{equation}
    \langle \delta\gamma^2 \rangle^{1/2}
          \sim 0.1 \left( { 10 \Sigma_{sat} \over \Sigma_c } \right)^{1/2}
               \left( { 10^3 b \over b_0 }\right)^{1/4} \left( { \ln \Lambda \over 10 }\right)^{1/2},
\end{equation}
where $\ln\Lambda=\ln(a/s)$ is a Coulomb logarithm required to make the integral converge
at small separations, is significantly larger.  The effects of substructure gain on those
from the primary lens as we move to quantities requiring more derivatives of the 
potential because the substructure has less mass but shorter length scales.  For
example most astronomical objects have masses and sizes that scale with internal
velocity $\sigma_v$ as $M \propto \sigma_v^4$ and $R\propto \sigma_v^2$.  So time
delays, which depend on the (two-dimensional) potential $\Psi \propto M \propto \sigma_v^4$, will 
be completely unaffected by substructure.  Deflections, which require one spatial
derivative of the potential, $\alpha \propto \Psi/R \propto \sigma_v^2$, are 
affected only be the more massive substructres.  Magnifications, which require
two spatial derivatives of the potential, 
$\kappa \sim \gamma \sim \Psi/R^2 \propto \sigma_v^0$, are affected equally by
all mass scales provided the Einstein radius of the object is larger than the
characteristic size of the source.
Substructure will also affect
brighter images more than fainter images because the magnifications of
the brighter images are more unstable to small perturbations.  Recall that
the magnification $\mu = (\lambda_+\lambda_-)^{-1}$ where one of the eigenvalues
$\lambda_\pm=1-\kappa\pm\gamma$, usually $\lambda_-$, is small for a highly magnified
image.  If we now add a shear perturbation $\delta\gamma$, the perturbation to the
magnification is of order $\delta\gamma/\lambda_-$ so you have a bigger fractional
perturbation to the magnification for the same shear perturbation if the image is
more highly magnified.  The last important effect from substructure, for which I know
of no simple, qualitative explanation, is that substructure discriminates between
saddle points and minima when it is a small fraction of the total surface density
(Schechter \& Wambsganss~\cite{Schechter2002p685}, Keeton~\cite{Keeton2003p664}).   In this regime, the magnification distributions for the saddle points
develop an extended tail toward demagnification that is not present for the 
minima.

\begin{figure}[p]
\centerline{\psfig{figure=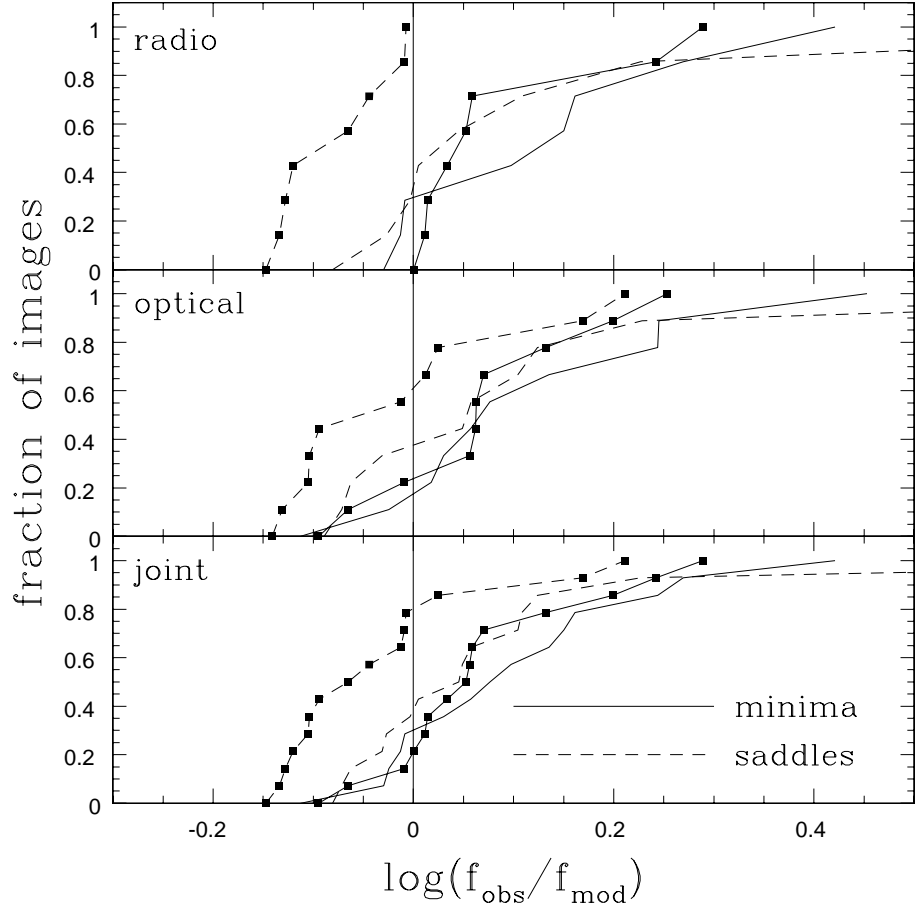,width=5.0in}}
\caption{ Saddle point suppression in lenses.  The three panels show the 
  cumulative distributions of model flux residuals, $\log(f_{obs}/f_{mod})$,
  in the real data, assuming constant fractional flux errors for each image.  The solid
  (dashed) lines are for minima (saddle points), with squares (no squares)
  for the distribution corresponding to the most (least) magnified image.
  From top to bottom the distributions are shown for samples of 8 radio,
  10 optical or 15 total four-image lenses.  If the flux residuals are
  created by propagation effects we would not expect the distributions
  to depend on the image parity or magnification, while if they are due
  to low optical depth substructure we would expect the distribution for
  the brightest saddle points to be shifted to lower observed fluxes.
   }
\labelprint{fig:substruc3}
\end{figure}

It turns out that anomalous flux ratios are very common -- a fact which had been staring 
us in the 
face but was ignored because most people (including the author!) were mainly just annoyed 
that the flux ratios could not be used to constrain the potential of the primary lens so as
to determine the radial mass profile.  When Dalal \& Kochanek (\cite{Dalal2002p25}) collected the
available four-image radio lenses to estimate the abundance of substructure, they found
that 5 of 6 systems showed anomalies.  In order to estimate the abundance of
substructure Dalal \& Kochanek (\cite{Dalal2002p25}) developed a Bayesian Monte Carlo method which
estimated the likelihood that adding substructure would significantly improve models
of seven four-image lenses including the fact that the model for the primary lens
would have to be adjusted each time any substructure was added.  Under the assumption
that the uncertainties in flux measurements (systematic as well as statistical)
were 10\%, they found a substructure mass fraction of $0.006 < f_{sat} < 0.07$
(90\% confidence) with a median estimate of $f_{sat}=0.02$.  This is consistent
with expectations from CDM simulations, including estimates of the destruction of
the satellites in the inner regions of galaxies (Zentner \& Bullock~\cite{Zentner2003p49},
Mao et al.~\cite{Mao2004p5}), and 
too high to be explained by normal satellite populations.  Because the result is
driven by the flux anomalies, which do not depend on the mass of the substructures,
rather than astrometric anomalies, which do depend on the mass, the results had
almost no ability to estimate the mass scale associated with the substructure.

While substructure with approximately the surface density expected from CDM is
consistent with the data, it is worth examining other possibilities.  We would
expect any effect from the ISM to be strongly frequency dependent (whether
in the radio or in the optical).  At least for radio lenses, Kochanek \& Dalal
(\cite{Kochanek2004}) found that the optical depth function needed to explain the radio flux
anomalies would have to be gray, ruling out all the standard radio suspects.
We would also expect propagation effects at radio frequencies to preferentially
affect the faintest images because they have the smallest angular sizes -- 
remember that more magnified images are always bigger even if
you cannot resolve the change in size.  The ISM also cannot discriminate between
images based on parity -- the ISM is a local property of the lens and the
parity is not, so they cannot show a correlation.  Hence, if radio propagation
effects created the anomalies they should be the same for minima and saddle
points and more important for the fainter than the brighter images.
Fig.~\ref{fig:substruc3} shows the cumulative distributions of flux residuals
for radio, optical and combined four-image lens samples from Kochanek \& Dalal
(\cite{Kochanek2004}).  The bright saddle point images clearly have a different distribution
in each case, as we would expect for substructure but not for the ISM.  The
Kolmogorov-Smirnov test significance of the differences between the most
magnified saddle points and the other three types of images (brightest minimum,
faintest minimum, faintest saddle) is 0.04\%, 5\% and 0.3\% for the
radio, optical and joint samples respectively.  The next most discrepant
image is the brightest minimum, also as expected for substructure, but with
less significance.  Various statistical games (bootstrap resampling methods
of estimating significance or testing for anomalies) always give the same
results.  Thus, the ISM is ruled out as an explanation.

Even though simple Taylor series arguments make it unlikely that changes
to the central potential are a solution (see \S\ref{sec:massquad}), 
it still has its advocates (Evans \& 
Witt~\cite{Evans2003p1351}, Quadri et al.~\cite{Quadri2003p659}, M\"oller,
Hewitt \& Blain~\cite{Moller2003p1}, Kawano et al.~\cite{Kawano2004}).
The basic answer is that it is possible to create flux anomalies by making
the deviations of the central potential from ellipsoidal sufficiently large
for the angular structure of the potential to change
rapidly enough between nearby images to produce the necessary magnification
changes.  There are three basic problems with this solution (see \S\ref{sec:modelfit}
as well).  

The first problem is that the required deviations from an ellipsoidal profile
far too large.  This is
true even though the biggest survey of such models allowed image positions 
to shift by approximately 10 times their actual uncertainties in order to 
alter the image fluxes (Evans \& Witt \cite{Evans2003p1351}) --  had they forced the 
models to match the true astrometric uncertainties they would have needed 
even larger perturbations.  Kochanek \& Dalal (\cite{Kochanek2004}) found that models
fitting the flux anomalies required $|a_4| \gg 0.01$ compared to the 
typical values observed for galaxies 
and simulated halos $|a_4|\sim0.01$ (see \S\ref{sec:massquad}).
 It is fair to say, however, that the quantitative
results on the multipole structure of simulated halos are limited.     

The second problem is that when we test these solutions in lenses for
which we have additional model constraints, the models are forced back
toward the standard ellipsoidal models.  The basic problem,
as Evans \& Witt (\cite{Evans2003p1351}) show, is that the problem of fitting 
image positions and
fluxes with potentials of the form $r F(\theta)$ can be reduced a a problem
in linear algebra if $F(\theta)$ is expanded as a multipole series -- by 
adding enough terms it is possible to fit any four-image lens exactly.
The reasons go back to the lack of constraints we discussed in \S\ref{sec:modelfit}.  
Fig~\ref{fig:substruc4} illustrates this point using the lens B1933+503.
Kochanek \& Dalal (\cite{Kochanek2004})
first fit the four compact images with a model including deviations
from an ellipsoidal surface density.  With sufficiently strong deviations
there were models that could eliminate the flux anomalies in this system.
However, this lens, B1933+503, actually has three components to its source --
a compact core forming the four-image system with the anomaly but also to
radio lobes lensed into another four-image system and a two-image system
for 10 images in all (Fig.~\ref{fig:b1933merlin}).  
When we add the constraints from these other images
the model is forced back to being a standard ellipsoidal model with a 
flux ratio anomaly.  In the future, the degree to which lens galaxy
potentials are ellipsoidal could be thoroughly tested in the lenses with
Einstein ring images of their host galaxies. 

 The third problem with 
using the central potential to produce flux ratio anomalies is that it
does not lead to the discrimination between saddle points and minima 
shown in Fig.~\ref{fig:substruc3}.  Kochanek \& Dalal (\cite{Kochanek2004}) demonstrate
this with Monte Carlo simulations, but the basic reason is simple.  
Consider a lens like PG1115+080 with two images merging at a saddle
point.  The sense with which the saddle point and minima are perturbed
depends on the phase of the higher order multipoles relative to the 
images and the critical line, but for any fixed lens potential, that
phase varies depending on the source position, so the average effect cannot
make the bright saddle points show a significantly different
set of properties from the bright minima.  Every observed flux anomaly
could be explained by adding complex angular structures to the main lens,
but the inability of these models to differentiate between saddle points
and minima would still rule them out.

\begin{figure}[t]
\centerline{\psfig{figure=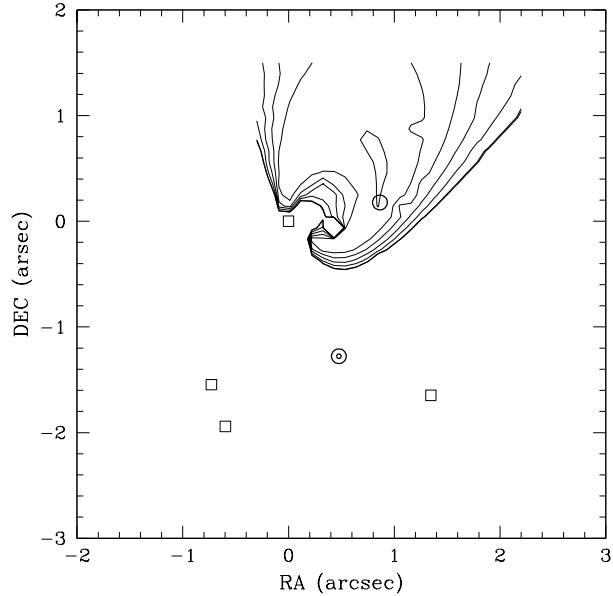,width=3.3in}}
\caption{The improvement in the fit to the Ros et al.~(\cite{Ros2000p845}) VLBI
   data on MG0414+0534 from adding an additional lens with a Einstein radius
   15\% that of the primary lens galaxy as a function of its position.  The
   squares show the location of the quasar images, the central circles mark
   the position of the main lens galaxy and the single circle marks the 
   position of object X (see Fig.~\ref{fig:mg0414h}).  The heavy contour 
   has the same $\chi^2=123$ as single component models, and they then drop
   a factor of 0.2 per lighter contour to a minimum of $\chi^2=0.6$ almost
   exactly at the position of Object X.
   }
\labelprint{fig:satchi}
\end{figure}

\begin{figure}[p]
\centerline{\psfig{figure=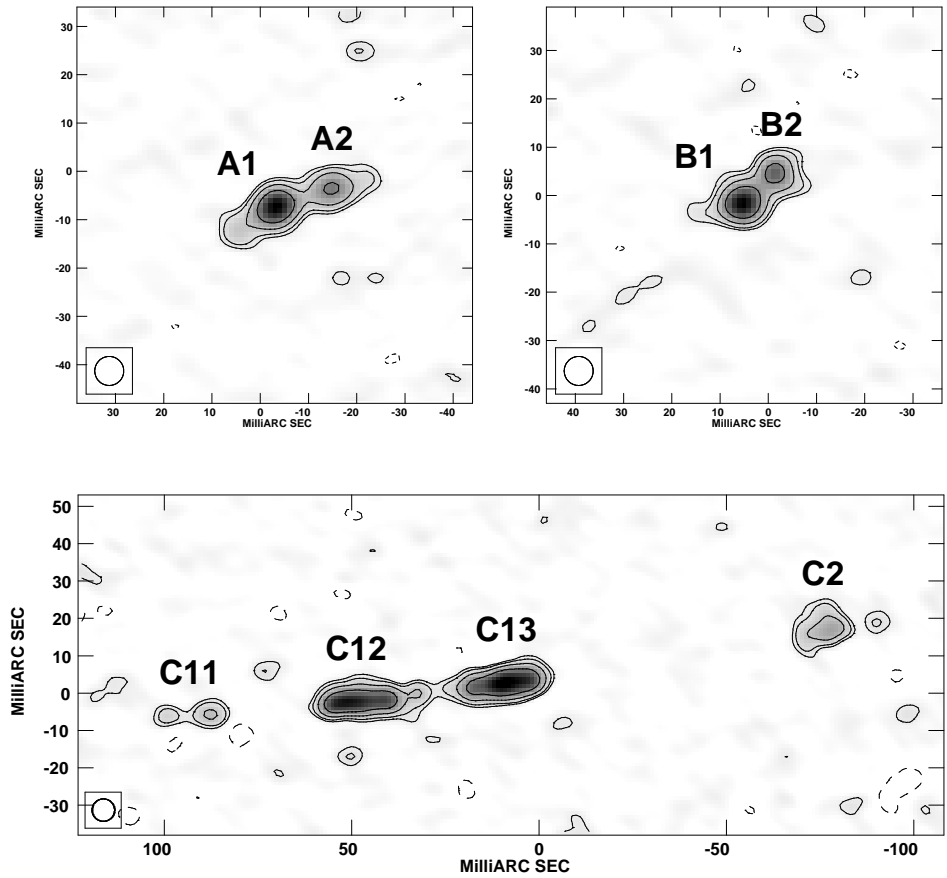,width=5.0in}}
\caption{VLBI maps of MG2016+112 (Koopmans et al.~\cite{Koopmans2002p39}).  
  The large difference in the
  C$_{11}$/C$_{12}$ separation as compared to the 
  C$_{13}$/C$_2$ separation is the clearest example of an 
  ``astrometric'' anomaly in a lens.
  The critical line passes between C$_{12}$ and C$_{13}$ and by
  symmetry we would expect the separations of the subcomponents
  on either side of the critical line to be similar.  In this
  case the cause of the asymmetry seems to be a galaxy D
  about 0\farcs8 South of the C image (see Fig.~\ref{fig:mg0414h}).
  Galaxy D has the same redshift as the primary lens 
  (Koopmans \& Treu~\cite{Koopmans2002p5}).  
   }
\labelprint{fig:astromsub}
\end{figure}

For the moment there are two barriers to improving estimates of the
substructure mass fraction.  First, radio lens surveys have run out
of sources bright enough to conduct efficient surveys.  This will only
change as upgrades to existing radio arrays are completed.  
The proposed Merlin and VLA upgrades will provide both sensitivity
and resolution improvements that will make
the next generation of radio lens surveys easier than the last.  Second,
searches for substructure using optical quasars need to separate the
effects of microlensing and substructure.  With simple imaging this 
can be done by finding parts of the quasar which are sufficiently extended
to avoid significant contamination from microlensing.  Emission line
(e.g. Moustakas \& Metcalf~\cite{Moustakas2003p607})
and dust emission regions should both be large enough to filter out
the effects of the stars.  Studying emission line ratios is now 
relatively easy because of the new generation of small-pixel integral
field spectrographs on 8m-class telescopes.  Mid-infrared flux ratios
for the dusty regions remain difficult, but the have been obtained for
one lens (Q2237+0305, Agol et al.~\cite{Agol2000p657}) and could be extended to several
more.  

The gold standard, however, would be astrometric detection of dark
substructure so that we would obtain a direct, mass estimate.  In
all the present analyses, the most massive substructures were 
included as part of the model.  They were not, however,
dark substructures because they matched to satellites visible in 
HST images of the lenses.  For example,  Object X in MG0414+0534
(Fig.~\ref{fig:mg0414h}) has effects on the image positions that are virtually
impossible to reproduce with changes in the potential of the central
lens galaxy (Trotter, Winn \& Hewitt~\cite{Trotter2000p671}), while
models with it easily fit the data (Ros et al.~\cite{Ros2000p845}).
Fig.~\ref{fig:satchi} shows the dependence of the goodness of fit to MG0414+0534
on the location of an additional lens component, with a deep minimum
located at the observed position of Object X.  The deflections produced
by an object of mass $M$ generally scale as $M^{1/2}$, so    
it is relatively easy to detect the deflection perturbations
from objects only 1\% the mass of the primary lens.  One approach
is to search lenses with VLBI structures for signs of perturbations.
This has been attempted for B1152+199 by Metcalf (\cite{Metcalf2002p696}), but the case for
substructure is not very solid given the limited nature of the data.
The cleanest example of astrometric detection of something small,
but sadly not dark, is in the VLBI structure of image C in 
MG2016+112 (Koopmans et al.~\cite{Koopmans2002p39}).  The asymmetry in the 
VLBI component separations of image C on either side of the critical line 
(see Fig.~\ref{fig:astromsub}) is due to a very faint galaxy 
0\farcs8 South of the image with a deflection scale $\sim 10$\% of
the primary lens (see Fig.~\ref{fig:mg0414h}).  
This is in reasonable agreement with the prediction from the
H-band magnitude difference of 4.6~mag and the (lens) Faber-Jackson 
relation between magnitudes
and deflections.  In this case, we even know that the satellite
is at the same redshift as the lens because Koopmans \& Treu (\cite{Koopmans2002p5}) 
accidentally measured its redshift in the course of their observations to measure
the velocity dispersion of the lens galaxy.  

\subsection{Low Mass Dark Halos \labelprint{sec:smallhalo}}

When we are examining a particular lens, almost all the substructure will consist of 
satellites associated with the lens, with only a $\sim 10$\% contamination from
other small halos along the line-of-sight to the source 
(Chen, Kravtsov \& Keeton~\cite{Chen2003p24}). However, the excess of
low mass halos in CDM mass functions relative to visible galaxies is a much more
general problem because the low mass CDM satellites should exist everywhere,
not just as satellites of massive galaxies.  Crudely, luminosity
functions diverge as $dn/dL \sim 1/L \sim 1/M$ while CDM mass functions
diverge as $dn/dM \sim M^{-1.8}$ so the fraction of low mass halos
that must be dark increases $\sim M^{-0.8}$ at low masses.
Fig.~\ref{fig:cdmvfunc} illustrates this assuming that all low mass halos have
baryons which have cooled (e.g. Gonzalez et al.~\cite{Gonzalez2000p145}, 
Kochanek~\cite{Kochanek2003p139}).  In the context of CDM, the solution to
this general problem is presumably the same as for the satellites responsible
for anomalous flux ratio -- they exist but lost their baryons before they
could form stars.  Such processes are implicit in semianalytic models which
can reproduce galaxy luminosity function (e.g. Benson et al.~\cite{Benson2003p38})
but can be modeled empirically in much the same way was employed for the
break between galaxies in clusters in \S\ref{sec:cluster} (e.g. 
Kochanek~\cite{Kochanek2003p139}).  In any model, the probability of the
baryons cooling to form a galaxy has to drop rapidly for halo masses below
$\sim 10^{11}M_\odot$ just as it has to drop rapidly for halo masses above
$\sim 10^{13}M_\odot$.  Unlike groups and clusters, where we still expect to
be able to detect the halos from either their member galaxies or X-ray 
emission from the hot baryons trapped in the halo, these low mass halos
almost certainly cannot be detected in emission.

We can only detect isolated, low-mass dark halos if they multiply image
background sources.  For SIS lenses the distribution of image separations
for small separations ($\Delta\theta/\Delta\theta_* \ll 1$, Eqn.~\ref{eqn:sepdist})
scales as
\begin{equation}
         { d \tau_{SIS} \over d \Delta\theta } \propto 
           \Delta\theta^{1+\gamma_{FJ}(1+\alpha)/2}
\end{equation}
where $\alpha$ describes the divergence of the mass/luminosity function at low
mass and $\gamma_{FJ}$ is the conversion from mass to velocity dispersion
(see \S\ref{sec:statgals}). For the standard parameters of galaxies,
$\alpha \simeq -1$ and $\gamma_{FJ} \simeq 4$, the separation distribution
is $d\tau_{SIS}/d\Delta\theta \propto \Delta\theta$.  In practice we do 
not observe this distribution because the surveys have angular selection
effects that prevent the detection of small image separations (below
$0\farcs25$ for the radio surveys), so the observed distributions show a
much sharper cutoff (Fig.~\ref{fig:basic1}).  Even without a cutoff, there
would be few lenses to find -- the CLASS survey found 9 lenses between 
$0\farcs3 \leq \Delta\theta \leq 1\farcs0$ in which case we expect only
one lens with $\Delta\theta < 0\farcs3$ even in the absence of any angular
selection effects.  A VLBI survey of 3\% of the CLASS sources
with milli-arcsecond resolution found no lenses (Wilkinson et al.~\cite{Wilkinson2001p37}),
nor would it be expected to for normal galaxy populations.  Our non-parametric
reconstruction of the velocity function including selection effects confirms 
that the existing lens samples are consistent with this standard model 
(Fig.~\ref{fig:nonpar}).

The result is very different if we extrapolate to low mass with the $\alpha \simeq -1.8$
slope of the CDM halo mass function.  The separation distribution becomes integrably
divergent, $d\tau_{SIS}/d\Delta\theta \propto \Delta\theta^{-0.6}$, and we would
expect 15 lenses with  $\Delta\theta < 0\farcs3$ given 9 between 
$0\farcs3 \leq \Delta\theta \leq 1\farcs0$.  Unfortunately, the 
Wilkinson et al.~(\cite{Wilkinson2001p37}) VLBI survey is too small
to rule out such a model.  A larger VLBI survey could easily do so,
allowing the lenses to confirm the galaxy counting argument for the
existence of second break in the density structure of halos at low
mass (Kochanek~\cite{Kochanek2003p139}, Ma~\cite{Ma2003p4}) similar
to the one between galaxies and high mass halos (\S\ref{sec:cluster}).
If the baryons in the low mass halos either fail to cool, or cool and are then ejected
by feedback, then their density distributions should revert to those of their
CDM halos.  If they are standard NFW halos, Ma~(\cite{Ma2003p4}) shows that such
low mass dark lenses will be very difficult to detect even in far larger surveys
than are presently possible.  Nonetheless, improving the scale of searches for 
very small separations from the initial attempt by  
Wilkinson et al.~(\cite{Wilkinson2001p37})
would provide valuable limits on their existence.

The resulting small, dark lenses would be the same as the dark lenses we discussed
in \S\ref{sec:binaryqsos} for binary quasars and explored by Rusin~(\cite{Rusin2002p705}).
They will also create the same problems about proving or disproving the lens hypothesis
as was raised by the binary quasars with the added difficulty that they will be far
more difficult to resolve.  Time delays, while short enough to be easily measured,
will also be on time scales where quasars show little variability.  Confirmation of
any small dark lens will probably requires systems with three or four images,
rather than two images, and the presence of resolvable (VLBI) structures.

\section{The Optical Properties of Lens Galaxies \labelprint{sec:optical} }

The optical properties of lens galaxies and the properties of their interstellar
medium (ISM) are important for two reasons.  First, statistical calculations
such as those in \S\ref{sec:stat} rely on lens galaxies obeying the same 
scaling relations as nearby galaxies and the selection effects depend
on the properties of the ISM.  Thus, measuring the scaling relations of the observed
lenses and the properties of their ISM are an important part of validating these
calculations.  Second, lenses have a unique advantage for studying the evolution
of galaxies because they are the only sample of galaxies selected based on mass
rather than luminosity, surface brightness or color.  Evolution studies using
optically-selected samples will always be subject to strong biases arising from
the difficulty of matching nearby galaxies to distant galaxies.  Selection by
mass rather than light makes the lens samples almost immune to these biases.

Most lens galaxies are early-type galaxies with relatively red colors and few signs
of significant on-going star formation (like the $3727$\AA\ or $5007$\AA\ Oxygen lines).
The resulting need to measure absorption line redshifts is one of the reasons that the
completeness of the lens redshift measurements is so poor.  Locally, early-type 
galaxies follow a series of correlations which also exist for the lens galaxies and
have been explored by Im, Griffiths \& Ratnatunga~(\cite{Im1997p457}),
Keeton, Kochanek \& Falco~(\cite{Keeton1998p561}), Kochanek
et al.~(\cite{Kochanek2000p131}), Rusin et al.~(\cite{Rusin2003p143}), Rusin,
Kochanek \& Keeton~(\cite{Rusin2003p29}), van de Ven, van Dokkum \& 
Franx~(\cite{vandeven2003p924}), Rusin \& Kochanek~(\cite{Rusin2004p1}).

The first, crude correlation is the Faber-Jackson relation between velocity 
dispersion and luminosity used in most lens statistical calculations.
 A typical local relation is that from \S\ref{sec:statgals} and shown in
Fig.~\ref{fig:kinematic}.  Most lenses lack directly measured velocity dispersions, 
but all lenses have a well-determined image separation $\Delta\theta$.  
For specific mass models the image separation can be converted
into an estimate of a velocity dispersion, such as the $\Delta\theta=8\pi(\sigma_v/c)^2 D_{ds}/D_s$ 
relation of the SIS, but the precise relationship depends on the mass distribution,
the orbital isotropy, the ellipticity and so forth (see  \S\ref{sec:dynamics}).
For the lenses, there is a close
relationship between the Faber-Jackson relation and aperture mass-to-light
ratios.  The image separation, $\Delta\theta$, defines the aperture mass interior 
to the Einstein ring,
\begin{equation}
    M_{ap} = {\pi \over 4 } \Sigma_c \Delta\theta^2
\end{equation}
where $\Sigma_c=c^2 D_s/4\pi G D_{ds} D_d$ is the critical surface density.  By
image separation we usually mean either twice the mean distance of the images 
from the lens galaxy or twice the critical radius of a simple lens model rather
than a directly measured image separation because these quantities will be less
sensitive to the effects of shear and ellipticity.  If
we measure the luminosity in the aperture $L_{ap}$ using (usually) HST, then we
know the aperture mass-to-light ($M/L$) ratio $\Upsilon_{ap}=M_{ap}/L_{ap}$. 

\begin{figure}[ph]
\begin{center}
\centerline{\psfig{figure=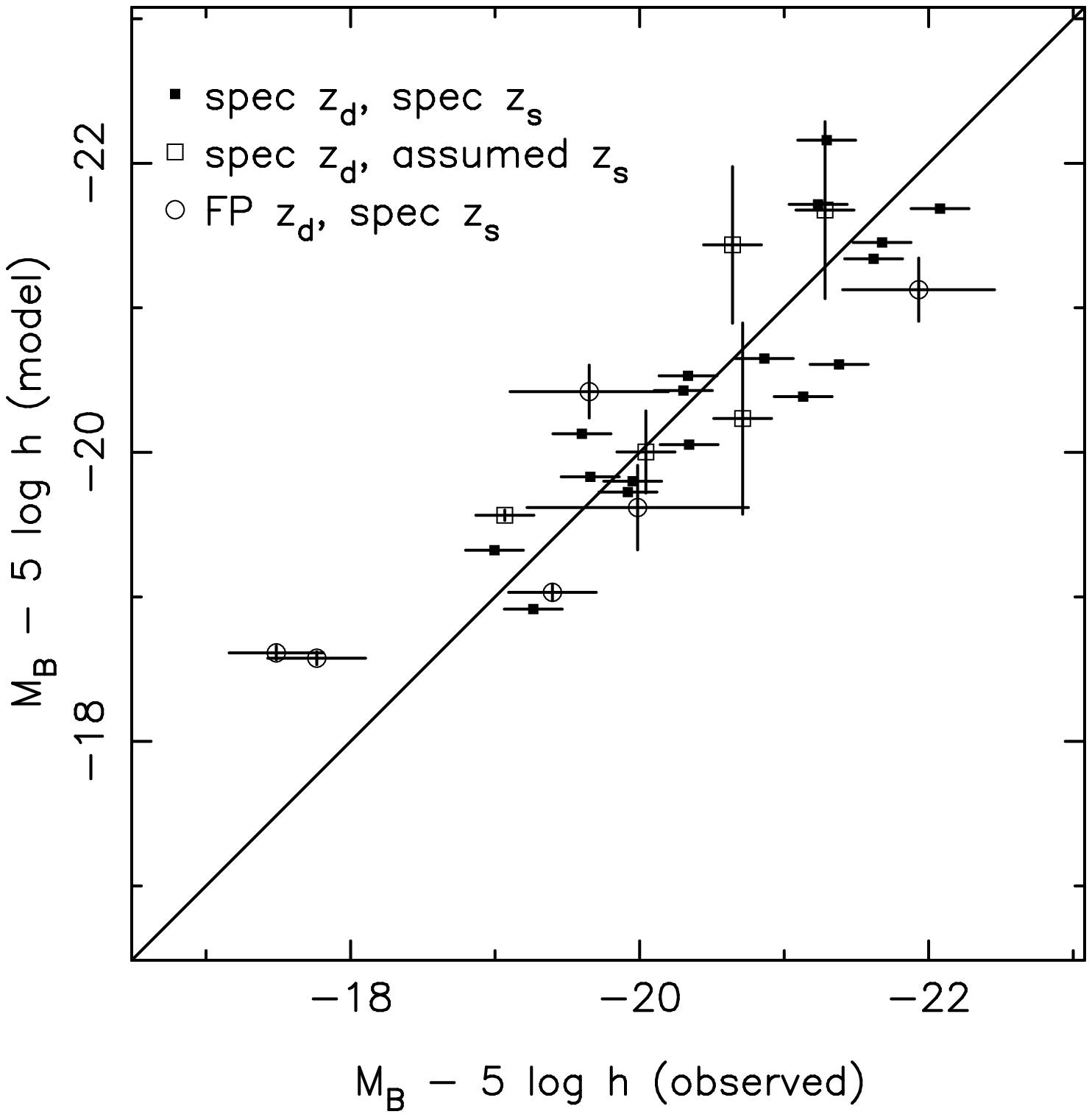,width=3.0in}}
\end{center}
\begin{center}
\centerline{\psfig{figure=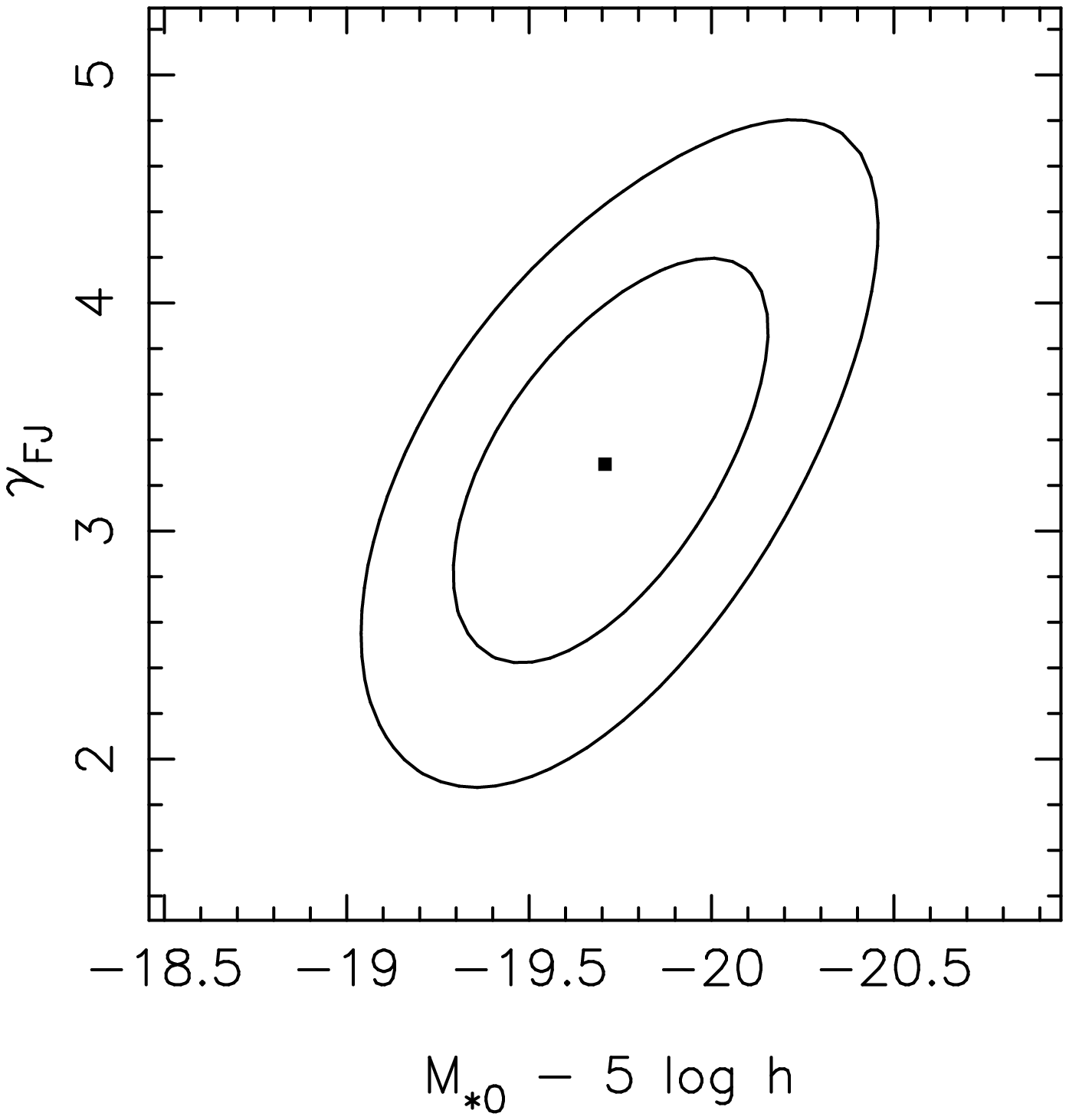,width=3.0in}}
\end{center}
\caption{ (Top)
   The ``Faber-Jackson'' relation for gravitational lenses.  The figure compares
   the observed absolute B magnitude corrected for evolution to that predicted
   from the equivalent of the Faber-Jackson relation for gravitational lenses 
   (Eqn.~\ref{eqn:lensfj}).  The different point styles indicate whether the
   lens and source redshifts were directly measured or estimated. From
   Rusin et al.~(\cite{Rusin2003p143}).
   }
\labelprint{fig:lensfj}
\caption{ (Bottom)
  The redshift zero absolute B-band magnitude and effective exponent of the 
  ``Faber-Jackson'' relation $L \propto \Delta\theta^{\gamma_{FJ}/2}$
   for gravitational lenses.
   }
\labelprint{fig:lensfjparams}
\end{figure}

If the mass-to-light ratio varies with radius or with mass, then to compare values of
$\Upsilon_{ap}$ from different lenses we must correct them to a common
radius and common mass.  If these scalings can be treated as power laws, then we can
define a corrected aperture mass to light ratio $\Upsilon_*=\Upsilon_{ap}(D_d^{ang}\Delta\theta/2R_0)^x$
where $R_0$ is a fiducial radius and $x$ is an unknown exponent, and we would
expect to find a correlation of the form
\begin{equation}
     \log \Upsilon_* = 2(1+a)\log \Delta\theta + 0.4 M_{abs} + \hbox{constant}
\end{equation}
where $M_{abs}$ is the absolute magnitude of the lens (in some band) and 
a value $a \ne 0$ indicates that the mass-to-light ratio varies either with
mass or with radius.  We can then rewrite this in a more familiar form as
\begin{equation}
     M_{abs} = M_{abs,0} + \gamma_{EV} z_l - 
          1.25 \gamma_{FJ} \log \left( { \Delta\theta \over \Delta\theta_0 } \right)
        \labelprint{eqn:lensfj}
\end{equation}
where $\Delta\theta_0$ sets an arbitrary separation scale, $\gamma_{EV}$ (or
a more complicated function) determines the evolution of the luminosity with
redshift, and $\gamma_{FJ}=4(1+a)$ sets the scaling of luminosity
with normalized separation defined so that for an SIS lens (where 
$\Delta\theta \propto \sigma_v^2$) the exponent $\gamma_{FJ}$ will match
the index of the Faber-Jackson relation (Eqn.~\ref{eqn:kbandfj}).
Fig.~\ref{fig:lensfj} shows the resulting relation converted to the rest frame B band at
redshift zero.  The relation is slightly tighter than local estimates of the
Faber-Jackson relation, but the scatter is still twice that expected from
the measurement errors.  The best fit exponent $\gamma_{FJ}=3.29\pm0.58$
(Fig.~\ref{fig:lensfjparams})
is consistent with local estimates and implies a scaling exponent $a=-0.18\pm0.14$
that is marginally non-zero.  If the mass-to-light ratio of early-type
galaxies increases with mass as $\Upsilon \propto M^x$, then $x=-a=0.18\pm0.14$ 
is consistent with estimates from the fundamental plane that more massive early-type
galaxies have higher mass-to-light ratios.  The solutions also require 
evolution with $\gamma_{EV}=-0.41\pm0.21$, so that early-type galaxies were
brighter in the past.  These scalings can also be done in terms of observed
magnitudes rather than rest frame magnitudes to provide simple estimation
formulas for the apparent magnitudes of lens galaxies in various bands as a
function of redshift and separation to an rms accuracy of approximately 
0.5~mag (see Rusin et al.~\cite{Rusin2003p143}).

\begin{figure}[ph]
\begin{center}
\centerline{\psfig{figure=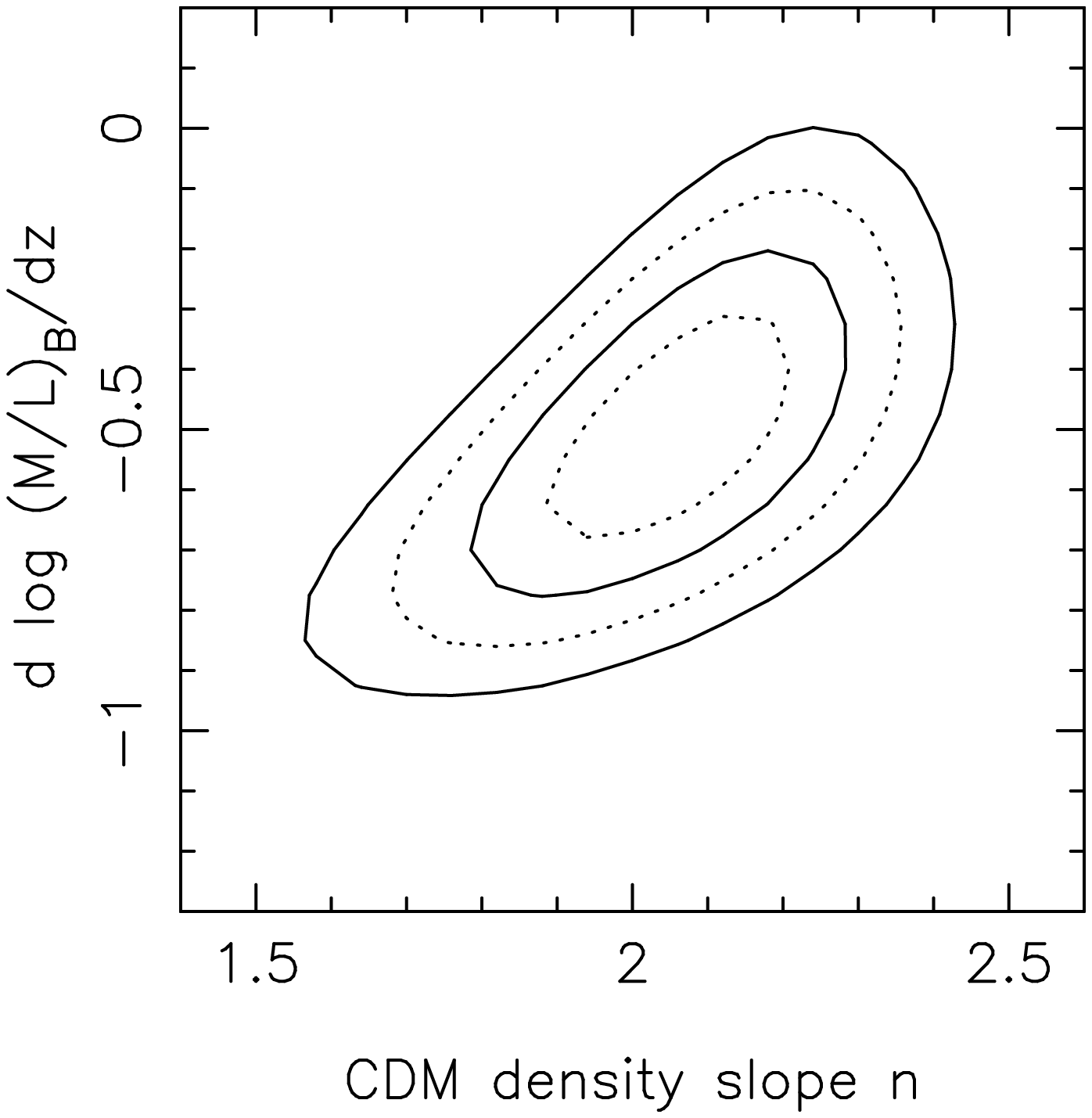,width=2.9in}}
\end{center}
\begin{center}
\centerline{ \psfig{figure=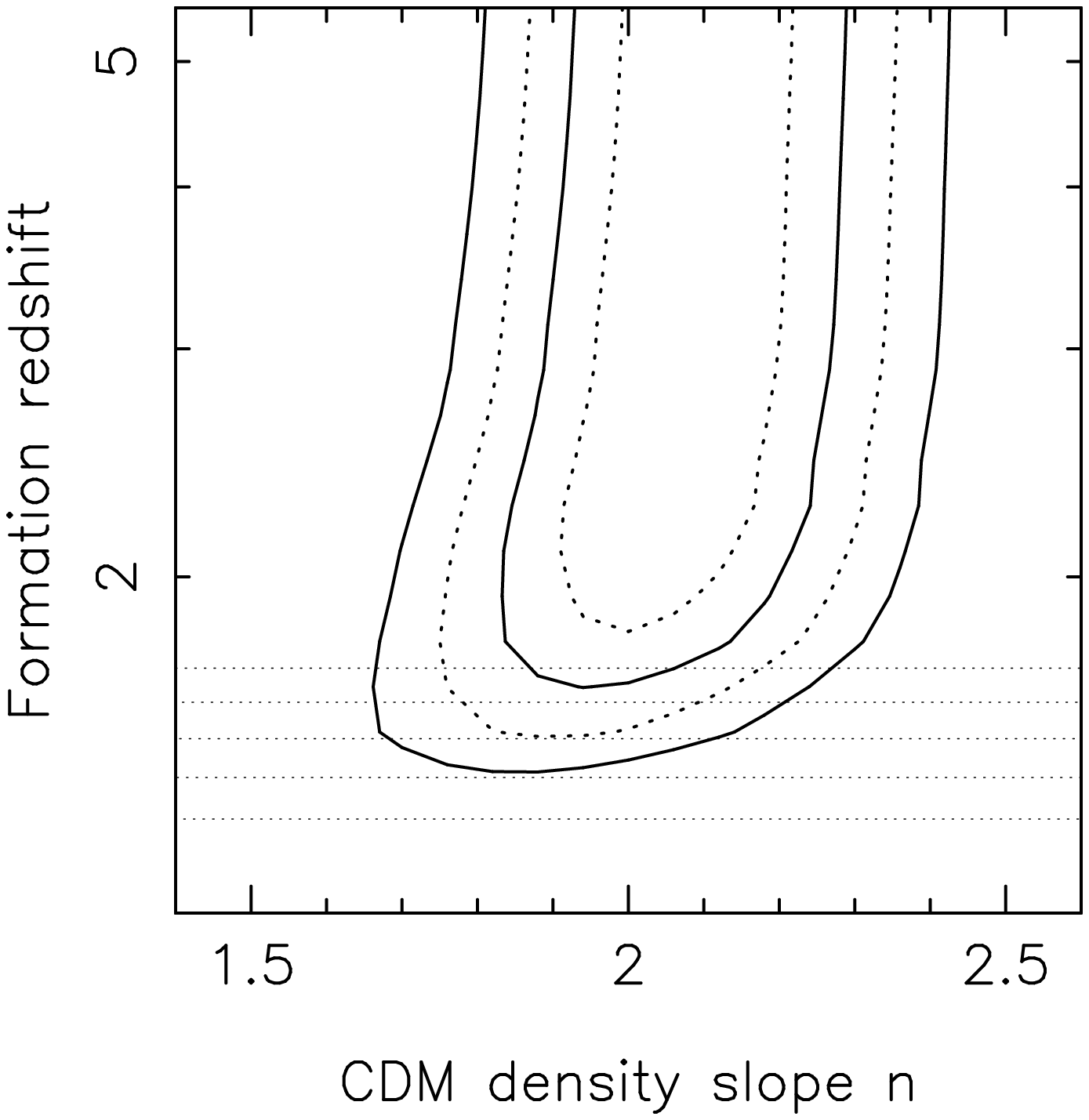,width=2.9in}}
\end{center}
\caption{ (Top)
   Constraints on the B-band luminosity evolution rate $d\log(M/L)_B/dz$ as a function
   of the logarithmic density slope $n$ ($\rho \propto r^{-n}$)
   of the galaxy mass distribution.  Solid (dashed)
   contours are the 68\% and 95\% confidence limits on two parameter (one parameter).
   These use the self-similar mass models of Eqn.~\ref{eqn:selfsimmass} and
   are closely related to the fundamental plane.
   From Rusin \& Kochanek~(\cite{Rusin2004p1}).
   }
\labelprint{fig:levolve1}
\caption{ (Bottom)
   Constraints on the mean star formation epoch $\langle z_f\rangle$ as  a function
   of the logarithmic density slope $n$ ($\rho \propto r^{-n}$) of the galaxy mass 
   distribution.  Solid (dashed)
   contours are the 68\% and 95\% confidence limits on two parameter (one parameter).
   The horizontal dotted lines mark $\langle z_f \rangle =1.3$, $1.4$, $1.5$, $1.6$
   and $1.7$.  The lens sample favors older stellar populations with 
   $\langle z_f \rangle > 1.5$ at 95\% confidence.
   These use the self-similar mass models of Eqn.~\ref{eqn:selfsimmass} and
   are closely related to the fundamental plane.
   From Rusin \& Kochanek~(\cite{Rusin2004p1}).
   }
\labelprint{fig:levolve2}
\end{figure}

The significant scatter of the Faber-Jackson relation makes it a crude tool.  
Early-type galaxies also follow a far tighter correlation known as the 
fundamental plane (FP, Dressler et al.~\cite{Dressler1987p42}, Djorgovski \& 
Davis~\cite{Djorgovski1987p17}) between the central, stellar velocity dispersion 
$\sigma_c$, the effective radius $R_e$ and the mean surface brightness inside the 
effective radius $\langle\hbox{SB}_e\rangle$ of the form
\begin{equation}
   \log\left( { R_e \over h^{-1}\hbox{kpc} }\right) =
       \alpha\log \left( { \sigma_c \over \hbox{km s}^{-1} } \right)
       +\beta \left( { \langle\hbox{SB}_e \rangle \over \hbox{mag arcsec}^{-2} }\right)
       +\gamma
            \labelprint{eqn:fp1}
\end{equation}
where the slope $\alpha$ and the zero-point $\gamma$ depend on wavelength but
the slope $\beta\simeq 0.32$ does not (e.g. Scodeggio et al.~\cite{Scodeggio1998p2738},  
Pahre, de Carvalho \& Djorgovski~\cite{Pahre1998p1606}).
Local estimates for the rest frame B-band give $\alpha=1.25$ and 
$\gamma_0 = -8.895 -\log(h/0.5)$ (e.g. Bender et al.~\cite{Bender1998p529}).
In principle both the zero points and the slopes may evolve with redshift, but all
existing studies have assumed fixed slopes and studied only the evolution of the
zero point with redshift.  For galaxies with velocity dispersion measurements,
the basis of the method is that measurement of $R_e$ and $\sigma_v$ provides an
estimate of the surface brightness the galaxy will have at redshift zero.  The
difference between the measured surface brightness at the observed redshift and
the surface brightness predicted for $z=0$ measures the evolution of the stellar
populations between the two epochs as a shift in the zero-point $\Delta\gamma$.  
The change in the zero-point is related to the change in the luminosity by
$\Delta L = -0.4 \Delta \hbox{SB}_e = \Delta\gamma/(2.5\beta)$.  While these
estimates are always referred to as a change in the mass-to-light ratio, no
real mass measurement enters operationally.
If, however, we assume a non-evolving virial mass estimate $M = c_M\sigma_v^2 R_e/G$
for some constant $c_M$, then
the FP can be rewritten in terms of a mass-to-light ratio,
\begin{equation}
   \log \Upsilon = \log\left( { M \over L} \right) \propto
      \left( { 10\beta -2 \alpha \over 5\beta }\right) \log \sigma_c
          + \left( { 2 - 5\beta \over 5\beta } \right) \log R_e 
           -  {\gamma \over 2.5\beta }
\end{equation}
so that if both $\alpha$ and $\beta$ do not evolve, the evolution of the 
mass-to-light ratio is $d\log \Upsilon/dz = - (d\gamma/dz)/(2.5\beta)$.  
Either way of thinking about the FP, either as an empirical estimator of the
redshift zero surface brightness or an implicit estimate of the virial mass,
leads to the same evolution estimates but alternate ways of thinking about
potential systematic errors.

Confusion about applications of lenses to the FP and galaxy evolution usually
arise because most gravitational lenses lack direct measurements of the central 
velocity dispersion.  Before addressing this problem, it is worth considering
what is done for distant galaxies with direct measurements.  The central
dispersion appearing in the FP has a specific definition -- usually either
the velocity dispersion inside the equivalent of a $3\farcs0$ aperture in
the Coma cluster or the dispersion inside $R_e/8$.  Measurements for 
particular galaxies almost never exactly match these definitions, so empirical
corrections are applied to adjust the velocity measurements in the observed
aperture to the standard aperture.  As we explore more distant galaxies, 
resolution problems mean that the measurement apertures become steadily larger
than the standard apertures.  The corrections are made with a single, average
local relation for all galaxies -- implicit in this assumption is that the 
dynamical structure of the galaxies is homogeneous and non-evolving.  This
seems reasonable since the minimal scatter around the FP seems to require
homogeneity, but says nothing about evolution.  These are also the same
assumptions used in the lensing analyses.

If early-type galaxies are homogeneous and have mass distributions that
are homologous with the luminosity distributions, then there is no difference
between the lens FP and the normal kinematic FP, independent of the actual mass
distribution of the galaxies (Rusin \& Kochanek~\cite{Rusin2004p1}).  If the mass
distributions are homologous, then the mass and velocity dispersion are related
by $M = c_M \sigma_c^2 R_e/G$ where $c_M$ is a constant, $\sigma_c$ is the 
central velocity dispersion (measured in a self-similar aperture like the
$R_e/8$ aperture used in many local FP studies), and $R_e$ is the effective
radius.  If we allow the mass-to-light ratio to scale with luminosity as
$\Upsilon \propto L^x$, then the normal FP can be written as
\begin{equation}
    \log R_e = { 2 \over 2 x + 1 } \log \sigma_c + { 0.4(x+1) \over 2 x +1 } 
    \langle SB_e \rangle + { \log c_M \over 2 x + 1},
   \labelprint{eqn:fp2}
\end{equation}
which looks like the local FP (Eqn.~\ref{eqn:fp1}) if $\alpha=2/(2x+1)$ and
$\beta=0.4(x+1)/(2x+1)$ (see Faber et al.~\cite{Faber1987p175}).  Thus, the
lens galaxy FP will be indistinguishable from the FP provided early-type
galaxies are homologous and the slopes can be reproduced by a scaling
of the mass-to-light ratio (as they can for $x\simeq 0.3$ given 
$\alpha \simeq 1.2$ and $\beta \simeq 0.3$, e.g., 
Jorgensen, Franx \& Kjaergaard~\cite{Jorgensen1996p167} or
Bender et al.~\cite{Bender1998p529}).  All the details about the mass
distribution, orbital isotropies and the radius interior to which the
velocity dispersion is measured enter only through the constant $c_M$
or equivalently from differences between the FP zero point $\gamma$ 
measured locally and with gravitational lenses.  In practice, Rusin
\& Kochanek~(\cite{Rusin2004p1}) show that the zero point must be measured
to an accuracy significantly better than $\Delta\gamma=0.1$ before there
is any sensitivity to the actual mass distribution of the lenses from the
FP.  Thus, there is no difference between the aperture mass estimates 
for the FP and its evolution and the normal stellar dynamical approach
unless the major assumption underlying both approaches is violated.  It
also means, perhaps surprisingly, that measuring central velocity dispersions
adds almost no new information once these conditions are satisfied.

Rusin \& Kochanek~(\cite{Rusin2004p1}) used the self-similar models we described
in \S\ref{sec:statfit} to estimate the evolution rate and the
star formation epoch of the lens galaxies while simultaneously estimating
the mass distribution.  Thus, the models for the mass include the uncertainties
in the evolution and the reverse.  Fig.~\ref{fig:levolve1} shows the estimated
evolution rate, and Fig.~\ref{fig:levolve2} shows how this is related to a limit
on the average star formation epoch $\langle z_f \rangle$ based on 
Bruzual \& Charlot~(\cite{Bruzual1993p538}, BC96 version) population
synthesis models.  This estimate is consistent with the earlier estimates
by Kochanek et al.~(\cite{Kochanek2000p131}) and 
Rusin et al.~(\cite{Rusin2003p143}) which used only isothermal lens models,
as we would expect. 
Van de Ven, van Dokkum \& Franx~(\cite{vandeven2003p924}) found a somewhat
lower star formation epoch ($\langle z_f \rangle = 1.8_{-0.5}^{1.4}$) when
analyzing the same data, which can be traced to  differences in the
analysis.  First, by weighting the galaxies by their measurement errors
when the scatter is dominated by systematics and by dropping two higher
redshift lens galaxies with unknown source redshifts, van de Ven
et al.~(\cite{vandeven2003p924}) analysis reduces the weight of the higher redshift
lens galaxies, which softens the limits on low $\langle z_f\rangle$.
Second, they used a power law approximation to the stellar evolution tracks
which underestimates the evolution rate as you approach the star formation
epoch, thereby allowing lower star formation epochs.  These two effects
leverage a small difference in the evolution rate\footnote{
   Rusin \& Kochanek~(\cite{Rusin2004p1}) obtained $d\log(M/L)_B/dz=-0.50\pm0.19$
   including the uncertainties in the mass distribution,
   Rusin et al.~(\cite{Rusin2003p143}) obtained $-0.54\pm0.09$ for a fixed
   SIS model, and van de Ven et al.~(\cite{vandeven2003p924}) obtained
   $-0.62\pm0.13$ for a fixed SIS model.}
into a much more dramatic difference in the estimated star formation epoch.
These evolution rates are consistent with estimates for cluster or
field ellipticals by (e.g. van Dokkum et al.~\cite{van_Dokkum1996p985}, 
\cite{van_Dokkum2001p39}, van Dokkum \& Franx~\cite{van_Dokkum2001p90},
van Dokkum \& Ellis~\cite{van_Dokkum2003p53},
Kelson et al.~\cite{Kelson1997p13}, \cite{Kelson2000p184}), 
and inconsistent with the much faster evolution
rates found by Treu et al.~(\cite{Treu2001p237}, \cite{Treu2002p13})
or Gebhardt et al.~(\cite{Gebhardt2003p1}).  

\begin{figure}[p]
\begin{center}
\centerline{\psfig{figure=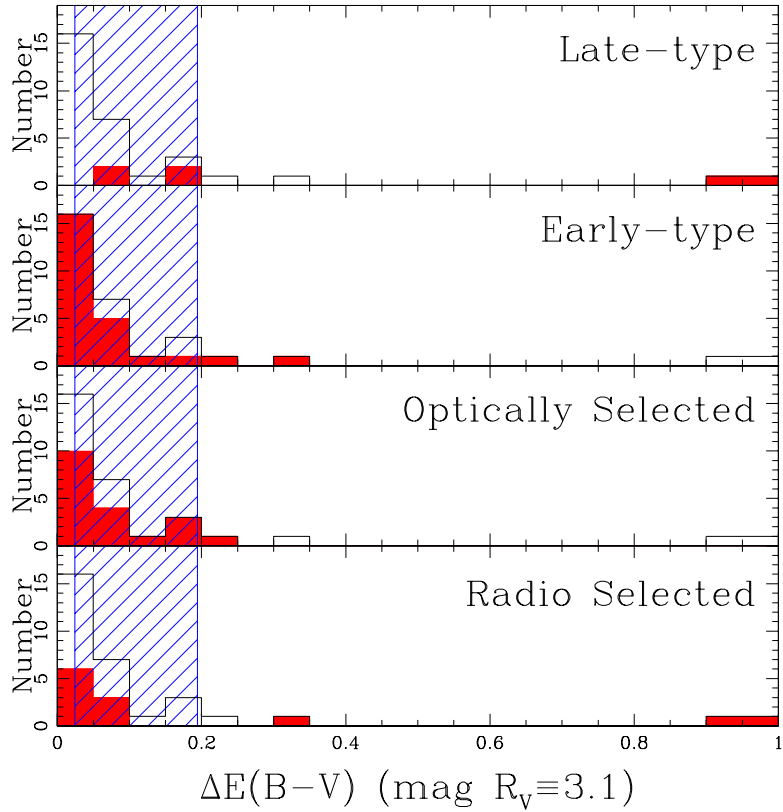,width=4.5in}}
\end{center}
\caption{ 
   Histograms of the differential extinction in various lens subsamples from
   Falco et al.~(\cite{Falco1999p617}).  In each panel the solid histogram
   shows the full sample of 37 differential extinctions measured in 23 lenses
   while the shaded histogram shows the distributions for different selection
   methods (radio/optical) or galaxy types (early/late).  The hatched region  
   shows the extinction range consistent with the Falco, Kochanek \& Mu\~noz
   (\cite{Falco1998p47}) analysis of the difference between the statistics
   of radio-selected and optically-selected lens samples (see \S\ref{sec:magbias}).
   Note that the most highly extincted systems, PKS1830--211 and B0218+357, are
   both radio-selected and late-type galaxies.  The lowest differential 
   extinction bins are contaminated by the effects of finite measurement
   errors. 
   }
\labelprint{fig:exthist}
\end{figure}

\subsection{The Interstellar Medium of Lens Galaxies \labelprint{sec:ism} }

As well as studying the emission by the lens galaxy we can study its 
absorption of emission from the quasar as a probe of the interstellar
medium (ISM) of the lens galaxies.  The most extensively studied effect
of the ISM is dust extinction because of its effects on estimating the
cosmological model from optically-selected lenses and because it allows
unique measurements of extinction curves outside the local Group.  
There are also broad band effects on the radio continuum due to 
free-free absorption, scatter broadening and Faraday rotation. While
all three effects have been observed, they have been of little 
practical importance so far.  Finally, in both the radio and the
optical, the lens can introduce narrow absorption features.  While
these are observed in some lenses, observational limitations have
prevented them from being as useful as the are in other areas of
astrophysics. 

As we mentioned in \S\ref{sec:stat}, extinction is an important systematic
problem for estimating the cosmological model using the statistics of 
optically selected lenses.  It modifies the results by changing the effective
magnification bias of the sample because it can make
lensed quasars dimmer than their unlensed counterparts.  
Because we
see multiple images of the same quasar, it is relatively easy to estimate
the differential extinction between lensed images under the assumption
that the quasar spectral shapes are not varying on the time scale 
corresponding to the time delay between the images and that microlensing
effects are not significantly changing the slope of the quasar continuum.
The former is almost certainly valid, while for the latter we simply lack
the necessary data to check the assumption (although we have a warning sign
from the systems where the continuum and emission line flux ratios
differ, see \partmicro).  Under these assumptions, the magnitude
difference at wavelength $\lambda$ between two images A and B 
\begin{equation}
     m_A(\lambda) - m_B(\lambda) = -2.5 \log\left| { \mu_A \over \mu_B }\right|
    +  R\left( { \lambda \over 1+z_l } \right) \Delta E(B-V)
     \labelprint{eqn:dust}
\end{equation}
depends on the ratio of the image magnifications $\mu_A/\mu_B$, the differential
extinction $\Delta E(B-V)=E_A-E_B$ between the two images and the extinction
law $R(\lambda/(1+z_l))$ of the dust in the rest frame of the dust.  We have the additional
assumption that either the extinction law is the same for both images or that
one image dominates the total extinction (Nadeau et al.~\cite{Nadeau1991p430}). 
Because it is a purely differential measurement that does not depend on knowing
the intrinsic spectrum of the quasar, it provides a means of determining 
extinctions and extinction laws that is otherwise only achievable locally where
we can obtain spectra of individual stars (the pair method, e.g. Cardelli,
Clayton \& Mathis~\cite{Cardelli1989p245}).  The total extinction cannot
be determined to any comparable accuracy because estimates of the total
extinction require an estimate of the intrinsic spectrum of the quasar.
Fig.~\ref{fig:exthist} shows the distribution of differential extinctions found
in the Falco et al.~(\cite{Falco1999p617}) survey of extinction in 23 
gravitational lenses.  Only 7 of the 23 systems had colors consistent with
no extinction, and after correcting for measurement errors and excluding the
two outlying, heavily extincted systems the data are consistent with a 
one-sided Gaussian distribution of extinctions starting at 0 and 
with a dispersion of $\sigma_{\Delta E}\simeq 0.1$~mag.  The two 
outlying systems, B0218+357 and PKS1830--211, were both radio-selected
and both have one image that lies behind a molecular cloud of a late
type lens galaxy (see below).

\begin{figure}[t]
\begin{center}
\centerline{\psfig{figure=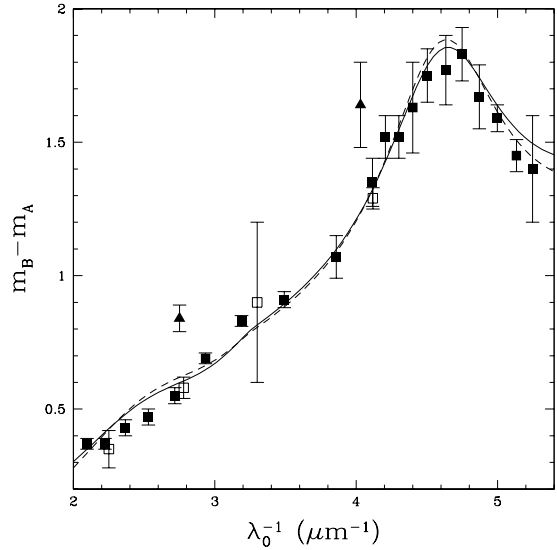,width=3.0in}}
\end{center}
\caption{
   The extinction curve of the dust in SBS0909+532 at $z_l=0.83$ 
   by Motta et al.~(\cite{Motta2002p719}).  The solid squares 
   show the magnitude difference as a function of inverse rest
   wavelength derived from integral field spectra of the continuum
   of the quasars.  The open squares are broad band measurements
   from earlier HST imaging and the filled triangles are the 
   flux ratios in the quasar emission lines.   The solid  
   curve shows the best fit $R_V=2.1\pm0.9$ Cardelli, Clayton 
   \& Mathis~(\cite{Cardelli1989p245}) extinction curve while
   the dashed curve shows a standard $R_V=3.1$ curve.  The offset
   between the continuum and emission line flux ratios seems not
   to depend on wavelength and is probably due to microlensing.   
   }
\labelprint{fig:extcurve}
\end{figure}

For lenses that have the right amount of dust, so that the image flux ratio
can be measured accurately over a broad range of wavelengths, it is
possible to estimate the extinction curve $R(\lambda/(1+z_l))$ of the dust
 (Nadeau et al.~\cite{Nadeau1991p430}) or to estimate the dust redshift
under the assumption that the extinction curve is similar to those measured
locally (Jean \& Surdej~\cite{Jean1998p729}).  Starting with 
Nadeau et al.~(\cite{Nadeau1991p430}), there have been many estimates
of extinction curves in lens galaxies (Falco et al.~\cite{Falco1999p617},
Toft, Hjorth \& Burud~\cite{Toft2000p115}, Motta et al.~\cite{Motta2002p719},
Mu\~noz et al.~\cite{Munoz2004p614}).  The most interesting of these are
for systems where the region near the 2175\AA\ extinction feature is
visible.  This requires source and lens redshifts that put the feature
at long enough wavelengths to be easily observed (i.e. higher lens
redshifts) with a quasar UV continuum extending to shorter wavelengths
(i.e. lower source redshifts).  Motta et al.~(\cite{Motta2002p719}) 
achieved the first cosmological detection of the feature in the 
$z_l=0.83$ lens SBS0909+532, as shown in Fig.~\ref{fig:extcurve}.  The
overall extinction curve is marginally consistent with a standard
Galactic R$_V=3.1$ extinction curve.  Other cosmologically distant
extinction curves are very different from normal Galactic models
ranging for an anomalously low $R_V$ curve in MG0414+0534 at $z_l=0.96$
(Falco et al.~\cite{Falco1999p617}), probably an SMC extinction
curve in LBQS1009--252 at an estimated redshift of $z_l\simeq 0.88$
(Mu\~noz et al.~\cite{Munoz2004p614}), and a anomalously high 
$R_V$ extinction curve for the dust in the molecular cloud of the
$z_l=0.68$ lens galaxy in B0218+357.  The Jean \& Surdej~(\cite{Jean1998p729})
idea of using the shape of the extinction curve to estimate the redshift
of the dust also seems to work given a reasonable amount of dust and
wavelength coverage (see Falco et al.~\cite{Falco1999p617}, Mu\~noz 
et al. \cite{Munoz2004p614}), but too few lenses with unknown redshifts
satisfy the requirements for widespread use of the method.

For broad band radio emission from the source, the three observed propagation
effects are free-free
absorption, scatter broadening and Faraday rotation.  For example, in
PMNJ1632--0033, the candidate third image of the lens (C) has the same
radio spectrum as the other two images except at the lowest frequency
observed (1.4~GHz) where it is fainter than expected. This can be
interpreted as free-free absorption by electrons at the center of the
lens galaxy but the interpretation needs to be confirmed by measurements
at additional frequencies to demonstrate that the dependence of the optical
depth on wavelength is consistent with the free-free process
(Winn, Rusin \& Kochanek~\cite{Winn2004p613}).  Scatter broadening
is observed in many radio lenses (e.g. PMN0134--0931, Winn et al.~\cite{Winn2003p26};
B0128+437, Biggs et al.~\cite{Biggs2004p1}; 
PKS1830--211, Jones et al.~\cite{Jones1996p23}; B1933+503, Marlow
et al.~\cite{Marlow1999p15}) primarily as changes in the fluxes of 
images between high resolution VLBI observations and lower resolution
VLA observations or apparently finite sizes for compact source components
in VLBI observations.  
In the presence of a magnetic field, the scattering medium will also
rotate polarization vectors (e.g. MG1131+0456, Chen \& Hewitt~\cite{Chen1993p1719}).
This is only of practical importance if maps which depend on the polarization
vector are used to constrain the lens potential.  In short, these effects
are observed but have so far been of little practical consequence.    

More surprisingly, absorption by atoms and molecules has also been of little
practical import for lens physics as yet.  Wiklind \& Alloin~(\cite{Wiklind2002p124})
provide an extensive review of molecular absorption and emission in gravitational
lenses.   The two systems with the strongest absorption systems are 
B0218+357 and PKS1830--211 (see Gerin et al.~\cite{Gerin1997p31} and
references therein) where one of the two images lies behind a molecular
cloud of the spiral galaxy lens.  These two systems also show the highest
extinction of any lensed images (Falco et al.~\cite{Falco1999p617}).
Molecular absorption systems can be used to determine time delays
(Wiklind \& Alloin~\cite{Wiklind2002p124}), measure the redshift of lens galaxies 
(the lens redshift in PKS1830-211 is measured using molecular absorption lines, 
Wiklind \& Combes \cite{Wiklind1996p139}), and potentially to determine the
rotation velocity of the lens galaxy (e.g. Koopmans \& de 
Bruyn~\cite{Koopmans2003p1}).  These studies at centimeter and millimeter
wavelengths are heavily limited by the resolution and sensitivity of existing
instruments, and the importance of these radio absorption features will probably
rise dramatically with the completion of the next generation of 
telescopes (e.g. ALMA, LOFAR, SKA).  

Similar problems face studies of
metal absorption lines in the optical.  Since most lenses are at modest
redshifts, the strongest absorption lines expected from the lens galaxies
tend to be observable only from space because they lie at shorter wavelengths
than the atmospheric cutoff.   For most lenses only the MgII (2800\AA) lines 
are observable from the ground since you only require a lens redshift
$z_l\gtorder 0.26$ to get the redshifted absorption lines longwards of 3500\AA.
The other standard metal line, CIV (1549\AA), is only visible for $z_l\gtorder 1.25$,
and we have no confirmed lens redshifts in this range.  Spectroscopy with HST
can search for metal lines in the UV, but the integration times tend to be
prohibitively long unless the quasar images are very bright.  Thus, while 
absorption lines either associated with the lens galaxy or likely to be
associated with the lens galaxy are occasionally found (e.g. SDSS1650+4251,
Morgan, Snyder and Reens~\cite{Morgan2003p2145}; or HE1104--1805, 
Lidman et al.~\cite{Lidman2000p62}), there have been no systematic studies
of metal absorption in gravitational lenses.  Nonetheless, some very bright
quasar lenses are favored targets for very high dispersion studies of they
Ly$\alpha$ forest, particularly the four-image lens B1422+231 and the three
image lens  APM08279+5255, because the lens magnification makes these systems
anomalously bright for quasars at $z_s > 3$.

\section{Extended Sources and Quasar Host Galaxies \labelprint{sec:hosts} }

As we saw in Figs.~\ref{fig:basic4a}, \ref{fig:basic4b}, and \ref{fig:mg1131}, 
we frequently see lensed emission from extended components
of the source.  
These arcs and rings are important because they can supply the extra constraints 
needed to determine the radial mass distribution that we lack in a simple
two-image of four-image lens (\S\ref{sec:massmono}).  The magnification
produced by gravitational lensing also allows us to study far fainter 
quasar host galaxies than is otherwise possible.  Comparisons of the 
luminosities and colors of high and low redshift host galaxies and the
relative luminosities of the host and the quasar are important for 
understanding the growth of supermassive black holes and their relationships
with their parent halos. 

Modeling extended emission is more difficult than modeling point sources largely
because of the complications introduced by the finite resolution of the observations.
In this section we first discuss a simple theory of Einstein ring images, then
some methods for modeling extended emission, and finally some results about the
mass distributions of lenses and the properties of quasar host galaxies.
All models of extended lenses sources start from the fact that lensing preserves
the surface brightness of the source -- what we perceive as magnification is only
an artifact of the finite resolution of our observations.  This can be modified
by absorption in the ISM of the lens galaxy 
(e.g. see, Koopmans et al.~\cite{Koopmans2003p70}), but we will neglect this 
complication in what follows.  We start with a simple analytic model for the
formation of Einstein rings, then discuss numerical reconstructions of lensed
sources and their ability to constrain mass distributions, and end with a 
survey of the properties of quasar host galaxies.

\begin{figure}[t]
\centerline{\psfig{figure=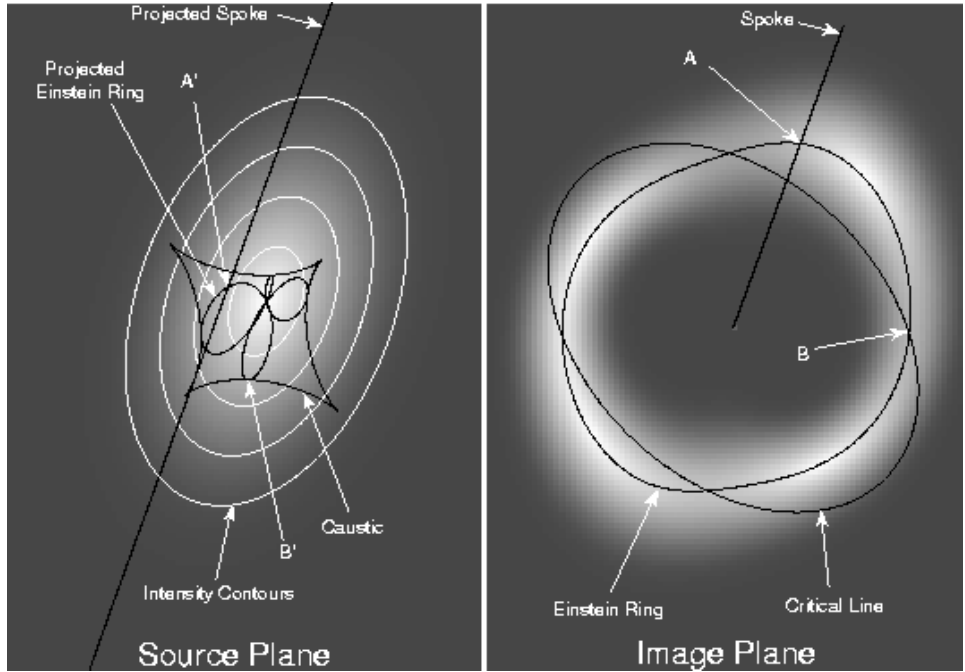,width=5.0in}}
\caption{An illustration of ring formation by an SIE lens.  An ellipsoidal
  source (left gray-scale) is lensed into an Einstein ring (right gray-scale).
  The source plane is magnified by a factor of 2.5 relative to the image
  plane.  The tangential caustic (astroid on left) and critical line 
  (right) are superposed.  The Einstein ring curve is found by looking for the
  peak brightness along radial spokes in the image plane.  For example, the
  spoke in the illustration defines point A on the ring curve.  The long 
  line segment on the right is the projection of the spoke onto the source
  plane.  Point A corresponds to point A' on the source plane where the 
  projected spoke is tangential to the intensity contours of the source.  The
  ring in the image plane projects into the four-lobed pattern on the source
  plane.  Intensity maxima along the ring correspond to the center of the source.
  Intensity minima along the ring occur where the ring crosses the critical
  curve (e.g. point B).  The corresponding points on the source plane (e.g. B')
  are where the astroid caustic is tangential to the intensity contours. 
   }
\labelprint{fig:host1}
\end{figure}

\begin{figure}[ph]
\begin{center}
\centerline{\psfig{figure=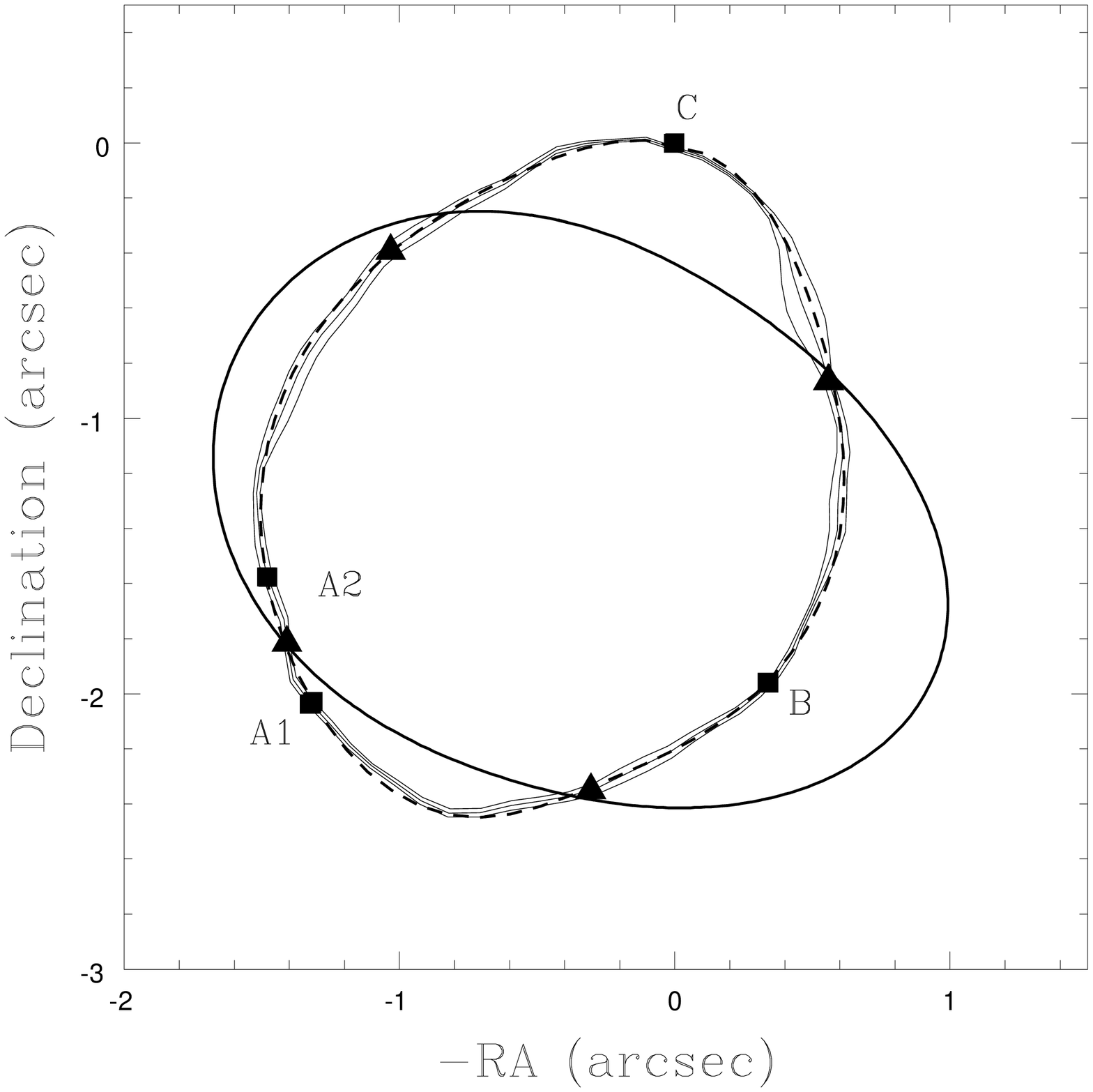,width=3.2in}}
\end{center}
\vspace{-0.25in}
\begin{center}
\centerline{ \psfig{figure=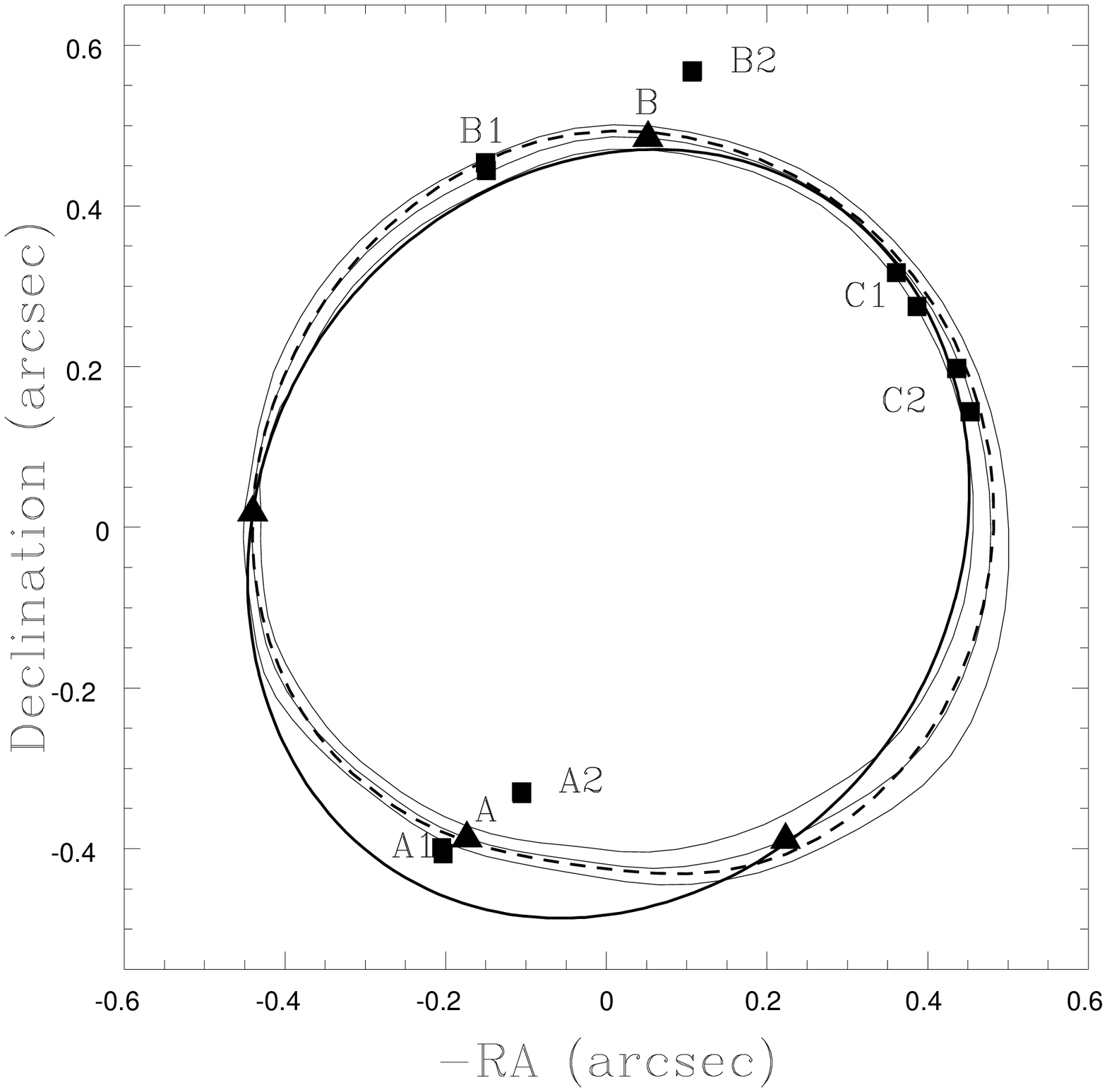,width=3.2in}}
\end{center}
\caption{The Einstein ring curves in PG1115+080 (top) and B1938+666 (bottom).  
  The black squares mark the lensed quasar or compact radio sources.  The light 
  black lines show the ring curve and its uncertainties.  The black triangles show the
  intensity minima along the ring curve (but not their uncertainties).  
  The best fit model ring curve is shown by the dashed curve, and the
  heavy solid curve shows the critical line of the best fit model.  The
  model was not constrained to fit the critical line crossings. }
\labelprint{fig:host2}
\end{figure}

\subsection{An Analytic Model for Einstein Rings \labelprint{sec:analring}}

Most of the lensed extended sources we see are dominated by an Einstein ring --
this occurs when the size of the source is comparable to the size of the astroid
caustic associated with producing four-image lenses. 
When the Einstein ring is fairly thin, there
is a simple analytic model for the formation of Einstein rings (Kochanek, Keeton \& McLeod
\cite{Kochanek2001p50}).
The important point to understand is that the ring is a pattern rather than a simple
combination of multiple images.  Mathematically, what we identify as the
ring is the peak of the surface brightness as a function of angle around the lens
galaxy.  We can identify the peak by finding the maximum intensity $\lambda(\chi)$
along radial spokes in the image plane, 
$\vec{\theta}(\lambda)=\vec{\theta}_0+\lambda(\cos\chi,\sin\chi)$.  At
a given azimuth $\chi$ we find the extremum of the surface brightness of the
image $f_D(\vec{\theta})$ along each spoke, and these lie at the solutions of 
\begin{equation}
      0 = \partial_\lambda f_D(\vec{\theta})=\nabla_{\vec{\theta}}f_D(\vec{\theta})
           \cdot { d \vec{\theta} \over d\lambda }.
\end{equation} 
The next step is to translate the criterion for the ring location onto the source
plane.  In real images, the observed image $f_D(\vec{\theta})$ is related to 
the actual surface density $f_I(\vec{\theta})$ by a convolution with the beam
(PSF), $f_D(\vec{\theta})=B * f_I(\vec{\theta})$, but for the moment we will
assume we are dealing with a true surface brightness map.  Under this assumption
$f_D(\vec{\theta})=f_I(\vec{\theta})=f_S(\vec{\beta})$  because of surface
brightness conservation.  When we change variables, the criterion for the
peak brightness becomes
\begin{equation}
      0 = \nabla_{\vec{\beta}}f_S(\vec{\beta}) \cdot M^{-1} 
           \cdot { d \vec{\theta} \over d\lambda }
   \labelprint{eqn:host1}
\end{equation} 
where the inverse magnification tensor $M^{-1}=d\vec{\theta}/d\vec{\beta}$ is
introduced by the variable transformation.  Geometrically we must find the point
where the tangent vector of the curve, $M^{-1}\cdot  d \vec{\theta}/d\lambda$
is perpendicular to the local gradient of the surface brightness 
$\nabla_{\vec{\beta}}f_S(\vec{\beta})$.  These steps are illustrated in
Fig.~\ref{fig:host1}.  

This result is true in general but not very useful.  We next assume that the
source has ellipsoidal surface brightness contours, $f_S(m^2)$, with 
$m^2=\Delta\vec{\beta}\cdot S \cdot \Delta\vec{\beta}$ where 
$\Delta\vec{\beta}=\vec{\beta}-\vec{\beta}_0$ is the distance from the 
center of the source, $\vec{\beta}_0$,
and the matrix $S$ is defined by the axis ratio $q_s=1-\epsilon_s \leq 1$ 
and position angle $\chi_s$ of the source.  We must assume that the surface
brightness declines monotonically, $df_s(m^2)/dm^2 < 0$, but require no
additional assumptions about the actual profile.  With these assumptions
the Einstein ring curve is simply the solution of
\begin{equation}
    0 = \Delta\vec{\beta} \cdot S \cdot \mu^{-1} \cdot { d \vec{\theta} \over d\lambda }.
\end{equation}
The ring curve traces out a four (two) lobed cloverleaf pattern when projected
on the source plane if there are four (two) images of the center of the source
(see Fig.~\ref{fig:host1}).  These lobes touch the tangential caustic at their
maximum ellipsoidal distance from the source center, and these cyclic variations
in the ellipsoidal radius produce the brightness variations seen around the
ring.  The surface brightness along the ring is defined by $f_I(\lambda(\chi),\chi)$ 
for a spoke at azimuth $\chi$ and distance  $\lambda(\chi)$ found by solving Eqn.~\ref{eqn:host1}.
The extrema in the surface brightness around the ring are located at the points
where $\partial_\chi f_I(\lambda(\chi),\chi)=0$, which occurs only at extrema
of the surface brightness of the source (the center of the source, 
$\Delta\vec{\beta}=0$ in the ellipsoidal model), or when the 
ring crosses a critical line of the lens and the magnification tensor is
singular ($|M|^{-1}=\mu^{-1}=0$) for the minima.   These are general results that
do not depend on the assumption of ellipsoidal symmetry.

For an SIE lens in an external shear field we can derive some simple properties 
of Einstein rings to lowest order in the various axis ratios.
  Let the SIE have critical radius $b$, axis ratio $q_l=1-\epsilon_l$ and put
its major axis along $\theta_1$. Let the external shear have amplitude $\gamma$ and 
orientation $\theta_\gamma$.  We let the source be an ellipsoid with axis ratio $q_s=1-\epsilon_s$
and a major axis angle $\chi_s$ located at position $(\beta\cos\chi_0,\beta\sin\chi_0)$
from the lens center.  The tangential critical line of the lens lies at radius
\begin{equation}
    r_{crit}/b = 1 + { \epsilon_l \over 2 } \cos 2 \chi - \gamma \cos 2(\chi-\chi_\gamma)  
\end{equation}
while the Einstein ring lies at
\begin{equation}
   { r_E \over b} = 1 + { \beta \over b }\cos(\chi-\chi_0)-{ \epsilon_l \over 6 } \cos 2 \chi 
           +  \gamma \cos 2(\chi-\chi_\gamma).  
\end{equation}
At this order, the Einstein ring is centered on the source position rather than the
lens position.  The orientation of the ring is generally perpendicular to that of
the critical curve, although it need not be exactly so when the SIE and the shear
are misaligned due to the differing coefficients of the shear and ellipticity terms
in the two expressions.  These results lead to a false impression that the results
do not depend on the shape of the source.  In making the expansion we assumed
that all the terms were of the same order ($\beta/b \sim \gamma \sim \epsilon_l \sim \epsilon_s$),
but we are really doing an expansion in the ellipticity of the potential of the
lens $e_\Psi \sim e_l/3$ rather than the ellipticity of the density distribution
of the lens, so second order terms in the shape of the source are as important as
first order terms in the ellipticity of the potential.  For example in a circular
lens with no shear ($\epsilon_l=0$, $\gamma=0$) the ring is located at
\begin{equation}
   { r_E \over b} = 1 + { \beta \over b } 
    { (2-\epsilon_s)\cos(\chi-\chi_0) + \epsilon_s \cos(2\chi_s-\chi-\chi_0) \over
               2 - \epsilon_s + \epsilon_s \cos 2(\chi_s-\chi) }
\end{equation}
which has only odd terms in its multipole expansion and converges slowly for 
flattened sources.  In general, the ring shape is a weak function of the source
shape only if the potential is nearly round and the source is almost centered on
the lens.  The structure of the lens potential dominates the even multipoles of
the ring shape, while the structure of the source dominates the odd multipoles.

In fact, the shape of the ring can be used to simply ``read off'' the amplitudes
of the higher order multipoles of the lens potential.  This is nicely illustrated
by an isothermal potential with arbitrary angular structure, $\Psi= r b F(\chi)$
with $\langle F(\chi)\rangle = 1$ (see Zhao \& Pronk~\cite{Zhao2001p401}, 
Witt et al.~\cite{Witt2000p98},
Kochanek et al.~\cite{Kochanek2001p50}, Evans \& Witt~\cite{Evans2001p1260}) in the 
absence of any shear.  The tangential critical line of the lens is
\begin{equation}
    { r_{crit} \over b } = F(\chi) + F''(\chi).
\end{equation}
If $\hat{e}_\chi$ and $\hat{e}_\theta$ are radial and tangential unit vectors
relative to the lens center and $\vec{\beta}_0$ is the distance of the source from 
the lens center, then the Einstein ring curve is
\begin{equation}
    { r_E \over b } = F(\chi) + 
     F'(\chi) { \hat{e}_\chi \cdot S \cdot \hat{e}_\theta \over 
                \hat{e}_\theta \cdot S \cdot \hat{e}_\theta }
    + { \vec{\beta}_0 \cdot S \cdot \hat{e}_\theta \over 
                \hat{e}_\theta \cdot S \cdot \hat{e}_\theta }
   \rightarrow F(\chi) + \vec{\beta}_0 \cdot \hat{e}_\theta
\end{equation}
with the limit showing the result for a circular source.  

Thus, by analyzing
the multipole structure of the ring curve one can deduce the multipole structure
of the potential.  While this has not been done non-parametrically, the ability
of standard ellipsoidal models to reproduce ring curves strongly suggests that
higher order multipoles cannot be significantly different from the ellipsoidal
scalings.   Fig.~\ref{fig:host2} shows two examples of fits to the ring curves
in PG1115+080 and B1938+666 using SIE plus external shear lens models.  The
major systematic problem with fitting the real data are that bright quasar images
must frequently be subtracted from the image before the ring curve can be
extracted, and this can lead to artifacts like the wiggle in the curve 
between the bright A$_1$/A$_2$ images of PG1115+080.  Other than that, the
accuracy with which the ellipsoidal (plus shear) models reproduce the curves
is consistent with the uncertainties.  In both cases the host galaxy is
relatively flat ($q_s=0.58\pm0.02$ for PG1115+080 and $0.62\pm0.14$ for
B1938+666).  The flatness of the host explains the ``boxiness'' of the 
PG1115+080 ring, while the B1938+666 host galaxy shape is poorly constrained
because the center of the host is very close to the center of the lens
galaxy so the shape of the ring is insensitive to the shape of the source.  
Unless the source is significantly offset from the center of the lens,
as we might see for the host galaxy of an asymmetric two-image lens, it 
does not constrain the radial density profile of the lens very well -- 
after considerable algebraic effort you can show that the dependence on the 
radial structure scales as $|\Delta\vec{\beta}|^4$.  It can, however, help
considerably in this circumstance because it eliminates the angular degrees
of freedom in the potential that make it impossible for two-image lenses
to constrain the radial density profile at all.  

\subsection{Numerical Models of Extended Lensed Sources \labelprint{sec:numring}}

Obviously the ring curve and its extrema are an abstraction of the real
structure of the lensed source.  Complete modeling of extended sources 
requires a real model for the surface brightness of the source.  In
many cases it is sufficient to simply use a parameterized model for the
source, but in other cases it is not.  The basic idea in any non-parametric
method is that there is an optimal estimate of the source structure for
any given lens model.  This is most easily seen if we ignore the smearing
of the image by the beam (PSF) and assume that our image is a surface
brightness map.  Since surface brightness is conserved by lensing,
$f_I(\vec{\theta})=f_S(\vec{\beta})$.  For any lens model with 
parameters $\vec{p}$, the lens equations define the source position
$\vec{\beta}(\vec{\theta},\vec{p})$ associated with each image position.  
If we had only single images of each source point, this would be useless
for modeling lenses.  However, in a multiply imaged region, more than 
one point on the image plane is mapped to the same point on the source
plane.  In a correct lens model, all image plane points mapped to the
same source plane position should have the same surface brightness, while
in an incorrect model, points with differing surface brightnesses will be
mapped to the same source point.  This provided the basis for the first
non-parametric method, sometimes known as the ``Ring Cycle'' method
(Kochanek et al.~\cite{Kochanek1989p43}, Wallington, Kochanek \& Koo~\cite{Wallington1995p58}).
Suppose source plane pixel $j$ is associated with image plane pixels
$i=1 \cdots n_j$ with surface brightness $f_i$ and uncertainties $\sigma_i$.
The goodness of fit for this source pixel is
\begin{equation}
        \chi^2_j = \sum_{i=1}^{n_j} \left( { f_i - f_s \over \sigma_i }\right)^2
\end{equation}
where $f_s$ will be our estimate of the surface brightness on the source
plane.  For each lens model we compute $\chi^2(\vec{p})=\sum\chi^2_j$
and then optimize the lens parameters to minimize the surface brightness
mismatches.

The problem with this algorithm is that we never have images that are
true surface brightness maps -- they are always the surface brightness
map convolved with some beam (PSF).  We can generalize the simple
algorithm into a set of linear equations.  Although the source and
lens plane are two-dimensional, the description is simplified if we
simply treat them as a vector $\vec{f}_S$ of source plane surface
brightness and a vector $\vec{f}_I$ of image plane flux densities
(i.e. including any convolution with the beam).  The two images are
related by a linear operator $A(\vec{p})$ that depends on the 
parameters of the current lens model and the PSF.  In the absence of a lens,
$A$ is simply the real-space (PSF) convolution operator.  In either
case, the fit statistic 
\begin{equation}
      \chi^2 =  { |\vec{f}_I - A(\vec{p})\vec{f}_S|^2 \over \sigma^2 }
\end{equation}
(with uniform uncertainties here, but this is easily generalized) must
first be solved to determine the optimal source structure for a given
lens model and then minimized as a function of the lens model.  The
optimal source structure $d\chi^2/d\vec{f}_S=0$ leads to the equation
that $\vec{f}_S = A^{-1}(\vec{p}) \vec{f}_I$.  The problem, which is
the same as we discussed for non-parametric mass models in \S\ref{sec:nonparam}, is that a
sufficiently general source model when combined with a PSF will lead
to a singular matrix for which $A(\vec{p})^{-1}$ is ill-defined --
physically, there will be wildly oscillating source models for which
it is possible to obtain $\chi^2(\vec{p})=0$.  

Three approaches have been used to solve the problem.  The first is
LensClean (Kochanek \& Narayan~\cite{Kochanek1992p461}, Ellithorpe, 
Kochanek \& Hewitt~\cite{Ellithorpe1996p556}, Wucknitz~\cite{Wucknitz2004p1}), 
which is based on the Clean algorithm of radio
astronomy.  Like the normal Clean algorithm, LensClean is a non-linear
method using a prior that radio sources can be decomposed into point
sources for determining the structure of the source.  The second
is LensMEM (Wallington, Kochanek \& Narayan~\cite{Wallington1996p64}), 
which is based on the Maximum Entropy Method
(MEM) for image processing.  The determination of the source structure
is stabilized by minimizing $\chi^2 + \lambda \int d^2\beta f_S\ln (f_S/f_0)$  
while adjusting the Lagrange multiplier $\lambda$ such that at the
minimum $\chi^2 \sim N_{dof}$ where $N_{dof}$ is the number of degrees
of freedom in the model.  Like Clean/LensClean, MEM/LensMEM is a
non-linear algorithm in which solutions must be solved iteratively.
Both LensClean and LensMEM can be designed to produce only positive-definite
sources.  The third approach is linear regularization where the source 
structure is stabilized by minimizing $\chi^2 +\lambda \vec{f}_S \cdot H \cdot \vec{f}_S$
(Warren \& Dye~\cite{Warren2003p673}, Koopmans et al.~\cite{Koopmans2003p70}).
The simplest choice for the matrix $H$ is the identity matrix, in which
case the added criterion is to minimize the sum of the squares of the source
flux.  More complicated choices for $H$ will minimize the gradients or
curvature of the source flux.  The advantage of this scheme is that the
solution is simply a linear algebra problem with 
$(A^T(\vec{p})A(\vec{p})+\lambda H) \vec{f}_S = A^T(\vec{p})\vec{f}_I$.

In all three of these methods there are two basic systematic issues 
which need to be addressed.  First, all the methods have some sort
of adjustable parameter -- the Lagrange multiplier $\lambda$ in LensMEM
or the linear regularization methods and the stopping criterion in the
LensClean method.  As the lens model changes, the estimates of the
parameter errors will be biased if the treatment of the multiplier
or the stopping criterion varies with changes in the lens model in some
poorly understood manner.  Second, it is difficult to work out the 
accounting for the number of degrees of freedom associated with the
model for the source when determining the significance of differences
between lens models.  Both of these problems are particularly
severe when comparing models where the size of the multiply imaged
region depends on the lens model.  Since only multiply imaged regions
supply any constraints on the model, one way to improve the goodness of
fit is simply to shrink the multiply imaged region so that there are
fewer constraints.  Since changes in the radial mass distribution have
the biggest effect on the multiply imaged region, this makes estimates
of the radial mass distribution particularly sensitive to controlling 
these biases.  It is fair to say that all these algorithms lack a
completely satisfactory understanding of this problem.  For radio
data there are added complications arising from the nature of interferometric
observations, which mean that good statistical models must work with
the raw visibility data rather than the final images (see Ellithorpe et 
al.~\cite{Ellithorpe1996p556}).  

\begin{figure}[t]
\begin{center}
\centerline{\psfig{figure=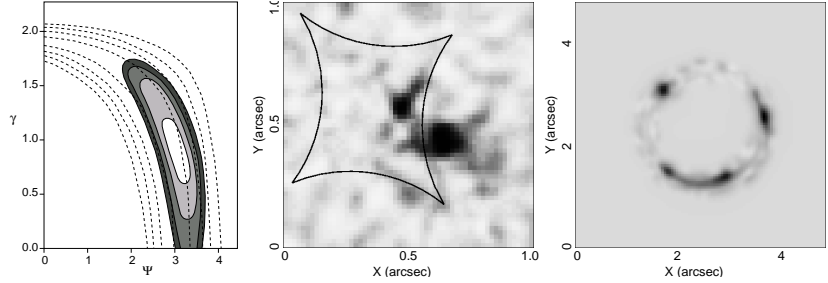,width=5.0in}}
\end{center}
\caption{Models of 0047--2808 from Dye \& Warren~(\cite{Dye2003p1}).  
  The right panel shows the lensed image of
  the quasar host galaxy after the foreground lens has been subtracted.
  The middle panel shows the reconstructed source and its position relative
  to the tangential (astroid) caustic.  The left panel shows the resulting
  constraints on the central exponent of the dark matter halo ($\rho \propto r^{-\gamma}$)
  and the stellar mass-to-light ratio of the lens galaxy. The dashed contours
  show the constraints for the same model using the central velocity
  dispersion measurement from Koopmans \& Treu~(\cite{Koopmans2003p606}).
  }
  \labelprint{fig:dye0047}
\end{figure}

\begin{figure}[t]
\begin{center}
\centerline{\psfig{figure=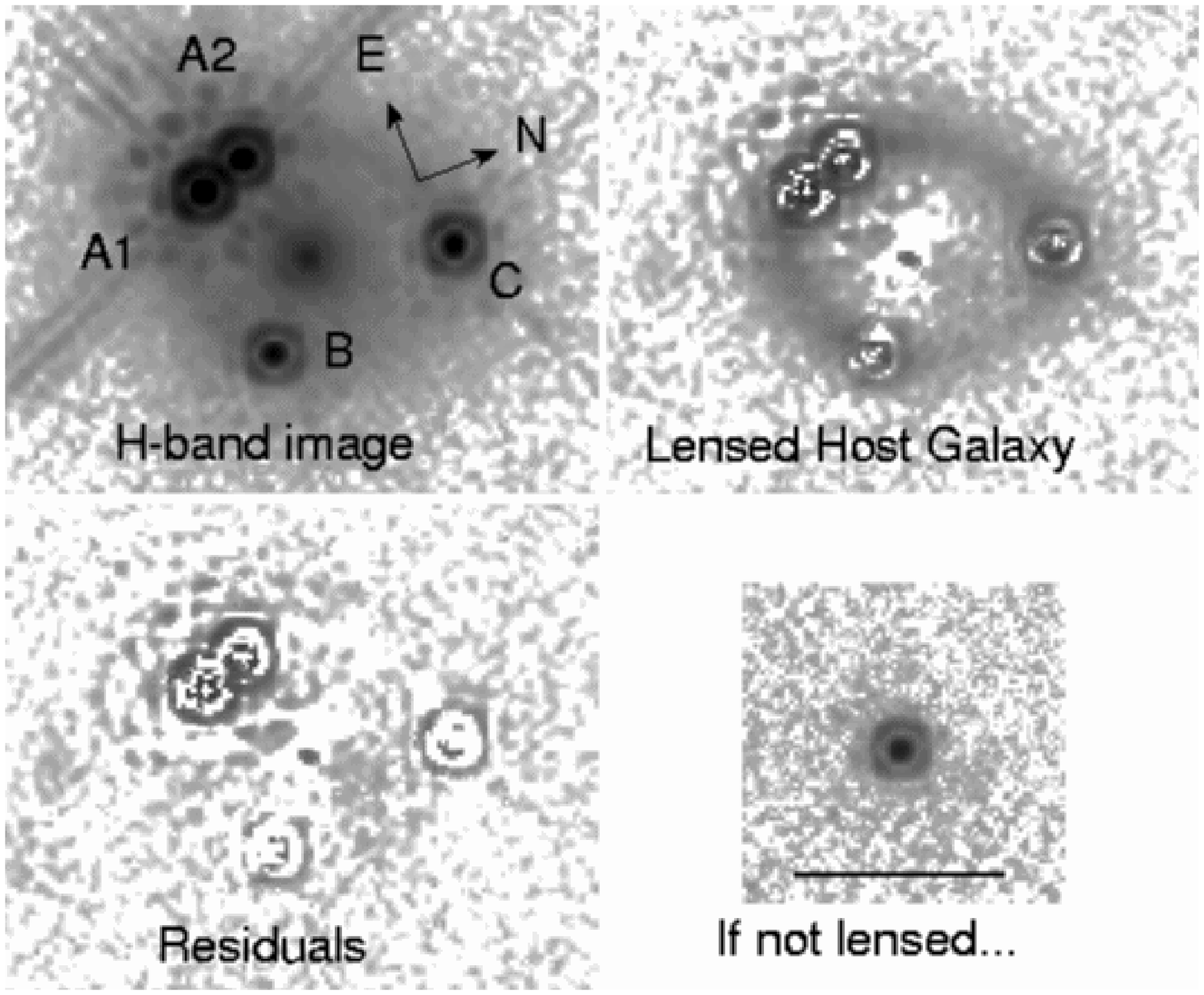,width=4.5in}}
\end{center}
\vspace{-0.25in}
\caption{The host galaxy in PG1115+080.  The top left panel
  shows the 1-orbit NICMOS image from Impey et al.~(\cite{Impey1998p551}).
  The top right panel shows the lensed host galaxy after subtracting
  the quasar images and the lens galaxy,  The lower left panel shows the
  residuals after subtracting the host as well.  For comparison, the
  lower right panel shows what an image of an unlensed PG1115+080
  quasar and host would look like in the same integration time and 
  on the same scales.  The host galaxy
  is an H$=20.8$~mag late-type galaxy (Sersic index $n=1.4$) with
  a scale length of $R_e=1.5h^{-1}$~kpc.  The demagnified magnitude
  of the quasar is H$=19.0$~mag. The axis ratio of the source,
  $q_s=0.65\pm 0.04$ is consistent with the estimate of $q_s=0.58\pm0.02$
  from the simpler ring curve
  analysis (\S\ref{sec:analring}, Fig.~\ref{fig:host2}, Kochanek
  Keeton \& McLeod~\cite{Kochanek2001p50}).
   }
\labelprint{fig:pengpg1115}
\end{figure}

These methods, including the effects of the PSF, have been applied to 
determining the mass distributions in 0047--2808 (Dye \& Warren~\cite{Dye2003p1}),
B0218+357 (Wucknitz, Biggs \& Browne
\cite{Wucknitz2004p14}), MG1131+0456 (Chen, Kochanek \& Hewitt~\cite{Chen1995p62}, and
MG1654+134 (Kochanek~\cite{Kochanek1995p559}).  We illustrate them with the
Dye \& Warren~(\cite{Dye2003p1}) results for 0047--2808 in Fig.~\ref{fig:dye0047}.
The mass distribution consists of the lens galaxy and a cuspy dark matter halo,
where Fig.~\ref{fig:dye0047} shows the final constraints on the mass-to-light
ratio of the stars in the lens galaxy and the exponent of the central dark
matter density cusp ($\rho \propto r^{-\gamma}$).  The allowed parameter 
region closely resembles earlier results using either statistical constraints
(Fig.~\ref{fig:selfsim1}) or stellar dynamics (Fig.~\ref{fig:vdisp}).  In fact,
the results using the stellar dynamical constraint from Koopmans \& 
Treu~(\cite{Koopmans2003p606}) are superposed on the constraints from the
host in Fig.~\ref{fig:dye0047}, with the host providing a tighter constraint
on the mass distribution than the central velocity dispersion.  The one problem
with all these models is that they have too few degrees of freedom in their
mass distributions by the standards we discussed in \S\ref{sec:modelfit}.  In
particular, we know that four-image lenses require both an elliptical lens
and an external tidal shear in order to obtain a good fit to the data 
(e.g. Keeton, Kochanek \& Seljak~\cite{Keeton1997p604}), while none of these
models for the extended sources allows for multiple sources of the angular
structure in the potential.  In fact, the lack of an external shear probably
drives the need for dark matter in the 0047--2808 models.  Without dark matter,
the decay of the stellar quadrupole and the low surface density at the Einstein
ring means that the models generate too small a quadrupole moment to fit the
data in the absence of a halo.  
The dark matter solves the problem both through its own ellipticity and
the reduction in the necessary shear with a higher surface density near the 
ring (recall that $\gamma \propto 1-\langle \kappa \rangle$).  
Again see the need for a greater focus on the angular structure of the
potential.

\subsection{Lensed Quasar Host Galaxies \labelprint{sec:lensedhosts} }

One advantage of studying lensed quasars is that the lens magnification
enormously enhances the visibility of the quasar host.  A typical HST
PSF makes the image of a point source have a mean surface brightness
that declines as $R^{-3}$ with distance $R$ from the quasar.  Compared
to an unlensed quasar, the host galaxy of a lensed quasar is stretched
along the Einstein ring leading to an improvement in the contrast 
between the host in the quasar of $\mu^2$ for an image
magnified by $\mu$ -- you gain $\mu^3$ by stretching the host away from 
the quasar and lose $\mu$ because the quasar is magnified.  Perpendicular
to the Einstein ring, the contrast becomes a factor of $\mu$ worse   
than for an unlensed quasar.  Since the alignment of the
magnification tensor relative to the host changes with each image, the
segment of the host where contrast is lost will correspond to a segment
where it is gained for another image leading to a net gain for almost
all parts of the source when you consider all the images.  The distortions
produced by lensing also mean the host structure is more easily 
distinguished from the PSF.  In a few cases, like SDSS0924+0219 in
Fig.~\ref{fig:substruc0}, microlensing or substructure may provide
a natural coronograph that supresses the flux from the quasar but
not that from the host.
Despite naive expectations (and TAC comments),
the distortions have little consequence for understanding the structure of 
the host even though a lens model is required to produce a photometric
model of the host. 

The only extensive survey of lensed quasar hosts is that of 
Peng~(\cite{Peng2004}).  Fig.~\ref{fig:pengpg1115} shows the
example of PG1115+080, a $z_s=1.72$ radio-quiet (RQQ) quasar.
The Einstein ring image is easily visible even in a short, 
one-orbit exposure.  For comparison, we also took the final
model for the quasar and the its host and produced the image 
that would be obtained in the same time if we observed the
quasar in the absence of lensing.  It is quite difficult to see
the host, and this problem will carry through in any numerical
analysis.

\begin{figure}[t]
\begin{center}
\centerline{\psfig{figure=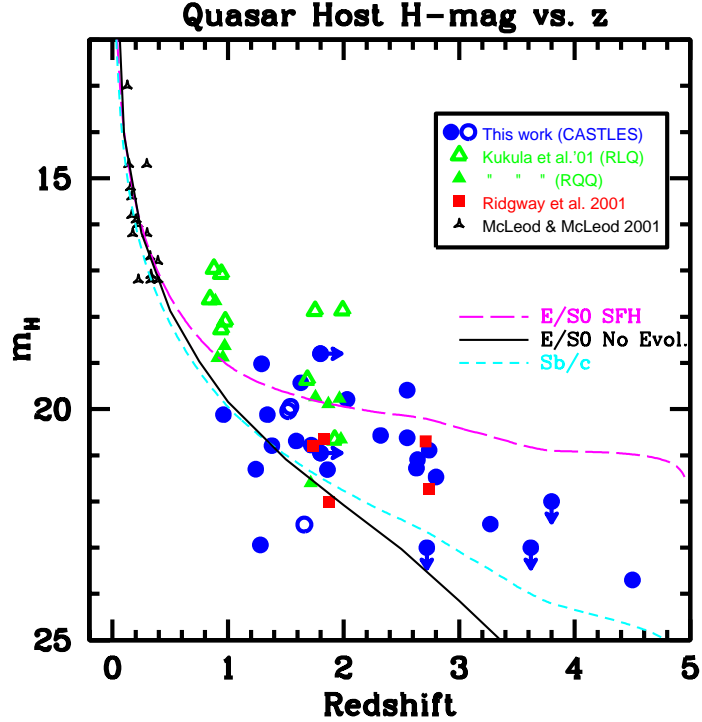,width=3.75in}}
\end{center}
\vspace{-0.25in}
\caption{Observed H-band magnitudes of quasar host galaxies.  The solid (open) circles
  are secure (more questionable) hosts detected in the CASTLES survey of  
  lensed hosts.   The low redshift points are from McLeod \& McLeod~(\cite{McLeod2001p782}).
  All the Ridgway et al.~(\cite{Ridgway2001p122}) systems are radio quiet. 
  For comparison, we superpose the evolutionary tracks for a non-evolving 
  E/S0 galaxy (solid curve), an evolving E/S0 galaxy which stars forming
  stars at $z_f=5$ with a 1~Gyr exponentially decaying star formation rate
  (long dashed line) and a star forming Sb/c model (short dashed line). 
  The evolution models are matched to the luminosity of an $L_*$ early-type
  galaxy at redshift zero.
  The CASTLES observations can reliably detect hosts about 4 magnitudes fainter
  than the quasar. From Peng~(\cite{Peng2004}).
  }
\labelprint{fig:hostmag}
\end{figure}

At low redshifts ($z<1$), quasar host galaxies tend to be massive early-type
galaxies (e.g. McLure et al.~\cite{Mclure1999p377},   
Dunlop et al.~\cite{Dunlop2003p1095}).  Over 80\% of quasars brighter
than $M_V<-23.5$~mag are in early-type galaxies with $L\gtorder 2L_*$
and effective radii of $R_e \sim 10$~kpc for $z \ltorder 0.5$.  Radio
quiet quasars (RQQ) tend to be in slightly lower luminosity hosts
than radio loud quasars (RLQ) but only by factors of $\sim 2$ at 
redshift unity.  Far fewer unlensed host galaxies have been detected
above redshift unity (e.g. Kukula et al.~\cite{Kukula2001p1533},  
Ridgway et al.~\cite{Ridgway2001p122}) with the surprising result
that the host galaxies are 2--3~mag brighter than the typical host
galaxy at low redshift and corresponded to $\sim 4 L_*$ galaxies.
  Given that the low redshift hosts were 
already very massive galaxies, it was expected that higher redshift
hosts would have lower masses because they were still in the process
of being assembled and forming stars (e.g. Kauffmann \& Haehnelt
\cite{Kauffmann2000p576}).  One simple explanation was that by 
selecting from bright radio sources, these samples picked
quasars with more massive black holes as the redshift increased, 
creating a bias in favor of more massive hosts.  The key to checking
for such a bias is to be able to detect far less luminous hosts, and
the improved surface brightness contrast provided by lensing the host
galaxies provides the means.  

\begin{figure}[t]
\begin{center}
\centerline{\psfig{figure=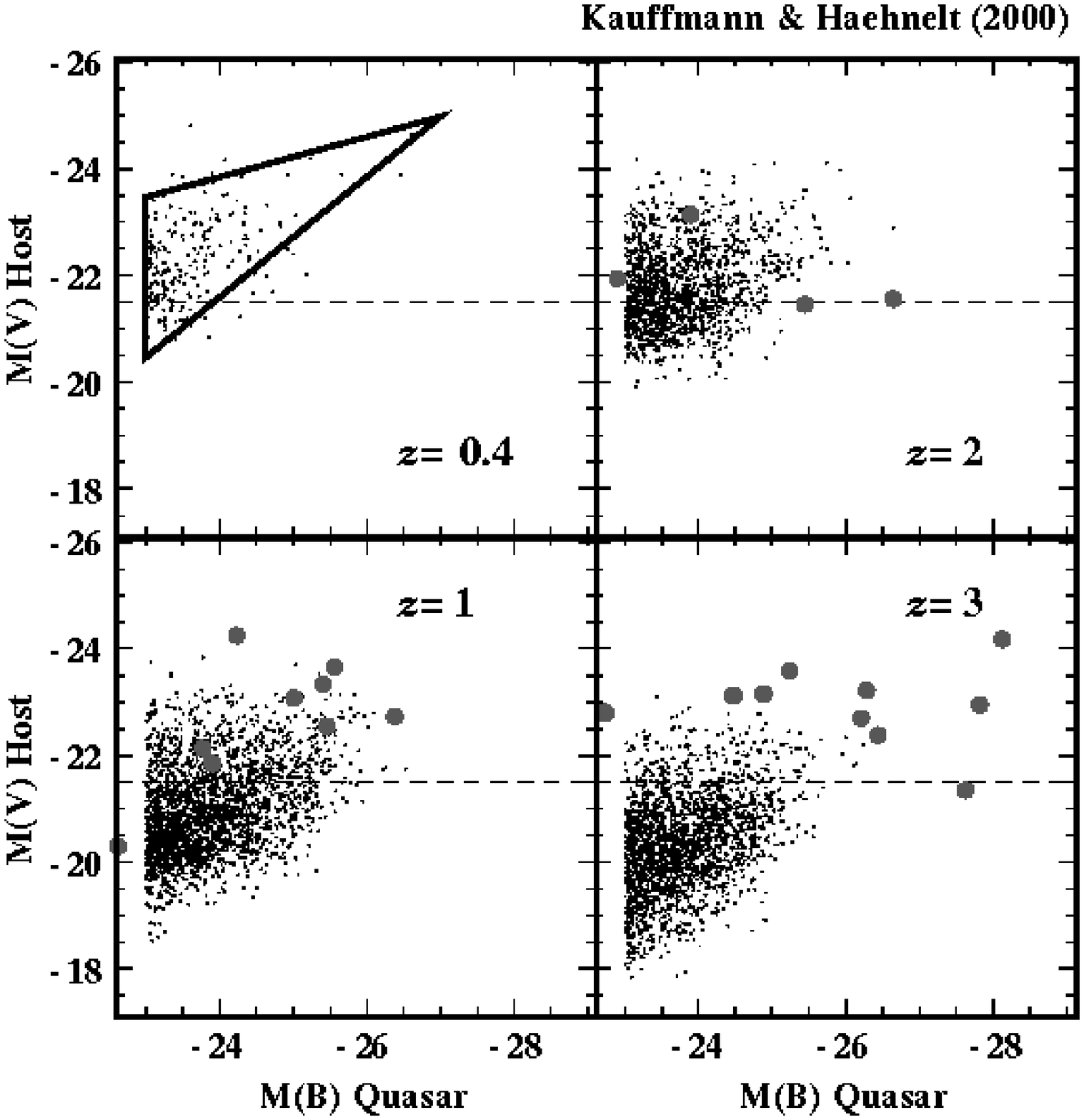,width=3.75in}}
\end{center}
\vspace{-0.25in}
\caption{
    A comparison of the estimated rest frame absolute magnitudes 
    of the quasars and hosts as compared to the theoretical models
    for the evolution of galaxies and the growth of black holes as
    a function of redshift by
    Kauffmann \& Haehnelt~(\cite{Kauffmann2000p576}).  The low redshift
    quasars from  McLeod \& McLeod~(\cite{McLeod2001p782}) occupy the
    triangle in the upper left panel.  At intermediate redshift the lensed host
    galaxies occupy a region similar to the models, but the two distributions
    are nearly disjoint by $z \simeq 3$.  Both the hosts and the quasars
    are significantly more luminous than predicted.   The horizontal line
    marks the luminosity of an $L_*$ galaxy at $z=0$.  From Peng~(\cite{Peng2004}).  
    }
\labelprint{fig:hosttheory}
\end{figure}

Fig.~\ref{fig:hostmag} shows the observed H-band magnitudes of the lensed
hosts as compared to low redshift host galaxies and other studies of high
redshift host galaxies.  Although 30\% of the lensed quasars are radio-loud,
they have luminosities similar to the lensed (or unlensed) radio-quiet 
hosts.  There are no hosts as bright as the Kukula 
et al.~(\cite{Kukula2001p1533}) radio-loud quasar hosts.  Once the
luminosities of the quasar and the host galaxy are measured we can
compare them to the theoretical expectations (Fig.~\ref{fig:hosttheory}).
While the models agree with the data at low redshift, they are nearly
disjoint by $z\sim 3 $ in the sense that the observed quasars and hosts
are significantly more luminous than predicted.  The same holds for the
Kukula et al.~(\cite{Kukula2001p1533}) and Ridgway et 
al.~(\cite{Ridgway2001p122}) samples, suggesting that black holes masses
grow more rapidly than predicted by the theoretical models or that accretion
efficiencies were higher in the past.  Vestergaard~(\cite{Vestergaard2004p676})
makes a similar argument based on estimates of black hole masses from 
emission line widths.

\section{Does Strong Lensing Have A Future? \labelprint{sec:future} }

Well, you can hardly expect an answer of ``No!" at this point, can you? 
Since we have just spent nearly 170 pages on the astrophysical uses of
lenses, there is no point in reviewing all the results again here.
Instead I suggest some goals for the future.

Our first goal is to expand the sample of lenses from $\sim 100$ to
$\sim 1000$.  While 80 lenses
seems like a great many compared to even a few years ago, it is still
too few to pursue many interesting questions.  The problem worsens
if the analysis must be limited to lenses meeting other criteria
(radio lenses, lenses found in a well-defined survey, lenses outside
the cores of clusters $\cdots$) or if the sample must be subdivided
into bins (redshift, separation, luminosity $\cdots$).  For example, 
one of the most interesting applications of lenses will be to map out 
the halo mass function.   This is difficult to do with any other
approach because no other selection method works homogeneously on
dark low-mass halos, galaxies of different types, groups and
clusters.  Unlike any other sample in astronomy, gravitational lenses
are selected based on mass rather than luminosity, so the same
search method works for all halos -- the separation distribution
of lenses is a direct mapping of the halo mass function.  It is
not a trivial mapping because the structure of halos changes with
mass, but the systematics are far better than those of any other
approach.  The fact that lenses are mass-selected also gives them an
enormous advantage in studying the evolution of galaxies with redshift
over optically-selected samples where it will be virtually impossible
to select galaxies in the same manner at both low and high redshift.
There is no shortage of detectable lenses in the universe -- it is 
simply a question of imaging enough of the sky at high angular 
resolution.
The upgraded VLA and Merlin radio arrays are the most promising
tools for this objective.

Our second goal is to systematically monitor the variability
of as many lenses as possible.  Time delays, if measured in large
numbers and measured accurately, can resolve most of the remaining
issues about the mass distributions of lenses.  This is true even
if you regard the $H_0$ as unmeasured or uncertain -- the Hubble
constant is the same number for all lenses, so as the number of time delay
systems increases, the contribution of the actual value of the
Hubble constant to constraining the mass distribution diminishes.
At the present time, we are certain that the typical early-type
galaxy has a substantial dark matter halo, but we are uncertain
how it merges with the luminous galaxy.  Steady monitoring of 
microlensing of the source quasars by the stars in the lens galaxy
will also help to resolve this problem because the patterns of the
microlensing variability constrain both the stellar surface density
near the lensed images and the total density (\partmicro, 
Schechter \& Wambsganss~\cite{Schechter2002p685}).  The constraints
from time delays and microlensing will be complemented by the
continued measurement of central velocity dispersions.

Our third goal should be to obtain ultra-deep, high resolution radio
maps of the lenses to search for central images in order to measure
the central surface densities of galaxies and to search for 
supermassive black holes.  
Keeton~(\cite{Keeton2003p17}) showed that the dynamic ranges of 
the existing radio maps of lenses are 1--2 orders of magnitude
too small to routinely detect central images given the expected
central surface densities of galaxies.  Only very asymmetric 
doubles like PMN1632--0033, where Winn et al.~(\cite{Winn2004p613})
have detected a central image, are likely to show central images
with the present data.  Once we reach the sensitivity needed to
detect central images, we will also either find central black
holes or set strict limits on their existence (Mao, Witt \&
Koopmans~\cite{Mao2001p301}).  This is the only approach that
can directly detect even quiescent black holes and determine
their masses at cosmological distances.  The existing limits 
could be considerably improved simply by co-adding the existing
radio maps either for individual lenses or even for multiple
lenses in order to obtain statistical limits. 

Our fourth goal should be to unambiguously identify a ``dark'' satellite
of a lens galaxy.  For starters we need to conduct complete
statistical analyses of lens galaxy satellites in general, 
by determining their mass functions and radial distributions.
As part of such an analysis we can obtain upper bounds on the
number of dark satellites.  Then, with luck, we will find an
example of a lens that requires a satellite at a specific location
for which there is no optical counterpart.  This may be
too conservative a condition. For example, Peng~(\cite{Peng2004})
argues that much of the flux of Object X in MG0414+0534 (Fig.~\ref{fig:mg0414h})
is actually coming from lensed images of the quasar 
host galaxy rather than the satellite.

Finally, lens magnification already means that it is far easier to
do photometry of a lensed quasar host galaxy than an unlensed galaxy. 
The next frontier is to measure the kinematics of cosmologically
distant host galaxies.  This might already be doable for the host
galaxy of Q0957+561 at $z_s=1.41$, but will generally require either
JWST or the next generation of ground based telescopes.  With larger
lens samples we may also find more cases like SDSS0924+0219 where
gravitational lensing provides a natural coronograph for the quasar.

\bigskip
\bigskip

\noindent Acknowledgments:  
These lectures are dedicated to Bohdan Paczynski at a very difficult time for
a man who made enormous contributions to this field both through his own work
and his support for the work of others.
I would like to thank E.E. Falco, C.R. Keeton, L.V.E. Koopmans,
D. Maoz, J. Mu\~noz, D. Rusin, D.H. Weinberg and J.N. Winn for 
commenting on this manuscript, and S. Dye, C.D. Fassnacht, D.R. Marlow, J.L. Mitchell,
 C.Y. Peng, D. Rusin and J.N Winn for supplying figures.  G. Meylan showed
considerable patience with the author while waiting for these lectures to
be delivered.
This research has been supported by the NASA ATP grant NAG5-9265, and 
by grants HST-GO-7495, 7887, 8175, 8804, and 9133 from the Space Telescope 
Science Institute.  The continuing HST observations of gravitational lenses
are an absolutely essential part of converting gravitational lenses from
curiosities into astrophysical tools.

\end{document}